\documentclass[article]{JHEP3}

\usepackage{graphicx}
\usepackage{amsmath,epsfig}
\usepackage{amssymb,amsfonts}
\usepackage{latexsym}
\usepackage{epstopdf}  

\relax
\renewcommand{\theequation}{\arabic{section}.\arabic{equation}}

\def\be{\begin{equation}}
\def\ee{\end{equation}}

\newcommand{\la}{\lambda}
\newcommand{\rc}{\nonumber\\}

\newcommand{\morder}[1]{\mathcal{O}\left(#1\right)}

\allowdisplaybreaks[2]

\newcommand{\bear}{\begin{align}}
\newcommand{\eear}{\end{align}}
\newcommand{\bea}{\begin{align}}
\newcommand{\eea}{\end{align}}
\newcommand{\nn}{\nonumber}
\newcommand{\xcfb}{x_c{}^{-}}

\def\hri#1#2{\href{http://arxiv.org/abs/#1}{[ArXiv:#1]#2}}
\def\hre#1#2{\href{http://arxiv.org/abs/#1/#2}{[ArXiv:#1/#2]}}

\newbox\pippobox

\def\II{\relax{\rm I\kern-.18em I}}

\def\e{\epsilon}
\def\l{\lambda}
\def\m{\mu}
\def\n{\nu}
\def\r{\rho}

\def\pa{\partial}
\def\sp{\;\;\, ,\;\;\;}

\def\z{\zeta}
\def\a{\alpha}

\def\tr{\ensuremath{\mathrm{Tr}}}

\def\t{\tau}

\def\h{\kappa}
\def\gf{w}
\def\gtx{\tilde g_{xx}}
\def\gtz{\tilde g_{rr}}
\def\nb{\nonumber}
\def\pfi{\partial_\Phi}
\def\ptt{\partial_\tau}
\def\pdfi{\partial_\Phi^2}

\def\nw{w_p}
\def\Awf{{A}} 
\def\G{G} 
\def\R{\mathfrak{R}}
\def\Q{\mathcal{Q}}

\title{The discontinuities of conformal transitions and mass spectra of  V-QCD  }

\author{Daniel Are\'an$^a$, Ioannis Iatrakis$^{c,e}$, Matti J\"arvinen$^c$ and Elias Kiritsis$^{b,c,d}$\\
~\\
$^a$  \href{http://www.ictp.it/}{International Centre for Theoretical Physics (ICTP)}
and INFN - Sezione di Trieste \\
Strada Costiera 11; I 34014 Trieste, Italy \\
~\\
$^b$ \href{http://www.apc.univ-paris7.fr}{APC, Universit\'e Paris 7, Diderot},\\
 CNRS/IN2P3, CEA/IRFU, Obs. de Paris, Sorbonne Paris Cit\'e, \\
  B\^atiment Condorcet, F-75205, Paris Cedex 13, France (UMR du CNRS 7164)\\
~\\
$^c$ \href{http://hep.physics.uoc.gr}{Crete Center for Theoretical Physics},
Department of Physics, University of Crete, 71003 Heraklion, Greece\\
~\\
$^d$\href{http://wwwth.cern.ch/}{Theory Group, Physics Department, CERN}, CH-1211, Geneva 23, Switzerland\\
~\\
$^e$ \href{http://www.physics.sunysb.edu/Physics/}{Department of Physics and Astronomy, Stony Brook University}, Stony Brook, New York 11794-3800, USA 
}

\preprint{CCTP-2013-7}

\abstract{Zero temperature spectra of mesons and glueballs are analyzed in a class of holographic bottom-up models for QCD in the Veneziano limit,
$N_c\to\infty$, $N_f\to \infty$, with  $x=N_f/N_c$ fixed (V-QCD). The backreaction of flavor on color is fully included.
It is found that spectra are discrete and gapped (modulo the pions) in the QCD regime, for $x$ below the critical value $x_c$ where the conformal transition takes place. The masses uniformly converge to zero in the walking region $x\to \xcfb$ due to Miransky scaling. All the  ratios of masses asymptote to non-zero constants as $x\to \xcfb$ and therefore there is no ``dilaton" in the spectrum.
The S-parameter is computed and found to be of $\mathcal{O}(1)$ in units of $N_f N_c$  in the walking regime,  while it is always an increasing function of  $x$. This indicates the presence of a subtle discontinuity of correlation functions across the conformal transition at $x=x_c$.
}

\begin{document}

\def\g{\gamma}
\def\go{\g_{00}}
\def\gi{\g_{ii}}

\maketitle 

\section{Introduction and Outlook}

The gauge/gravity duality has been widely used to describe the physics of strongly coupled gauge theories. The perturbative expansion of  Yang-Mills (YM) theory in the ('t Hooft) large $N_c$ limit, $N_c \to \infty$ and $\lambda = g^2_{YM} N_c=\text{finite}$, suggests that gauge theories have a dual string theory description.
We will consider QCD with gauge group $SU(N_c)$ and with $N_f$ quarks in the fundamental representation of $SU(N_c)$. In the limit of large $N_c$, the effects of flavor degrees of freedom are suppressed.  Veneziano introduced an alternative topological expansion of QCD, via the so-called Veneziano limit
\begin{equation}
N_c \rightarrow \infty \,\,,\,\,\, N_f \rightarrow \infty \,\, ,\,\,\, x={N_f \over N_c}=\text{fixed}  \,\,,\,\,\, \lambda = g_{YM}^2 N_c=\text{fixed} \,;
\label{venli}
\end{equation}
in this limit we may study phenomena depending on the flavor degrees of freedom such as hadronic multiparticle production, \cite{veneziano} and the $U(1)$ problem in QCD, \cite{vu1}.

By studying the QCD $\beta$-function,
\be \label{QCDbetaVen}
\l\equiv {g^2 N_c}\sp \dot \l=-b_0\l^2+b_1\l^3+{\cal O}(\l^4)
\ee
with
\be
b_0={2\over 3}{(11-2x)\over (4\pi)^2}+{\cal O}(N_c^{-2})\sp {b_1\over b_0^2}=-{3\over
2}{(34-13x)\over (11-2x)^2}+{\cal O}(N_c^{-2})
\ee
it is evident that the theory in the  Veneziano limit exhibits several interesting features.

 \begin{itemize}

\item For $x> 11/ 2$ the theory is not asymptotically free but IR free, and therefore the IR physics is simple to assess.

\item At values $x<11/2$, there is also an IR fixed point of the two-loop $\beta$-function. This region of $x$ where the theory has an IR fixed point is called ``conformal window''. For $x$ close to $11/2$ (Banks-Zaks region) the IR fixed point is at weak coupling,  \cite{bankszaks}.
In the Banks-Zaks region,  perturbation theory is trustworthy at all energy scales. As $x$ decreases, the IR fixed point coupling increases and perturbative methods cannot be applied. QCD with $x$ in the conformal window has unbroken chiral symmetry.

\item The theory is expected to reduce (at qualitative level) to standard QCD for small $N_f$ and to pure Yang-Mills for $N_f=0$. Therefore,  there should be a critical value $x=x_c<11/2$ where there is a transition from chirally symmetric theories in the conformal window to confining theories with broken chiral symmetry in the IR.
In the region below but close to $x_c$ the theory is expected to exhibit ``walking'' behavior. There is an approximate IR fixed point in the RG flow, which dominates the behavior of the theory for a large range of energies.
    It is eventually  missed by the RG flow and the theory ends up in the IR regime  of broken chiral symmetry. In particular, the coupling constant varies slowly for a long energy range, \cite{jk}.

\end{itemize}

The standard picture is that the transition at $x_c$ is a phase transition of the Berezinskii-Kosterlitz-Thouless (BKT) type, \cite{bkt,miransky} and is known as a conformal phase transition, \cite{miransky}.
It has been proposed to be caused by the annihilation of an IR and a UV fixed point, \cite{son}. The dimensionful observables of the theory  scale as the condensate in the  BKT transition in two dimensions, (also known in this context  as Miransky scaling). The value of $x_c$ is difficult to be determined since the dynamics of the theory is strongly coupled in this region.
It is also difficult to disentangle this transition in lattice calculation due to finite volume limitations. There have, however, been many  efforts along various directions in order to find the critical $N_f$ for finite $N_c$ where the transition takes place, see \cite{ds}, \cite{cw}, \cite{antipin} and \cite{lattice}.

Nearly conformal gauge theories are of great importance in models for the physics beyond the Standard Model, \cite{holdom}, \cite{walk2}. In particular, they are an important ingredient of technicolor models where electroweak symmetry breaks spontaneously through the same mechanism as chiral symmetry in QCD. However, in technicolor the breaking happens at a higher energy scale and in a new gauge sector, \cite{tech}, \cite{review}.
The model includes ``technifermions'' transforming in some representation of the technicolor  group whose condensate breaks dynamically the electroweak symmetry.

In order to generate the correct masses of the Standard model fermions, technicolor is generalized to extended technicolor which has a larger hidden gauge group.
It follows that the lepton and quark masses depend on the dimension of the scalar $\bar qq$ operator that condenses. For phenomenological purposes (related to the size of the lepton and quark masses)  the dimension of this operator should be away  from three, which is its free field theory value, and should be closer to two. This can be achieved in a ``walking'' theory, \cite{ds}, \cite{Holdom2} and \cite{fukano}.
A typical issue with technicolor models is that they add relatively large contributions to the S-parameter, which was first defined in \cite{peskin}. Such contributions  create tension with the electroweak precision measurements. Walking theories have been conjectured to have reduced, or even arbitrarily small S-parameter, \cite{as}.

 Several efforts have been made to study gauge theories including flavor degrees of freedom in the context of AdS/CFT. Most of the research has been concentrated in the case of the quenched approximation, $N_f \ll N_c$, where flavor branes are introduced in the color background geometry without backreaction, \cite{topdown}.
In order to study the Veneziano limit of gauge theories holographically, the backreaction of flavor to the color action should  be taken into account. A possible way of attacking such a problem is to consider the action of type IIA/B supergravity with smeared flavor branes, \cite{Bigazzi:2005md,Casero:2006pt}. Alternatively, the flavor sources may be localized, so their charge density is a sum of delta functions. From the string theory viewpoint, such systems correspond to intersecting branes.
Some well known examples are the D3-D7 intersections, \cite{Kehagias:1998gn}, D2-D6, \cite{Itzhaki:1998uz}, D4-D8, \cite{Nastase:2003dd}, and D4-D8-$\overline{D8}$, \cite{Burrington:2007qd}.

Bottom-up holographic models, which are based on the hard-wall AdS/QCD model \cite{Erlich:2005qh}, have been constructed to  describe walking field theories, \cite{butc,hong}. Top-down approaches for describing field theories with walking behavior are \cite{nunez} and \cite{kutasov}.

In \cite{jk}, a bottom-up class of holographic models was built for QCD in the Veneziano limit, called V-QCD.  The $SU(N_c)$ sector of QCD was described by the Improved Holographic QCD (IHQCD) model, \cite{ihqcd}, whose action is the Einstein-dilaton one with a specific, non-trivial dilaton potential. The dilaton is the field dual to the scalar ${\mathbb Tr} [F^2]$ operator of YM.
The near-boundary asymptotics of the potential are chosen in order to match the perturbative $\beta$-function of Yang-Mills coupling and the IR asymptotics are such that reproduce features as confinement, asymptotic linear glueball trajectories and a mass gap. By tuning two phenomenological parameters of the potential, the model  agrees with lattice data both at zero and finite temperature, \cite{ihqcd2}, \cite{data}. For a review of lattice studies of large $N_c$ gauge theories see \cite{Lucini:2012gg}.
There have been  efforts to model walking behavior in the context of a single scalar  model by adjusting appropriately the $\beta$-function (potential), \cite{antipin}, \cite{Jarvinen:2009fe}. However, in those cases no
flavor degrees of freedom were considered.

In \cite{jk} fully backreacting flavor degrees of freedom were included  in the IHQCD backgrounds. The framework for this class of models  model was first studied in \cite{Bigazzi:2005md,ckp}. Its flavor action is the low-energy effective action of $N_f$ brane-antibrane pairs. This action was first proposed by Sen, \cite{sen} around flat spacetime.
The fields of the model include the complex tachyon which corresponds to the lightest state of the string stretching among branes and antibranes, as well as a left and a right gauge field dual to the left and right flavor QCD currents, respectively. The tachyon field is dual to the quark mass operator whose non-trivial vacuum expectation value causes chiral symmetry breaking $U(N_f)_L  \times U(N_f)_R \to U(N_f)_V$.
The model was found to reproduce various low energy features of the QCD meson sector, \cite{ikp}. Tachyon condensation has also been studied in the Sakai-Sugimoto model, \cite{sstachyon}. A lattice study of meson physics of large $N_c$ gauge theories was published recently in \cite{Bali:2013kia}.

V-QCD is therefore created by the fusion of IHQCD and tachyon dynamics, as modeled by generalizations of the Sen action. The dilaton potential is taken to have the same form as in the IHQCD setup. The tachyon potential must have two basic properties; to vanish exponentially in the IR in order to have the brane-antibrane pair annihilate and to give the right UV mass dimension to the dual operator of the tachyon.

The vacuum saddle point  of the theory is determined by the non-trivial profiles of the  metric, dilaton and tachyon fields. The left and right gauge fields are trivial in the vacuum solution.  Making  a few reasonable assumptions,  the model produces a phase diagram which has similar structure as that expected from QCD in the Veneziano limit and does not depend on the details of the potentials.

In the range $x_c<x<11/2$, where the theory has an IR fixed point, the IR dimension of the chiral condensate can be determined. It is found to decrease as a function of $x$
and the point where it becomes 2, determines  $x_c$, as has been argued in  \cite{son}.
Upon matching the $\beta$-function of the Yang-Mills coupling and the anomalous dimension of $\overline{q} q$  to the QCD result in the UV, $x_c$ is found to be close to 4,
\be
 3.7 \lesssim ~~x_c~~ \lesssim 4.2 \, ,
\ee
which agrees with other estimates, \cite{ds,cw,antipin}. A similar phase transition has also been found in more simplified holographic models for QCD, which were likewise matched properties of perturbative QCD near the boundary, \cite{Alvares:2012kr}.

V-QCD models have been analyzed in detail at finite temperature in \cite{alho}. Generically they exhibited a non-trivial chiral-restoration transition in the QCD phase and no transition in the conformal window. Depending on the model class, they might exhibit more than one phase transitions especially in the walking region.

In the present article, we study the quadratic fluctuations of V-QCD for zero quark masses. Some of the main results first appeared in the short publication \cite{letter}. Among other issues, there are two relevant questions for walking gauge theories which we will answer here:

\begin{itemize}

\item Is there a light dilaton\footnote{Note that this dilaton is different from the bulk scalar $\phi$ that we also call the (string theory) dilaton.
 Here dilaton stands for one of the bound states of the QCD spectrum that is hierarchically lighter than all others states, and can be interpreted as the Goldstone boson of (spontaneously) broken conformal invariance.} state in the spectrum,  due to the approximate scale invariance, \cite{walk2}?

\item  Is the S-parameter strongly suppressed as expected on general grounds, \cite{as}?

\end{itemize}

Various holographic models have been proposed which explore the above questions. In  \cite{dilaton} and \cite{kutasov}  the lightest state is found to be a scalar. But its identification with the dilaton seems to be model dependent.
 The S-parameter has also been calculated in holographic bottom-up, \cite{hong}, and Sakai-Sugimoto
 like models, \cite{cobi}. In \cite{rubakov}, it was argued that the S-parameter
 is bounded below for a specific class of holographic models.

We consider small fluctuations of the fields which are involved in the action. The glue sector has the metric, the dilaton and the (closed string) axion. Their normalizable fluctuations correspond to $J^{PC}=0^{++},\; 0^{-+},\; 2^{++}$ glueballs, where $J$ is the spin, $P$ is parity and $C$ the charge conjugation of the state. The flavor action includes the complex tachyon and the left and right gauge fields. These give rise to  $J^{PC}=1^{++},\; 1^{--},\; 0^{++},\;  0^{-+}$ towers of mesons.
The states above are separated in two distinct symmetry classes, the flavor non-singlet and the flavor singlet states. Singlet or non-singlet fluctuations with different $J^{PC}$ have an infinite tower of excited states. The flavor singlet states which are in the meson and in the glueball sectors mix at leading order in $1/N_c$, since we explore the Veneziano limit (such a mixing is ${\cal O}(N_c^{-2})$ in the 't Hooft limit).

As mentioned above, the dilaton and the tachyon potentials are constrained by QCD properties, like confinement, glueball spectra, anomalies etc. Meson spectra which are calculated in this work also set some constraints on the potentials which are used in the action, see section \ref{subsec:conmsp}.
We explore the correlation between different IR asymptotics of the various potential functions that enter the holographic V-QCD action, and properties of the glueball and meson spectra. Once this is done we can pin down the asymptotics that provide the correct expected gross properties.
The main requirements in the flavor non-singlet sector are that
all meson towers have linear asymptotic (radial) trajectories.
In the flavor singlet case, mesons and glueballs decouple at large excitation number and the mesons have the same asymptotics as the flavor non-singlet meson with the same $J^{PC}$. The glueball trajectories are linear. Even if the above requirements do not set tight constraints on the potentials appearing in the action, they constrain the region of their parameters, see section \ref{subsec:conmsp}.

An interesting related issue is whether the slopes of the linear  trajectories for the axial vector and the vector mesons are the same.
It has been pointed out in \cite{ckp} that, due to chiral symmetry breaking, they might differ. Indeed in the models discussed in \cite{ckp} and \cite{ikp} these slopes were different.
Despite an ensuing  debate in the literature, (see \cite{sv}-\cite{espriu} and references therein),
what happens in QCD remains an open issue. We will also characterize this possibility in terms of the IR asymptotics of the flavor potential functions.

The numerical solution of the fluctuation equations for zero quark mass produces the spectrum of the model, as described in  section \ref{sec:results}.
The analysis of the spectrum was done for two different classes of tachyon potentials, potentials I and potentials II. Potentials I reproduce well the physics of real QCD in the Veneziano limit. For instance, it has been checked that the finite temperature phase diagram has the correct structure, \cite{alho}. The asymptotic meson trajectories are also linear but with possible logarithmic corrections. Potentials II are a bit further from the detailed  QCD behavior.
We have investigated them in order to study the robustness of our results against changes in the asymptotic form of the potential.
 Potentials II have quadratic asymptotic trajectories for mesons.

 \subsection{Results}

We summarize below our results.

\begin{enumerate}

\item The main generic properties of the spectra are as follows.

\begin{itemize}
 \item Below the conformal window, in the chirally broken phase with $x<x_c$, the spectra are discrete and gapped. The only exception are the $SU(N_f)$ pseudoscalar pions
that are massless, due to chiral symmetry breaking.

\item In the conformal window, $x_c<x<11/2$, all spectra are continuous and gapless.

\item All masses in the Miransky scaling region (aka ``walking region'') are obeying Miransky scaling
$m_n\sim \Lambda_{\rm UV} \exp({-{\kappa \over \sqrt{x_c-x}}})$. This is explicitly seen in the case of the $\rho$ mass in figure
\ref{fnss3}.
\end{itemize}

\item The non-singlet fluctuations include the L and R vector meson fluctuations, packaged into an axial and
vector basis, $V_{\mu},A_{\m}$, the pseudoscalar mesons (including the massless pions), and the scalar mesons.
Their second order equations are relatively simple. Our main results for the non-singlet sector are:

\begin{itemize}
\item The mass spectra of the low lying mesons can be seen in figure \ref{fnss1} for potentials I and in figure \ref{fnss2} for potentials II (note that the left hand plots have their vertical axis in logarithmic scale). The lowest masses of the mesons vary little with $x$ until we reach the walking region. There, Miransky scaling takes over and the masses dip down exponentially fast.
The $\Lambda_\mathrm{UV}$ scale is extracted as usual from the logarithmic running of $\l$ in the UV.

\item
The mass ratios asymptote to  finite and ${\cal O}(1)$ constants as $x \to x_c$.
\end{itemize}

\item The singlet fluctuations
include the $2^{++}$ glueballs, the $0^{++}$ glueballs and scalar
mesons that mix to leading order in $1/N_c$ in the Veneziano limit, and the $0^{-+}$ glueballs and the $\eta'$ pseudoscalar tower.
Although the spin-two fluctuation equations are always
simple, summarized by the appropriate Laplacian,
the scalar and pseudoscalar equations are very involved. Our main results for the singlet sector are:

\begin{itemize}
\item
The $U(1)_A$ anomaly appears
at leading order and the mixture of the $0^{-+}$ glueball and the $\eta'$ has a mass of ${\cal O}(1)$.

\item In figures \ref{fss1} and \ref{fss2} we present the results for the singlet scalar meson spectra. The dependence on $x$ is qualitatively similar as in the non-singlet sector.

\item In the scalar sector, for small $x$, where the mixing between glueballs and mesons is small, the lightest state is a meson, the next lightest state is a glueball, the next a meson and so on. However, with increasing $x$, non-trivial mixing sets in and level-crossing seems to be generic.
    This can be seen in figure \ref{fss2}.

\item All singlet mass ratios asymptote to constants as $x\to x_c$ (see figure~\ref{fss1}). The same holds for mass ratios between the flavor singlet and non-singlet sectors, as confirmed numerically in figure~\ref{fratios}. There seems to be no unusually light state (termed the ``dilaton'') that reflects the nearly unbroken scale invariance in the walking region.
The reason is a posteriori simple: the nearly unbroken scale invariance is reflected in the {\em whole} spectrum of bound states scaling exponentially to zero due to Miransky scaling. The breaking however of the scale invariance is {\em not spontaneous.}
\end{itemize}

\item The asymptotics of the spectra at high masses is in general a power-law with logarithmic corrections, with the powers depending on the potentials. The trajectories are approximately linear ($m_n^2\sim c\, n$) for type I potentials
  and quadratic ($m_n^2\sim c\, n^2$) for type II potentials. There is the possibility, first seen in \cite{ckp} that the proportionality coefficient $c$ in the linear case is different between axial and vector mesons.

\item There are several dilaton-dependent functions that enter the V-QCD action, which can be constrained by using various known properties of QCD. They include $V_g(\l)$ the dilaton potential in the glue sector action in (\ref{vg}), as well as the tachyon potential $V_f(\tau,\l)$, the kinetic function  for the tachyon, $\kappa(\l)$, and the kinetic function for the gauge fields $w(\l)$ that appear in the flavor action (\ref{generalact}).
Moreover, the tachyon potential function, motivated by flat space string theory,  is parametrized as in (\ref{vf})
    \be
    V_f(\tau,\l)=V_{f0}(\l)e^{-a(\l)\tau^2}
    \ee
  in terms of two extra functions of the dilaton, $V_{f0}(\l)$ and $a(\l)$. We find that the functions can be constrained as follows.

\begin{itemize}
 \item
  The UV asymptotics of these functions can be fixed from perturbation theory. The IR asymptotics are more obscure.
   The pure glue potential has been established in previous works, \cite{ihqcd,data}.

\item
   We have parametrized the $\l\to\infty$ IR asymptotics as
\begin{align} \label{par}
 \h(\l) &\sim \l^{-\h_p}(\log\l)^{-\h_\ell}\,,& a(\l) &\sim \l^{a_p}(\log\l)^{a_\ell}\,,&& \nn\\
 \gf(\l) &\sim \l^{-\gf_p}(\log\l)^{-\gf_\ell}\,,& V_{f0}(\l) &\sim \l^{v_p} \,,& (\l &\to \infty) \, .
\end{align}
$V_{f0}(\l)$ is constrained indirectly so that $V_g(\l)-x V_{f0}(\l)$ has a non-trivial IR fixed point for a range of $x$.
The others however are severely constrained. By asking various generic criteria to be satisfied as well as requiring the existence of
asymptotically linear meson towers, we obtain that
 \be \label{finalchoice1}
 \h_p=\frac{4}{3}\,,\quad \h_\ell=-\frac{1}{2}\sp a_p=0  \sp  \quad a_\ell=0\, .
\ee
and
\be \label{wconstraint1}
 \frac{\h(\l)}{\gf(\l)} \to 0 \,, \qquad ( \l \to \infty)\,.
\ee
If on the other hand  $\h(\l)/\gf(\l) \to $ constant, then the axial and vector towers have different slopes.
In view of this we opt for
\be
\gf_p={4\over 3}\sp \gf_\ell<-{1\over 2}\,.
\ee
These parameter values are essentially those of potentials~I.

{\em It is curious to note that all the functions, at large $\l$ behave (modulo the logs) like in standard non-critical tree-level string theory, (in Einstein frame).}
\end{itemize}

\item In the region $0<x <x_c$ and for zero quark mass, there are infinite subdominant saddle points, that we called Efimov solutions and which are
labeled by a natural enumeration $n=1,2, \ldots $,  that indicates the number of zeros of the tachyon solution, \cite{jk}.
We have verified numerically and analytically\footnote{The analytic check is doable in the limit $x \to \xcfb$ or at asymptotically large $n$. In the former case, scalar singlet and non-singlet sectors both contain $n$ tachyonic modes with anomalously large $-m^2$, as also observed for a different model in \cite{kutasov}.} that these saddle points are perturbatively unstable. Tachyonic fluctuation modes are seen in the scalar singlet and non-singlet towers.

\item The behavior of correlation functions across the conformal transition turned out to be interesting and in part different from previous expectations.
  We have computed the two-point functions of several operators including the axial and vector currents.
We focus on the two-point function of the vector and axial currents which can be written in momentum space as
\be
\langle V_{\m}^{a}(q) V_{\n}^{b}(p) \rangle= \Pi^{ab}_{\mu \nu,V}(q,p)=-(2 \pi)^4 \delta^4 (p+q)\, \delta^{ab}
\left( q^2\eta_{\m\n}-{q_{\m} q_{\n}}\right)\Pi_{V} (q) \,,
\ee
and similarly for the axial vector. We have  the decomposition
\be
V_{\m}(x)=\int {d^4 q \over (2\pi)^4} e^{i q x} V_{\m}^{a}(q) t^a \, \psi_V(r)\;,
\ee where $t^a$, $a=1,\ldots, N_f^2-1$ are the flavor group generators and $\psi_V(r)$ are the radial wave-functions.

\begin{itemize}
 \item
Using the expansions
\be
\Pi_A = {f_{\pi}^2 \over q^2} +\sum_{n} {f_n^2 \over q^2 +m_n^2 - i \epsilon}\,,\qquad
\Pi_V=\sum_{n} {F_n^2 \over q^2 +M_n^2 - i \epsilon}\,,
\label{2int}
\ee
we determine $f_{\pi}$ as
\be
f_{\pi}^2= -{N_{c} N_{f} \over 12 \pi^2 } \left. {\partial_{r} \psi^A
    \over r} \right|_{r=0,\, q=0}\,,
\ee
where the normalization was fixed by matching the UV limit of the two-point functions to QCD.
The dependence of $f_{\pi}$ on $x$ is shown in figure \ref{fpif}.
The pion decay constant changes smoothly for most $x$, but is affected directly by Miransky scaling which makes it vanish exponentially in the walking regime.

\item The S-parameter is defined as
\be
S\equiv 4 \pi {d \over dq^2}\left[q^2 (\Pi_V - \Pi_A)\right]_{q=0} =-{N_c N_f \over 3 \pi} {d \over dq^2}\left. \left( {\partial_r \psi^V (r) \over
    r}-{\partial_r \psi^A (r) \over r} \right) \right|_{r=0,\, q=0}
\label{3int}\ee
$$
=4\pi\sum_{n}\left({F_n^2\over M_n^2}-{f_n^2\over m_n^2}\right) \ .
$$

As both masses and decay constants in (\ref{2int},\ref{3int}) are affected similarly by Miransky scaling, the S-parameter is insensitive to it (Miransky-scaling-invariant).
Therefore its value cannot be predicted by Miransky scaling alone.
Our results show that generically the S-parameter (in units of $N_fN_c$) remains finite in the QCD regime, $0<x<x_c$ and asymptotes to a finite constant at
$x_c$ (see figure \ref{sparf}). The S-parameter is identically zero inside the conformal window (massless quarks) because of unbroken chiral symmetry. This suggests a subtle discontinuity of correlators across the conformal transition.
We have also found choices of potentials
where the S-parameter becomes very large as we approach $x_c$.
Our most important result is that generically the S-parameter is an increasing function of $x$, reaching it highest value at $x_c$ contrary to previous expectations, \cite{sannino}.

\item We have also calculated the next derivative of the difference, (related to the X-parameter of~\cite{Barbieri:2004qk}) as
\be \label{Spdefi}
 S' \equiv -2\pi\frac{d^2}{(dq^2)^2} \left[q^2\left(\Pi_V(q^2)-\Pi_A(q^2)\right)\right]_{q=0}
\ee
so that
\be
 q^2\left(\Pi_A(q^2)-\Pi_V(q^2)\right) = f_\pi^2 - \frac{S}{4\pi} q^2 + \frac{S'}{4\pi}q^4 + \cdots\,.
\ee
This parameter is shown for both potentials~I and~II in Fig.~\ref{spparf}. The $x$ dependence (in IR units) is qualitatively rather similar to the S-parameter so that the values typically increase with $x$, and approach fixed values as $x \to x_c$. However, unlike for the S-parameter, there is also a region with decreasing values near $x=x_c$.

\end{itemize}

\end{enumerate}

\subsection{Outlook}

Our exhaustive analysis of the class of V-QCD models and the results obtained paint a
reasonably clear  holographic picture for the behavior of QCD in the Veneziano limit.
Although V-QCD should be considered as a toy model for QCD in the Veneziano limit, there are two facts that give substantial weight to our findings.

\begin{itemize}

\item The ingredients of the holographic models follow as closely as possible what we know from string theory about the dynamics of the dilaton and  open-string tachyons. This is treated in more detail in section \ref{sec:string}.

\item We have explored parametrizations of the functions and potentials that enter the holographic action, especially in the IR. General qualitative
guidelines suggest that these functions are the same as in (naive) string theory corrected by logarithms of the string coupling.
This was first seen in \cite{ihqcd} where the dilaton potential behaves as $V\sim \l^{4\over 3}\sqrt{\log \l}$ as $\l\to\infty$.
Note that in the Einstein frame the noncritical dilaton potential in five dimensions is proportional to $\l^{4\over 3}$.

Moreover the power  of the subleading log was fixed at the time in order for the glueball radial trajectories to be linear.
It was later realized that only asymptotic potential of the form $V\sim \l^{4\over 3}~(\log \l)^a$ lead to non scale-invariant IR asymptotics, \cite{blaise}.
It was also independently found, \cite{Gursoy},  that the power $a={1\over 2}$ was also responsible for providing the well known power of $T^2$ in the free energy just above the deconfinement phase transition, \cite{2}.

\end{itemize}

Despite the success of the framework, there are several conceptual issues that remain to be addressed successfully.

\begin{itemize}

\item The effects of loops of the non-singlet mesons are not suppressed, as the large number of flavors compensates for the large $N_c$ suppression of interactions.

\item The CP-odd sector requires further attention as the $x\to 0$ limit in that sector seems to not be smooth.

\end{itemize}

To this we add two obvious open problems that involve the understanding of phase diagram at finite temperature and density, and the construction of a well-tuned model to real QCD. All of the above are under current scrutiny.

\section{V-QCD}

The complete action for the V-QCD model can be written as
\be
 S = S_g + S_f + S_a
\ee
where $S_g$, $S_f$, and $S_a$ are the actions for the glue, flavor and CP-odd sectors, respectively. We will define these three terms separately below. As discussed in \cite{jk}, only the first two terms contribute in the vacuum structure of the theory if the phases of the quark mass matrix and the $\theta$ angle vanish.
The full structure of $S_f$ and $S_a$  was not detailed in \cite{jk}, as this was  not necessary in order to study the vacuum structure of the model. However, the extra terms do contribute to the spectrum of fluctuations and will therefore be discussed in detail below.

\subsection{The glue sector} \label{sec:VQCDglue}

The glue action was introduced in \cite{ihqcd},
\be
S_g= M^3 N_c^2 \int d^5x \ \sqrt{-g}\left(R-{4\over3}{
(\partial\lambda)^2\over\lambda^2}+V_g(\lambda)\right) \, .
\label{vg}\ee
Here $\l=e^\phi$ is the exponential of the dilaton. It is dual to the ${\mathbb Tr} F^2$  operator, and its background value is identified as the 't Hooft coupling. The Ansatz for the vacuum solution of the metric is
\be
ds^2=e^{2 \Awf(r)} (dx_{1,3}^2+dr^2)\,,
\label{bame}
\ee
where the warp factor $A$ is identified as the logarithm of the energy scale in field theory.
 Our convention will be that the UV boundary lies at $r=0$ (and $A\to\infty$), and the bulk coordinate therefore runs from zero to infinity. The metric will be close to the AdS one except near the IR singularity at $r=\infty$. Consequently, $A \sim -\log(r/\ell)$, where $\ell$ is the (UV) AdS radius. In the UV $r$ is therefore identified roughly as the inverse of the energy scale of the dual field theory.

\subsection{The flavor sector} \label{sec:VQCDflavor}

The flavor action is the generalized Sen's action,
\be
S_f= - \frac{1}{2} M^3 N_c  {\mathbb Tr} \int d^4x\, dr\,
\left(V_f(\l,T^\dagger T)\sqrt{-\det {\bf A}_L}+V_f(\l, TT^\dagger)\sqrt{-\det {\bf A}_R}\right)\,,
\label{generalact}
\ee
where the quantities inside the square roots are defined as
\begin{align}
{\bf A}_{L\,MN} &=g_{MN} + \gf(\l,T) F^{(L)}_{MN}
+ {\h(\l,T) \over 2 } \left[(D_M T)^\dagger (D_N T)+
(D_N T)^\dagger (D_M T)\right] \,,\nonumber\\
{\bf A}_{R\,MN} &=g_{MN} + \gf(\l,T) F^{(R)}_{MN}
+ {\h(\l,T) \over 2 } \left[(D_M T) (D_N T)^\dagger+
(D_N T) (D_M T)^\dagger\right] \,,
\label{Senaction}
\end{align}
with the covariant derivative
\be
D_M T = \partial_M T + i  T A_M^L- i A_M^R T\,.
\ee
The fields  $A_{L}$, $A_{R}$ as well as $T$ are $N_f \times N_f$ matrices in the flavor space. We also define
 \be
 x\equiv {N_f\over N_c}\;.
 \ee
 It is not known, in general, how the determinants over the Lorentz indices in~\eqref{generalact} should be defined when the arguments~\eqref{Senaction} contain non-Abelian matrices in flavor space. However, for our purposes such definition is not required: our background solution will be proportional to the unit matrix $\mathbb{I}_{N_f}$, as the quarks will be all massless or all have the same mass $m_q$.
In such a case, the fluctuations of the Lagrangian are unambiguous up to quadratic order.

The form of the tachyon
potential that we will use for the derivation of  the spectra is
\be
V_f(\l,TT^\dagger)=V_{f0}(\l) e^{- a(\l) T T^\dagger} \, .
\label{tachpot}
\ee
This is the string theory tachyon potential where the constants have been allowed to depend on the dilaton $\la$.
For the vacuum solutions (with flavor independent quark mass) we will have $T =\tau(r) \mathbb{I}_{N_f}$ where $\tau(r)$ is real, so that
\be
 V_f(\l,T)=V_{f0}(\l) e^{- a(\l)\tau^2} \, .
\label{vf}\ee
The coupling functions $\h(\l,T)$ and $\gf(\l,T)$ are allowed in general to depend on $T$,  through such combinations that the expressions~\eqref{Senaction} transform covariantly under flavor symmetry. In this paper, we will take them to be eventually independent of $T$, emulating the known string theory results.

We discuss how the $\l$-dependent functions $V_{f0}(\l)$, $a(\l)$, $\h(\l)$, and $\gf(\l)$ should be chosen  in Section~\ref{sec:constraints}.

\subsection{The CP-odd sector and the closed string axion}

Here we follow \cite{ckp} in order to discuss the coupling of the closed string axion to the phase
 of the bifundamental tachyon, dual to the quark mass operator and the axial U(1)$_A$
gauge boson.
This discussion adapted to 5d holographic QCD is as follows.

We start with a three-form RR axion  $C_{\m\n\r}$, with field strength, $H_4=dC_3$ and
\be
S_a=S_\mathrm{closed}+S_\mathrm{open}\sp
S_\mathrm{closed}=-{M^3\over2}\int d^5x\, \sqrt{g}\,{|H_4|^2\over Z(\l)}\sp H_4=dC_3
\ee
and
\begin{align}
S_\mathrm{open}&=i\int C_3\wedge\, \Omega_2=i\int C_3\wedge \,d\Omega_1\, ,\qquad
&\Omega_1&=i\,N_f\left[2V_a(\l,T)\,A-\theta\, dV_a(\l,T)\right]\, ,\nn\\
A_{M}&={A_M^{L}-A_{M}^{R}\over 2}\,.&
\end{align}
Here $\theta$ is the overall phase of the tachyon, $T=\tau e^{i\theta}\cdot \mathbb{I}_{N_f}$.
In flat-space tachyon condensation $V_a(\l,T)$ is independent of the dilaton,
and is the same as the potential that appears in the tachyon DBI, \cite{ckp}. In our case it may be different in principle. However,
it must have the same basic properties; in particular it becomes a field-independent
 constant (related to the anomaly) at $T=0$, and vanishes exponentially at $T=\infty$.
We may dualize the three-form to a pseudo-scalar axion field $a$ by solving  the equations of motion as
\be
{H_4 \over  {Z(\l)}}={^*}\left(d\tilde a+i\,\Omega_1\right)\,.
 \ee
Therefore, the dual action takes the form
\be
S_a=-{M^3\,N_c^2\over2}\int d^5x\, \sqrt{g}\,Z(\l)\left[da-x\left(2V_a(\l,T)\,A-\theta\, dV_a(\l,T)
\right)\right]^2 
\label{samain}
\ee
in terms of the QCD axion $a=\tilde a/N_c$.

This is
normalized so that $a$ is dual to  $\theta/N_c$ with $\theta$ being the standard $\theta$-angle of QCD.
The coupling to the axial vector,  $A$, reflects the axial anomaly in QCD
\be
A_{\m}\to A_{\m}+\pa_{\m}\e\sp \theta\to \theta-2\e\sp a\to a+2x\,V\e
\label{u1transf}
\ee
with $V_a(\l,T=0)=1$, which gives the correct U(1)$_A$ anomaly.
This normalization is correct when $A_{\mu}$ is normalized so that the two-point function of the dual
current
$$\langle J^5_\m J^5_{\n}\rangle \sim N_fN_c~~~{\rm  with}~~~
J^5_{\m}\simeq \sum_{i=1}^N\bar\psi^i\gamma^5\gamma_{\m}\psi^i\,.$$
{}From the coupling between source and operator, $\sum_{ij}J^{ij}_{\mu}{A^{ij}}^{\mu}$,  with
$J^{ij}_{\mu}=\bar\psi^i\gamma^5\gamma_{\m}\psi^j$
we obtain the parametrization $A^{ij}_\mu=A_{\mu}\delta_{ij}+{\rm traceless}$.

The terms above mix the axion both with $\theta$ and the longitudinal part of $A_{\m}$.
As we will see, for $A_{\m}$ and $\theta$ there are other terms coming from the DBI action.
In the 't Hooft limit, $x\sim {1/ N_c}$ and the flavor corrections are subleading,
whereas in the Veneziano limit $x\sim {\cal O}(1)$, and the corrections are important.

The natural form for the function $V_a$ is to keep the tachyon exponential without the $V_{0f}(\l)$
term, i.e., $V_a(\l,\t)=\exp\left[-a(\l)\t^2\right]$.
We will work out the fluctuation problem however  for general $V_a(\l,\t)$.

Since we  have $A_{\m}=0$ in the background, we must first solve the ${\cal O}(N_c^2)$ action $S_g+S_f$
to determine $g_{\m\n},\l,$ and $\tau$.
In the quenched limit, the calculation we have done is enough to leading order in $1/N_c$.
In the Veneziano limit the full second order fluctuation system must be derived.

\subsection{The relation to string theory models} \label{sec:string}

We would like here to compare the class of models we are studying in this paper to expectations from string theory and QCD.

The main ingredients associated with the pure glue part of the model has been studied extensively and motivated from string theory.
These discussions can be found in the original papers, \cite{ihqcd}. It should be noted that the important region for such models is in the IR.
Although a gravity description is not expected in the UV (due to the weak coupling),  the approach taken in \cite{ihqcd} and here is that of  matching the gravitational theory to reproduce perturbative $\beta$-functions in the UV. This guarantees the correct UV boundary conditions for the (more interesting) IR
  theory.

In the flavor sector, the main ingredient is the Sen-like action for open string tachyons of unstable branes and brane-antibrane pairs in flat space, \cite{sen}.
Although this action has not been tested in all possible contexts it has passed a lot of non-trivial tests in the past.

However, it is not immediately obvious how the action should be written down in the case of a curved background and a running dilaton that is relevant here. There are nevertheless some simplifications:

\begin{itemize}

\item All YM-like geometries in the IR that have been studied in the past, \cite{ihqcd}, are nearly flat {\em in the string frame}.
This is non-trivial and we have no deep understanding of this fact. There are hints though: the asymptotic potentials in the Einstein frame
\be
V_n(\l)\sim \lambda^{4\over 3}\log^{n\over 2}\l
\ee
become $\tilde V_n(\l)\sim \log^{n\over 2}\l$ in the string frame. The case $n=0$ corresponds to the non-critical string dilaton potential that is constant
in the string frame. It gives rise to the well-known linear dilaton solution where the dilaton runs but the string metric remains flat.
The solutions for positive $n$ are similar. The metric is almost flat and the dilaton runs as the $(n+1)$ power of the radial coordinate.
$n=1$ is the case relevant for QCD, as it is the only one that gives linear asymptotic trajectories for glueballs, and the appropriate temperature scaling of the free energy just above the phase transition.

\item  Given this,  it is a natural choice to use an {\em adiabatic Ansatz} for the flavor action: to make all constants in the Sen action dilaton dependent.
This is precisely what we do.

\item Very little is known about the non-abelian tachyon action. This however is not limiting our analysis provided (a) we do not break the vector
$SU(N_f)$ symmetry, and (b) we are studying effects up to second order in the fluctuations.
If the vector symmetry is not broken by the quark masses, the vacuum expectation value for the tachyon is proportional to the unit matrix, and therefore
non-abelian subtleties are absent.
Similarly, up to quadratic order in fluctuations,  the non-abelian subtleties do not arise if we use for example a symmetric prescription.

\item The Sen's action, like any other DBI-like action, has its limitations; it does not include ``acceleration corrections'' (terms $\sim |D^2 T|^2$). These are well defined in the abelian (flavor) case, but
apply also to the non-abelian case relevant for our purposes.
It is therefore important to check that these corrections are not important for our solutions.

Indeed an analysis directly in the string frame of all tachyon solutions for all potentials used here,  shows they have ``small accelerations'' in the IR regime.
This might not be so in the middle of the geometry. However that regime is not important for the physics.
The only exception to the above is the ``walking regime'', for $x$ smaller and close to $x_c$; yet we have checked that for that case too,  the accelerations remain small.

\item Unlike the 't Hooft limit, where the loops of singlet bulk fields are suppressed by extra powers of $N_c$, in the Veneziano limit, there is no such
suppression for the non-singlet flavor fields. Therefore, we would have to think of our bottom-up action as the Wilsonian effective action that includes the effects of integrating out massive modes. In the QCD phase,  it therefore does not contain the contributions of the pions which are the only massless states.
Although some contributions may be relevant in some cases, we do not expect them to modify substantially neither the vacuum structure, nor the finite temperature and density structure. This however needs further analysis.

In the conformal window all modes are gapless. However, we do not expect instabilities of the Wilsonian action beyond the one specified by the only relevant operator (quark mass).

The only subtle region is the one near the conformal transition. In that region qualitative changes may happen because of the quantum effects. This issue requires further analysis but we will not pursue it in this paper.

\item
Within the framework set by the general principles above,
we have analyzed many different types of IR actions (especially their dependence on the dilaton).
We have mapped their IR landscape onto important phenomenological properties of the theory. This has been analyzed in section \ref{irs} and appendices \ref{app:bgUVIR}, \ref{app:Regge} and \ref{app:wfasympt}.

The remarkable conclusion is that, like the glue part of the theory, the dilaton functions, apart from logarithmic corrections, should have in the IR the same values as in standard non-critical string theory around flat space. For the dilaton functions $\h(\l), \gf(\l), V_{f0}(\l)$ and $a(\l)$,
defined in (\ref{Senaction}) and (\ref{vf}) we obtain from (\ref{par}) (and dropping the logarithms)
\begin{align} \label{par1}
 \h(\l) &\sim \l^{-{4\over 3}}\,,& a(\l) &\sim \l^{0}\,,&& \nn\\
 \gf(\l) &\sim \l^{-{4\over 3}}\,,& V_{f0}(\l) &\sim \l^{{7\over 3}} \,,& (\l &\to \infty) \, .
\end{align}
and after transforming to the string frame they become
\begin{align} \label{par2}
 \h(\l) &\sim \l^{0}\,,& a(\l) &\sim \l^{0}\,,&& \nn\\
 \gf(\l) &\sim \l^{0}\,,& V_{f0}(\l) &\sim \l^{-1} \,,& (\l &\to \infty) \,,
\end{align}
in agreement with tree level string theory results.

It should be also mentioned that the parameter least sensitive to phenomenological
 constraints is the power $v_p$ of $ V_{f0}(\l) \sim \l^{v_p} $. Indeed our potentials~I, which were constructed such that all constraints from QCD are satisfied at qualitative level, implement the choice of~\eqref{par1} except for the value of $v_p$ (see subsection~\ref{sec:potentials}). We chose a polynomial $V_{f0}$ and therefore $v_p=2$ for simplicity, but modifying the power to $7/3$ would not result in any qualitative changes.

\item Another issue is the justification of the use of the DBI form of
the tachyon action even after integrating out non-singlet degrees of freedom.
This issue turns out to not be important in the intermediate regions of the geometry, as the results depend very little on this region.

The only regime where this issue is of importance is when the kinetic term of the tachyon,  $|DT|^2$, diverges.
This is happening only in the IR part of the geometry, if the tachyon condenses and if $a(\lambda)$ is constant (as is the case for potentials~I).

We have analyzed various asymptotic powers in the actions different from the square root characteristic of DBI by parametrizing the tachyon action as
as $V(T,\lambda)(\det(g+DTDT^{\dagger}))^b$ with $b\not={1/2}$. This is discussed in Appendix~\ref{app:modDBI}.
We have found that $b={1/ 2}$ is a ``critical value'' for the exponent where many cancelations are operative in the equations of motion.
For $b<{1/ 2}$ there are no regular solutions to the equations of motion, whereas for $b>{1/2}$ the diverging tachyon solutions are powerlike rather than exponential. Meson trajectories in this case cannot be linear.

These are indirect but convincing arguments for the use of tachyon DBI action, but a first principles derivation in string theory is however desirable.

\end{itemize}

To conclude this section, V-QCD is a toy model for real QCD but it seems to
have many features that suggest it belongs to the correct universality class.

\subsection{The background solutions} \label{sec:bg}

We will discuss some general features of the background solutions of the V-QCD models. We restrict first to the standard case, which has a phase diagram similar to what is usually expected to arise in QCD. In Sec.~\ref{sec:constraints} we shall discuss which V-QCD models fall in this category.

To find the background, we consider $r$-dependent Ans\"atze for  $\l$, and $\Awf$. Assuming that the quark mass is flavor independent, we further take $T=\t(r)\mathbb{I}_{N_f}$, set all other fields to zero, and look for solutions to the equations of motion. The models are expected to have two types of (zero temperature) vacuum solutions \cite{jk}:
\begin{enumerate}
 \item Backgrounds with nontrivial $\l(r)$, $\Awf(r)$ and with zero tachyon $\t(r)=0$. These solutions have zero quark mass and intact chiral symmetry.
 \item Backgrounds with nontrivial $\l(r)$, $\Awf(r)$ and $\t(r)$. These solution have broken chiral symmetry. As usual, the quark mass and the chiral condensate are identified as the coefficients of the normalizable and non-normalizable tachyon modes in the UV (see Appendix~\ref{app:bgUVIR}).
\end{enumerate}
In the first case, the equations of motion can be integrated analytically into a single first order equation, which can easily be solved numerically. In the second case, we need to solve a set of coupled differential equations numerically. At the UV boundary and at the IR singularity, analytic expansions can be found (see Appendix~\ref{app:bgUVIR}).

We constrain the ratio $x=N_f/N_c$ to the range $0 \le x <11/2$ where the upper bound was normalized to the Banks-Zaks (BZ) value in QCD, where the leading coefficient of the $\beta$-function turns positive. The standard phase diagram at zero quark mass has two phases separated by a phase transition at some $x=x_c$ within this range.
\begin{itemize}
 \item When $x_c\le x<11/2$, chiral symmetry is intact. The dominant vacuum solution is of the first type with the tachyon vanishing identically.
 \item When $0<x<x_c$, chiral symmetry is broken. The dominant vacuum therefore has nonzero tachyon even though the quark mass is zero.
\end{itemize}

Interestingly, the phase transition at $x=x_c$ (which is only present at zero quark mass) involves BKT \cite{bkt} or Miransky \cite{miransky} scaling. The order parameter for the transition, the chiral condensate $\sigma \sim \langle \bar q q \rangle$ vanishes exponentially,
\be \label{condscaling}
 \sigma \sim \exp\left(-\frac{2 \hat K}{\sqrt{x_c-x}}\right)
\ee
as $x \to x_c$ from below. Here the constant $\hat K$ is positive.
When $x \ge x_c$, $\sigma$ is identically zero as chiral symmetry is intact. The Miransky scaling is linked to the ``walking'' behavior of the coupling constant: the field $\l(r)$ takes an approximately constant value $\l_*$ for a wide range of $r$, and the length of this region obeys the same scaling as (the square root of) the condensate in~\eqref{condscaling}.
The walking behavior is connected to the IR fixed point which is found for $x \ge x_c$: then $\l(r) \to \l_*$ as $r \to \infty$, and the geometry becomes asymptotically AdS also in the IR.

The Miransky scaling can also be discussed in terms of the energy scales of the theory. We may define the UV and IR scales, denoted by $\Lambda_\mathrm{UV}=\Lambda$ and $\Lambda_\mathrm{IR} =1/R$ (see Appendix~\ref{app:bgUVIR} for details). When $x<x_c$ and $x_c-x$ is not small, the V-QCD models involve only one scale, reflecting the behavior of ordinary QCD.
We therefore have $\Lambda_\mathrm{IR}/\Lambda_\mathrm{UV} = \morder{1}$. When $x \to x_c$, the two scales become distinct, and their ratio obeys Miransky scaling:
\be \label{scalescal}
 \frac{\Lambda_\mathrm{UV}}{\Lambda_\mathrm{IR}} \sim  \exp\left(\frac{\hat K}{\sqrt{x_c-x}}\right)\,.
\ee
The scale $\Lambda_\mathrm{UV}$ continues to be the one where the coupling constant $\l$ becomes small even as $x \to x_c$. The coupling walks for $\Lambda_\mathrm{UV}^{-1}\ll r \ll \Lambda_\mathrm{IR}^{-1}$, and starts to diverge at $r \sim \Lambda_\mathrm{IR}^{-1}$. In terms of the two scales, the chiral condensate behaves as $\sigma \sim \Lambda_\mathrm{UV} (\Lambda_\mathrm{IR})^2$.
The result~\eqref{condscaling} is therefore understood to hold when $\sigma$ is measured in the units of $\Lambda_\mathrm{UV}$.

 We recall how the constant $\hat K$ can be evaluated (see Sec.~10 of \cite{jk} for details). Let $m_\mathrm{IR}$ be the bulk mass of the tachyon $\t$ and $\ell_\mathrm{IR}$ the AdS radius, both evaluated at the IR fixed point. We then have
\be \label{hatKres}
 \hat K = \frac{\pi}{\sqrt{\frac{d}{dx}\left(m_\mathrm{IR}^2 \ell_\mathrm{IR}^2\right)_{x=x_c}}}\,.
\ee

The models may also have subdominant vacua. Including the solutions with finite quark mass, the generic structure is as follows.
\begin{itemize}
 \item When $x_c\le x<11/2$, only one vacuum exists, even at finite quark mass.
 \item When $0<x<x_c$ and the quark mass is zero, there is an infinite tower of (unstable) Efimov vacua in addition to the standard, dominant solution.\footnote{We are assuming here for simplicity that the IR effective potential of~\eqref{Veff} has an extremum, corresponding to an IR fixed point, for all positive values of $x$, and the the Breitenlohner-Freedman bound is violated at the fixed point all the way down to $x=0$ (see the discussion in subsection~\ref{subsec:BF}).}
 \item When $0<x<x_c$ and the quark mass is nonzero, there is an even number (possibly zero) of Efimov vacua. The number of vacua increases with decreasing quark mass for fixed $x$.
\end{itemize}

The infinite tower of Efimov vacua, which appears at zero quark mass, admits a natural enumeration $n=1,2,3,\ldots$ (where $n$ is the number of tachyon nodes of the background solution). A generic feature of these backgrounds is, that they ``walk'' more than the dominant, standard vacuum, so that the scales $\Lambda_\mathrm{UV}$ and $\Lambda_\mathrm{IR}$ become well separated for all $0<x<x_c$ when $n$ is large enough. It is possible to show that
\be \label{scalescalEf1}
  \frac{\Lambda_\mathrm{UV}}{\Lambda_\mathrm{IR}} \sim  \exp\left(\frac{\pi n}{k}\right)\,,\qquad (n\to\infty)\,,
\ee
for any $0<x<x_c$. Here $k$ can also be computed analytically (see Appendix~F in \cite{jk}). On the other hand, as $x \to x_c$ we find that
\be \label{scalescalEf2}
 \frac{\Lambda_\mathrm{UV}}{\Lambda_\mathrm{IR}} \sim  \exp\left(\frac{\hat K(n+1)}{\sqrt{x_c-x}}\right)\,
\ee
for any value of $n$. In particular, $n=0$ corresponds to the standard solution of~\eqref{scalescal}. We also found a similar scaling result for the free energies of the solutions as $x \to x_c$ in \cite{jk}, therefore proving that the Efimov vacua are indeed subdominant, and verified this numerically for all $0<x<x_c$. In this article we shall show in Sec.~\ref{sec:results} that the Efimov vacua are perturbatively unstable (again analytically as $x \to x_c$, and numerically for all $0<x<x_c$).

\section{Quadratic fluctuations} \label{quadfluct}

In order to compute the spectrum of mesons and glueballs we need to study the fluctuations of all the fields of V-QCD.
In the glue sector, the relevant fields are the metric $g_{\m\n}$, the dilaton $\phi$ and the QCD axion $a$. Their normalizable
fluctuations correspond to glueballs with $J^{PC}=0^{++},\; 0^{-+},\; 2^{++}$, where $J$ stands for the spin and $P$ and
$C$ for the field properties under parity and charge conjugation respectively.
In the meson
 sector,  one has the tachyon $T$, and the gauge fields $A_\m^{L/R}$; their normalizable fluctuations
corresponding to mesons with $J^{PC}=1^{++},\; 1^{--},\; 0^{++},\;  0^{-+}$.

The fluctuations fall into two classes according to their transformation properties under the flavor group: flavor non-singlet modes (expanded in terms of the generators of $SU(N_f)$)
and flavor singlet
modes. The glue sector contains only flavor singlet modes, whereas each fluctuation in the meson sector can be divided into flavor singlet and non-singlet terms. Those (flavor singlet) modes which are present in both sectors will mix.
Since we are in the Veneziano limit,
the mixing takes place at leading order in
$1/N_c$: the $0^{++}$ glueball mixes with the $0^{++}$ flavor singlet $\sigma$-meson, and the pseudoscalar $0^{-+}$
flavor singlet meson mixes with the $0^{-+}$ glueball due to the axial anomaly (realized by the CP-odd sector).
All classes, with various $J^{PC}$ and transformation properties under the $U(N_f)$ group, contain an infinite discrete tower of excited states once we are below the conformal window.

To compute the masses of the different glueballs and mesons we expand the action up to quadratic order in the fluctuations and
separate the fluctuations into the different decoupled sectors described above.
We postpone the technical details of this analysis to Appendix~\ref{app:quadfluctdet} and in the rest of this section present the
basic structure of each sector.

We  start by defining the vector and axial vector combinations of the gauge fields:
\be
V_M = \frac{A_M^L + A_M^R}{2}\,,\qquad
A_M = \frac{A_M^L - A_M^R}{2}\,.
\label{VAdefsmain}
\ee
They will appear both in the singlet and non-singlet flavor sectors that we describe in the following.
We also write the complex tachyon field as
\be \label{Tfluctdef}
T(x^\mu,r)=\left[\tau(r)+s(x^\mu,r)+\mathfrak{s}^a(x^\mu,r)\,t^a\right]\,\exp\left[i\theta(x^\mu,r)+i\,\pi^a(x^\mu,r)\,t^a\right]\,,
\ee
where $t^a$ are the generators of $SU(N_f)$, $\tau$ is the background solution, $s$ ($\theta$) is the scalar (pseudoscalar) flavor singlet fluctuation, and $\mathfrak{s}^a$ ($\pi^a$) are the scalar (pseudoscalar) flavor non-singlet fluctuations.

\subsection{The flavor non-singlet sector}

The class of flavor non-singlet fluctuations involves the $SU(N_f)$ part of the vector, axial vector, pseudoscalar and scalar mesons.
The relative fields are split as
\be
\begin{split}
V_\mu^F (x^\mu, r) &=  \psi_V(r)\, {\cal V}^a_\mu (x^\mu)\,t^a\,, \\
A_\m^F (x^\mu, r)
&=\psi_A(r) {\cal A}_\m^a(x^\m)\,t^a
-\psi_L(r) \partial_\m({\cal P}^a(x^\m))\,t^a\,, \\
\pi^a (x^\mu, r) &= 2\,  \psi_P(r)\, {\cal P}^a(x^\mu)\,, \\
\mathfrak{s}^a(x^\mu,r) &= \psi_S (r) \,{\cal S}^a(x^\mu)\,,
\end{split}
\ee
where the superscript $F$ denotes that we are only considering the traceless terms of the fluctuations $V_\mu$ and $A_\mu$.
The gauge $V_r=A_r=0$ is chosen and ${\cal V}^a_\mu$ and ${\cal A}^a_\mu$ are transverse, $\partial^{\m} {\cal V}^a_\m = \partial^{\m}{\cal A}^a_\mu=0$. In addition, $\mathfrak{s}^a$ and $\pi^a$ are the fluctuations of the tachyon modulus and phase, respectively, as defined in~\eqref{Tfluctdef}.

The fluctuation equations are
derived in Appendix~\ref{app:SUNfluct}.
The result for the vector and the transverse axial vector wave functions reads
\be
\begin{split}
& \frac{1}{V_f(\l,\t)\, \gf(\l,\t)^2\, e^{\Awf}\,\G}\partial_r \left( V_f(\l,\t)\, \gf(\l,\t)^2 \,e^{\Awf}\,\G^{-1}\, \partial_r \psi_V \right)
+m_V^2\, \psi_V  = 0\,, \\
& \frac{\partial_r \left( V_f(\l,\t)\, \gf(\l,\t)^2\, e^{\Awf}\,
\G^{-1}\, \partial_r \psi_A \right)}{V_f(\l,\t)\, \gf(\l,\t)^2\, e^{\Awf}\,\G  }
-4 {\t^2\, e^{2 \Awf} \over \gf(\l,\t)^2}\,\h(\l,\t)\, \psi_A  +m_A^2\, \psi_A = 0\,,
\end{split}
\label{vecaxeommain}
\ee
where we introduced a shorthand notation for
\be \label{Gdeftext}
 \G(r) = \sqrt{1 + e^{-2A(r)}\h(\l,\tau) (\partial_r \t(r))^2}\,,
\ee
and the various potentials were defined in sections~\ref{sec:VQCDglue} and~\ref{sec:VQCDflavor}.
Notice that the two equations differ by a mass term which comes from the coupling of the axial vectors to the tachyon. This term implements the effect of chiral symmetry breaking, sourcing the differences between the spectra of vector and axial vector mesons.

The fluctuation equations for the non-singlet pseudoscalars and scalars are given in Eqs.~\eqref{PSflucts} and~\eqref{flscalflucts}, respectively, in Appendix~\ref{app:SUNfluct}. The pseudoscalar fluctuations also contain the pions which become massless as the quark mass tends to zero, and obey the Gell-Mann-Oakes-Renner relation for small but finite quark mass.

\subsection{The flavor singlet sector}

We  first consider the pseudoscalar fluctuations which give rise to the $0^{-+}$ glueballs and the $\eta'$ mesons.
The flavor singlet pseudoscalar
degrees of freedom correspond to gauge invariant combinations of the longitudinal part of the flavor singlet axial vector fluctuation $A^{\lVert S}_\m$, the pseudoscalar phase of the tachyon $\theta$ and the axion
field $a$. These fields are decomposed as
\begin{align}
A^{\lVert S}_\m(x^\mu, r) &= -\varphi_L(r)\,\partial_\m({\cal T}(x^\m))\,,\nonumber\\
\theta(x^\mu, r) &=2\varphi_\theta(r)\,{\cal T}(x^\m)\,,\nonumber\\
a(x^\mu, r) &=2\varphi_\mathrm{ax}(r)\,{\cal T}(x^\m)\,.
\label{defsamain}
\end{align}
The gauge invariant combinations of the above fields
are
\begin{align}
P(r)&\equiv\varphi_\theta(r)-\varphi_L(r)\,,\nonumber\\
Q(r)&\equiv\varphi_\mathrm{ax}(r)+x\,V_a(\l,\t)\,\varphi_L(r)\,,
\label{ginvpsmain}
\end{align}
which correspond to the pseudoscalar glueball ($0^{-+}$) and $\eta'$ meson towers. These combinations satisfy the coupled
equations (\ref{cpsys1}, \ref{cpsys2}), reflecting the expected mixing of the glueballs with the mesons.

The scalar fluctuations
should realize the $0^{++}$ glueball and the flavor singlet $\sigma$ meson.
The contributing fields come from the expansion of the tachyon modulus ($s$), of the dilaton ($\chi$) and the metric.
The only scalar metric fluctuation which remains after eliminating the nondynamical degrees of freedom is $\psi$, which appears as the coefficient of the flat Minkowski metric $\eta^{\mu\nu}$ (see~\eqref{fluctdef} and~\eqref{lorentzf2} in Appendix~\ref{app:quadfluctdet}).
The combinations
\be
\zeta=\psi-{\Awf'\over\Phi'}\,\chi\, ,\qquad \xi=\psi-{\Awf'\over\tau'}\,s\;
\label{xizetdefmain}
\ee
are invariant under bulk diffeomorphisms (see section \ref{app_singlet_scalar}) and correspond to scalar glueballs and mesons which
mix at finite $x$, see (\ref{zeteq}, \ref{xieq}). As was first pointed out in \cite{ihqcd} they correspond to RG invariant operators in the dual theory.

For the flavor singlet spin-one states there are no glueballs and hence no mixing. The singlet vectors have the same spectrum as the non-singlet ones. The fluctuation equation for the singlet axial vector fluctuations differs from the one for the non-singlet axials in~\eqref{vecaxeommain} by a positive mass term, coming from the CP-odd part of the action, see Eq.(\ref{axvectoreom2}). Therefore,  the singlet axial vector states have generically higher masses than the non-singlet ones.

Finally, the singlet sector includes a traceless rank-2 tensor fluctuation, which satisfies the scalar Laplacian  equation, (\ref{tenseom}), in the 5-dimensional background (\ref{bame}). This fluctuation generates the tower of
$2^{++}$ glueballs.

\section{Constraining the action} \label{sec:constraints}

Agreement with the dynamics of QCD sets various requirements on the potential functions  ($V_g(\l)$, $V_f(\l,T)$, $\kappa(\l,T)$, $w(\la,T)$)
of V-QCD. In particular, both the UV and IR asymptotics of these functions  need to fulfill constraints, which have been analyzed in part in \cite{ihqcd,jk,alho}.
In this article, we perform a more detailed analysis of the IR structure than was done before. In addition, we study the constraints arising from the meson spectra. These constraints also apply to the function $\gf(\l,T)$, which has not been discussed in earlier work. We will first discuss generic features of the potentials, list the detailed constraints and give some examples below.

The asymptotics of the dilaton potential $V_g(\l)$, which governs the glue dynamics, has been analyzed in detail in \cite{ihqcd}. An overall fit to Yang-Mills data was done in \cite{data}.

There are other undetermined functions in the flavor action which can also depend on the tachyon field $T$. When the quark mass is flavor independent, the background solution is of the form $T = \tau(r) \mathbb{I}_{N_f}$. Evaluated on the background, the potentials
($V_f(\l,\tau)$,
$\kappa(\l,\tau)$,
$w(\la,\tau)$) must satisfy the following generic requirements \cite{jk}:
\begin{itemize}
\item[(a)] There should be two extrema in the potential for  $\tau$: an unstable maximum at $\tau=0$ with chiral symmetry intact and a minimum at $\tau=\infty$ with chiral symmetry broken.
\item[(b)] The full dilaton potential at $\tau=0$, namely $V_\mathrm{eff}(\la)=V_g(\la)-xV_{f}(\la,\tau=0)$, must have a nontrivial IR extremum at $\la=\la_*(x)$ that moves from $\la_*=0$
at $x={11/2}$ to large values as $x$ is lowered.
\end{itemize}
In \cite{jk,alho}, the flavor potential was parametrized as $V_{f}(\l,\tau)= V_{f0}(\l) \exp(-a(\l)\tau^2)$ in order to satisfy the first requirement (a). This is apparently the simplest Ansatz which works, is motivated by string theory and we will restrict to this form here. More general Ans\"atze could also be considered, for example quartic terms in the tachyon \cite{parnachev}.
In V-QCD it is, however, more essential to include the $\l$-dependence in $V_{f0}(\l)$ (and possibly in $a(\l)$), as discussed in subsection~\ref{sec:string}.
This is also natural in order to reproduce the ``running'' of the coupling (and the quark mass) of QCD, (see \cite{jk}, and the next subsection).

The second requirement (b) is necessary in order for the phase diagram to have the desired structure as $x=N_f/N_c$ is varied. Assuming the parametrization discussed above, $V_{f}(\l,\tau)= V_{f0}(\l) \exp(-a(\l)\tau^2)$, the existence of the extremum of
\be \label{Veff}
V_\mathrm{eff}(\la)=V_g(\la)-xV_{f}(\la,\tau=0) = V_g(\la)-xV_{f0}(\l)
\ee
is guaranteed  in the BZ region ($x \to ({11/2})^-$) if the $\l$-dependence of $V_g(\l)$ and $V_{f0}(\l)$ is matched with the $\beta$-function of QCD. On the field theory side the extremum is mapped to a (perturbative) IR fixed point. For generic values of $x$ the existence of the fixed point is nontrivial, and its existence needs to be studied case by case as the phase diagram may be affected.

The simplest Ansatz for the remaining functions $\h$ and $\gf$ of the flavor action is  to take them to be functions of $\l$ only: $\h=\h(\l)$ and $\gf=\gf(\l)$. Again the $\l$-dependence of $\gf$ is necessary to reproduce the running of the quark mass in QCD. As both functions appear as couplings under the square root in the DBI action, it is natural to expect that they have qualitatively similar functional form.

The CP-odd action $S_a$ contains two additional functions $Z(\l)$ and $V_a(\l,\t)$. The form of $Z(\l)$ can be constrained by studying Yang-Mills theory. $Z(\l)$ should go to  constant in the UV ($\l \to 0$) and diverge as $\sim \l^4$ in the IR ($\l \to \infty$) \cite{ihqcd}. Further constraints have been discussed recently in \cite{cs}.  The standard Ansatz is therefore
\be
 Z(\l) = Z_0 (1+c_a\l^4)\,,
\ee
where the constant $Z_0$ can be matched to the topological susceptibility of Yang-Mills, and $c_a$ can be fitted to the spectrum of glueballs (see \cite{data} for details). Notice that the value of $c_a$ also depends on the choice for $V_g$.

\subsection{UV structure}
\label{uvstruct}

The UV properties of most of the functions have been discussed in detail in \cite{jk,alho}. We repeat here the main points for completeness. As it turns out, the functions $V_g(\l)$, $V_{f0}(\l)$, $a(\l)$ and $\h(\l)$ must be analytic in the UV, $\l \to 0$. Therefore we may expand them as
\begin{align}  \label{VUVexps}
 V_{g}(\l) &= V_0(1+V_1\l+V_2\l^2+\cdots)\,,& \qquad V_{f0}(\l) &=  W_0(1+W_1 \l +W_2 \l^2 + \cdots)\,, \\
 a(\l) &= a_0(1+a_1 \l +a_2 \l^2+ \cdots)\,,& \qquad \ \  \h(\l) &= \h_0(1+\h_1 \l +\h_2 \l^2 + \cdots)\, .
\end{align}

 We first explain how the leading coefficients are fixed.
The leading coefficient of the effective potential $V_\mathrm{eff} = V_g - x V_{f0}$ is related to the UV AdS radius of the metric $\ell$ by
\be
 \frac{12}{\ell^2} = V_0 -x W_0 \, .
\ee
Changes in $\ell$ can be absorbed in redefining the fields, but one physically meaningful free parameter remains, which we shall choose to be $W_0$. Below we will fix  $V_0=12$ such that $\ell(x=0)=1$. As $\ell^2$ is positive, $W_0$ is then constrained by $0<W_0< 12/x$ \cite{jk}, so that the largest possible constant value is $24/11$ (which is the upper limit at the BZ point $x=11/2$).
One can also choose an $x$ dependent $W_0$ such that the thermodynamic functions automatically take the Stefan-Boltzmann form at large temperatures (see \cite{alho}).

Requiring the UV dimensions of the quark mass and the chiral condensate to be correct further sets
\be
 \frac{\h_0}{a_0} = \frac{2\ell^2}{3} \, .
\ee
The remaining free parameter (the normalization of $a$ and $\h$) can be eliminated by rescaling the tachyon field. Therefore we can choose, e.g., $\h_0=1$ without qualitatively affecting the background.

 We now discuss how the UV expansions can be related to perturbative QCD. First, we can use the identification of the field $\l(r)$ as the 't Hooft coupling in QCD, and $\Awf(r)$ as the logarithm of the energy scale, and require that $\l'(r)/\Awf'(r)$ agrees with the QCD $\beta$-function perturbatively at small $\l(r)$.
The higher order coefficients $V_1,V_2,\ldots$ of the dilaton potential $V_g$ are then mapped to the coefficients of the perturbative $\beta$-function of Yang-Mills theory \cite{ihqcd}. Further, the coefficients $W_1,W_2,\ldots$ of the flavor potential are fixed in terms of the coefficients of the QCD $\beta$-function in the Veneziano limit. Actually, it is convenient to use the coefficients of the effective potential  $V_\mathrm{eff} = V_g - x V_{f0}$ (see Appendix~\ref{app:bgUVIR}).

Second, we can require that the energy dependence of the (common) quark mass agrees with perturbative QCD at small $\l$ for the background solutions with a finite quark mass. The running quark mass can be identified as the coefficient of the linear term $\propto r$ of the tachyon UV asymptotics (see Appendix~\ref{app:bgUVIR}).
This results in a relation between the coefficients $a_1,a_2\,\ldots$, $\h_1,\h_2,\ldots$ in~\eqref{VUVexps} and the perturbative anomalous dimension of the quark mass in QCD. More precisely, the higher order coefficients of the ratio $a(\l)/\h(\l)$ are in one-to-one correspondence with the coefficients of the anomalous dimension. Therefore, the expansions of $a$ and $\h$ also contain a set of coefficients which are not fixed by the matching.

In practice, we require that the UV expansions of the potentials agree with the two-loop QCD $\beta$-function and the one-loop anomalous dimension, which are scheme independent.

In addition to the potentials in~\eqref{VUVexps}, the fluctuations of V-QCD depend on the function $w(\l)$, i.e., the coupling of the bulk gauge fields. We expect that it has a similar UV expansion as the other potentials:
\be
 w(\l) = w_0  (1+w_1 \l +w_2 \l^2+ \cdots)\,.
\ee
By analyzing the UV behavior of the two-point correlators, we show in Appendix~\ref{app:2ptf} that the leading coefficient in this expansion is related to $\h_0$ in~\eqref{VUVexps} via
\be \label{kappawrelation}
{\ell^4 \h_0 \over \gf_0^2}={3 \over 2}\,.
\ee

\subsection{IR structure\label{irs}}

The asymptotic IR structure of V-QCD is rather involved, and has been discussed in \cite{ihqcd,jk,alho}. Here we refine this analysis. Most of the constraints arise from the background solution with nonzero tachyon profile, which corresponds to a phase with chiral symmetry breaking on the field theory side.
The key requirement, which is necessary in order to reproduce QCD like physics in the IR, is that the tachyon potential $V_f(\l,\t)$ vanishes fast enough in the IR \cite{ckp}. Therefore, the flavor dynamics decouples asymptotically from the glue. Therefore, we may proceed in two steps:
\begin{itemize}
 \item We first solve the asymptotics of $\l(r)$ and $\Awf(r)$ from the glue action $S_g$.
 \item Inserting the results for $\l(r)$ and $\Awf(r)$, we solve the asymptotics of $\t(r)$ from the flavor action $S_f$.
\end{itemize}

In the first step the solutions are determined by the asymptotics of the dilaton potential $V_g$ only, and the requirements are the same as for IHQCD \cite{ihqcd}.
We need to ensure that the IR singularity is of the ``good'' kind, i.e., fully repulsive\footnote{Here ``repulsive'' means that all perturbations around the acceptable background solution, which keep the quark mass fixed, grow rapidly toward the IR.} \cite{gubser}. In addition, requiring the glueball spectra to be linear, and the theory to be confining fixes the asymptotics of $V_g$ to
\be \label{VgIR}
 V_g(\l) \sim \l^{4/3}(\log\l)^{1/2}\,,\qquad (\l \to \infty)\, .
\ee

In the second step, the tachyon solution depends on the asymptotics of the various potentials of the flavor action. We need to require that the tachyon potential $V_f(\l,\t)$ indeed vanishes in the IR, and the IR singularity remains ``good'', i.e., repulsive. We will call backgrounds which satisfy these criteria ``acceptable''.\footnote{Notice that the ``acceptable'' backgrounds will be further constrained by the asymptotics of the spectra as discussed in Sec.~\ref{subsec:conmsp}.}
The background is most sensitive to the asymptotics of $\h(\l)$ and $a(\l)$, but also the asymptotics of $V_{f0}$ affects the analysis. Because the gauge fields are zero in the background, their coupling $\gf(\l)$ does not appear here, but it will be constrained by the spectra as we shall discuss below in subsection~\ref{subsec:conmsp}. We parametrize
\begin{align} \label{potasympt}
 \h(\l) &\sim \l^{-\h_p}(\log\l)^{-\h_\ell}\,,& a(\l) &\sim \l^{a_p}(\log\l)^{a_\ell}\,&& \nn\\
 \gf(\l) &\sim \l^{-\gf_p}(\log\l)^{-\gf_\ell}\,,& V_{f0}(\l) &\sim \l^{v_p} \,,& (\l &\to \infty) \, .
\end{align}
The dependence of the asymptotics on $\h_p$, $a_p$, and $v_p$ was analyzed in \cite{jk}, but it turns out that the logarithmic corrections to $\h(\l)$ and $a(\l)$ may also be important, in analogy with the logarithmic corrections to the asymptotics of $V_g$.

\begin{figure}
\centerline{
\includegraphics[width=.45\textwidth]{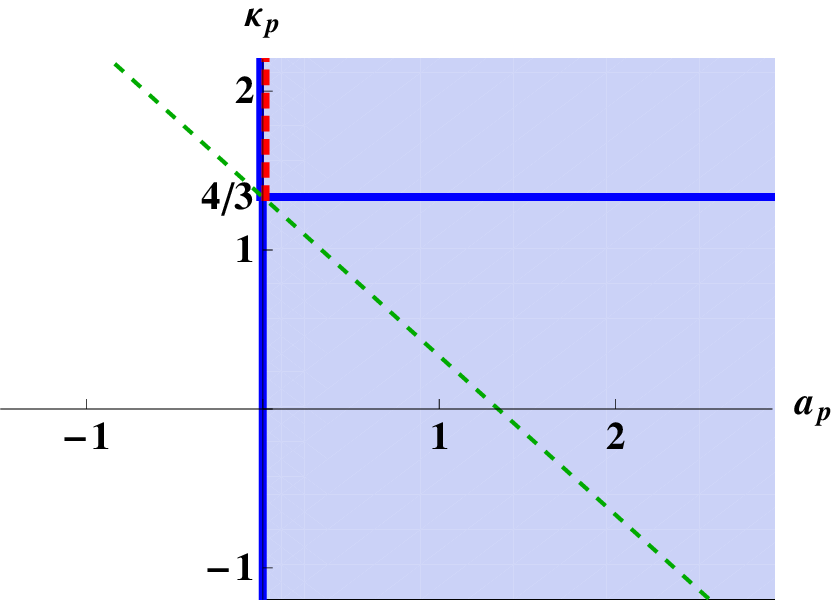}\hspace{0.05\textwidth}%
\includegraphics[width=.45\textwidth]{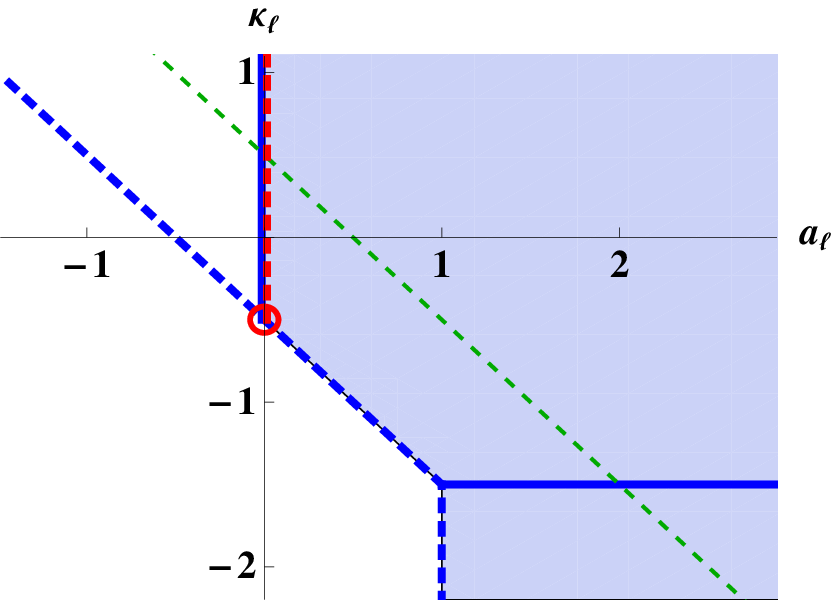}}
\caption{
Map of the acceptable IR asymptotics of the functions $\h(\l)$ and $a(\l)$. Left: qualitatively different regions of tachyon asymptotics as a function of the parameters $\h_p$ and $a_p$ characterizing the power-law asymptotics of the functions. Right: regions of tachyon asymptotics at the critical point $\h_p =4/3$, $a_p=0$ as a function of the parameters $\h_\ell$ and $a_\ell$ characterizing the logarithmic corrections to the functions.
In each plot, the shaded regions have acceptable IR behavior, and the thick blue lines denote changes in the qualitative IR behavior of the tachyon background. On the solid blue lines good asymptotics can be found, whereas on the dashed lines such asymptotics is absent. The thin dashed green line shows the critical behavior where the BF bound is saturated as $x \to 0$. Potentials above this line are guaranteed to have broken chiral symmetry at small $x$.
Finally, on the red dashed lines the asymptotic meson mass trajectories are linear with subleading logarithmic corrections.
The red circle shows the single choice of parameters where the logarithmic corrections are absent.
}\label{fig:IRmaps}
\end{figure}

We have analyzed all the asymptotics having the form~\eqref{potasympt} (see Appendix~\ref{app:tachyonIR} for details). The main results are presented in Fig.~\ref{fig:IRmaps}. For the dependence on the exponents $\h_p$ and $a_p$, Fig.~\ref{fig:IRmaps} (left), is as in \cite{jk}.
Acceptable solutions (shaded regions in the plot) are found for $a_p \ge 0$, while for $a_p<0$ either solutions with good IR singularities do not exist, or such solution do exist but the tachyon potential does not vanish in the IR. We can identify critical values $a_p=0$ and $\h_p=4/3$, where the tachyon asymptotics of the acceptable solutions change
qualitatively. Notice that these values match with those motivated by string theory, see subsection~\ref{sec:string} above. In many cases there is also an upper bound for $v_p$, see Appendix~\ref{app:tachyonIR}.

When $a_p$ and/or $\h_p$ take their critical values, the logarithmic corrections, characterized by the exponents $a_\ell$ and $\h_\ell$, are also important. In particular, the critical point $a_p=0$, $\h_p=4/3$ includes various qualitatively different cases depending on the values of $a_\ell$ and $\h_\ell$. The main features are shown in Fig.~\ref{fig:IRmaps} (right).
Again in the shaded regions acceptable solutions can be found. Moreover, the solid blue lines also have acceptable solutions, while the dashed lines do not.

In summary, the acceptable solutions are those shown by the shaded regions and solid blue lines in Fig.~\ref{fig:IRmaps}.\footnote{The case $a_p=0$ and $\h_p \ne 4/3$ was not included in the plots. For these parameter values the result depends on $a_\ell$ as follows. For $\h_p>4/3$, acceptable solutions exist when $a_\ell \ge 0$. For $\h_p<4/3$, acceptable solutions exist when $a_\ell>1$.}

\subsection{The IR fixed point and the BF bound}  \label{subsec:BF}

We already pointed out above that the effective potential
\be
 V_\mathrm{eff}(\l)=V_g(\l)- x V_{f0}(\l)
\ee
should admit an extremum $\l_*(x)$ which tends to zero as $x \to 11/2$ from below, and moves to larger values of the coupling when $x$ is decreased. On the field theory side, $\l_*(x)$ marks the position of a nontrivial IR fixed point (unless the fixed point is screened by the tachyon). The extremum of the effective potential might exist all the way to $x=0$ or only above some critical value $x=x_*$.
Its existence near $x=11/2$ (the BZ region) is guaranteed by the matching of the effective potential with the perturbative QCD $\beta$-function.

It is essential to study when the tachyon mass at the IR fixed point satisfies the Breitenlohner-Freedman (BF) bound,
\be
 - m_\mathrm{IR}^2 \ell_\mathrm{IR}^2 < 4
\ee
where $\ell_\mathrm{IR}$ is the IR AdS radius. It is not difficult to understand why this is important. For $x$ close to $11/2$ (and for zero quark mass) the dominant background has no chiral symmetry breaking, and the tachyon vanishes identically. The holographic RG flow runs from the UV fixed point at $\l=0$ to the IR fixed point at $\l=\l_*(x)$.
As $x$ is decreased, if the BF bound is violated at the IR fixed point, the background becomes unstable against fluctuations of the tachyon in the IR.
Then the dominant background solution has nonzero tachyon and therefore also chiral symmetry is broken. Therefore, there is a phase transition at the value $x=x_c$ where the BF bound is saturated, \cite{son}. If the BF bound cannot be saturated at the fixed point, the model is likely not to have chiral symmetry breaking at all.
When the UV expansions of the potentials are fixed in terms of the two-loop QCD $\beta$-function and one-loop anomalous dimension of the quark mass, a critical value $x_c\simeq 4$ exists, unless the
potentials
are modified by large higher order terms \cite{jk}.

The existence of the fixed point can also be linked to the asymptotics of the potentials. Recall that the glue potential had the asymptotics $V_g(\l) \sim \l^{4/3}\sqrt{\log\l}$. Matching with UV behavior of QCD requires that $V_\mathrm{eff}(\l)$ increases with $\l$ for small $\l$. Therefore, taking $V_{f0}(\l)\sim \l^{v_p}$,
\begin{itemize}
 \item When $v_p>4/3$, the effective potential has at least one extremum (and in general an odd  number of
 extrema) for all $0<x<11/2$. For low enough $x$, there is a single extremum which tends to $\infty$ as $x \to 0$.
 \item When $v_p<4/3$ the effective potential has an even number extrema, or no extrema at all. When $x$ is
 decreased down from $x=11/2$, the perturbative BZ fixed point disappears at some $x=x_*$ by joining another
 fixed point.
\end{itemize}
In the critical case $v_p=4/3$, there are three possibilities, depending on higher order terms:
the behavior can be as in either of the two above cases, or the fixed point disappears by running to $\l=\infty$ at a finite
$x=x_*$ \cite{alho}.

\begin{figure}
\centerline{
\includegraphics[width=.45\textwidth]{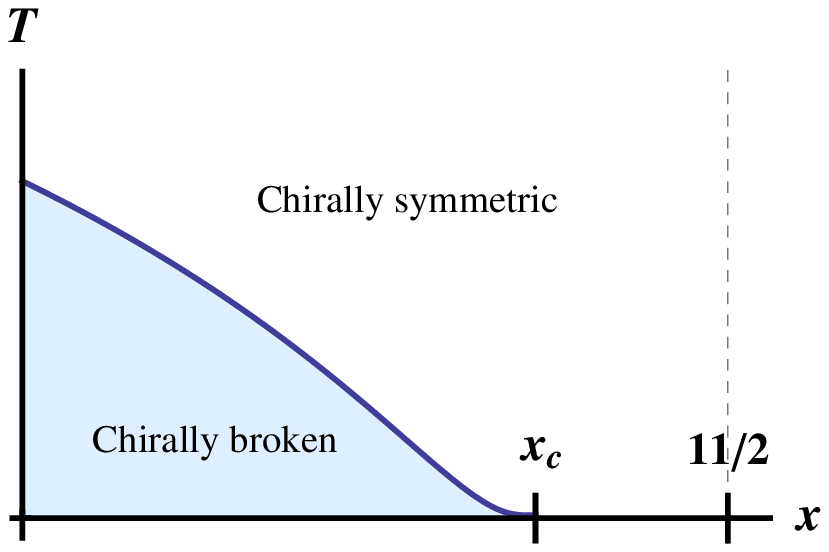}\hspace{0.05\textwidth}%
\includegraphics[width=.45\textwidth]{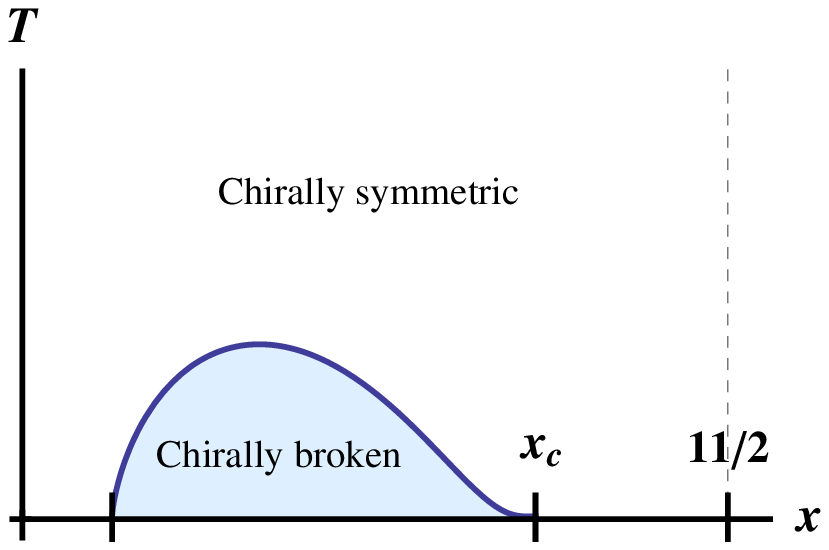}}
\caption{A qualitative sketch of the different small-$x$ behaviors of the phase diagram on the $(x,T)$-plane. Left: The ``standard'' case, where the low temperature chirally broken phase exists down to $x=0$. Right: The ``exceptional'' case, where the theory is chirally symmetric at all temperatures when $x$ lies below certain critical value. The latter case is only possible if the BF bound is satisfied at the IR fixed point at low $x$.
}\label{fig:finiteTpd}
\end{figure}

All potentials which we study here belong to the class with $v_p>4/3$. In this case, it is possible to calculate the tachyon mass from the asymptotics of the potentials as $x \to 0$ (and therefore $\l_* \to \infty$) \cite{alho}. A straightforward calculation yields
\be
 - m_\mathrm{IR}^2 \ell_\mathrm{IR}^2 = \frac{24 a(\l_*)}{\h(\l_*)V_\mathrm{eff}(\l_*)} \, .
\ee
Therefore,
\be
 - m_\mathrm{IR}^2 \ell_\mathrm{IR}^2 \sim \l_*^{a_p+\h_p-4/3}(\log\l_*)^{a_\ell+\h_\ell-1/2}\,, \qquad (\l_* \to \infty)\,,
\ee
where we used the fact that $V_\mathrm{eff}(\l_*)\sim V_g(\l_*)$. It was observed in \cite{alho} that potentials for which the tachyon mass tends to zero as $x \to 0$ may have an exceptional phase diagram, where chiral symmetry is preserved at small $x$. In Fig.~\ref{fig:finiteTpd} (right) we show a sketch of the phase diagram on the ($x,T$)-plane for such a case.\footnote{There might also be additional finite structure in addition to the main features shown in Fig.~\ref{fig:finiteTpd}. These have been discussed in detail in~\cite{alho}.} The standard behavior (i.e., chiral symmetry breaking) at small $x$ is guaranteed, if we require that the tachyon mass diverges as $x \to 0$. Then the phase diagram is qualitatively similar to that shown on Fig.~\ref{fig:finiteTpd} (left).
The tachyon mass diverges if $a_p+\h_p>4/3$, or possibly if $a_p+\h_p=4/3$, in which case we also need to require $a_\ell+\h_\ell>1/2$. The edge of this region is marked as the thin dashed green line in Fig.~\ref{fig:IRmaps}.

\subsection{Constraints from the meson spectra}
\label{subsec:conmsp}

The asymptotic behavior of the meson masses at large excitation number depends strongly on the IR asymptotics of the tachyon. It therefore allows to further narrow down the choices for physically relevant functions, and, in particular, also set constraints on the IR behavior of $\gf(\l)$. The asymptotics has been analyzed in detail in Appendix~\ref{app:Regge}.  The generic findings are as follows:
\begin{itemize}
 \item The asymptotics of each tower has the generic form
     \be \label{genasympt}
        m_n^2 \sim n^P (\log n)^Q\, , \qquad (n \to \infty) \, .
     \ee
 \item In the flavor non-singlet sectors there are two possibilities for all potentials classified as acceptable above, depending on the asymptotics of $\gf(\l)$:
 \begin{enumerate}
  \item  All the towers (vectors, axials, scalars and pseudoscalars) have the same asymptotic behavior (including the slope, i.e., the proportionality constant in~\eqref{genasympt}).
  \item The vector and scalar towers have different asymptotics from the axial and pseudoscalar towers. In several cases, the only difference is the slope.
 \end{enumerate}
 \item In the flavor singlet towers (scalars and pseudoscalars) the glueball and meson ($\bar q q$) excitations decouple at large $n$. The asymptotics for the meson asymptotics are exactly the same for the flavor non-singlet mesons with the same parity. All glueball towers (including the spin-two glueballs) have linear asymptotics $m_n^2 \sim n$ with the same slope. This is because the glueball asymptotics only depends on the potentials $V_g(\l)$ and $Z(\l)$ which were fixed as in IHQCD.
\end{itemize}
We have also computed the asymptotics of the vector and axial vector decay constants in Appendix~\ref{app:asdecconn}.

We require that all the meson towers have the same asymptotics. This happens for all potential choices, unless $\gf(\l)$ vanishes too fast in the IR.
The rule of thumb is that $\gf(\l)$ should vanish slower than $\h(\l)$ in the IR, i.e., the ratio $\h(\l)/\gf(\l)$ should vanish in the IR (see Appendix~\ref{app:Regge} for details):
\be \label{wconstraint}
 \frac{\h(\l)}{\gf(\l)} \to 0 \,, \qquad (\l \to \infty)\,.
\ee
When this is the case
the leading meson asymptotics depend on the choice of $a(\l)$ as follows (dropping logarithmic corrections and only including potentials which produce an acceptable background in the IR):
\begin{itemize}
 \item When $a_p$ of~\eqref{potasympt} is positive, we find that $m_n^2 \sim n^2$.
 \item When $a_p=0$ but $a_\ell>0$, we find that $m_n^2 \sim n^P$ with $1<P<2$.
 \item When $a(\l)$ is constant (such that $a_p=0=a_\ell$), we find that $m_n^2 \sim n$.
\end{itemize}
There is a single exception to the above rules: taking the critical choice $\h_p=4/3$ and $a_p=0$, and with $a_\ell=1$ and $\h_\ell=-3/2$ (which is a meeting point of lines in Fig.~\ref{fig:IRmaps} (right)), the background is acceptable and the asymptotics is $m_n^2 \sim n \log n$.

Requiring that the leading meson trajectories are linear therefore leaves us with the choices where $a(\l)$ is constant, which are marked with the red dashed line in Fig.~\ref{fig:IRmaps}.\footnote{It is also possible to construct potentials with linear asymptotics with nonconstant $a(\l)$ if the requirement on how $V_f(\l,\t)$ vanishes in the IR is modified, see Appendix~\ref{app:Regge}.}
Interestingly, further requiring that the subleading logarithmic corrections to the asymptotics of $m_n^2$ vanish, leaves us with a single\footnote{There is also another special choice which has linear trajectories but requires more delicate tuning of parameters. This choice has also the critical power law $a_p=0$, $\h_p=4/3$, but now $a_\ell+\h_\ell=-1/2$ with $0<a_\ell<1$, and $a_\ell$ should be sufficiently close to $a_\ell=1$.
This case was not listed as acceptable in Fig.~\ref{fig:IRmaps} because the flavor potential $V_f$ does not vanish for generic values of $a_\ell$ in the range $0<a_\ell<1$.
However in the limit $a_\ell \to 1$ the background should
be fine, as discussed in Appendix~\ref{app:Regge}.} choice of asymptotics of $a(\l)$ and $\h(\l)$ marked as the red circle in Fig.~\ref{fig:IRmaps}:
\be \label{finalchoice}
 \h_p=\frac{4}{3}\,,\quad a_p=0\,,\quad \h_\ell=-\frac{1}{2}\,,\quad a_\ell=0\, .
\ee
Further, all meson trajectories have the same slopes if the constraint~\eqref{wconstraint} for $\gf(\l)$ is satisfied.
This choice is a slight modification of Potentials I which were studied extensively in \cite{jk,alho} (and have instead $\kappa_\ell=0$). In this case, there is a further requirement related to the IR asymptotics of the tachyon, which is a power-law,
\be
 \t(r) \sim \t_0 r^C \, , \qquad (r \to \infty)\, .
\ee
In order for the tachyon potential $V_f(\l,\t)$ to vanish in the IR, and therefore to ensure the decoupling of the tachyon in the IR, we need that $C>1$, which can be satisfied by varying additional parameters in the potential: see the discussion in the next subsection.
Notice that the slopes of the meson and glueball trajectories might be different even in this case.

Interestingly, the asymptotics~\eqref{finalchoice} which has exactly linear trajectories is somewhat analogous to the choice of $V_g$ in~\eqref{VgIR}: in both cases we find power laws in $\l$ inspired by string theory, modified by logarithmic corrections, as we already discussed above in subsection~\ref{sec:string}.
The power $4/3$ in $V_g(\l)$ and in $\kappa(\l)$ arises from the transformation from the string frame to the Einstein frame, see Eqs.~\eqref{par1} and~\eqref{par2}.

The only constraint from the considerations above on the coupling factor $\gf(\l)$ of the gauge fields in the DBI action at intermediate ($\l = \morder{1}$) and large ($\l \to\infty$) values of the coupling is given in~\eqref{wconstraint}. The choice of this potential, however, does have
a strong effect on the qualitative behavior on the spectra for low excitation numbers. It is expected to affect strongly the masses of the vector and axial mesons which are identified with the fluctuation modes of the bulk gauge fields.
In particular, the lightest meson is usually either a vector or a scalar depending on the choice of $\gf(\l)$.
This kind of properties depend on the behavior of the potentials at $\l = \mathcal{O}(1)$ and $r= \mathcal{O}(1)$, where the solutions are not analytically tractable, and therefore need to be analyzed numerically.
The natural expectation, which can be confirmed by numerics, is that when $\gf(\l)$ and $\h(\l)$ have qualitatively similar $\l$-dependence, then the spectra of the vector and the scalar mesons look qualitatively similar.
In practice this means that we will choose the string-motivated value for the power laws of both coupling $\h(\l)$ and $\gf(\l)$, i.e., $\h_p=4/3=\gf_p$.

To conclude, the only choice of potentials that results in exactly linear trajectories is given by~\eqref{finalchoice}. Further, the meson trajectories have the same slopes if $\gf_p<\h_p=4/3$, or also if we have the critical power law $\gf_p=\h_p=4/3$ and in addition $\gf_\ell<\h_\ell=-1/2$.

\subsection{Examples of potentials} \label{sec:potentials}

In~\cite{jk} and in~\cite{alho} we discussed two classes of potentials $V_g$, $V_{f0}$, $\h$, and $a$, which we called potentials I and potentials II. They can be defined as follows.
\begin{itemize}
 \item \textbf{Both Potentials I \& II.}
  \begin{align} \label{potIandIIcommon}
    V_{g}(\l)  & = V_0\left[1+V_1 \l + V_2 \l^2 \frac{\sqrt{1+\log(1+\frac{\lambda}{\l_0})}}{\left(1+\frac{\lambda}{\l_0}\right)^{2/3}}\right]\,, \nonumber\\
    V_{f0}(\l) & = W_0\left[1+W_1 \l + W_2 \l^2\right]\,.
  \end{align}
 \item \textbf{Potentials I.}
  \be \label{potIdefs}
    a(\l)   = a_0\,,\qquad   \h(\l)  = \frac{1}{\left(1+\frac{3a_1}{4}\l\right)^{4/3}}\,.
  \ee
 \item \textbf{Potentials II.}
  \be \label{potIIdefs}
    a(\l)   = a_0\,\frac{1+a_1 \l + \frac{\l^2}{\l_0^2}}{\left(1+\frac{\l}{\l_0}\right)^{4/3}}\,, \qquad    \h(\l)  = \frac{1}{\left(1+\frac{\l}{\l_0}\right)^{4/3}}\,.
  \ee
\end{itemize}

Here the coefficients are fixed by matching to perturbative QCD as discussed above, except for $W_0$, which remains as a free parameter. We also set $\ell(x=0)=1$, and choose the parameter $\l_0$, which only affects the higher order coefficients of the UV expansions, such that the higher order coefficients have approximately the same relative size as with standard scheme choices in perturbative QCD. Explicitly, the coefficients satisfy
\begin{align}
 V_0 &= 12\, , \qquad V_1 = \frac{11}{27 \pi^2}\,,\qquad  V_2= \frac{4619}{46656 \pi ^4}\, ; \nonumber \\
 W_1 &= \frac{24+(11-2 x) W_0}{27 \pi ^2 W_0}\,,\qquad W_2 = \frac{24 (857-46 x)+\left(4619-1714 x+92 x^2\right) W_0}{46656 \pi ^4 W_0}\,; \nonumber\\
 a_0 &= \frac{12-x W_0}{8} \, , \qquad a_1 =  \frac{115-16 x}{216 \pi ^2} \, , \qquad \l_0 = {8 \pi^2} \,.
\end{align}

As we discussed above, we consider two qualitatively different choices for $W_0$: either constant $W_0$, which satisfies
\be \label{W0range}
0<W_0<24/11\,,
\ee
or $W_0$ fixed such that the pressure agrees with the Stefan-Boltzmann (SB) result at high temperatures \cite{alho} (without the need to include $x$ dependence in the normalization of the action). The latter option gives (when $\ell(x=0)=1$)
\be
 W_0 = \frac{12}{x}\left[1-\frac{1}{(1+\frac{7}{4}x)^{2/3}}\right]\qquad \textrm{(Stefan-Boltzmann)}\,,
\ee
so that the AdS radius is
\be
 \ell(x) = \sqrt[3]{1+\frac{7}{4}x}\,.
\ee
The finite temperature phase diagram is of the exceptional type of Fig.~\ref{fig:finiteTpd} (right) for potentials I if $W_0$ is large or SB normalized,~\cite{alho}, so that there is a chirally symmetric phase at small $x$, as discussed above. An acceptable value of $W_0$ for potentials~I is therefore, e.g., $W_0=3/11$, which is relatively close to the lower limit of the range~\eqref{W0range}. For potentials II all choices produce the standard phase diagram of Fig.~\ref{fig:finiteTpd} (left).

Based on the earlier discussion in this section we notice that
\begin{itemize}
 \item \textbf{Potentials I} were chosen such that the power behavior of $\h(\l)$ and $a(\l)$ is the critical one, $\h_p=4/3$ and $a_p=0$. These potentials admit a regular IR solution with exponential tachyon, $\t \sim \t_0 e^{C r}$, where $C$ can be computed in terms of the potentials and $\t_0$ is an integration constant (see Appendix~\ref{app:tachyonIR} for details). The asymptotic trajectories of masses in all towers are linear but have logarithmic corrections.
 \item \textbf{Potentials II} have instead $\h_p=4/3$ and $a_p=2/3$. These potentials admit a regular IR solution with $\t \sim \sqrt{C r +\t_0}$, and the asymptotic trajectories of masses in all towers are quadratic.
\end{itemize}
We also see that potentials I can be quite easily modified so that the asymptotic trajectories are exactly linear and that logarithmic corrections are absent: we need to add a critical logarithmic correction to $\h(\l)$ such that $\h_\ell = -1/2$ and we are sitting at the red circle of Fig.~\ref{fig:IRmaps} (right). An explicit choice is
\be \label{logmodk}
\h(\l)  = \frac{1}{\left(1+\frac{3a_1}{4}\l\right)^{4/3}}\sqrt{1+\frac{1}{D}\log\left[1+\left(\frac{\lambda} {\lambda_0}\right)^2\right]} \,.
\ee
There is however the following observation. For these potentials we find that the regular solution has the tachyon IR asymptotics
\be
 \t(r) \sim \t_0 r^C\,,
\ee
where the coefficient
\be
C = \frac{27\times 3^{1/3} \sqrt{D}\, (115-16 x)^{4/3} \left(12-x\, W_0\right)}{295616\times 2^{1/6}}
\ee
must be larger than one. For this to happen for all reasonable $W_0$ and for all values of $x$ up to $x_c$  we need a relatively large $D$, e.g., $D=200$. This means that the logarithmic modification in~\eqref{logmodk} sets in only at very high values of $\l$, i.e., only close to the IR singularity. Therefore the logarithmic correction term is expected to only cause minor changes to observables such as the finite temperature phase diagram and low lying masses.

Finally we also need to choose the function $w(\l)$. As argued above, it should not vanish too fast in the IR, and should have qualitatively similar $\l$ dependence as $\h(\l)$. An intriguing choice would be $w(\l) = w_0 \h(\l)$, where $w_0 = \ell^2 \sqrt{2}/\sqrt{3}$ due to~\eqref{kappawrelation} as we have chosen $\h_0=1$.
However, as detailed in Appendix~\ref{app:Regge}, such a  choice of $w(\l)$ is the critical one which would often make the slopes of, e.g., the asymptotic vector and axial vector trajectories to be different. If we want the slopes to be the same, $w(\l)$ should vanish slightly slower in the IR than $\h(\l)$. Therefore a reasonable choice, which would work with potentials I and also together with $\h(\l)$ of~\eqref{logmodk}, would be
\be \label{logmodw}
 w(\l) =\frac{w_0}{\left(1+\frac{3a_1}{4}\l\right)^{4/3}}\left\{1+\frac{1}{D}\log\left[1+\left(\frac{\lambda} {\lambda_0}\right)^2\right]\right\}\,.
\ee

We have done the numerical analysis of the next two sections by using the following choices:
\begin{itemize}
 \item \textbf{Potentials I with $W_0=3/11$ and $w(\l)=\h(\l)$.} The motivation for this choice is to mimic (at qualitative level, without fitting any of the numerical results to QCD data) the physics of real QCD in the Veneziano limit. We have checked that the finite temperature phase diagram has the standard structure of Fig.~\ref{fig:finiteTpd} (left) when $W_0=3/11$.
Notice that we did not implement the logarithmic corrections of~\eqref{logmodk} and~\eqref{logmodw}, but as we have argued, these factors only affect slightly the numerical results.

 \item \textbf{Potentials II with SB normalized $W_0$ and $w(\l)=1$.} This choice might not model QCD as well as the first one, but the motivation is merely to pick a background with different IR structure in order to see how model dependent our results are.
\end{itemize}

\section{Spectra: numerical results} \label{sec:results}

Here we present the results of the numerical solution of the fluctuation equations for two different classes of potentials (I and II) as specified above in Sec.~\ref{sec:potentials}. We compute the spectrum of all excitation modes as a function of $x$ and for zero quark mass. To find the mass spectrum one has to require normalizability of the wave functions of the fluctuation modes both in the IR and in the UV.
Then, the numerical integration of the fluctuation equations leads to discrete towers of masses corresponding to physical states.

In practice, the computation proceeds as follows. We choose a set of potentials (I or II) and a value of $x$ below the critical one $x_c$. The dominant background, which has a nontrivial tachyon profile, is then constructed~\cite{jk} by shooting from the IR and matching to the IR expansions of the various fields given in Appendix~\ref{app:bgUVIR}.
The coefficients of the fluctuation equations, which are discussed in Sec.~\ref{quadfluct} and in Appendix~\ref{app:quadfluctdet}, are then evaluated on the background.
After this the fluctuation equations are solved by shooting from the IR and matching to the IR expansions of the IR-normalizable fluctuation modes given in Appendix~\ref{app:wfasympt}. For the non-singlet sector it is easy to identify the meson masses by varying the bound state mass and reading off the values where the solution becomes normalizable in the UV. We have also computed the mass spectrum for the singlet scalars, which requires analyzing two coupled fluctuation equations, as explained in
Appendix~\ref{app:nummet}. All computations were done by using $A$ as the coordinate instead of $r$, because for this choice, all fields behave smoothly even when the background is close to having a fixed point.

Some general features of both singlet and non-singlet sectors are
\begin{itemize}
\item For $x< x_c$ all spectra are discrete and gapped except for the non-singlet pseudoscalars (pions) which are the massless Goldstone bosons of chiral symmetry breaking.
\item All the mass spectra are continuous in the conformal region, $x_c\le x <11/2$.
\item
In the walking region, as $x \rightarrow x_c$, all masses follow Miransky scaling \be m_n \sim \Lambda_\mathrm{UV} \exp \left( -{\kappa \over \sqrt{x_c -x}} \right)\,.\ee
\item All mass ratios asymptote to finite constants for $x \rightarrow x_c$.
\end{itemize}

\begin{figure}[t]
\begin{center}
\includegraphics[width=0.49\textwidth]{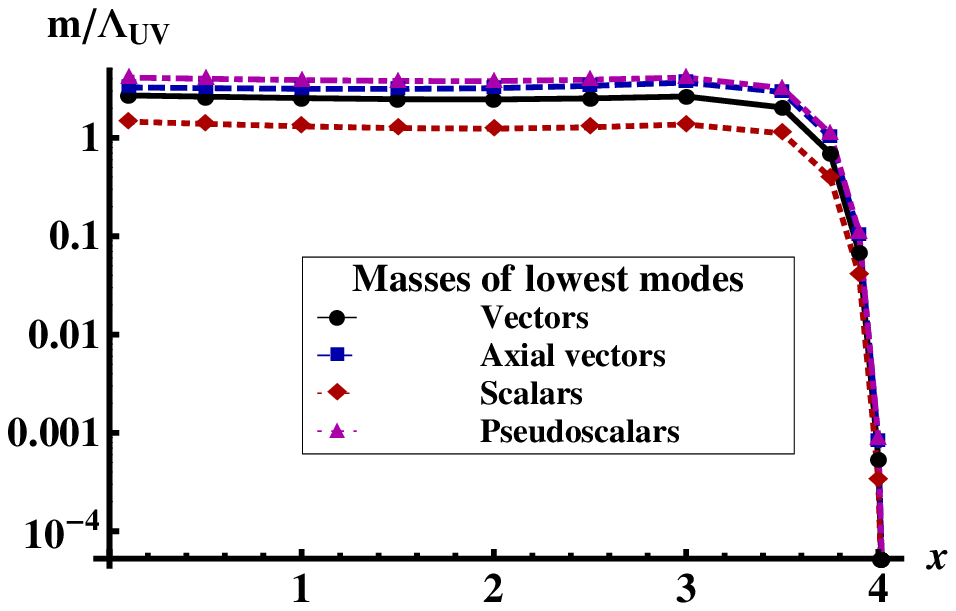}\hfill
\includegraphics[width=0.49\textwidth]{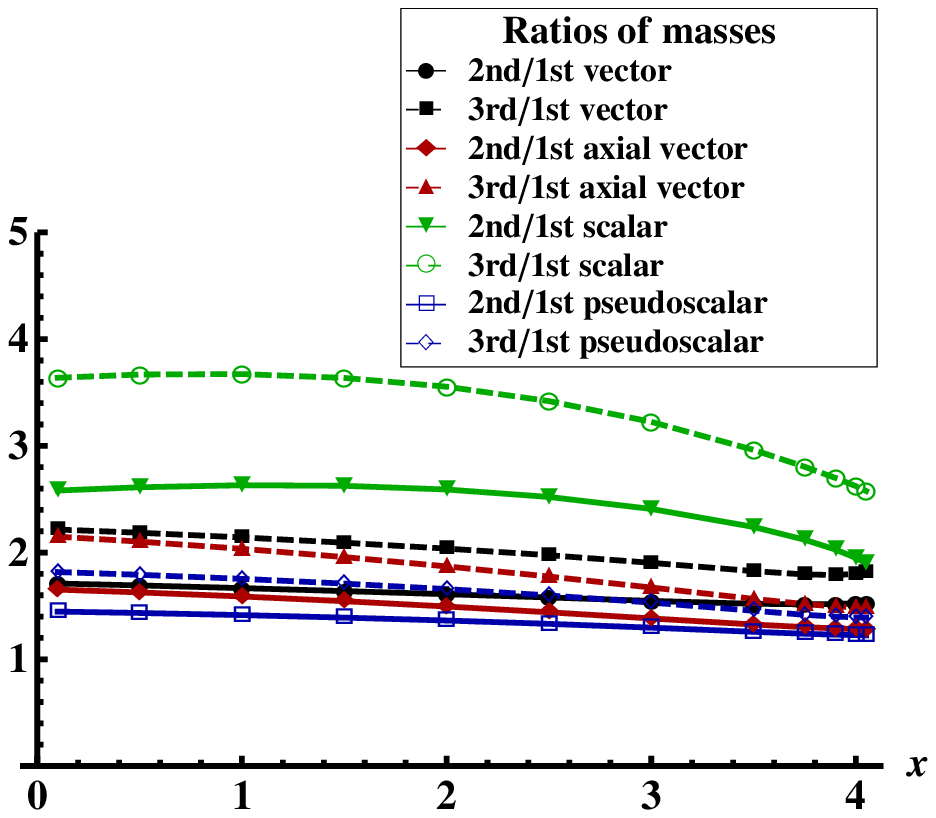}
\end{center}
\caption{Non-singlet meson spectra in the potential I class ($W_0=3/11$), with $x_c \simeq 4.0830$. Left: the lowest nonzero masses of all four towers of mesons, as a function of $x$,
in units of $\Lambda_{\rm UV}$, below the conformal window. Right: the ratios of masses of up to the third massive states as a function of $x$.
}
\label{fnss1}\end{figure}

\begin{figure}[t]
\begin{center}
\includegraphics[width=0.49\textwidth]{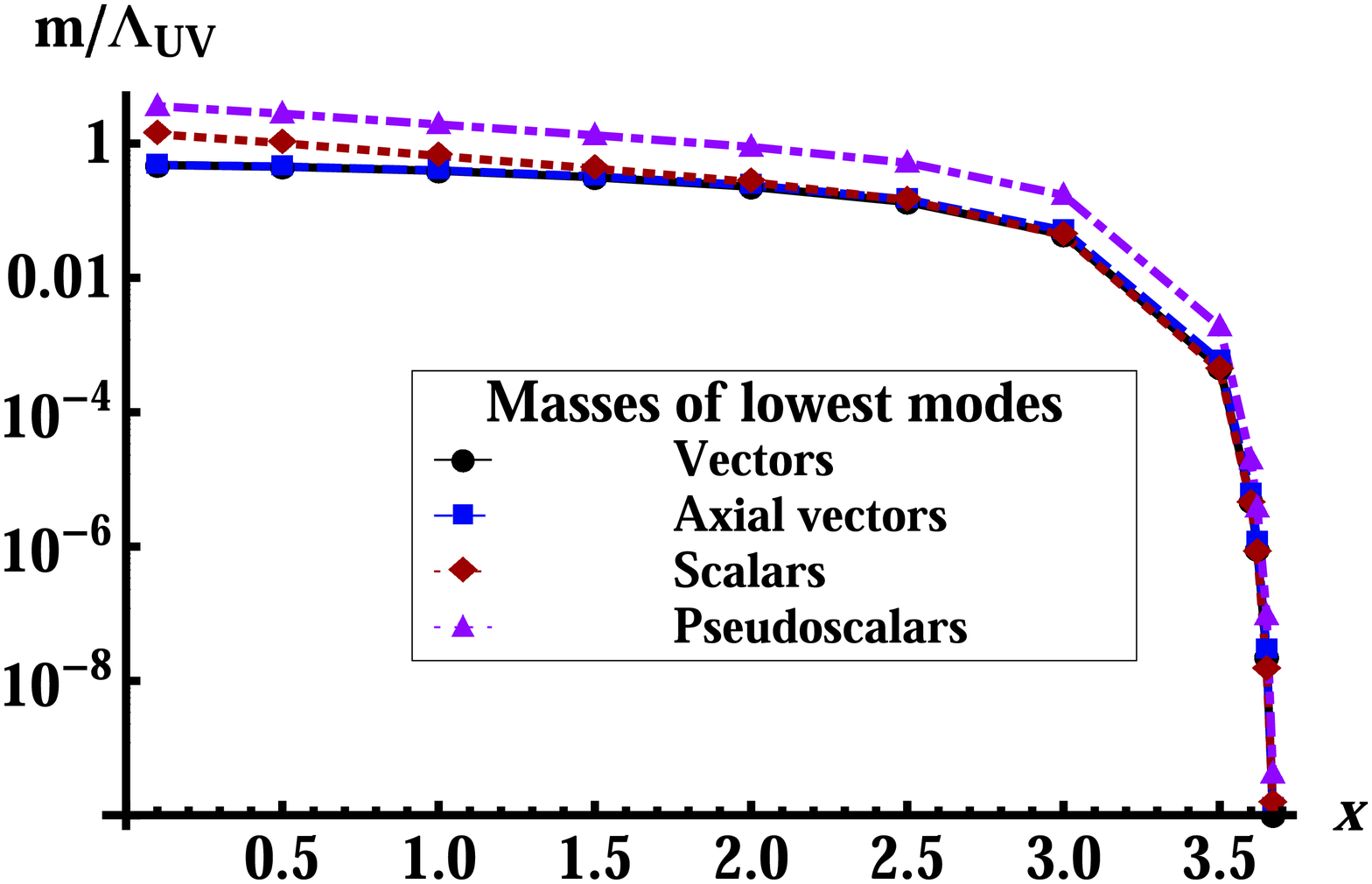}\hfill
\includegraphics[width=0.49\textwidth]{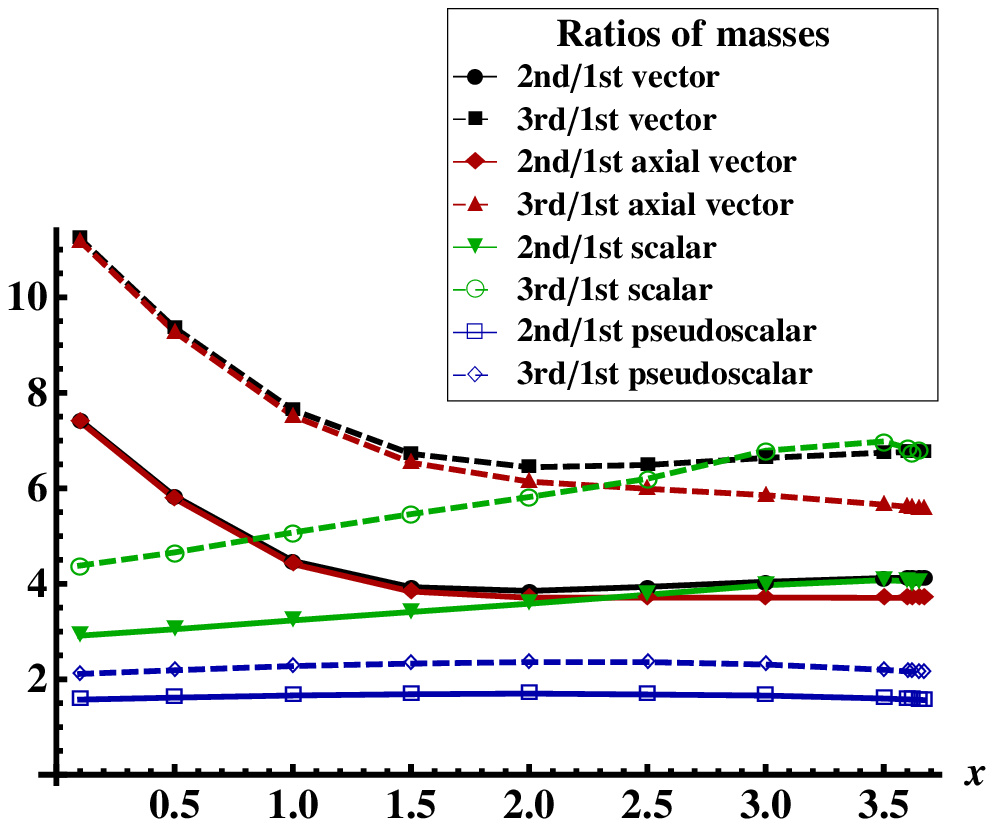}
\end{center}
\caption{Non-singlet meson spectra in the potential II class with SB normalization for $W_0$ (so that $x_c \simeq 3.7001$). Left: the lowest nonzero masses of all four towers of mesons, as a function of $x$, in units of
$\Lambda_{\rm UV}$, below the conformal window. Right: the ratios of masses of up to the third massive states as a function of $x$.
}
\label{fnss2}\end{figure}

\begin{figure}[!tb]
\begin{center}
\includegraphics[width=0.49\textwidth]{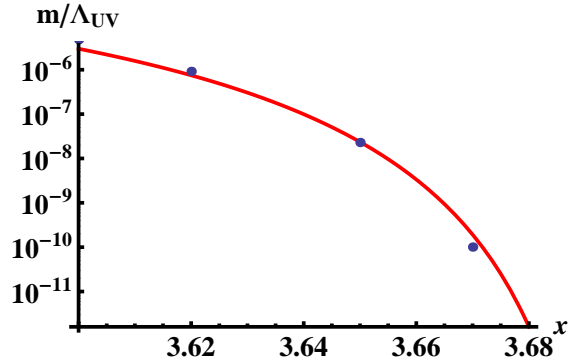}\hfill
\end{center}
\caption{A fit of the $\rho$ mass to the Miransky scaling factor, for Potential II with SB normalization for $W_0$.}
\label{fnss3}\end{figure}

\subsection{The flavor non-singlet sector}

The results for the mass spectrum of all the flavor non-singlet excitation modes, vectors, axial vectors, pseudo-scalars and scalars are shown in Figs.~\ref{fnss1} and~\ref{fnss2}, for potentials~I with $W_0=3/11$ and for potentials~II with SB normalized $W_0$, respectively.
The left hand plots depict the mass of the lowest mode of each excitation tower (excluding the massless Goldstones of the pseudoscalar tower) as a function of $x$ in units of $\Lambda_\mathrm{UV}$. Notice that the vertical axis is in logarithmic scale. For potentials~II the lowest massive mode is the vector and for potentials~I the scalar.

The dependence of masses on $x$ is rather mild below the walking region, i.e., when $0<x<x_c$ and $x_c - x = \morder{1}$. There are, however, some tendencies depending on the potentials. For potentials~II (Fig.~\ref{fnss2}), the vectors have qualitatively similar $x$ dependence as the axial vectors, but it  is somewhat different from that of the scalars and pseudo-scalars, as best seen at small $x$. For potentials~I, however, masses in all towers have qualitatively similar $x$ dependence.
This reflects our choices for the functions $\h(\l)$ and $w(\l)$ appearing as couplings in~\eqref{Senaction}: As $w(\l)$ is the coupling for the gauge fields, it only affects the fluctuation equations for the spin-one modes.
For potentials~I we chose $\h(\l)=w(\l)$, and therefore the masses of spin-zero and spin-one states were expected to have qualitatively similar behavior as we discussed in Sec.~\ref{sec:constraints}, which is confirmed by the numerics in Fig.~\ref{fnss1}. For potentials~II we chose $\h(\l)$ as
in~\eqref{potIIdefs}, but we took $w(\l)=1$ so that the $\l$-dependence of $\h(\l)$ differs qualitatively from that of $w(\l)$, which explains the qualitative differences between the states of different spins in this case.

As it is shown explicitly for the $\rho$ mass in figure \ref{fnss3}, the masses approach zero exponentially in the walking region (as $x \to x_c$ from below), following Miransky scaling. Finally, the right hand plots in figures~\ref{fnss1} and~\ref{fnss2}, show how the ratios of the mass of the second and third excitation modes over the first  one depend on $x$.
All of these approach finite numbers close to $x_c$. These findings appear to be in qualitative agreement with earlier analysis based on Dyson-Schwinger and Bethe-Salpeter equations~\cite{Harada:2003dc}.

 \begin{figure}[!tb]
\begin{center}
\includegraphics[width=0.49\textwidth]{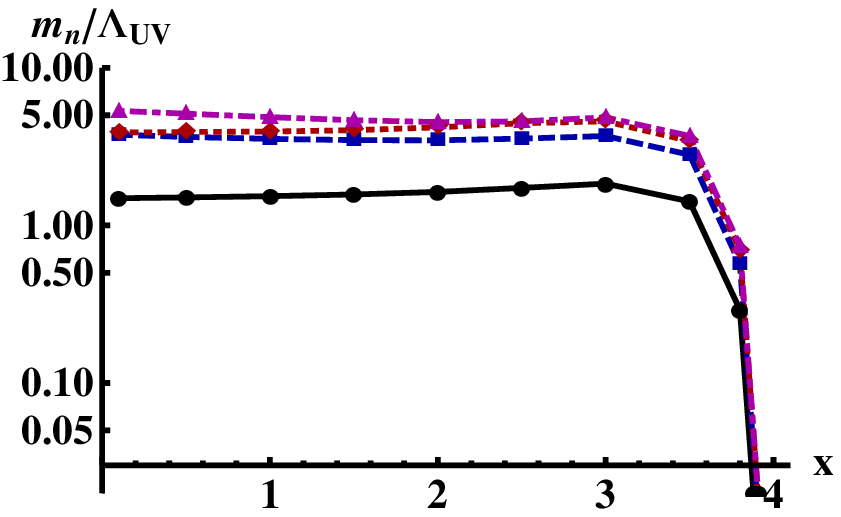}\hfill
\includegraphics[width=0.49\textwidth]{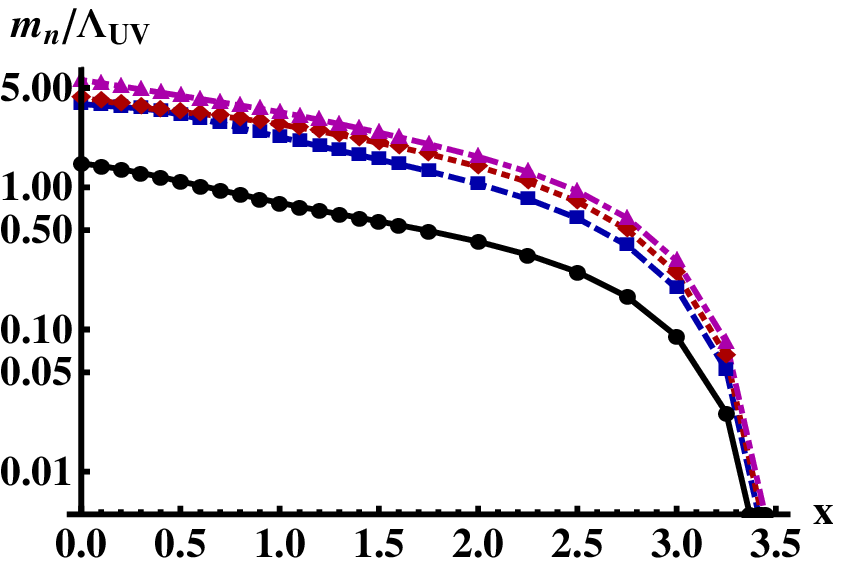}
\end{center}
\caption{Singlet scalar meson spectra.
The plots contain the four lowest masses as a function of $x$ in units
of $\Lambda_{\rm UV}$, including two $0^{++}$ glueballs and two singlet $0^{++}$
 mesons that mix here at leading order. Left: potentials~I with $W_0=3/11$. Right: potentials~II with SB normalized $W_0$.
}
\label{fss1}\end{figure}

 \begin{figure}[!tb]
\begin{center}
\includegraphics[width=0.49\textwidth]{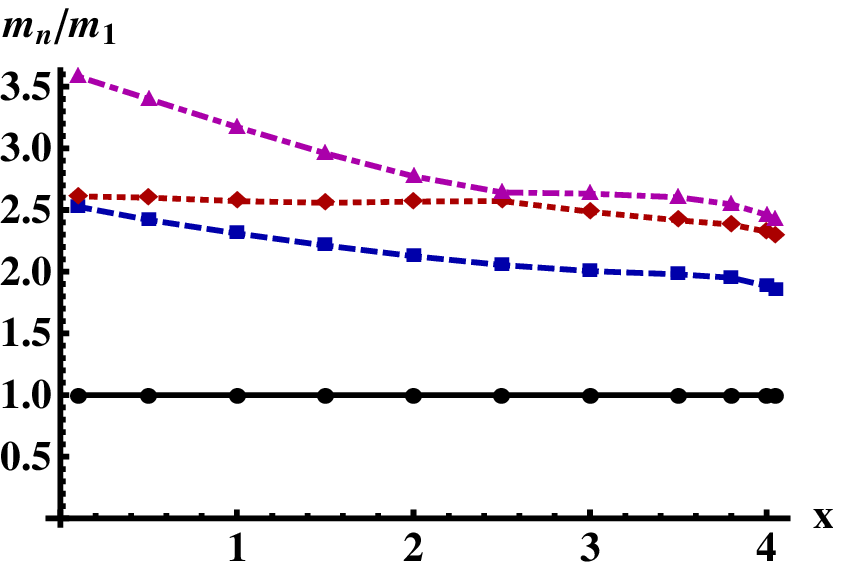}\hfill
\includegraphics[width=0.49\textwidth]{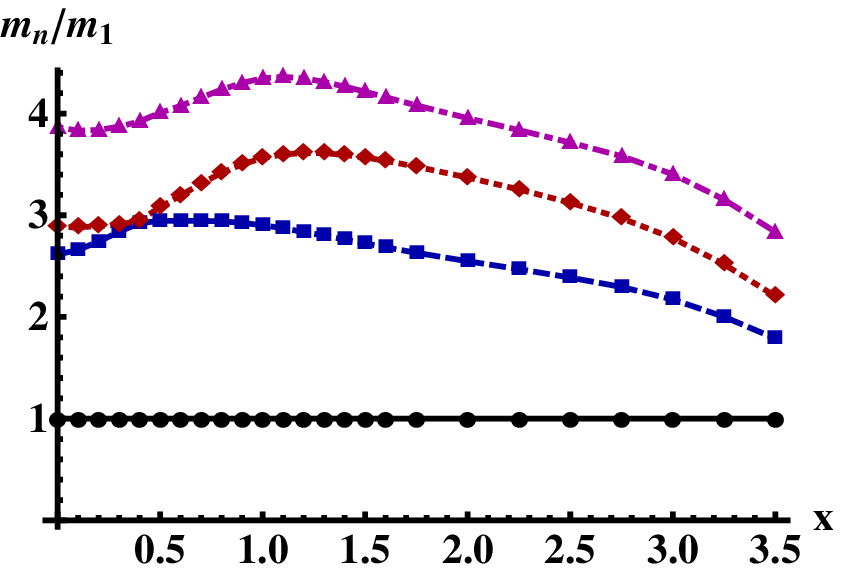}
\end{center}
\caption{The ratios of the singlet scalar masses of up to the fourth massive states as a function of $x$.   Left: potentials~I with $W_0=3/11$. Right: potentials~II with SB normalized $W_0$.
}
\label{fss2}\end{figure}

\subsection{The flavor singlet sector}

The flavor singlet scalar and pseudoscalar spectra are interesting due to the nontrivial mixing between the glueballs and meson states. We consider here the singlet scalar spectrum, whereas the pseudoscalar spectrum will be studied in a future publication. Note that the ``dilaton'', the alleged Goldstone mode due to the almost unbroken conformal symmetry~\cite{walk2}, should appear in the singlet scalar spectrum as $x \to x_c$ if such a state exists.
The possibility of a light dilaton, in QCD and similar theories with a walking regime, has been studied extensively both by field theoretical methods~\cite{dilaton2} and more recently by using holography~\cite{dilaton}, with various results.

In Fig.~\ref{fss1}, we plot the masses of the four lowest singlet scalar states as functions of $x$ for both potentials~I and~II. The behavior is qualitatively similar to the non-singlet $SU(N_f)$ states. In Fig.~\ref{fss2}, the ratios of the masses up to the fourth state are shown. In the limit  $x \rightarrow 0$
the mixing between  mesons and glueballs
vanishes. In this regime  we can identify the states as glueballs or mesons.

 It is clear that there is mixing of states and level crossings at finite $x$. Moreover, it is seen that as $x$ approaches $x_c$ all the ratios of singlet state masses asymptote to finite numbers. There is no single state becoming much lighter than the rest, which could be interpreted as the dilaton.
We will discuss this in more detail below and in Appendix~\ref{app:xtoxc}. Instead, the restoration of the conformal symmetry as $x \to x_c$ is reflected in all masses falling exponentially to zero, as specified by Miransky scaling. Therefore, {\it the conformal symmetry in the walking region is NOT spontaneously broken}.

\begin{figure*}[!tb]
\begin{center}
\includegraphics[width=0.49\textwidth]{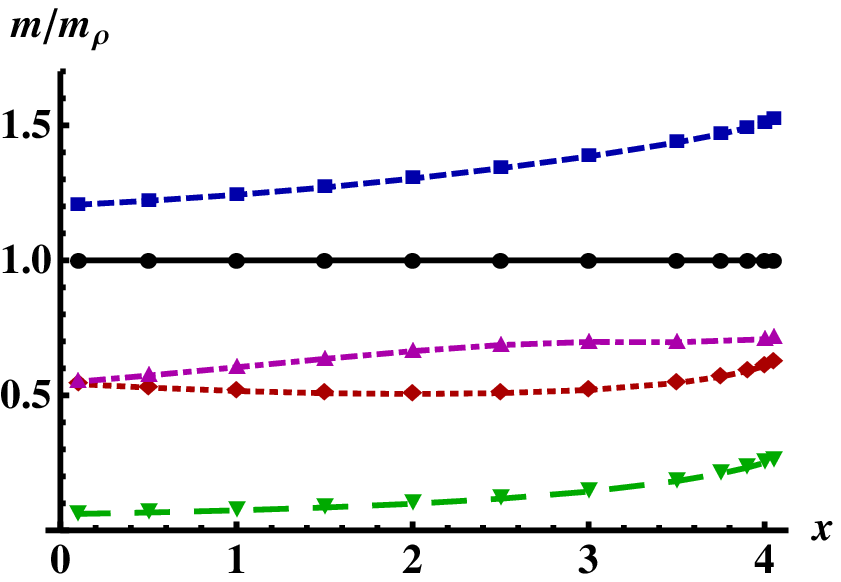}\hfill
\includegraphics[width=0.49\textwidth]{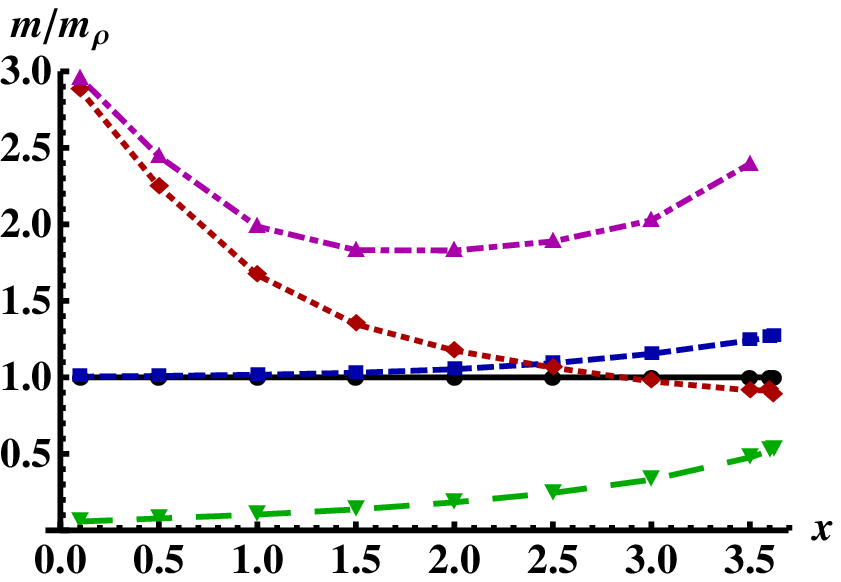}
\end{center}
\caption{
The masses of the lightest states of various towers, and $f_{\pi}/\sqrt{N_fN_c}$ as a function of $x$ in units of the $\rho$ mass. Left: potentials~I with $W_0=3/11$. Right: potentials~II with SB normalized $W_0$. Solid black, dashed blue, dotted red,  and dotdashed magenta curves show the masses of the lightest vector, axial, flavor non-singlet scalar, and flavor singlet scalar states, respectively, while the long-dashed green curve is $f_{\pi}/\sqrt{N_fN_c}$.}
\label{fratios}\end{figure*}

\subsection{Behavior as $x \to \xcfb$ and the dilaton state} \label{sec:xtoxclimit}

We verify that all mass ratios tend to constants as $x \to x_c$ from below for both potentials~I and~II in Fig.~\ref{fratios}. In fact, the same applies to other dimensionful quantities such as decay constants, as can be checked  for $f_\pi$ (the long-dashed green curves) in Fig.~\ref{fratios}.
An analytical explanation~\cite{alho} of these results is sketched in Appendix~\ref{app:xtoxc}. The main point is as follows. As $x \to x_c$ from below, the model develops an approximate fixed point as we have discussed in Sec.~\ref{sec:bg}. The background can be divided into an ``IR piece'', which describes the RG flow from the singularity at $\l=\infty$ to the fixed point at $\l=\l_*$, and an ``UV piece'', which describes the flow from $\l=\l_*$ to $\l=0$.
It is then possible to show that masses and decay constants only depend on the IR piece of the background. This piece is characterized by a single energy scale, and therefore these observables tend to constants when expressed in units of this scale as we take the limit $x \to x_c$.

The scalar sector shows interesting behavior as $x \to x_c$, related to the possibility of having a dilaton in the spectrum, which is also explained in more detail in Appendix~\ref{app:xtoxc}. As it turns out, the Schr\"odinger potential of the flavor non-singlet scalars has a negative dip in the ``walking'' regime, i.e., when $\l \simeq \l_*$ and $x_c-x\ll 1$. The potential has approximately the critical behavior~\cite{son} in this region:
\be \label{eq1o4}
 V_S(r) \sim -\frac{1}{4r^2} \,.
\ee
It can be checked that the singlet scalar fluctuations are similarly almost critical in the vicinity of the fixed point, but in this case a Schr\"odinger potential cannot be easily defined due to mixing of the glue and flavor fluctuations.
The  dip
has therefore just the critical width, so that it is not clear without doing the numerics if there is an unstable state or possibly a light dilaton state.
As we have seen above, numerical analysis  proves that the dilaton (as well as tachyonic states) are absent from the spectrum in the end. As seen from Fig.~\eqref{fratios}, the lightest flavor singlet meson is heavier than its non-singlet counterpart, and it is also heavier than the spin-one states in the case of potentials~II.

 We now discuss the behavior of the $\beta$-function near the fixed point as $x \to x_c$, which is usually understood to be intimately tied to the existence of the dilaton. Indeed the simplest estimates state that the squared dilaton mass should be proportional to the minimum of (the absolute value of) the $\beta$-function at the approximate fixed point.
Our model provides an explicit prediction for the ``walking'' RG flow of the coupling in this regime. The result is obtained by using the expressions for the behavior of $\l(r)$ near the fixed point given in Appendices~D.2 and~E.3 of~\cite{jk}. It reads
\be \label{laflow}
 \l \simeq \l_* - c_1 \left(r \Lambda_\mathrm{UV}\right)^{-\delta} + c_2(\log r) \left(r \Lambda_\mathrm{IR}\right)^4\,,\qquad (\Lambda_\mathrm{UV}^{-1} \ll r \ll \Lambda_\mathrm{IR}^{-1})\,,
\ee
where $c_1$ is a positive $\morder{1}$ constant which can be computed numerically, $c_2(\log r)$ is a positive, regular $\morder{1}$ function, and\footnote{It is understood that the definition~\eqref{deltadef} should be evaluated by using the subdominant solution where the tachyon vanishes for all $r$, and therefore the fixed point is reached.  For the potentials used here $\delta$ is close to the value of the derivative of the two-loop $\beta$-function of QCD in the Veneziano limit.}
\be \label{deltadef}
 \delta = \lim_{\l\to \l_*} \frac{1}{\l-\l_*} \frac{\l'(r)}{A'(r)} > 0
\ee
is the derivative of the $\beta$-function at the fixed point.
Notice that there is a simple interpretation for the terms in~\eqref{laflow}: the second term is the flow induced by the ``perturbative'' (zero-tachyon) $\beta$-function with a fixed point, whereas the last term is the perturbation due to non-zero tachyon which will eventually drive the coupling away from the fixed point.

We make some observations:
\begin{itemize}
 \item The result~\eqref{laflow} is, at the qualitative level, independent of the details of the Lagrangian.
 \item The coupling only depends on $x_c-x$ through the scales $\Lambda_\mathrm{UV}$ and $\Lambda_\mathrm{IR}$ near the fixed point, which are related by Miransky scaling of~\eqref{scalescal}.
 \item The (absolute value of the) $\beta$-function reaches its minimum at the intermediate point $r\Lambda_\mathrm{UV} \sim (\Lambda_\mathrm{UV}/\Lambda_\mathrm{IR})^{4/(4+\delta)}$. The minimum is given by
\be
 -\frac{d\la}{dA} \sim \left(\frac{\Lambda_\mathrm{IR}}{\Lambda_\mathrm{UV}}\right)^\frac{4\delta}{4+\delta} \sim \exp\left[-\frac{4\delta\hat K}{(4+\delta)\sqrt{x_c-x}}\right] \,,
\ee
and is therefore suppressed by Miransky scaling.
\end{itemize}

To conclude,
the value of the minimum of the $\beta$-function suggests that if there was an anomalously light state in the model, its mass should be suppressed by Miransky scaling with respect to the other masses as $x \to x_c$. There is no reason to expect a state whose mass would be suppressed, e.g., by a power law.
However, the numerical study of the spectrum shows that there is no anomalously light state at all.

\subsection{(In)stability of the Efimov solutions}
\label{subsec:efimov}

Recall from Sec.~\ref{sec:bg} that when $0<x<x_c$ and the quark mass is zero, there can be
subdominant vacuum solutions, indexed by the number $n$ of zeros of the tachyon solution. If the effective potential $V_\mathrm{eff}=V_g-xV_{f0}$ has an extremum, signaling the presence of a fixed point, and the BF bound is violated at the fixed point, there is an infinite number of Efimov vacua.
This is the case for potentials~II in the whole range $0<x<x_c$, and for potentials~I (with $W_0=3/11$) within the interval $x_*<x<x_c$, where $x_*\simeq 1.0005$,~\cite{alho}. When there is no fixed point or the BF bound is satisfied, there may still be a finite number of Efimov-like vacua.

By extending the above argument of the scalar sector to these vacua, it is possible to show that they are, quite in general, unstable as seen in a similar case already in \cite{kutasov}.

We first discuss what happens in the non-singlet scalar tower as $x \to x_c$.
The length of the walking regime is longer for the Efimov solutions than for the standard one, as seen from~\eqref{scalescalEf2}. If we assume a small scalar mass $m \sim \Lambda_\mathrm{IR}$, the fluctuation wave function has $n$ nodes in the walking region located approximately at (see the results~\eqref{Vschsc} and~\eqref{scalflu} in Appendix~\ref{app:xtoxc})
\be
  r \sim \Lambda_\mathrm{IR}^{-1} \exp\left(-\frac{\hat K l}{\sqrt{x_c-x}}\right)\,,\qquad l=1,2,\ldots,n\,.
\ee
This signals the presence of $n$ tachyonic scalar states. Their masses can be estimated by computing the value of the Schr\"odinger potential at the nodes, which gives
\be
 m_l^2 \sim - \Lambda_\mathrm{IR}^{2} \exp\left(\frac{2 \hat K l}{\sqrt{x_c-x}}\right)\,,\qquad l=1,2,\ldots,n\,.
\ee
The singlet scalar sector can also be checked to have similar states. The Efimov vacua are therefore highly unstable in the sense that the scale of the masses of these tachyonic modes is exponentially enhanced with respect to the mass scale of the spectra of the standard ($n=0$) background solutions, which is $\sim \Lambda_\mathrm{IR}$.

When $x_c-x$ is not small, and there is a fixed point where the BF bound is violated so that a full Efimov tower is present, we can argue analytically that the vacua with high enough $n$ are unstable. Given any fixed $x$ between zero and $x_c$,~\eqref{scalescalEf1} implies that the coupling walks for high enough $n$,
and
further that the walking regime can be made long enough for instabilities to appear by increasing $n$.

For generic values of $x$ and for finite $n$, the stability of these saddle points can be checked numerically. For the potentials
used here
already the first Efimov vacuum is always unstable.  Apart from the tachyonic modes, the Efimov vacua also admit a spectrum of modes similar to that of the standard vacuum, i.e., there are states at positive squared mass $m^2 \sim \Lambda_\mathrm{IR}^2$.
The unstable mode with the smallest $m^2$ is typically found in the non-singlet sector.

\section{Two-point functions and the S-parameter} \label{sec:Sparam}
In this section we will study the non-singlet vector, axial vector, and scalar two-point functions. This will allow us to compute the pion decay constant, and especially the S-parameter. Of special interest is the study of the S-parameter in the walking region $x\to \xcfb$. We will find it to be discontinuous at $x=x_c$, which will prompt us to study the behavior of the correlators for $x\to \xcfb$.



\subsection{The S-parameter and the decay constants}

The normalization of the decay constants of meson states and the S-parameter can be fixed by matching the bulk vector and scalar two-point functions to the corresponding quantum field theory correlators in the limit of large Euclidean momentum \cite{Erlich:2005qh}, \cite{ikp}.
 We define\footnote{The factors of $N_f/2$ are necessary in order to have similar normalization of $\Pi_V$ and $\Pi_S$ in the flavor singlet and non-singlet sectors.}
\begin{align}
 \langle J_{\mu}^{a\,(V)}(q) J_{\n}^{b\,(V)}(p) \rangle &=\Pi^{ab}_{\mu \nu}(q,p)=-(2 \pi)^4 \delta^4 (p+q) \frac{2\delta^{ab}}{N_f}
\left( q^2 \eta_{\m\n}-{q_{\m} q_{\n}}\right)\Pi_V (q) &\\
 \langle J^{a\,(S)}(q) J^{b\,(S)}(p) \rangle &=\Pi^{ab}_{\phantom{\mu \nu}}(q,p) = (2 \pi)^4 \delta^4 (p+q) \frac{2\delta^{ab}}{N_f} \Pi_S (q)\, .
\end{align}
We calculate the UV limit of $\Pi_V (q)$ and $\Pi_S(q)$,
and match to the quantum field theory result in Appendix \ref{app:2ptf}. The matching results in the following expressions for the normalization constants of the bulk action:
\be
M^3 N_c N_f W_0 \, \gf_0^2 \ell={N_c N_f \over 6 \pi^2} \,,\qquad
M^3 N_c N_f W_0 \ell^5 \h_0= {N_c N_f \over 4 \pi^2} \, ,
\label{macond}
\ee
from which we conclude that ${\ell^4 \h_0 /\gf_0^2}={3/2}$.

\begin{figure}[!tb]
\begin{center}
\includegraphics[width=0.49\textwidth]{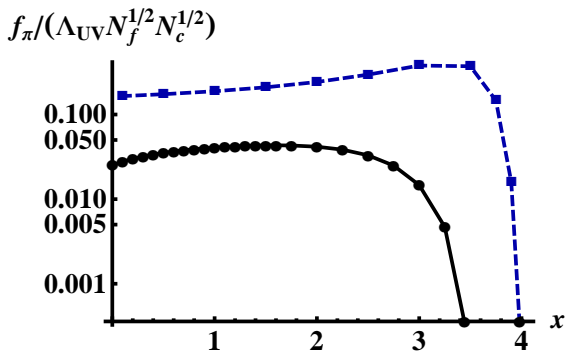}
\end{center}
\caption{ The pion decay constant $f_{\pi}$ as a function of $x$ in units of $\Lambda_{\rm UV}$. It vanishes near $x_c$ following again Miransky scaling. The dashed blue curve is the result for potentials~I with $W_0=3/11$, while the continuous black curve is for potentials~II with SB normalized $W_0$.}
\label{fpif}\end{figure}

Using the expansions
\be
\Pi_A = {f_{\pi}^2 \over q^2} +\sum_{n} {f_n^2 \over q^2 +m_n^2 - i \epsilon}\,,\qquad
\Pi_V=\sum_{n} {F_n^2 \over q^2 +M_n^2 - i \epsilon}\,,
\label{2}\ee
we can determine $f_{\pi}$ in terms of the bulk axial vector wavefunction
\be
f_{\pi}^2= -{N_{c} N_{f} \over 12 \pi^2 } \left. {\partial_{r} \psi^A
    \over r} \right|_{r=0,\, q=0} \,,
\ee
where the normalization was fixed by using (\ref{macond}) and we required $\psi^A(r=0)=1$. The pion decay constant is portrayed in figure \ref{fpif} for both potentials~I and~II. It has qualitatively similar $x$-dependence to the meson masses: it depends only weakly on $x$ for small $x$ and  decreases exponentially to zero close to $x_c$.

The S-parameter is defined as
\begin{align} \label{Sdef}
S=4 \pi {d \over dq^2}\left[q^2 (\Pi_V - \Pi_A)\right]_{q=0} &=-{N_c N_f \over 3 \pi} {d \over dq^2}\left. \left( {\partial_r \psi^V (r) \over
    r}-{\partial_r \psi^A (r) \over r} \right) \right|_{r=0,\, q=0}
 \\\nonumber
&=4\pi\sum_{n}\left({F_n^2\over M_n^2}-{f_n^2\over m_n^2}\right) \, .
\end{align}

\begin{figure}[!tb]
\begin{center}
\includegraphics[width=0.49\textwidth]{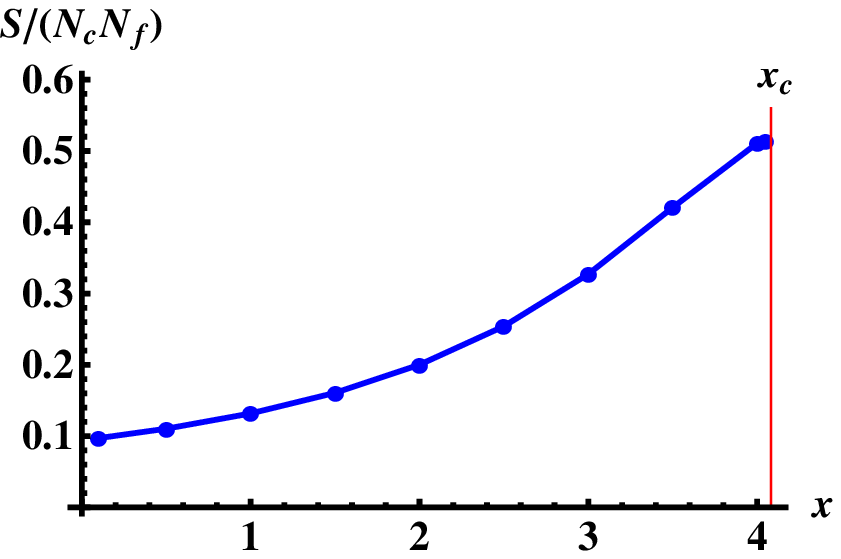}\hfill
\includegraphics[width=0.49\textwidth]{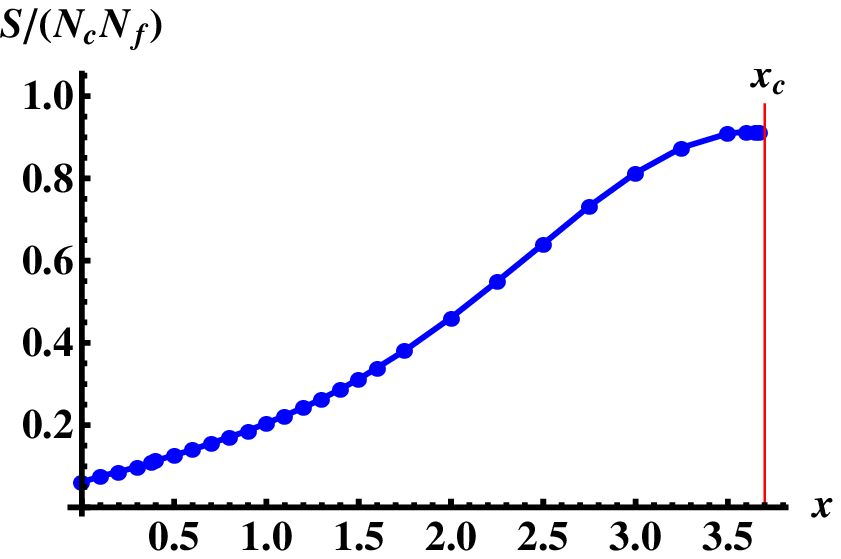}
\end{center}
\caption{ Left: The S-parameter as a function of $x$ for potential class I with $W_0={3/11}$.
  Right: The S-parameter as a function of $x$ for potential class II with SB normalization for $W_0$.
In both cases $S$ asymptotes to a finite value as $x\to x_c$.
}
\label{sparf}
\end{figure}

As both masses and decay constants in (\ref{2}) and (\ref{Sdef}) are affected similarly by Miransky scaling, the S-parameter is {\it invariant} under Miransky scaling.
Therefore its value in the limit $x \to x_c$ cannot be predicted by Miransky scaling alone.
Our numerical results show that generically the S-parameter (in units of $N_f N_c$) remains finite in the QCD regime, $0<x<x_c$ and asymptotes to a finite constant at
$x_c$, as seen in figure \ref{sparf}. The S-parameter is identically zero inside the conformal window (massless quarks) because of unbroken chiral symmetry. This suggests a subtle discontinuity of correlators across the conformal transition, which we shall discuss in more detail below.

\begin{figure}[!tb]
\begin{center}
\includegraphics[width=0.49\textwidth]{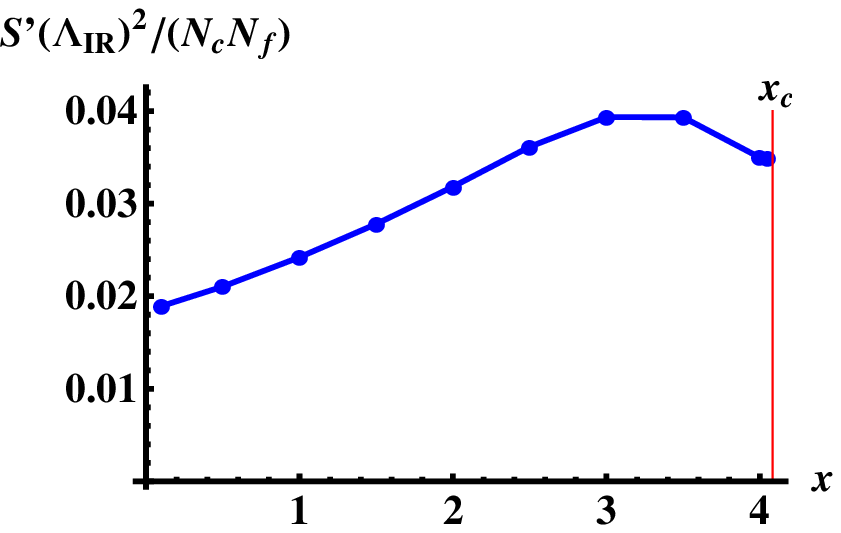}\hfill
\includegraphics[width=0.49\textwidth]{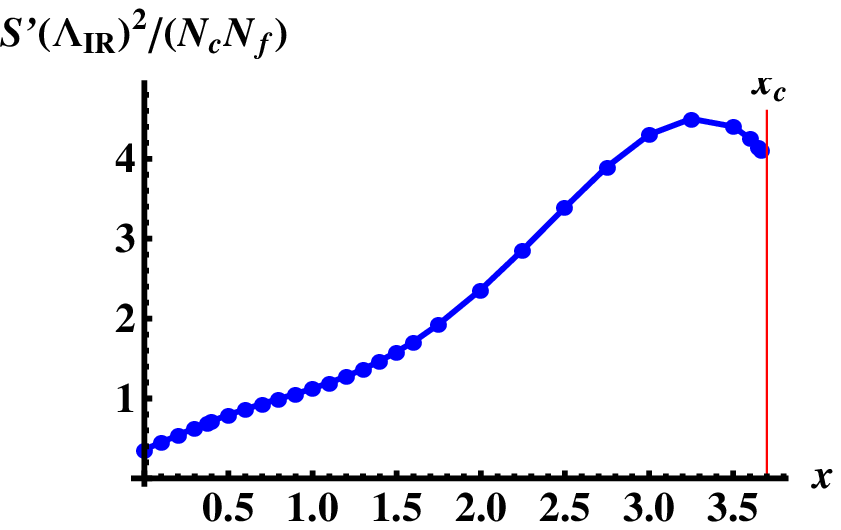}
\end{center}
\caption{ Left: The higher order coefficient $S'$ in units of $\Lambda_\mathrm{IR}$ as a function of $x$ for potential class I with $W_0={3/11}$.
  Right: The same parameter as a function of $x$ for potential class II with SB normalization for $W_0$.
}
\label{spparf}
\end{figure}

The discontinuous behavior of $S$ at the conformal phase transition is in agreement with estimates based on Dyson-Schwinger equations~\cite{SinDS} and the analysis of the BZ limit in field theory \cite{sannino}.
However, according to most estimates~\cite{walkingS} the S-parameter should be reduced in the walking regime, and therefore decrease with increasing $x$~\cite{SinDS,sannino}.
Our most important result is that generically the S-parameter is an increasing function of $x$, reaching it highest value at $x_c$, contrary to previous expectations. We have also found choices of potentials where the S-parameter becomes very large as we approach $x_c$.

We also define a higher order coefficient (related to the X-parameter of~\cite{Barbieri:2004qk}) as
\be \label{Spdef}
 S' \equiv -2\pi\frac{d^2}{(dq^2)^2} \left[q^2\left(\Pi_V(q^2)-\Pi_A(q^2)\right)\right]_{q=0}
\ee
so that
\be
 q^2\left(\Pi_A(q^2)-\Pi_V(q^2)\right) = f_\pi^2 - \frac{S}{4\pi} q^2 + \frac{S'}{4\pi}q^4 + \cdots\,.
\ee
This parameter is shown for both potentials~I and~II in Fig.~\ref{spparf}. The $x$ dependence (in IR units) is qualitatively rather similar to the S-parameter so that the values typically increase with $x$, and approach fixed values as $x \to x_c$. However, unlike for the S-parameter, there is also a region with decreasing values near $x=x_c$.

\subsection{Discontinuity at $x=x_c$} \label{sec:Sparamxxc}

In order to understand the $x$ dependence of the S-parameter and in particular the discontinuous behavior at $x=x_c$, it is useful to analyze the difference $q^2\left(\Pi_A(q^2)-\Pi_V(q^2)\right)$ of the vector-vector and axial-axial correlators at large $q^2$. We present here only rough estimates while a more precise analysis is done in Appendix~\ref{app:Pidiff}.

To start with, the equations~\eqref{vecaxeommain} for the vector and axial wave functions can be roughly approximated for $r\ll \Lambda_{IR}^{-1}$ as
\be \label{psiAVeqs}
 r\partial_r \left( r^{-1} \psi_V'(r)\right) - q^2 \psi_V(r) \simeq 0\,,\qquad  r\partial_r \left( r^{-1} \psi_A'(r)\right) - q^2 \psi_A(r)- H(r) \psi_A(r)\simeq 0\, ,
\ee
where
\be \label{Hdef}
 H(r) = H_A(r) = \frac{4 \t^2(r) e^{2 A(r)} \h(\l,\t)  }{\gf(\l,\t)^2}\,.
\ee
These equations are the leading order terms at fixed points, but give a reasonable estimate also for the behavior between the walking regime and the UV fixed point when $x_c-x$ is small. By using the standard relation between the correlators and bulk wave functions, the difference of the correlators becomes
\be \label{corrdiff}
 q^2\left(\Pi_A(q^2)-\Pi_V(q^2)\right) \propto - \lim_{r\to 0} \frac{\phi(r)}{r}\,,
\ee
where  $\phi = (\psi_A-\psi_V)/\psi_A$ and the proportionality constant is independent of $q$. The equations~\eqref{psiAVeqs} imply
\be \label{feq}
 r\partial_r \left( r^{-1} \phi'(r)\right)  + 2 \partial_r\left( \log \psi_A(r)\right) \phi'(r)- \left(1-\phi(r)\right)H(r) \simeq 0 \,.
\ee

 We then consider large momentum, $q \gg \Lambda_\mathrm{IR}$. As it turns out, we can approximate $1-\phi(r) \simeq 1$ in the last term for the region
of interest to us.  The relevant solutions of~\eqref{feq} are
\begin{align} \label{phiappsol}
 \phi(r) &\simeq b\, r^2 \,,& &(r \ll q^{-1})\,,\nonumber\\
 \phi(r) &\simeq -\frac{1}{2q} \int_0^r dr' H(r')\,,&(q^{-1}\ll\ &r \ll \Lambda_\mathrm{IR}^{-1})\, ,
\end{align}
where we used the UV boundary conditions to rule out the non-normalizable solution, and the fact that $\partial_r \log \psi_A(r)\simeq -q$ for $q^{-1}\ll r \ll \Lambda_\mathrm{IR}^{-1}$. Using~\eqref{corrdiff} and requiring continuity of~\eqref{phiappsol} at $r \sim q^{-1}$ gives
\be \label{corrdiff2}
  q^2\left(\Pi_A(q^2)-\Pi_V(q^2)\right) \sim -b \sim q \int_0^{1/q} dr H(r) \sim q \int_0^{1/q} dr \frac{\t(r)^2}{\ell^2r^2}\,,
\ee
where we dropped the slowly varying potential functions of~\eqref{Hdef} at the last step.

Finally, we write down the result~\eqref{corrdiff2} explicitly for zero quark mass in two regions of $x$.
\begin{enumerate}
 \item In the region having dynamics similar to ordinary QCD ($x<x_c$ but $x_c-x = \morder{1}$), there is only one scale $\Lambda_\mathrm{UV} \sim \Lambda_\mathrm{IR}$. The tachyon behaves as
\be
 \frac{\t(r)}{\ell} \sim \sigma r^3 \sim \Lambda_\mathrm{UV}^3 r^3
\ee
up to logarithmic corrections.
Therefore,  we find that
\be \label{corrdiffruns}
   \Pi_A(q^2)-\Pi_V(q^2) \sim \frac{\Lambda_\mathrm{UV}^6}{q^6}\,,\qquad (q \gg \Lambda_\mathrm{UV})\,.
\ee
This result agrees with earlier computations in the probe approximation, \cite{Son:2010vc}.

\item If $x_c-x \ll 1$, we have walking behavior and the scales $\Lambda_\mathrm{UV}$, $\Lambda_\mathrm{IR}$ are well separated. The tachyon behaves roughly as (see Section~\ref{sec:bg}, Appendix~\ref{app:bgUVIR} and~\cite{jk})
\begin{align}
 \frac{\t(r)}{\ell} &\sim \sigma r^3 \sim \Lambda_\mathrm{UV} \Lambda_\mathrm{IR}^2 r^3\,,& &(r\ll \Lambda_\mathrm{UV}^{-1})\,,\nonumber\\
 \frac{\t(r)}{\ell} &\sim \frac{\sigma}{\Lambda_\mathrm{UV}}  r^2 \sim \Lambda_\mathrm{IR}^2 r^2\,,& (\Lambda_\mathrm{UV}^{-1}\ll\ &r\ll \Lambda_\mathrm{IR}^{-1})\,.
\end{align}
Inserting these in~\eqref{corrdiff2}, we obtain
\begin{align} \label{corrdiffwalks}
  \Pi_A(q^2)-\Pi_V(q^2) &\sim \frac{\Lambda_\mathrm{UV}^2 \Lambda_\mathrm{IR}^4}{q^6}\,,&  &\!(\Lambda_\mathrm{UV} \ll q) \,,\nonumber\\
  \Pi_A(q^2)-\Pi_V(q^2) &\sim \frac{\Lambda_\mathrm{IR}^4}{q^4}\,,& (\Lambda_\mathrm{IR} \ll\ &q \ll \Lambda_\mathrm{UV})
\end{align}
when $x$ is close to $x_c$.
\end{enumerate}

\begin{figure}[!tb]
\begin{center}
\includegraphics[width=0.49\textwidth]{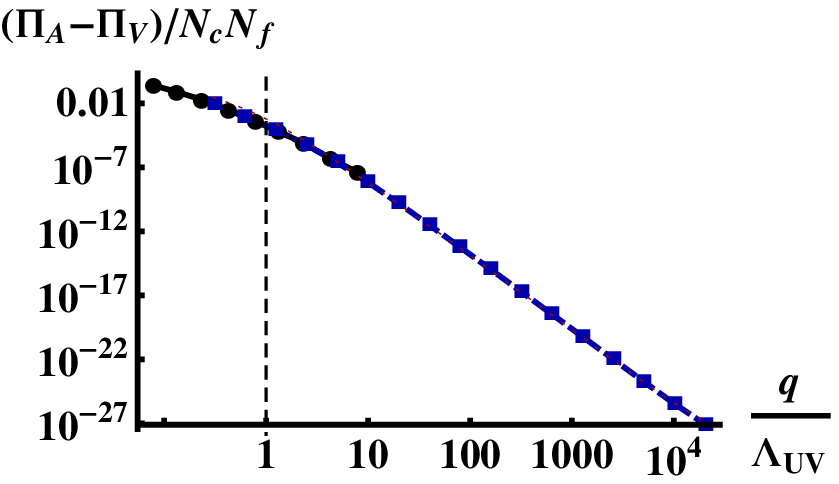}\hfill
\includegraphics[width=0.49\textwidth]{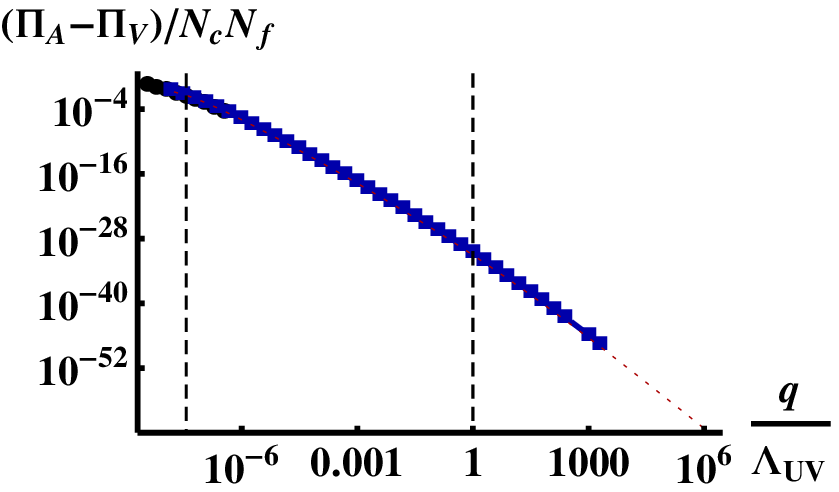}

\vspace{5mm}

\includegraphics[width=0.49\textwidth]{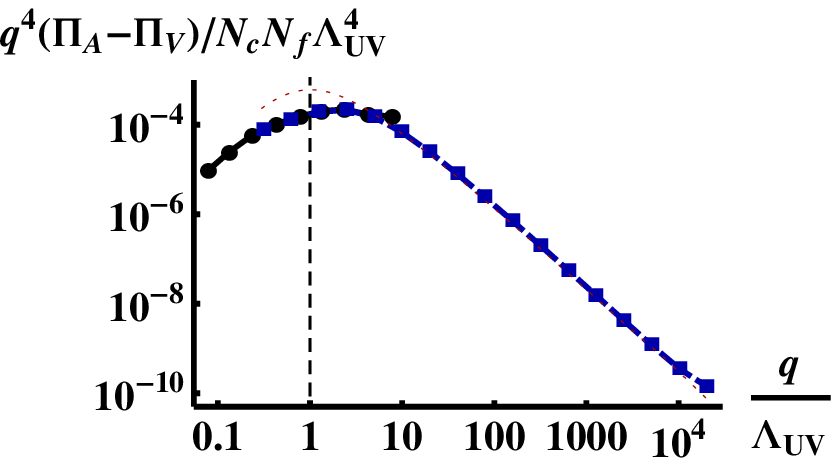}\hfill
\includegraphics[width=0.49\textwidth]{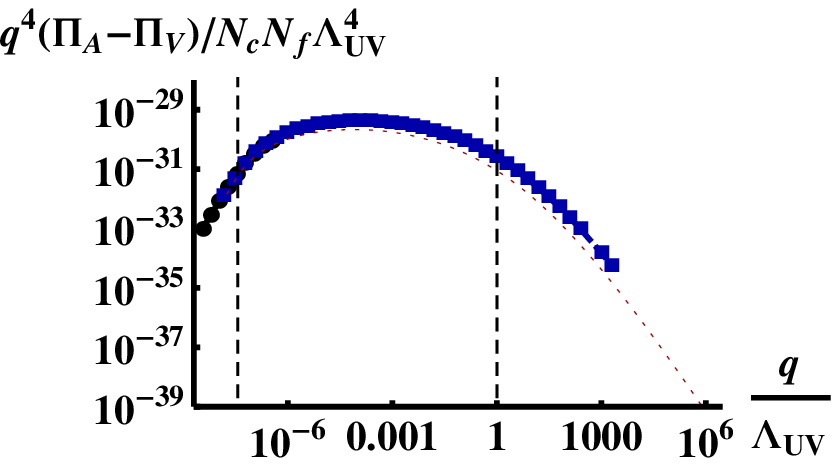}
\end{center}
\caption{The $q$-dependence of the difference $\Pi_A-\Pi_V$ for potentials II with SB normalization of $W_0$ (so that $x_c\simeq 3.7001$). The various curves show different estimates for $\Pi_A-\Pi_V$, see text for details. The vertical dashed lines show values of $q$ where qualitative changes in behavior are expected. Left column: a case with dynamics similar to ordinary QCD ($x=1$). Right column: a case with walking dynamics ($x=3.65$). Top row: $\Pi_A-\Pi_V$ normalized to $N_f N_c$ as a function of $q/\Lambda_\mathrm{UV}$. Bottom row: $q^4(\Pi_A-\Pi_V)$, in well-chosen units, as a function of $q/\Lambda_\mathrm{UV}$. Multiplication by $q^4$ makes the details of the $q$-dependence better visible.  
}
\label{pidifffig}
\end{figure}

Notice that the results~\eqref{corrdiffruns} and~\eqref{corrdiffwalks} are rough estimates of~\eqref{corrdiff2}, because the logarithmic RG flow of the chiral condensate, which is included in $\t(r)$, was neglected. However the integral in~\eqref{corrdiff2} is dominated by the region in the vicinity of its upper endpoint, and therefore we can write down a more generic estimate for the $q$-dependence.
Recall that we matched the RG flow of the chiral condensate to QCD such that (see Appendix~\ref{app:bgUVIR})
\be
 -\frac{d\t}{dA} \simeq \frac{d\t}{d\log r} \simeq 3 -\gamma
\ee
in the UV, where the anomalous dimension $\gamma = d\log m_q/d\log q$ in QCD. Therefore
\be
   \Pi_A(q^2)-\Pi_V(q^2) \sim \frac{1}{q} \int^{1/q}\frac{\t(r)^2}{r^2}\sim \frac{1}{q^{6-2 \gamma}}\,,
\ee
where we approximated the anomalous dimension to be constant. The result agrees with the one from the operator product expansions (see, e.g.,~\cite{SinDS}).

We plot $\Pi_A-\Pi_V$ as a function of the momentum in UV units, $q/\Lambda_\mathrm{UV}$, for Potentials II with SB normalized $W_0$ in Fig.~\ref{pidifffig}. The left hand plots are for ``running'' dynamics ($x=1$), and the right hand plots are for ``walking'' dynamics ($x=3.65$). In the latter case, we chose the value of $x$ very close to the critical one ($x_c=3.70001$), so that the walking regime can be clearly distinguished ($\Lambda_\mathrm{UV}/\Lambda_\mathrm{IR} \sim 10^7$). On the top row we plot  the ratio $(\Pi_A-\Pi_V)/N_fN_c$ which is finite in the Veneziano limit. On the bottom row we plot the same quantity multiplied by the factor $q^4/\Lambda_\mathrm{UV}^4$.
The various curves were computed as follows. The black continuous curve is the exact numerical result, which can be reliably extracted from~\eqref{corrdiff} for small values of $q$. The blue dashed curve is given by the estimate~\eqref{Pidiffexact} in Appendix~\ref{app:Pidiff}. The thin red dashed curve (which is best visible for high $q$ in the right hand plot) is given by the rough estimate~\eqref{corrdiff2}.

The $q$-dependence follows roughly the results~\eqref{corrdiffruns} and~\eqref{corrdiffwalks}. On the top-left plot, one can see kink at $q\sim \Lambda_\mathrm{UV}$ as the $q$-dependence changes from the low $q$ one ($\sim f_\pi/q^2$) to the UV one ($\sim 1/q^6$). The expected location of the kink, at $q \sim \Lambda_\mathrm{UV}$ is marked with the vertical dashed line. The change in behavior is more clearly visible in the bottom-left plot, where we multiplied $\Pi_A-\Pi_V$ by an additional factor of $q^4/\Lambda_\mathrm{UV}^4$ to suppress the overall strong power dependence on $q$. 
On the right hand plots there is, in addition, the regime reflecting the walking dynamics between the two vertical dashed lines at $q\sim \Lambda_\mathrm{IR}$ and at $q\sim \Lambda_\mathrm{UV}$, where the difference of correlators behaves roughly as $\sim 1/q^4$. There are now two (rather smooth) kinks at $q \sim \Lambda_\mathrm{IR} \sim 10^{-7} \Lambda_\mathrm{UV}$ and at $q\sim \Lambda_\mathrm{UV}$, best visible in the bottom-right plot.

We now  comment on the connection of the result~\eqref{corrdiffwalks} to Weinberg's sum rules. We first recall how the sum rules are derived. By applying the spectral decomposition~\eqref{2} to $\Pi_A-\Pi_V$, we see that the first two terms of the expansion at $q^2=\infty$ behave as $1/q^2$ and $1/q^4$. However, as seen from~\eqref{corrdiffruns} and~\eqref{corrdiffwalks},  $\Pi_A-\Pi_V$ vanishes as $1/q^6$ in the UV.
Therefore, the coefficients of the $\morder{q^{-2}}$ and $\morder{q^{-4}}$ terms in the series expansion have to vanish, which results in the sum rules. We notice that our result~\eqref{corrdiffwalks} is consistent with the analysis of~\cite{as}: the difference in the intermediate range of energies vanishes only as $1/q^4$, and the contribution from this range to the second sum rule may be viewed as a modification to it.

We then consider the limit $x \to \xcfb$ in order to analyze the discontinuous behavior of the S-parameter observed above. First, we notice that due to Miransky scaling in~\eqref{scalescal} the high momentum tail in~\eqref{corrdiffwalks} is exponentially suppressed. We can write down a similar limit as in~\eqref{lambdalims}:
\be \label{corrUVlim}
  \Pi_A(q^2)-\Pi_V(q^2) \to 0\qquad \textrm{as}\quad x \to x_c\quad \textrm{with}\quad  \frac{q}{\Lambda_\mathrm{UV}}\quad \textrm{fixed}\,.
\ee
The convergence is pointwise in $q/\Lambda_\mathrm{UV}$ but fast when $q/\Lambda_\mathrm{UV} = \morder{1}$: it obeys (quartic) Miransky scaling. This can also be seen from the top-right plot in Fig.~\ref{pidifffig} where the value of $\Pi_A-\Pi_V$ is highly suppressed at $q\sim \Lambda_\mathrm{UV}$. The result reflects the behavior of the background which was discussed in Sec.~\ref{sec:xtoxclimit} and in Appendix~\ref{app:xtoxc}.
For $q \sim \Lambda_\mathrm{UV}$ the correlators only depend on the UV piece of the background solution as $x \to x_c$. The UV piece approaches smoothly the background at $x=x_c$ which has an IR fixed point and is chirally symmetric (zero tachyon) so that $\Pi_A-\Pi_V=0$. This behavior is reflected in the vanishing of the difference of correlators in the limit of~\eqref{corrUVlim}.

Second, at low momenta ($q \sim \Lambda_\mathrm{IR}$) the difference of correlators is $\mathcal{O}(1)$, as can be seen from the top-row plots in Fig.~\ref{pidifffig}, and has nontrivial structure. If we take the limit~\eqref{corrUVlim} keeping $q/\Lambda_\mathrm{IR}$ fixed instead, the result is nonzero. This is consistent with the fact that $f_\pi/ \sqrt{N_f N_c}$, $S/(N_f N_c)$ and $S'/(N_f N_c)$ approach finite values when measured in units of $\Lambda_\mathrm{IR}$  as $x \to x_c$ from below (as seen from Figs.~\ref{fpif},~\ref{sparf} and~\ref{spparf}).
The situation is similar to what was found for the decay constants in Appendix~\ref{app:xtoxc}: for $q \sim \Lambda_\mathrm{IR}$ the difference of the correlator only depends on the IR piece of the background, up to a calculable $\mathcal{O}(1)$ correction factor. Therefore, we understand that the nonvanishing value of the S-parameter as $x \to x_c$ arises from the IR piece of the background, which includes a nonvanishing tachyon, and is absent for the solutions with $x_c \le x <11/2$.

In summary, the naive expectation that $\Pi_A-\Pi_V$ should vanish as $x \to \xcfb$ due to restoration of chiral symmetry in the conformal window, holds in the sense of~\eqref{corrUVlim}. The vanishing of  $\Pi_A-\Pi_V$ is not in contradiction with the finite value of the S-parameter in this limit, because the convergence to zero in~\eqref{corrUVlim} is not uniform in  $q/\Lambda_\mathrm{UV}$.
The finite value of the S-parameter reflects the structure of $\Pi_A-\Pi_V$ at values of $q$ which are suppressed with respect to $\Lambda_\mathrm{UV}$ by the Miransky scaling factor as $x \to \xcfb$.

 \addcontentsline{toc}{section}{Acknowledgments}
\acknowledgments

We would like to thank D.~D.~Dietrich, N.~Evans, K.~Kajantie, and K.~Tuominen  for helpful conversations and correspondence.
D.A. would like to thank the Crete Center for Theoretical Physics for hospitality and the FRont Of
pro-Galician Scientists for unconditional support.

This work was supported in part by grants PERG07-GA-2010-268246, PIF-GA-2011-300984, the EU program ``Thales'' and ``HERAKLEITOS II'' ESF/NSRF 2007-2013 and was also co-financed by the European Union (European Social Fund, ESF) and Greek national funds through the Operational Program ``Education and Lifelong Learning'' of the National Strategic Reference Framework (NSRF) under ``Funding of proposals that have received a positive evaluation in the 3rd and 4th Call of ERC Grant Schemes''.

\newpage

\appendix
\renewcommand{\theequation}{\thesection.\arabic{equation}}
\addcontentsline{toc}{section}{Appendices}
\section*{APPENDIX}

\section{Derivation on the quadratic fluctuation equations} \label{app:quadfluctdet}

We shall consider the following fluctuations of the metric, tachyon, dilaton and gauge fields:
\begin{align}
 g_{MN}&=g_{MN}^{(0)}+\hat g_{MN}\,,& \hat g_{MN}\,d\xi^M d\xi^N&=e^{2\Awf}\left(2\phi\,dr^2
+2\hat A_\mu\,dr\,dx^\mu+h_{\mu\nu}\,dx^\mu\,dx^\nu\right)\,,\nonumber \\
\Phi&=\Phi_0+\chi\,, &
T&=(\tau+s+\mathfrak{s}^at^a)e^{i\theta+i\,\pi^at^a}\,.
\label{fluctdef}
\end{align}
where $t^a$ are the generators of $SU(N_f)$. We are mostly interested in the standard vacuum for $0<x<x_c$ which is expected to have a nontrivial spectrum. Therefore, the background solution $\t(r)$ is nonzero and the phases $\theta$, $\pi^a$ in~\eqref{fluctdef} are well defined.

The vacuum with zero tachyon, which is the dominant one for $x_c\le x <11/2$, has continuous spectrum. This is clear as the coupling flows to an IR fixed point and the metric is asymptotically AdS in the IR. The Schr\"odinger potential vanishes as $r \to \infty$ in the non-singlet sector, as we shall demonstrate below.

We use the vector and axial combinations of the left and right gauge
fields
\be
V_M = \frac{A_M^L + A_M^R}{2}\, ,\qquad
A_M = \frac{A_M^L - A_M^R}{2}\,.
\label{VAdefstext}
\ee
The associated field strengths will be $V_{MN}$, $A_{MN}$. We choose the gauge $A_r = V_r = 0$.

The vector field in the bulk is written as
\be
V_\mu (x^\mu, r) =  \psi_V(r)\, {\cal V}_\mu (x^\mu)\,,
\ee
where ${\cal V}_\mu$ is transverse, $\partial^\m {\cal V}_\mu = 0$, and the longitudinal term can be set to zero.
As usual, to look for $4D$ mass eigenstates we insert a plane wave Ansatz ${\cal V}_\mu (x^\mu) = \exp(ip_\mu x^\mu)$ where $p_\mu p^\mu = -m_V^2$, so that
\be
\partial_\nu\partial^\nu\,{\cal V}_\mu = m_V^2\,{\cal V}_\mu\,,
\ee
and analogously for the rest of the fluctuations we study.
For the axial vectors, we first need to separate the transverse and longitudinal parts:
\be
A_\m (x^\mu, r)=A^\bot_\m (x^\mu, r)+ A^{\lVert}_\m(x^\mu, r) \,,
\ee
where $\partial^\n A^\bot_\n (x^\mu, r) = 0$ and the longitudinal term is the divergence of a scalar function.

For the vector modes it is not necessary to treat the flavor non-singlet and singlet terms separately, as the fluctuations modes are the same in both the sectors. However, for the axial vector modes we have to do that. Therefore we write
\begin{align}
  A^{\bot}_\m(x^\mu, r) &= A^{\bot F}_\m(x^\mu, r)+ A^{\bot S}_\m(x^\mu, r) =
\psi_A(r) {\cal A}_\m^a(x^\m)t^a + \varphi_A(r)\, {\cal X}_\m(x^\m) \\
  A^{\lVert}_\m(x^\mu, r) &= A^{\lVert F}_\m(x^\mu, r)+ A^{\lVert S}_\m(x^\mu, r) =
-\psi_L(r) \partial_\m({\cal P}^a(x^\m))t^a - \varphi_L(r)\, \partial_\m({\cal T}(x^\m)) \, .
\label{ansaxlong}
\end{align}
The flavor non-singlet terms from the pseudoscalar and scalar sectors can be expanded similarly:
\begin{align} \label{somedefsnorm}
\pi^a (x^\mu, r)&= 2\,  \psi_P(r)\, {\cal P}^a(x^\mu)\,, \rc
\mathfrak{s}^a(x^\mu,r) &= \psi_S (r) \,{\cal S}^a(x^\mu)\,. 
\end{align}
Notice that the $x^\mu$ dependence of the pion fluctuations $\pi^a$ and the flavored axial vector fluctuations are related. This is required in order to satisfy the fluctuation equations below. The flavor singlet components will be discussed in more detail later. Their analysis is more involved due to mixing terms between the various fields.

We also define a shorthand notation for the omnipresent factor
\be \label{Gdef}
 \G(r) = \sqrt{1 + e^{-2A(r)}\h(\l,\tau) (\partial_r \t(r))^2}
\ee
so that the ``effective'' metric factor, which often appears due to the nonzero tachyon background, reads
\be
 g_{rr} +\h(\l,\tau) (\partial_r \t(r))^2  =  e^{2A(r)} \G(r)^2 \, .
\ee

\subsection{Flavor non-singlet sector} \label{app:SUNfluct}

For the flavor non-singlet fluctuations the fluctuation analysis is rather straightforward as only the DBI action $S_f$ contributes. Therefore we will write down the quadratic actions for the vector, axial vector, pseudoscalar and scalar sectors directly below.

\subsubsection{Vector Mesons}
\label{app:NSvmesons}

For spin-one vector excitations, the spectra in the non-singlet and singlet sectors are identical.
The quadratic action for the vector mesons is
\be
S_V = -  {1\over2}\, M^3 N_c\,  {\mathbb Tr} \int d^4x\, dr
V_f(\l,\t)\, \gf(\l,\t)^2\, \G^{-1}\,e^{\Awf}
\left[ \frac12\, \G^2\, V_{\m\n}V^{\m\n} + \partial_r V_\m \partial_r V^{\m}
\right]\,,
\label{vectoracti}
\ee
where $V_{\m\n}=\partial_\m V_\n-\partial_\n V_\m$, and the trace is over the flavor indices.
The fluctuation equation therefore reads
\be
\frac{1}{V_f(\l,\t)\, \gf(\l,\t)^2\, e^{\Awf}\,\G}
\partial_r \left( V_f(\l,\t)\, \gf(\l,\t)^2 \,e^{\Awf}\,
\G^{-1}\, \partial_r \psi_V \right)
+m_V^2\, \psi_V  = 0 \, .
\label{vectoreom}
\ee

This equation can be transferred to Schr\"odinger form as shown in Appendix~\ref{app:schro}.
The Schr\"odinger functions for the vector meson equation are
$C_3(r)=M(r)=0$ and
\begin{equation}
C_1(r)=V_f(\l,\t)\, \gf(\l,\t)^2\, e^{\Awf(r)}\,\G(r)^{-1}\,,\qquad
C_2(r)=V_f(\l,\t)\, \gf(\l,\t)^2\, e^{\Awf(r)}\, \G(r) \,.
\label{ABdefs}
\end{equation}
Further defining
\be \label{XiHV}
 \Xi_V(r) = \left(C_1(r)C_2(r)\right)^{1/4} = \, \gf(\l,\t)\,\sqrt{V_f(\l,\t)\, e^{\Awf(r)}}\,,\qquad H_V(r) = \frac{M(r)}{C_2(r)} = 0\,,
\ee
the Schr\"odinger potential for the flavor non-singlet vectors reads
\be
 V_V(u) = \frac{1}{\Xi_V(u)}\frac{d^2\Xi_V(u)}{du^2} + H_V(u)\,.
\ee
Here the Schr\"odinger coordinate $u$ is defined by
\be \label{udefAppA}
 \frac{du}{dr} = \sqrt{\frac{C_2(r)}{C_1(r)}} = G(r)\,
\ee
and requiring that $u \to 0$ in the UV. $G$ is defined in~\eqref{Gdef}. The definition of the coordinate $u$ will be the same for all non-singlet meson towers, but the potential will change, as we shall see below.

By using the UV expansions of the potentials in~\eqref{VUVexps} and the expansions of the background from Appendix~\ref{app:bgUVIR}, we can compute the UV asymptotics of the potential $V_V$. The potential factors of $\Xi_V$ in~\eqref{XiHV} are almost constants, so that the leading contribution arises from the warp factor $e^{A/2} \sim 1/\sqrt{r}$, giving
\be \label{VVatUV}
 V_V(u)  = \frac{3}{4 u^2}\left[1 + \morder{\frac{1}{\left[\log (u \Lambda)\right]^2}}\right]\,,\qquad (u \to 0)\,.
\ee
At an IR fixed point, i.e., in the conformal window ($x_c\le x<11/2$) or below the conformal window ($0<x<x_c$) for the subdominant vacuum with $\t\equiv 0$, we can derive a similar result, as the potentials are again nearly constants and the metric is close to AdS:
\be \label{VVatIRFP}
 V_V(u)  = \frac{3}{4 u^2}\left[1 + \morder{\frac{1}{u^\delta}}\right]\,,\qquad (u \to \infty)\,,
\ee
as can be checked by using the results of Appendix~E.3 in~\cite{jk} where the constant $\delta$ is also computed. This result also applies when the system is very close to having a fixed point. This is the case for the dominant vacuum (with $\t \ne 0$) in the ``walking'' regime when $x<x_c$ and $x_c-x \ll 1$: Eq.~\eqref{VVatIRFP} holds if  we require $1/\Lambda_\mathrm{UV} \ll u \ll 1/\Lambda_\mathrm{IR}$ instead of $u \to \infty$.

\subsubsection{Axial Vector Mesons}

The action for the non-singlet and singlet sectors of the (transverse) axial vector modes differs by a term
coming from the CP-odd sector. The quadratic action for the $SU(N_f)$ sector of the axial vector meson
excitations reads
\be
\begin{split}
S_A = & - {1\over2}\, M^3 N_c\, {\mathbb Tr} \int d^4x\, dr
V_f(\l,\t)\, e^{\Awf}\, \G^{-1}
\bigg[ {1 \over 2}\, \G^2\, \gf(\l,\t)^2\, A_{\mu\nu}A^{\mu\nu} +
 \\
&+
 \gf(\l,\t)^2\, \partial_r A^{\bot F}_\mu \partial_r A^{\bot F\,\mu}+4\h(\l,\t)\, \tau^2\, e^{2 \Awf}\, \G^2\,
A^{\bot F}_\mu A^{\bot F\,\mu}
\bigg]
\label{axacti}
\end{split}
\ee
where  $A_{\m\n}=\partial_\m A_\n^{\bot F}-\partial_\n A_\m^{\bot F}$.
The fluctuation equation reads
\be
\frac{\partial_r \left( V_f(\l,\t)\, \gf(\l,\t)^2\, e^{\Awf}\,
\G^{-1}\, \partial_r \psi_A \right)}{V_f(\l,\t)\, \gf(\l,\t)^2\, e^{\Awf}\,\G  }
-4 {\t^2\, e^{2 \Awf} \over \gf(\l,\t)^2}\,\h(\l,\t)\, \psi_A  +m_V^2\, \psi_A = 0
\label{axvectoreom}
\ee

The Schr\"odinger functions are otherwise the same
as for vectors
but now
\be \label{MdefA}
M(r)=4 C_2(r)\, {\t(r)^2\, e^{2 \Awf(r)} \over \gf(\l,\t)^2}\,\h(\l,\t)
\ee
 is nonzero. Therefore we find that
\be \label{XiHA}
 \Xi_A(r) = \Xi_V(r)\,,\qquad H_A(r) =  {4 \t(r)^2\, e^{2 \Awf(r)} \over \gf(\l,\t)^2}\,\h(\l,\t)\,,
\ee
and the definition of $u$ is as in~\eqref{udefAppA}. The asymptotic results for the Schr\"odinger potential are the same as for the vectors,~\eqref{VVatUV} and~\eqref{VVatIRFP}, because the additional mass term $H_A$ does not contribute.

\subsubsection{Pseudoscalar Mesons}

The quadratic action reads:
\begin{align}
S\ =&- \frac{1}{2} M^3 N_c\,  {\mathbb Tr} \int d^4 x\, dr\, V_f(\l,\t)\, e^{\Awf}\,\G^{-1}
\Big[ \gf(\l,\t)^2 \, \left(\partial_{r}  A^{\lVert F}_\m\right)^{2}+\nonumber \\
&+\t^2\, e^{2 \Awf}
\,\h(\l,\t) \,\left(\partial_{r}\pi^at^a\right)^{2}+ \t^2\, e^{2 \Awf}\, \G^2\,\h(\l,\t) \,\left(\partial_{\mu}\pi^at^a+2  A^{\lVert F}_\m\right)^2 \Big]\,.
\label{pseudoac}
\end{align}
After substituting in (\ref{somedefsnorm}), the fluctuation equations for $\psi_P (r)$ and $\psi_L (r)$ are
\begin{align}
&{1 \over  V_f(\l,\t)\, e^{\Awf}\,\G\,\h(\l,\t)}
\partial_r  \left( V_f(\l,\t)\, e^{ \Awf }\, \G^{-{1}}\,
  \gf(\l,\t)^2\,\partial_r \psi_L \right)
\nonumber \\
&
-4 \t^2\, e^{2 \Awf}\,  (\psi_L-\psi_P )=0\,,
\label{pseom1}
\\
&4 \t^2\, e^{2 \Awf}\,\h(\l,\t)\, \partial_r \psi_P -m^2\,\gf(\l,\t)^2\, \partial_r \psi_L=0\,.
\label{pseom2}
\end{align}
These two equations can be combined into one by solving (\ref{pseom1}) for $\psi_P$ and inserting this into
(\ref{pseom2}).
\begin{align} \label{PSflucts}
&V_f(\l,\t)\, \t^{2}\, e^{3 \Awf }\, \G^{-{1}}\,\h(\l,\t)\,
\partial_{r} \left[ { 1 \over V_f(\l,\t)\, \t^2\, \h(\l,\t)\, e^{3 \Awf}\,\G}\partial_{r} \hat \psi_{P}\right] \nonumber \\
&-4 \t^2\, e^{2 \Awf}\, {\h(\l,\t)\over \gf(\l,\t)^2}\,\hat \psi_{P}+m^2\, \hat \psi_{P}=0\,,
\end{align}
where we have defined
\begin{equation}
\hat \psi_{P}(r)=-V_f(\l,\t)\, e^{\Awf (r)}\, \G(r)^{-1}\, \gf(\l,\t)^2\,\partial_r \psi_L(r)\,.
\end{equation}

When $m_q=0=m$ all solutions to~\eqref{PSflucts} are normalizable in the UV. Therefore, the IR normalizable solution is the pion mode, for which $\psi_P$ is constant and $\psi_L$ tends to (the same) constant in the UV. The standard calculation, \cite{ckp}, shows that the mass of the pion satisfies the Gell-Mann-Oakes-Renner relation when small $m_q$ is turned on.

The Schr\"odinger functions read
\be
\begin{split}
C_1(r)&=V_f(\l,\t)^{-1}\, \t(r)^{-2}\, e^{-3 \Awf(r)}\, \G(r)^{-1}\,\h(\l,\t)^{-1}
\,,\\
C_2(r)&=V_{f}(\l,\t)^{-1}\, \t(r)^{-2}\, e^{- 3 \Awf(r)}\, \G(r)\,\h(\l,\t)^{-1}\,,
\\
M(r)&= 4 C_2(r)\,\t(r)^2\, e^{2 \Awf(r)}\,\h(\l,\t)\, \gf(\l,\t)^{-2}\,,
\end{split}
\ee
and $C_3(r)=0$. Therefore,
\be \label{XiHP}
 \Xi_P(r) = \frac{1}{\t(r)\sqrt{V_f(\l,\t)\,\h(\l,\t)\, e^{3 \Awf(r)}}}\,,\qquad H_P(r) = \frac{4\t(r)^2\, e^{2 \Awf(r)}\,\h(\l,\t)}{ \gf(\l,\t)^{2}}\,.
\ee

As $\Xi_P$ depends strongly on the tachyon, the UV behavior of the Schr\"odinger potential $V_P$ differs from that of $V_V$ and $V_A$. By using the tachyon asymptotics from Appendix~\ref{app:bgUVIR} we find that
\begin{align}
\label{VPasUVFP1}
 V_P(u) &= \frac{15}{4 u^2}\left[1 + \morder{\frac{1}{\log (u \Lambda)}}\right]\,,& (u &\to 0\,,\quad m_q=0)\,,\\
\label{VPasUVFP2}
 V_P(u) &= -\frac{1}{4 u^2}\left[1 + \morder{\frac{1}{\log (u \Lambda)^2}}\right]\,,& (u &\to 0\,,\quad m_q\ne0)\,.
\end{align}
Notice that the UV behavior of the potential is the critical one when $m_q \ne 0$, \cite{son}. This is linked to the existence of the pion modes: the Schr\"odinger problem can have solutions with arbitrary low mass as $m_q \to 0$ only if the UV asymptotics is critical.\footnote{For exactly zero $m_q$, the potential does not need to have any specific asymptotics since, as we pointed out above, the wave function of the pion mode has free UV boundary conditions.} When $m_q$ is small but finite, \eqref{VPasUVFP2} holds for $u \ll \sqrt{m_q/\sigma}$ while \eqref{VPasUVFP1} holds for $\sqrt{m_q/\sigma} \ll u \ll 1/\Lambda_\mathrm{UV}$. 

The behavior at an IR fixed point is more tricky to calculate, because the tachyon cannot be set to zero in~\eqref{XiHP}. We need to introduce an ``infinitesimal'' tachyon that behaves as $\t \sim u^\Delta$, where $\Delta>0$ is the dimension of the quark mass at the fixed point. Then we find that
\be \label{VPasIRFP1}
 V_P(u) = \left(\frac{3}{2}-\Delta\right)\left(\frac{1}{2}-\Delta\right) \frac{1}{u^2}\left[1 + \morder{\frac{1}{u^\delta}}\right]\,,\qquad (u \to \infty)
\ee
with the understanding that the tachyon normalization is taken to zero before taking $u \to \infty$. Eq.~\eqref{VPasIRFP1} holds in the conformal window since we assumed that $\Delta$ is real. It can be extended to complex $\Delta$ and therefore for the IRFP below the conformal window (with $\t\equiv 0$) by taking the real part:
\begin{align} \label{VPasIRFP2}
 V_P(u) &= \mathrm{Re}\left[\left(\frac{3}{2}-\Delta\right)\left(\frac{1}{2}-\Delta\right)\right] \frac{1}{u^2}\left[1 + \morder{\frac{1}{u^\delta}}\right]&&\nn\\
&= \left(\frac{3}{4}-(\mathrm{Im}\Delta)^2\right) \frac{1}{u^2}\left[1 + \morder{\frac{1}{u^\delta}}\right]\,,&\qquad (u &\to \infty)\,.
\end{align}
Similarly as~\eqref{VVatIRFP}, this latter expression is also valid at the approximate fixed point which appears in the standard vacuum in the walking regime ($x \to \xcfb$).

\subsubsection{Scalar Mesons}

The quadratic action is
\be
\begin{split}
S&= -{x\over2}\, M^3 N_c^2\,  {\mathbb Tr} \int d^4 x\,  e^{3\Awf}\, G^{-1} \bigg[ V_{f}(\l,\t)\,G^{-2}\,\h(\l,\t)
  (\partial_r \mathfrak{s})^2  \\
&- \left( 2\, \partial_\t V_f(\l,\t)\, \h(\l,\t)   \t'-  (1+G^{-2}) \t'\,V_{f}(\l,\t)\, \partial_{\t}\h(\l,\t)
\right) \mathfrak{s} \partial_r \mathfrak{s} \\
&+ \left( \partial_\t^2 V_f(\l,\t)\, G^2\, e^{2 \Awf} + \partial_\t V_f(\l,\t)\,  \t'^2\, \partial_\t\h(\l,\t) -{\t'^4
    \over 4} G^{-2} e^{-2\Awf}\,V_{f}(\l,\t)\, (\partial_\t\h(\l,\t) )^2 \right.\\
& + \left. {1\over 2}
  \t'^2\,V_{f}(\l,\t)\, \partial^2_\t\h(\l,\t) \right) \mathfrak{s}^2 - V_{f}(\l,\t)\,\h(\l,\t) (\partial_{\mu}
  \mathfrak{s})^2 \bigg]\,,
\end{split}\ee
where $\mathfrak{s}=\mathfrak{s}^a t^a$. The fluctuation equation therefore becomes
\be \label{flscalflucts}
\psi_S''+ \partial_r (\log C_1(r)) \psi_S' - {M-{1\over 2} \partial_r C_3
  \over C_1} \psi_S + {C_2 \over C_1} m^2 \psi_S=0\,,
\ee
where the Schr\"odinger functions read
\begin{align}
C_1(r)&=V_{f}(\l,\t)  e^{3 \Awf} G^{-3}\h(\l,\t) \nn\\
C_2(r) &=V_{f}(\l,\t)  e^{3 \Awf} G^{-1}\h(\l,\t) \nn\\
C_3(r)&= e^{3 \Awf} G^{-1} \left( 2 \partial_\t V_{f}(\l,\t)\h(\l, \t) \t'+
 (1+G^{-2}) \t' V_{f}(\l,\t)   \partial_\t\h(\l, \t) \right) \nn\\
M(r)&=   G e^{5 \Awf} \partial_\t^2 V_{f}(\l,\t)+ G^{-1} e^{3 \Awf} \t'^2 \partial_\t V_{f}(\l,\t) \partial_\t\h(\l,\t)
 \nn\\
& - {\t'^4 \over 4} G^{-3} e^{\Awf} V_{f}(\l,\t) (\partial_\t\h(\l,\t))^2
  +{1\over 2} G^{-1} e^{3 \Awf} \t'^2 V_{f}(\l,\t) \partial^2_\t \h(\l,\t)  \, .
\end{align}

If $\h(\l,\t)$ is independent of $\t$ and $V_f(\l,\t)=V_{f0}(\l)\exp\left(-a(\l)\t^2\right)$, then
\be
\begin{split}
 {M(r)-{1\over 2} \partial_r C_3(r)}
&=-2
\h(\l)^{-1} C_2(r) \left[e^{2 \Awf(r)} a(\l) -\t(r) \t'(r) \l'(r) {d
  a(\l)\over d\l } \h(\l) \right]\,, \\
{C_2(r) \over C_1(r)} &= G(r)^2 = 1 + e^{-2\Awf(r)} \h(\l,\t) \t'(r)^2 \, ,
\end{split}
\ee
and we obtain
\be\begin{split}
 \Xi_S(r) &= \frac{1}{G(r)}\sqrt{V_{f}(\l,\t)\,\h(\l,\t)\, e^{3 \Awf} }\,,\\
 H_S(r) &=-\frac{2}{\h(\l)}  \left[e^{2 \Awf(r)} a(\l) -\t(r) \t'(r) \l'(r) {d
  a(\l)\over d\l } \h(\l) \right]\,.
\end{split}\ee

The UV behavior of the potential is
\be \label{VSUVFP}
 V_S(u)=\frac{3}{4 u^2}\left[1 + \morder{\frac{1}{\log (u \Lambda)}}\right]\,,\qquad (u \to 0)\,,
\ee
where also the term $H_S$ contributes. At an IR fixed point we find that
\begin{align} \label{VSIRFP}
 V_S(u) &= \left[\frac{15}{4}-\Delta(4-\Delta)\right]\frac{1}{u^2}\left[1 + \morder{\frac{1}{u^\delta}}\right] \nn\\
  &=  \left(\frac{5}{2}-\Delta\right)\left(\frac{3}{2}-\Delta\right)\frac{1}{u^2}\left[1 + \morder{\frac{1}{u^\delta}}\right]\,,\qquad (u \to \infty)\,.
\end{align}
Here we used
\be
 \Delta(4-\Delta) = -m_\mathrm{IR}^2 \ell_\mathrm{IR}^2 = \frac{24\, a(\l_*)}{\h(\l_*)V_\mathrm{eff}(\l_*)}\,,
\ee
where $m_\mathrm{IR}$ is the tachyon mass at the fixed point, and $\ell_\mathrm{IR}$ is the IR AdS radius. The result~\eqref{VSIRFP} applies at real and complex values of $\Delta$ and is therefore valid for all solutions with vanishing tachyon, both in and below the conformal window. It also holds near the approximate fixed point of the standard vacuum in the walking regime.

\subsection{Flavor singlet sector: Lagrangian terms}

For flavor singlet (scalar and pseudoscalar) fluctuations the contributions from the different terms $S_g$, $S_f$ and $S_a$ in the action are coupled. Therefore, the analysis is somewhat more involved than for the non-singlet sector. We use the notation for the fluctuations defined in (\ref{fluctdef}).

\subsubsection{Terms from the flavor action}

The quantities ${\bf A}_{(i)MN}$ appearing in the DBI action can
be written as
\be
{\bf A}_{(i)MN}=\tilde g_{MN}+\hat g_{MN}+G_{MN}+\gf (\Phi,T)\,F^{(i)}_{MN}\,,
\label{h1}\ee
where $G_{MN}$ is defined in terms of the symmetric\footnote{We will use the convention $M_{(ab)}=M_{ab}+M_{ba}$.} part of $(D_M T)^\dagger D_NT$ as
\begin{align}
&\h(\Phi,T)\,(D_M T)(D_N T)^\dagger=\h_0\,\partial_M\tau\,\partial_N\tau +G_{MN}+H_{MN}\,,
\quad{\rm where}\;\; H_{MN}=-H_{NM}\,,\nonumber \\
&G_{MN}=\h_1\,\partial_M\tau\,\partial_N\tau+\h_0\,\partial_{(M}\tau\,\partial_{N)}s+\h_2\,
\partial_M\tau\,\partial_N\tau +\h_1\,\partial_{(M}\tau\,\partial_{N)}s +\h_0\,\partial_M S\,
\partial_N S +\nonumber \\
&\qquad\quad\;+\h_0\,\tau^2\left(\partial_M\theta+(A_M^{(L)}-A_M^{(R)})\right)
\left(\partial_N\theta+(A_N^{(L)}-A_N^{(R)})\right)\,,
\end{align}
$\hat g_{MN}$ is defined in (\ref{fluctdef}) and we have introduced the following useful quantity:
\be
\tilde g_{MN}=g_{MN}^{(0)}+\h_0\,\partial_M\tau\,\partial_N\tau\,,
\label{t1}\ee
and expanded $\h(\Phi,T)$ in terms of the fluctuations as
\begin{align}
\h(\Phi,T)&=\h_0+\h_1+\h_2+\mathcal{O}(s^2,\chi^2,\chi\,s)\,,\nonumber \\
\h_0&=\h(\Phi^{(0)},\tau)\,,\quad \h_1={\partial \h\over\partial\tau}\,s+
{\partial \h\over\partial\Phi}\,\chi\,,\quad \h_2={\partial^2 \h\over\partial\Phi\partial\tau}\,
\chi\,s+{1\over2}{\partial^2 \h\over\partial\Phi^2}\,\chi^2+{1\over2}
{\partial^2 \h\over\partial s^2}\,s^2\,.\nonumber \\
\end{align}
We will need to do the same for the potential $V_f(\Phi,T)$:
\begin{align}
V_f(\Phi,T)=V_f(\Phi^{(0)},\tau)+\left({\partial V_f\over\partial\Phi}\,\chi+
{\partial V_f\over\partial\tau}\,s\right)+\left({\partial^2 V_f\over\partial\Phi\partial\tau}\,
\chi\,s+{1\over2} {\partial^2 V_f\over\partial\Phi^2}\,\chi^2
+{1\over2} {\partial^2 V_f\over\partial\tau^2}\,s^2\right)\,.\nonumber \\
\end{align}
For notational simplicity we shall drop below the superscripts and subscripts $(0)$, $0$. It is understood that $\Phi$ stands
for $\Phi_0$ and the potentials are always evaluated on the vacuum solutions.
Moreover, to make lengthy expressions more compact, in the rest of this section we will not be writing explicitly the functional
dependence of the different potentials and functions: we will write $V_f$ instead of $V_f(\Phi,\t)$ and $\tau$ instead of $\tau(r)$
and so on.

Now we can expand the determinants appearing in (\ref{generalact}) up to second order in the
fluctuations by means of
\be
\sqrt{\det(\mathbb{I}+X)}=1+{1\over2}\tr{X}-{1\over4}\tr{X^2}+{1\over8}(\tr{X})^2+\mathcal{O}(X^3)\,.
\ee
In our case $X$ is given by
\be
X^a_b=\tilde g^{ac}\,\hat g_{cb}+\tilde g^{ac}\, G_{cb}+\gf(\Phi,T)\,\tilde g^{ac}\,F^{(i)}_{cb}\,.
\ee
where $\hat g$ and $\tilde g$ are defined in (\ref{h1}) and (\ref{t1}) respectively.

We shall now write the action (\ref{generalact}) up to second order in the fluctuations. The result
is quite voluminous. We will therefore first write the part of the action corresponding to the
left and right gauge fields which only mix with the fluctuations of the phase of the
tachyon $\theta$ (and as we have seen also with the QCD dilaton through the CP-odd sector). It takes the form
\begin{align}
{\cal L}_{\rm VA}\ =&-{x\over2}\,V_f\,\tilde g_{rr}^{1/2}\,\tilde g_{xx}^2\,
\bigg[ {1\over2}\gf^2\,\tilde g_{xx}^{-2}\,V^{\mu\nu}V_{\mu\nu}+\gf^2\,\tilde g_{rr}^{-1}\,\tilde g_{xx}^{-1}
\left(V_\mu ' \right)^2+ {1\over2}\gf^2\,\tilde g_{xx}^{-2}\,A^{\mu\nu}A_{\mu\nu}+\nonumber \\
&+\gf^2\,\tilde g_{rr}^{-1}\tilde g_{xx}^{-1}\left(A_\mu ' \right)^2+\h\,\tilde g_{xx}^{-1}\,\tau^2\,
\left(\partial_\mu\theta+2A_\mu\right)^2+\h\,\tilde g_{rr}^{-1}\,\tau^2\,\theta'^2\bigg]\,,
\label{lagvec}
\end{align}
in terms of the combinations defined in (\ref{VAdefstext}) (and in the gauge $A_r=V_r=0$).

The rest of the fluctuations mix with each other and the Lagrangian resulting from the quadratic
DBI reads
\begin{align}
{\cal L}_{\rm DBI}^{\rm mix}\ =&-\frac{1}{2}x\,V_f\,\tilde g_{rr}^{1/2}\,\tilde g_{xx}^2\,\Bigg\{-{1\over2}(h_{\mu\nu})^2
-\tilde g_{rr}^{-1}\,\tilde g_{xx}\,(\hat A_\mu)^2-2\tilde g_{rr}^{-1}\h\,\tau'\,\hat A^\mu\,
\partial_\mu s+{1\over4}\,(h^\mu_\mu)^2+\nonumber\\
&+\tilde g_{rr}^{-1}\,\tilde g_{xx}\,\phi\,h^\mu_\mu+\tilde g_{rr}^{-1}\,\h\,\tau'\,h^\mu_\mu\,s'
+\tilde g_{rr}^{-2}\,\tilde g_{xx}\,\h\,s'^2+\tilde g_{rr}^{-1}\,\h\,(\partial_\mu s)^2
-2\tilde g_{rr}^{-2}\,\tilde g_{xx}\,\h\,\tau'\,\phi\,s'+\nonumber\\
&-\tilde g_{rr}^{-2}\,\tilde g_{xx}^2\,\phi^2
+\left[2\tilde g_{rr}^{-1}\,\tilde g_{xx} (V_f)^{-1}
{\partial V_f\over\partial\Phi}-\tilde g_{rr}^{-2}\,\tilde g_{xx}\,\tau'^2\,{\partial \h\over
\partial\Phi}\right]\chi\,\phi+\nonumber\\
&+\left[2\tilde g_{rr}^{-1}\,\tilde g_{xx} (V_f)^{-1}{\partial V_f\over\partial\tau}
-\tilde g_{rr}^{-2}\,\tilde g_{xx}\,\tau'^2\,{\partial \h\over\partial\tau}\right]s\,\phi
+\left[ (V_f)^{-1}{\partial V_f\over\partial\Phi}+{1\over2}\tilde g_{rr}^{-1}\,\tau'^2
{\partial \h\over\partial\Phi}\right]\chi\,h^\mu_\mu+\nonumber\\
&+\left[(V_f)^{-1}{\partial V_f\over\partial\tau}+{1\over2}\tilde g_{rr}^{-1}\,\tau'^2\,
{\partial \h\over\partial\tau}\right]s\,h^\mu_\mu+
\bigg[2(V_f)^{-1}{\partial^2 V_f\over\partial\tau\partial\Phi}+\tilde g_{rr}^{-1}\,
{\partial^2 \h\over\partial\tau\partial\Phi}\,\tau'^2+\nonumber\\
&+\tilde g_{rr}^{-1}\,\tau'^2\,(V_f)^{-1}\,{\partial V_f\over\partial\tau}\,
{\partial \h\over\partial\Phi}+\tilde g_{rr}^{-1}\,\tau'^2\,(V_f)^{-1}\,{\partial V_f\over\partial\Phi}\,
{\partial \h\over\partial\tau}-{1\over2}\,\tilde g_{rr}^{-2}\,\tau'^4\,{\partial \h\over\partial\tau}\,
{\partial \h\over\partial\Phi}\bigg]s\,\chi+\nonumber\\
&+\bigg[(V_f)^{-1}{\partial^2 V_f\over\partial\tau^2}+{1\over2}\,\tilde g_{rr}^{-1}\,\tau'^2\,
{\partial^2 \h\over\partial\tau^2}+\tilde g_{rr}^{-1}\,(V_f)^{-1}\,\tau'^2\,{\partial \h\over\partial\tau}\,
{\partial V_f\over\partial\tau}-{1\over4}\tilde g_{rr}^{-2}\,\tau'^4\left({\partial \h\over\partial\tau}\right)^2
\bigg]s^2
+\nonumber\\ &+
\bigg[(V_f)^{-1}\,{\partial^2 V_f\over\partial\Phi^2}+{1\over2}\tilde g_{rr}^{-1}\,\tau'^2
{\partial^2 \h\over\partial\Phi^2}+\tilde g_{rr}^{-1}\,(V_f)^{-1}\,\tau'^2
{\partial V_f\over\partial\Phi}\,{\partial \h\over\partial\Phi}-{1\over4}\tilde g_{rr}^{-2}\,\tau'^4
\left({\partial\h\over\partial\Phi}\right)^2\bigg]\chi^2+\nonumber\\
&+\bigg[\tilde g_{rr}^{-2}\,\tilde g_{xx}\,\tau'\,{\partial \h\over\partial\Phi}
+2\tilde g_{rr}^{-1}\,(V_f)^{-1}\,\tau'\,\h\,{\partial V_f\over\partial\Phi}
+\tilde g_{rr}^{-1}\tau'\,{\partial \h\over\partial\Phi}\bigg]\chi\,s'+\nonumber\\
&+\bigg[\tilde g_{rr}^{-2}\,\tilde g_{xx}\,\tau'\,{\partial \h\over\partial\tau}
+2\tilde g_{rr}^{-1}\,(V_f)^{-1}\,\tau'\,\h\,{\partial V_f\over\partial\tau}
+\tilde g_{rr}^{-1}\tau'\,{\partial \h\over\partial\tau}\bigg]s\,s'\,\Bigg\}\,.
\end{align}
We shall now write this in a shortened form:
\begin{align}
{\cal L}_{\rm DBI}^{\rm mix}=&-\alpha(x,r)\bigg[-{1\over2}(h_{\mu\nu})^2
-\tilde g_{rr}^{-1}\,\tilde g_{xx}\,(\hat A_\mu)^2-2\tilde g_{rr}^{-1}\h\,\tau'\,\hat A^\mu\,
\partial_\mu s+{1\over4}\,(h^\mu_\mu)^2+\tilde g_{rr}^{-1}\,\tilde g_{xx}\,\phi\,h^\mu_\mu+\nonumber\\
&+\tilde g_{rr}^{-1}\,\h\,\tau'\,h^\mu_\mu\,s'
+\tilde g_{rr}^{-2}\,\tilde g_{xx}\,\h\,s'^2+\tilde g_{rr}^{-1}\,\h\,(\partial_\mu s)^2
-2\tilde g_{rr}^{-2}\,\tilde g_{xx}\,\h\,\tau'\,\phi\,s'-\tilde g_{rr}^{-2}\,\tilde g_{xx}^2\,\phi^2+\nonumber\\
&+{\bold T}_1\,\chi\,\phi+{\bold T}_2\,s\,\phi+{\bold T}_3\,\chi\,h^\mu_\mu+{\bold T}_4\,s\,h^\mu_\mu
+{\bold T}_5\,s\,\chi+{\bold T}_6\,s^2+\nonumber\\
&+{\bold T}_7\,\chi^2+{\bold T}_8\,s'\,\chi+{\bold T}_9\,s'\,s\bigg]\,,
\label{lagdbi}
\end{align}
where we have defined
\begin{align}
&\alpha(x,r)=\frac{1}{2}x\,V_f\,\tilde g_{rr}^{1/2}\,\tilde g_{xx}^2\,,
\end{align}
\begin{align}
&{\bold T}_1=2\tilde g_{rr}^{-1}\,\tilde g_{xx} (V_f)^{-1}
{\partial V_f\over\partial\Phi}-\tilde g_{rr}^{-2}\,\tilde g_{xx}\,\tau'^2\,{\partial \h\over
\partial\Phi}\, ,\quad
{\bold T}_2=2\tilde g_{rr}^{-1}\,\tilde g_{xx} (V_f)^{-1}{\partial V_f\over\partial\tau}
-\tilde g_{rr}^{-2}\,\tilde g_{xx}\,\tau'^2\,{\partial \h\over\partial\tau}\,,\nonumber\\
&{\bold T}_3=(V_f)^{-1}{\partial V_f\over\partial\Phi}+{1\over2}\tilde g_{rr}^{-1}\,\tau'^2\,
{\partial \h\over\partial\Phi}\, ,\qquad\qquad\quad\;\;
{\bold T}_4=(V_f)^{-1}{\partial V_f\over\partial\tau}+{1\over2}\tilde g_{rr}^{-1}\,\tau'^2\,
{\partial \h\over\partial\tau}\,,\nonumber\\
&{\bold T}_5=2(V_f)^{-1}{\partial^2 V_f\over\partial\tau\partial\Phi}+\tilde g_{rr}^{-1}\,
{\partial^2 \h\over\partial\tau\partial\Phi}\,\tau'^2+\tilde g_{rr}^{-1}\,\tau'^2\,(V_f)^{-1}\,{\partial V_f\over\partial\tau}\,
{\partial \h\over\partial\Phi}+\tilde g_{rr}^{-1}\,\tau'^2\,(V_f)^{-1}\,{\partial V_f\over\partial\Phi}\,
{\partial \h\over\partial\tau}+\nonumber\\
&\quad\quad\;-{1\over2}\,\tilde g_{rr}^{-2}\,\tau'^4\,{\partial \h\over\partial\tau}\,
{\partial \h\over\partial\Phi}\,,\nonumber\\
&{\bold T}_6=(V_f)^{-1}{\partial^2 V_f\over\partial\tau^2}+{1\over2}\,\tilde g_{rr}^{-1}\,\tau'^2\,
{\partial^2 \h\over\partial\tau^2}+\tilde g_{rr}^{-1}\,(V_f)^{-1}\,\tau'^2\,{\partial \h\over\partial\tau}\,
{\partial V_f\over\partial\tau}-{1\over4}\tilde g_{rr}^{-2}\,\tau'^4\left({\partial \h\over\partial\tau}\right)^2\,,
\nonumber\\
&{\bold T}_7=(V_f)^{-1}{\partial^2 V_f\over\partial\Phi^2}+{1\over2}\,\tilde g_{rr}^{-1}\,\tau'^2\,
{\partial^2 \h\over\partial\Phi^2}+\tilde g_{rr}^{-1}\,(V_f)^{-1}\,\tau'^2\,{\partial \h\over\partial\Phi}\,
{\partial V_f\over\partial\Phi}-{1\over4}\tilde g_{rr}^{-2}\,\tau'^4\left({\partial \h\over\partial\Phi}\right)^2\,,
\nonumber\\
&{\bold T}_8=\tilde g_{rr}^{-2}\,\tilde g_{xx}\,\tau'\,{\partial \h\over\partial\Phi}
+2\tilde g_{rr}^{-1}\,(V_f)^{-1}\,\tau'\,\h\,{\partial V_f\over\partial\Phi}
+\tilde g_{rr}^{-1}\tau'\,{\partial \h\over\partial\Phi}\nonumber\\
&{\bold T}_9=\tilde g_{rr}^{-2}\,\tilde g_{xx}\,\tau'\,{\partial \h\over\partial\tau}
+2\tilde g_{rr}^{-1}\,(V_f)^{-1}\,\tau'\,\h\,{\partial V_f\over\partial\tau}
+\tilde g_{rr}^{-1}\tau'\,{\partial \h\over\partial\tau}\nonumber\\
&{\bold T}_{10}=2{\h\,\tau'^2\over\tilde g_{rr}}\,(V_f)^{-1}
{\partial V_f\over\partial\Phi}
+{\tau'^2\over\tilde g_{rr}^2}(\tilde g_{rr}+\tilde g_{xx})\,{\partial \h\over\partial\Phi}\,.
\label{tdefs}
\end{align}

\subsubsection{Terms from the glue action}

The action for the glue sector was expanded up to quadratic order in the fluctuations in ref. \cite{ihqcd,glue}.
In order to use the results of \cite{glue} we will first need to do some field redefinitions. The action for the
glue sector in our case \cite{jk} takes the form
\be
{\cal L}_{\rm glue}=\sqrt{-g}\left(R-{4\over3}(\partial\Phi)^2+V_g(\Phi)\right)\,.
\ee
Notice that our notation differs from that used in \cite{glue} by
\be
\Phi=\sqrt{3\over4}\tilde\Phi\, , \qquad V_g = -\tilde V_g\,, \qquad \Awf=-{\tilde \Awf\over2}\,,
\ee
where the tilded functions correspond to the ones appearing in \cite{glue}.

In the computation of \cite{glue} the background equations of motion where used in computing the
action of the fluctuations. There will now be some extra contributions since the background equations
of motion have new terms coming from the DBI sector. We can read the background Einstein and dilaton
EoMs from \cite{jk}. They take the form
\begin{align}
&6{b''\over b}=-\frac{4}{3}\Phi'^2+b^2\, V_g-x\,b^2\,V_f\,\sqrt{1+b^{-2}\,\h\,\tau'^2}\,,\label{ein1}\\
&12{b'^2\over b^2}=\frac{4}{3}\Phi'^2+b^2 V_g-x\,b^2{V_f\over \sqrt{1+b^{-2}\,\h\,\tau'^2}}\,,\label{ein2}\\
&3b^5{\partial V_g\over \partial\Phi}=8(b^3\Phi')'-3x\,b^5{\partial V_f\over \partial\Phi}
\sqrt{1+b^{-2}\,\h\,\tau'^2}- 2x\,b^3{\partial \h\over \partial\Phi}{V_f\,\tau'^2\over \sqrt{1+b^{-2}\,\h\,\tau'^2}}\,,
\label{bckeoms}
\end{align}
where
\be
b=e^{\Awf}\,,
\ee
Now (\ref{ein1}) and (\ref{ein2}) can be combined to obtain
\be
b^3\left(12{b'^2\over b^2}-\frac{4}{3}\Phi'^2\right)=(b^3)''+\frac{1}{2}x\,b^3V_f{\h\,\tau'^2\over
\sqrt{1+b^{-2}\,\h\,\tau'^2}}\,,
\ee
and
\be
b^3\left(12{b'^2\over b^2}-\frac{4}{3}\Phi'^2+b^2 \tilde V_g\right)=2 (b^3)''+x\,b^5 V_f
\sqrt{1+b^{-2}\,\h\,\tau'^2}\,.
\ee
Using these equations and proceeding as in the Appendix A of \cite{glue},
the action for the fluctuations at quadratic order is given by
\begin{align}
{\cal L}_{\rm glue}\ =&e^{3\Awf}\bigg[{\cal L}_{\rm ein}^{(2)}-{1\over4}(h_{\mu\nu}')^2+{1\over4}h'^2
-{1\over4}(\hat F_{\mu\nu})^2-\frac{4}{3}(\partial_\mu\,\chi)^2-\frac{4}{3}\chi'^2-{1\over2}e^{2\Awf}
(\partial_{\Phi}^2V_g)\chi^2
+\nonumber\\
&-\partial^\mu\phi(\partial^\nu h_{\mu\nu}-\partial_\mu h)+\frac{8}{3}\Phi'\,\phi'\,\chi+\frac{16}{3}\Phi'\,
\phi\,\chi'+\frac{4}{3}\Phi'\,h'\,\chi+\frac{8}{3}\Phi'\,\hat A^\mu\,\partial_\mu\chi\bigg]+\nonumber\\
&+\left(e^{3\Awf}\,
\hat A^\mu\right)'
(\partial^\nu h_{\mu\nu}-\partial_\mu h)-\left(e^{3\Awf}\right)'\left(2\hat A^\mu\,
\partial_\mu\phi+2\phi\,\phi'+\phi\,h'\right)+\nonumber\\
&+\alpha\left[-{1\over2}(h_{\mu\nu})^2+{1\over4}h^2-{\tilde g_{xx}\over\tilde g_{rr}}(\hat A_\mu)^2
{\bold T}_3\left(2\phi+h\right)\chi+\left(1-2{\tilde g_{xx}\over\tilde g_{rr}}\right)\phi^2
+{\tilde g_{xx}\over\tilde g_{rr}}\,\phi\,h
\right]\,,\nonumber\\
{\cal L}_{\rm ein}^{(2)}\ =&-{1\over4}\partial^\mu h_{\rho\sigma}\,\partial_\mu h^{\rho\sigma}+{1\over2}
\partial^\mu h_{\rho\mu}\,\partial_\nu h^{\rho\nu}-{1\over2}\partial_\mu h\,\partial_\rho h^{\rho\mu}
+{1\over4}(\partial_\mu h)^2\,.\label{lagglue}
\end{align}
Here and in what follows
\be
h^\mu_\mu\equiv h\,.
\ee
%
%
%

\subsubsection{Full Lagrangian, fluctuation equations and field decomposition}

Adding up (\ref{lagvec}), (\ref{lagdbi}), (\ref{lagglue}), and the contribution from the CP-odd sector we obtain the following final
Lagrangian for the fluctuations of the singlet sector:
\be
{\cal L}={\cal L}_{\rm VA}+{\cal L}_{\rm sg}+{\cal L}_a\,,
\ee
where ${\cal L}_{\rm VA}$ is given in (\ref{lagvec}), ${\cal L}_a$ can be read from (\ref{samain}) and ${\cal L}_{\rm sg}$ takes the form
\begin{align}
{\cal L}_{\rm sg}\ =\ &e^{3\Awf}\bigg[{\cal L}_{\rm ein}^{(2)}-{1\over4}(h_{\mu\nu}')^2+{1\over4}h'^2
-{1\over4}(\hat F_{\mu\nu})^2-\frac{4}{3}(\partial_\mu\,\chi)^2-\frac{4}{3}\chi'^2-{1\over2}e^{2\Awf}
(\partial_{\Phi}^2V_g)\chi^2+\nonumber\\
&-\partial^\mu\phi(\partial^\nu h_{\mu\nu}-\partial_\mu h)+\frac{8}{3}\Phi'\,\phi'\,\chi+\frac{16}{3}\Phi'\,
\phi\,\chi'+\frac{4}{3}\Phi'\,h'\,\chi+\frac{8}{3}\Phi'\,\hat A^\mu\,\partial_\mu\chi\bigg]+\nonumber\\
&+\left(e^{3\Awf}\,
\hat A^\mu\right)'
(\partial^\nu h_{\mu\nu}-\partial_\mu h)-\left(e^{3\Awf}\right)'\left(2\hat A^\mu\,
\partial_\mu\phi+2\phi\,\phi'+\phi\,h'\right)+\nonumber\\
&+\alpha\bigg[
2\tilde g_{rr}^{-1}\h\,\tau'\,\hat A^\mu\,\partial_\mu s+
-\tilde g_{rr}^{-1}\,\h\,\tau'\,h\,s'
-\tilde g_{rr}^{-2}\,\tilde g_{xx}\,\h\,s'^2
-\tilde g_{rr}^{-1}\,\h\,(\partial_\mu s)^2+\nonumber\\
&+2\tilde g_{rr}^{-2}\,\tilde g_{xx}\,\h\,\tau'\,\phi\,s'
+\tilde g_{rr}^{-2}\,\h\,\tau'^4\,\phi^2-{\bold T}_2\,s\,\phi-{\bold T}_4\,s\,h
{\bold T}_5\,s\,\chi-{\bold T}_6\,s^2+\nonumber\\
&-
{\bold T}_7\,\chi^2-
{\bold T}_8\,s'\,\chi-{\bold T}_9\,s'\,s+
{\bold T}_{10}\,
\chi\,\phi\bigg]\,.
\label{fluclag}
\end{align}

Notice that the fields appearing in ${\cal L}_{\rm sg}$ do not contribute to ${\cal L}_{\rm VA}$ and vice versa, so we can study
each piece of the Lagrangian separately. ${\cal L}_{\rm VA}$ has the same form as for the non-singlet sector, hence its
contributions to the fluctuation equations for the different fields are the same as in that sector. But those fluctuation equations
will receive extra contributions from  the CP-odd Lagrangian as we will see in the next section.
Instead, the contributions from ${\cal L}_{\rm sg}$ are decoupled from those of ${\cal L}_{\rm VA}$ and ${\cal L}_a$, and
therefore we can study it separately. In the rest of this section we will compute the fluctuation equations resulting from
${\cal L}_{\rm sg}$. We shall follow \cite{glue} and decompose $\hat A_\mu$ and $h_{\mu\nu}$ in irreducible representations
of the 4D Lorentz group as
\begin{align}
\hat A_\mu&=\partial_\mu W+\hat A_\mu^\bot \, ,\,\,\,\, \partial^\mu \hat A_\mu^\bot=0
\label{lorentzf1}
\\
h_{\mu\nu}&=2\eta_{\mu\nu}\,\psi+2\partial_\mu\partial_\nu E+2\partial_{(\mu}\hat V_{\nu)}^\bot
+h_{\mu\nu}^{\bot\bot} \, ,
\label{lorentzf2}
\end{align}
with $\partial^\mu \hat A_\mu^\bot=\partial^\mu \hat V_\mu^\bot=0$ and $\partial^\mu h_{\mu\nu}^{\bot\bot}=h_{\mu}^{\bot\bot\,\mu}=0$. The field equations resulting from  ${\cal L}_{\rm sg}$ read
\begin{align}
(h^{\mu\nu})\quad
&h^{\bot\bot''}_{\mu\nu}+3\Awf'\,h^{\bot\bot'}_{\mu\nu}+\Box\,h^{\bot\bot}_{\mu\nu}
-2e^{-3\Awf}\,\partial_{(\mu}\left[e^{3\Awf}\left(\hat A^\bot_{\nu)}-\hat V^{\bot'}_{\nu)}\right)\right]'+\nonumber\\
&-6\eta_{\mu\nu}\left[\psi''+3\Awf'\,\psi'-\Awf'\,\phi'-(3\Awf'^2+\Awf'')\phi+{4e^{-3\Awf}\over9}(e^{3\Awf}\,
\Phi'\,\chi)'\right]+\nonumber\\
&+2(\partial_\mu\partial_\nu-\eta_{\mu\nu}\Box)\left[\phi+2\psi-(W-E')'-3\Awf'(W-E')\right]+\nonumber\\
&-2\alpha\,e^{-3\Awf}\eta_{\mu\nu}\left({\bold T}_4\,s+{\h\,\tau'\over\tilde g_{rr}}\,s'\right)=0\,,\\
(\hat A^{\mu})\quad
&6\partial_\mu\left[\psi'-\Awf'\,\phi+{4\over9}\Phi'\,\chi\right]+\Box\left(\hat A_\mu^\bot-\hat V_\mu^{\bot'}\right)+\nonumber\\
&+2\alpha\,e^{-3\Awf}{\h\,\tau' \over \tilde g_{rr}}\partial_\mu s=0\,,\\
(\phi)\quad
&\Box\psi+4\Awf'\,\psi'-\left(\Awf''+3\Awf'^2\right)\phi-\Awf'\,\Box\left(W-E'\right)
+{4e^{-3\Awf}\over9}\left(e^{3\Awf}\,\Phi'\,\chi\right)'
-{8\over9}\Phi'\,\chi'+\nonumber\\
&-\alpha\,{e^{-3\Awf}\over6}\left[
{\bold T}_{10}\,\chi+2\left({\tau'^2\,\h\over\tilde g_{rr}}\right)^2\phi+2{\tilde g_{xx}\over\tilde g_{rr}^2}\h\,\tau's'
-{\bold T}_2\,s\right]=0\,,\\
(\chi)\quad
&\chi''+3\Awf'\chi'+\Box\chi-{3e^{2\Awf}\over8}(\partial_{\Phi}^2V_g)
\chi-2e^{-3\Awf}\left(e^{3\Awf}\,\Phi'\,\phi\right)'+\Phi'\,\phi'
+4\Phi'\,\psi'+\nonumber\\
&-\Phi'\,\Box(W-E')
+\alpha\,{3e^{-3\Awf}\over8}\left[{\bold T}_{10}\,\phi-{\bold T}_5\,s-2{\bold T}_7\,\chi-{\bold T}_8\,s'\right]=0\,,
\nonumber\\
(s)\quad
&-2\tilde g_{rr}\,\h\,\Box s+2\tilde g_{rr}\,\h\,\tau'\,\Box(W-E')-2\tilde g_{xx}\,\h\,s''-8\tilde g_{rr}\,\h\,\tau'\,\psi'+2\tilde g_{xx}\,\h\,\tau'\,\phi'+\nonumber\\
&\bigg[6(3\tilde g_{xx}-4\tilde g_{rr})\,\h\,\Awf'+(\tilde g_{xx}-3\tilde g_{rr})\,\Phi'(\partial_\Phi\h)+2(V_f)^{-1}
(2\tilde g_{xx}-3\tilde g_{rr})\,\h\,\Phi'\,(\partial_\Phi V_f)+\,\nonumber\\
&-2\tilde g_{xx}\,\tau'(\partial_\tau \h)+4\tilde g_{xx}\,(V_f)^{-1}\h\,\tau'(\partial_\tau V_f)\bigg]s'+\nonumber\\
&+\Big[-8\Awf'\,\tau'^3\h(\partial_\Phi \h)+2\tilde g_{xx}\,{\tau'\over\h}\,\Phi'\,(\partial_\Phi\h)\,(\partial_\tau\h)
+\tilde g_{xx}{\tau'^2\over \h}(\partial_\tau\h)^2 \rc
&-2(\tilde g_{xx})^2(V_f\,\h)^{-1}(\partial_\tau\h)(\partial_\tau V_f)+
\nonumber\\
&+2\gtx\,\gtz(V_f)^{-2}(\partial_\tau V_f)^2-2\h\,(V_f)^{-1}\,\tau'^3\,\Phi'(\partial_\tau\h)(\partial_\Phi V_f)+
\nb\\
&+2\gtz(V_f)^{-2}\,\h\,\tau'\,\Phi'\,(\partial_\tau V_f)(\partial_\Phi V_f)-(\gtx+\gtz)\,\tau'\,\Phi'\,
(\partial_\Phi\partial_\tau\h)-\gtx\,\tau'^2(\partial^2_\tau\h)+\nb\\
&+2\gtz\gtx\,(V_f)^{-1}(\partial^2_\tau V_f)-2\gtz\,\h\,(V_f)^{-1}\,\tau'\,\Phi'(\partial_\Phi\partial_\tau V_f)\Big]s
\nb\\
&+\Big[-8\Awf'\,\tau'^3\h(\partial_\Phi \h)+2{\gtx\over\h}\,\tau'\,\Phi'\,(\partial_\Phi\h)^2+2\gtz\,\h\,(V_f)^{-2}\,
\tau'\,\Phi'\,(\partial_\Phi V_f)^2+\nb\\
&-(\gtx+\gtz)\,\tau'\,\Phi'\,(\partial^2_\Phi\h)-2\gtz\,\h\,(V_f)^{-1}\,\tau'\,\Phi'\,(\partial^2_\Phi V_f)+{\gtx\over\h}\,
\tau'^2(\partial_\tau\h)(\partial_\Phi\h)+\nb\\
&-2{\gtx^2\over\h}\,(V_f)^{-1}\,(\partial_\Phi\h)(\partial_\tau V_f)-2\h\,(V_f)^{-1}\,\tau'^3\,\Phi'\,
(\partial_\Phi\h)(\partial_\Phi V_f)+\nb\\
&-2\gtx\,\gtz\,(V_f)^{-2}\,(\partial_\tau V_f)(\partial_\Phi V_f)
-\gtx\,\tau'^2\,(\partial_\tau\partial_\Phi \h)+2\gtx\,\gtz\,(V_f)^{-1}\,(\partial_\tau\partial_\Phi V_f)\Big]\chi+\nb\\
&-\Big[(\gtx+\gtz)(\partial_\Phi\h)\,\tau'+2\gtz\,(V_f)^{-1}\,\h\,\tau'\,(\partial_\Phi V_f)\Big]\chi'\nb\\
&+\big[16\Awf'(\h)^2\,\tau'^3+2\h\,\tau'^3\,\Phi'\,(\partial_\Phi\h)+4(\h)^2 (V_f)^{-1}\,\tau'^3\,\Phi'\,
(\partial_\Phi V_f)\rc
&+4\gtx\,(V_f)^{-1}\,(\partial_\tau V_f)\Big]\phi=0\,.\nb\\
\end{align}

These can be split into separate equations involving scalar, vector, and tensor modes only:
\begin{align}
&h^{\bot\bot''}_{\mu\nu}+3\Awf'\,h^{\bot\bot'}_{\mu\nu}+\Box\,h^{\bot\bot}_{\mu\nu}=0\,,\label{graveq}\\
\nb\\
&\left[e^{3\Awf}\left(\hat A^\bot_{\mu}-\hat V^{\bot'}_{\mu}\right)\right]'=0\,,\qquad
\Box\left(\hat A_\mu^\bot-\hat V_\mu^{\bot'}\right)=0\label{vceq}\,,\\
\nb\\
&\psi''+3\Awf'\,\psi'-\Awf'\,\phi'-(3\Awf'^2+\Awf'')\phi+{4e^{-3\Awf}\over9}(e^{3\Awf}\,
\Phi'\,\chi)'+\alpha\,{e^{-3\Awf}\over3}\left({\bold T}_4\,s+{\h\,\tau'\over\tilde g_{rr}}\,s'\right)=0\,,\nb\\
\label{sceq1} \\
&\phi+2\psi-(W-E')'-3\Awf'(W-E')=0\,\label{sceq2}\\
\nb\\
&\psi'-\Awf'\,\phi+{4\over9}\Phi'\,\chi+\alpha\,{e^{-3\Awf}\over3}\,{\h\,\tau' \over \tilde g_{rr}}\,s=0\,,\label{sceq3}\\
\nb\\
&\Box\psi+4\Awf'\,\psi'-\left(\Awf''+3\Awf'^2\right)\phi-\Awf'\,\Box\left(W-E'\right)
+{4e^{-3\Awf}\over9}\left(e^{3\Awf}\,\Phi'\,\chi\right)'
-{8\over9}\Phi'\,\chi'+\nonumber\\
&-\alpha\,{e^{-3\Awf}\over6}\left[
{\bold T}_{10}\,\chi+2\left({\tau'^2\,\h\over\tilde g_{rr}}\right)^2\phi+2{\tilde g_{xx}\over\tilde g_{rr}^2}\h\,\tau's'
-{\bold T}_2\,s\right]=0\,,\label{sceq4}\\
\nb\\
&\chi''+3\Awf'\chi'+\Box\chi-{3e^{2\Awf}\over8}(\partial_{\Phi}^2V_g)
\chi-2e^{-3\Awf}\left(e^{3\Awf}\,\Phi'\,\phi\right)'+\Phi'\,\phi'
+4\Phi'\,\psi'+\nonumber\\
&-\Phi'\,\Box(W-E')
+\alpha\,{3e^{-3\Awf}\over8}\left[{\bold T}_{10}\,\phi-{\bold T}_5\,s-2{\bold T}_7\,\chi-{\bold T}_8\,s'\right]=0\,,\label{sceq5}\\
\nb\\
&-2\tilde g_{rr}\,\h\,\Box s+2\tilde g_{rr}\,\h\,\tau'\,\Box(W-E')-2\tilde g_{xx}\,\h\,s''-8\tilde g_{rr}\,\h\,\tau'\,\psi'+2\tilde g_{xx}\,\h\,\tau'\,\phi'+\nonumber\\
&\bigg[6(3\tilde g_{xx}-4\tilde g_{rr})\,\h\,\Awf'+(\tilde g_{xx}-3\tilde g_{rr})\,\Phi'(\partial_\Phi\h)+2(V_f)^{-1}
(2\tilde g_{xx}-3\tilde g_{rr})\,\h\,\Phi'\,(\partial_\Phi V_f)+\,\nonumber\\
&-2\tilde g_{xx}\,\tau'(\partial_\tau \h)+4\tilde g_{xx}\,(V_f)^{-1}\h\,\tau'(\partial_\tau V_f)\bigg]s'+\nonumber\\
&+\Big[-8\Awf'\,\tau'^3\h(\partial_\Phi \h)+2\tilde g_{xx}\,{\tau'\over\h}\,\Phi'\,(\partial_\Phi\h)\,(\partial_\tau\h)
+\tilde g_{xx}{\tau'^2\over \h}(\partial_\tau\h)^2 \rc
&-2(\tilde g_{xx})^2(V_f\,\h)^{-1}(\partial_\tau\h)(\partial_\tau V_f)+
\nonumber\\
&+2\gtx\,\gtz(V_f)^{-2}(\partial_\tau V_f)^2-2\h\,(V_f)^{-1}\,\tau'^3\,\Phi'(\partial_\tau\h)(\partial_\Phi V_f)+
\nb\\
&+2\gtz(V_f)^{-2}\,\h\,\tau'\,\Phi'\,(\partial_\tau V_f)(\partial_\Phi V_f)-(\gtx+\gtz)\,\tau'\,\Phi'\,
(\partial_\Phi\partial_\tau\h)-\gtx\,\tau'^2(\partial^2_\tau\h)+\nb\\
&+2\gtz\gtx\,(V_f)^{-1}(\partial^2_\tau V_f)-2\gtz\,\h\,(V_f)^{-1}\,\tau'\,\Phi'(\partial_\Phi\partial_\tau V_f)\Big]s
\nb\\
&+\Big[-8\Awf'\,\tau'^3\h(\partial_\Phi \h)+2{\gtx\over\h}\,\tau'\,\Phi'\,(\partial_\Phi\h)^2+2\gtz\,\h\,(V_f)^{-2}\,
\tau'\,\Phi'\,(\partial_\Phi V_f)^2+\nb\\
&-(\gtx+\gtz)\,\tau'\,\Phi'\,(\partial^2_\Phi\h)-2\gtz\,\h\,(V_f)^{-1}\,\tau'\,\Phi'\,(\partial^2_\Phi V_f)+{\gtx\over\h}\,
\tau'^2(\partial_\tau\h)(\partial_\Phi\h)+\nb\\
&-2{\gtx^2\over\h}\,(V_f)^{-1}\,(\partial_\Phi\h)(\partial_\tau V_f)-2\h\,(V_f)^{-1}\,\tau'^3\,\Phi'\,
(\partial_\Phi\h)(\partial_\Phi V_f)+\nb\\
&-2\gtx\,\gtz\,(V_f)^{-2}\,(\partial_\tau V_f)(\partial_\Phi V_f)
-\gtx\,\tau'^2\,(\partial_\tau\partial_\Phi \h)+2\gtx\,\gtz\,(V_f)^{-1}\,(\partial_\tau\partial_\Phi V_f)\Big]\chi+\nb\\
&-\Big[(\gtx+\gtz)(\partial_\Phi\h)\,\tau'+2\gtz\,(V_f)^{-1}\,\h\,\tau'\,(\partial_\Phi V_f)\Big]\chi'\nb\\
&+\big[16\Awf'(\h)^2\,\tau'^3+2\h\,\tau'^3\,\Phi'\,(\partial_\Phi\h)+4(\h)^2 (V_f)^{-1}\,\tau'^3\,\Phi'\,
(\partial_\Phi V_f)+4\gtx\,(V_f)^{-1}\,(\partial_\tau V_f)\Big]\phi=0\,.
\label{sceq6}
\end{align}

\subsection{Flavor singlet sector: fluctuation equations}

In this section we will summarize the results of the fluctuation analysis for the different flavor singlet sectors.
Note that the action for the vector mesons is the same as for the non-singlet sector and therefore we will not repeat its
analysis (see section \ref{app:NSvmesons}).

\subsubsection{Axial-vector Mesons}

As mentioned above, the action for the singlet sector of the (transverse) axial vector modes has an
extra term coming from the CP-odd sector. It then takes the form
\begin{align}
S_A\ =& -{M^3 N_c^2\over2}\!\int\! d^4x\, dr
\bigg\{ x\,V_f(\l,\t)\, e^\Awf\, \G^{-1}\!\bigg[\,
{1 \over 2}\,\G^2 \, \gf(\l,\t)^2 A_{\mu\nu}\,A^{\mu\nu} +\nonumber\\
&+
\,\gf(\l,\t)^2\, \partial_r A^{\bot S}_\mu\, \partial_r A^{\bot S\,\mu}+4\h(\l,\t)\,\tau^2\, e^{2 \Awf}\,\G^{2}\,A^{\bot S}_\mu A^{\bot S\,\mu}\bigg]+ \nonumber\\
&+4x^2\,Z(\l)\,e^{3\Awf}\,V_a(\l,\t)^2\,A^{\bot S}_\mu A^{\bot S\,\mu}
\bigg\} \,,
\label{axacti2}
\end{align}
where $G$ was defined in~\eqref{Gdef}.
The fluctuation equation is given by
\begin{align}
&\frac{1}{V_f(\l,\t)\,\gf(\l,\t)^2\,e^\Awf \G  }\,
\partial_r \left( V_f(\l,\t)\, \gf(\l,\t)^2\, e^{\Awf}
\G^{-1} \partial_r \varphi_A \right)
+m_V^2 \varphi_A+\nonumber\\
&-4\left[x\,e^{2\Awf}{Z(\l)\,V_a(\l,\t)^2\over V_f(\l,\t)\,\G\,\gf(\l,\t)^2}+ {\t^2 e^{2 \Awf}
\over \gf(\l,\t)^2}\h(\l,\t)\right]\varphi_A= 0\,.
\label{axvectoreom2}
\end{align}

\subsubsection{Pseudoscalar Mesons}

The quadratic action has two contributions: $S_1$ coming from the DBI piece and $S_2$ from
the CP-odd sector. We write each separately:
\begin{align}\label{pscalu1s1}
S_1\ =&-M^3 N_c^2\,{x\over2} \int d^4 x\, dr\,V_f(\l,\t)\, e^{\Awf}\, G^{-1}
\\\nonumber
&\times\bigg[
\gf(\l,\t)^2\,\left(\partial_r A^{\lVert S}_\m\right)^2 + e^{2\Awf}\h(\l,\t)\,\t^2\,(\partial_r\theta)^2
+e^{2\Awf}\,\G^2\,\h(\l,\t)\,\t^2\left(\partial_\m\theta+2 A^{\lVert S}_\m\right)^2
\bigg]\,,
\end{align}
and
\be
S_2=-{M^3\, N_c^2\over2} \int d^4 x\, dr\,Z(\l)\,e^{3\Awf}\left[
\left(\partial_\m a-2x\,V_a(\l,\t)\, A^{\lVert S}_\m\right)^2
+\left(\partial_r a+x\,\partial_r V_a(\l,\t)\,\theta\right)^2
\right]\,,\label{pscalu1s2}
\ee
where as before we have set $A_r=0\,$.\footnote{Notice that the gauge choice $A_r=0$ should be imposed at the
level of the equations of motion. We have checked that, after setting $A_r=0$ the equation of motion for $A_r$ reduces
to the first order differential equation (\ref{pseudoscu1stode}), which follows from the other equations of motion.}
One can easily check that both pieces are invariant under the residual gauge transformation
(\ref{u1transf}) (where $\epsilon\neq\epsilon(r)$).
\newline
We  recall the Ansatz for  $A^{\lVert S}_\m$ in (\ref{ansaxlong}), and write down similar Ans\"atze for $\theta$ and
$a$:
\begin{align}
A^{\lVert S}_\m(x^\mu, r) &= -\varphi_L(r)\,\partial_\m({\cal T}(x^\m))\,,\nonumber\\
\theta(x^\mu, r) &=2\varphi_\theta(r)\,{\cal T}(x^\m)\,,\nonumber\\
a(x^\mu, r) &=2\varphi_\mathrm{ax}(r)\,{\cal T}(x^\m)\,.
\label{defsagain}
\end{align}
In terms of this new fields the gauge invariant combinations appearing in (\ref{pscalu1s1}) and
(\ref{pscalu1s2}) read
\begin{align}
\partial_\m\theta+2A^{\lVert S}_\m=2\left[\varphi_\theta(r)-\varphi_L(r)\right]\partial_\m({\cal T}(x^\m))
&\equiv 2\partial_\m({\cal T}(x^\m))\,P(r)\,,\nonumber\\
\partial_\m a-2x\,V_a(\l,\t)\,A^{\lVert S}_\m=2\left[\varphi_\mathrm{ax}(r)+x\,V_a(\l,\t)\,\varphi_L(r)\right]
\partial_\m({\cal T}(x^\m))&\equiv 2\partial_\m({\cal T}(x^\m))\,Q(r)\,,\nonumber\\
\partial_r a+x\,\partial_r V_a(\l,\t)\,\theta=2\left[\varphi_\mathrm{ax}'(r)+x\,\partial_r V_a(\l,\t)\,\varphi_\theta(r)\right]
{\cal T}(x^\m)&\equiv2{\cal T}(x^\m)\,R(r)\,.
\label{ginv}
\end{align}
where we have defined the new functions $P(r)$, $Q(r)$ and $R(r)$.

Computing the fluctuation equations for $A^{\lVert S}_\m$, $\theta$, and
$a$; and then substituting  in the Ansatz (\ref{defsagain}) one arrives at
\begin{align}
\partial_r\left(V_f(\l,\t)\,e^{\Awf}\,\G^{-1}\,\gf(\l,\t)^2\varphi_L'\right)
+4V_f(\l,\t)\,e^{3\Awf}\,\G \,\h(\l,\t)\,\t^2\,(\varphi_\theta-\varphi_L)+&\nonumber\\\nonumber
-4e^{3\Awf}\,Z(\l)\,V_a(\l,\t)\,(\varphi_\mathrm{ax}+x\,V_a(\l,\t)\,\varphi_L)&=0\,,
\\[2mm]
\partial_r\left(V_f(\l,\t)\,e^{3\Awf}\,\G^{-1}\,\h(\l,\t)\,\t^2\,\varphi_\theta'\right)
+m^2\,V_f(\l,\t)\,e^{3\Awf}\,\G\,\h(\l,\t)\,\t^2\,(\varphi_\theta-\varphi_L)+&\\\nonumber
-e^{3\Awf}\,Z(\l)\,\partial_r V_a(\l,\t)\,
\left(\varphi_\mathrm{ax}'+x\,\partial_r V_a(\l,\t)\,\varphi_\theta\right)&=0\,,\\[2mm] \nonumber
\partial_r\left[e^{3\Awf}\,Z(\l)\left(\varphi_\mathrm{ax}'+x\,\partial_r V_a(\l,\t)\,\varphi_\theta\right)\right]
+m^2\,e^{3\Awf}\,Z(\l)\,(\varphi_\mathrm{ax}+x\,V_a(\l,\t)\,\varphi_L)&=0\,.
\end{align}
We  now write these three equations in terms of the functions $P(r)$, $Q(r)$ and $R(r)$ defined in
(\ref{ginv}):
\begin{align}
\partial_r\left[V_f\,e^{\Awf}\,\G^{-1}\,\gf^2\left({V_a'\over V_a}\,P+{Q'-R\over x\,V_a}
\right)\right]
+4V_f\,e^{3\Awf}\,\G\,\h\,\t^2\,P-4e^{3\Awf}\,Z\,V_a\,Q&=0\,,
\label{amutheomgi}\\
\partial_r\left[V_f\,e^{3\Awf}\,\G^{-1}\,\h\,\t^2\,\left({V_a'\over V_a}\,P+P'
+{Q'-R\over x\,V_a}\right)\right]
+m^2\,V_f\,e^{3\Awf}\G\,\h\,\t^2\,P-e^{3\Awf}\,Z\,V_a'\,R&=0\,,\label{cptheomgi}\nonumber\\ \\
\partial_r\left[e^{3\Awf}\,Z\,R\right]+m^2\,e^{3\Awf}\,Z\,Q&=0\,.\label{cpaeomgi}
\end{align}
Here all primes denote derivatives with respect to $r$.
Combining these three equations one gets the following first order one:
\begin{align}
V_f\,e^{\Awf}\,\G^{-1}\,\gf^2\left({V_a'\over V_a}\,P+{Q'-R\over x\,V_a}\right)
&-{4\over m^2}\,V_f\,e^{3\Awf}\,\G^{-1}\,\h\,\t^2\,\left({V_a'\over V_a}\,P+P'
+{Q'-R\over x\,V_a}\right)+\nonumber\\
&+{4\over m^2}\,e^{3\Awf}\,Z\,V_a\,R=0\,.
\label{pseudoscu1stode}
\end{align}
{}From this equation we can solve for $R(r)$; substituting the result in (\ref{amutheomgi}) and
(\ref{cpaeomgi}) we obtain
\begin{align}
&\partial_r\left[V_f\,e^{\Awf}\,\G^{-1}\,\gf^2\left(
-4e^{2\Awf}\,{V_f\,\h\,\t^2\over N_a+N_b}\,P'+{V_a'\over V_a}\,{N_b\over N_a+N_b}\,P
+{N_b\over x\,V_a\,(N_a+N_b)}Q'\right)\right]+\nonumber\\
&+4V_f\,e^{3\Awf}\,\G\,\h\,\t^2\,P-4e^{3\Awf}\,Z\,V_a\,Q=0\,,\label{cpsys1}\\ \nonumber \\
&\partial_r\left[e^{3\Awf}\,Z\,\left(
4x\,e^{2\Awf}\,{V_a\,V_f\,\h\,\t^2\over N_a+N_b}\,P'+x\,{V_a'\,N_a\over N_a+N_b}\,P
+{N_a\over N_a+N_b}Q'\right)\right]+m^2\,e^{3\Awf}\,Z\,Q=0\,,\nonumber\\ \label{cpsys2}
\end{align}
where  $N_a$ and $N_b$ are given by the following expressions:
\be \label{NaNbdef}
N_a=V_f\left(4e^{2\Awf}\,\h\,\t^2-m^2\,\gf^2\right)\, ,\qquad
N_b=4x\,e^{2\Awf}\,Z\,V_a^2\,\G\,.
\ee

An alternative choice of variables is given by
\begin{align}
P(r)&\equiv\varphi_\theta(r)-\varphi_L(r)\,,\nonumber\\
Q(r)&\equiv\varphi_\mathrm{ax}(r)+x\,V_a(\l,\t)\,\varphi_L(r)\,,\nonumber\\
S(r)&\equiv \partial_r\varphi_\theta(r)\,.
\label{ginv2}
\end{align}
In terms of these, the Lagrangian (\ref{pscalu1s1}), (\ref{pscalu1s2}) reads
\begin{align}
S\ =&-{M^3 N_c^2\over2} \int d^4 x\, dr\,{\cal T}(x^\mu)^2\,\Bigg\{V_f(\l,\t)\,e^{\Awf(r)}\,\G(r)^{-1}\bigg[
4\,e^{2\Awf(r)}\,\G(r)^2\,\h(\l,\t)\,\t(r)^2\,S(r)^2+\nonumber \\
&-4m^2\,\h(\l,\t)\,\t(r)^2\,P(r)^2-m^2e^{2\Awf(r)}\,\tilde g_{rr}^{-\frac12}\,\gf(\l,\t)^2(S(r)-\partial_r P(r))^2\bigg]+\nonumber \\
&+4e^{3\Awf(r)}\,Z(\l)\,\bigg[-m^2\,Q(r)^2+\nonumber \\
&+\Big(\partial_r Q(r)+x\,V_a(\l,\t)\,\partial_r P(r)+x\,\partial_r V_a(\l,\t)\,P(r)
-x\,V_a(\l,\t)\,S(r)\Big)^2\bigg]\Bigg\}\,.
\label{lagginv2}
\end{align}
The fluctuation equations resulting from this Lagrangian reduce to the system (\ref{cpsys1}), (\ref{cpsys2}).

\subsubsection{Scalar mesons}
\label{app_singlet_scalar}
The flavor-singlet scalar mesons are described by the action ${\cal L}_{\rm sg}$ given by Eq. (\ref{fluclag}).
The resulting equations of motion are Eqs. (\ref{graveq} - \ref{sceq6}) and their analysis will run similarly to the one
carried out in the Appendix A of \cite{glue}. First, one should notice that the action (\ref{fluclag}) is invariant under
the 5d diffeomorphisms $\delta r=\xi^5$, $\delta x^\mu=\xi^\mu$, which act on the fields as \cite{glue}:
\begin{align}
&\delta h_{\mu\nu}=-\partial_\mu\xi_\nu-\partial_\nu\xi_\mu-2\eta_{\mu\nu}\,A'\,\xi^5\,,\quad
\delta A_\mu=-\xi_\mu'-\partial_\mu\xi^5\,,\nonumber\\
&\delta \phi=-\xi^{5'}-A'\,\xi^5\,,\quad
\delta\chi=-\Phi'\,\xi^5\,,\quad
\delta s=-\tau'\,\xi^5\,.
\label{5ddiff}
\end{align}
Therefore, the equations of motion must satisfy the two identities that follow from the equalities:
\be
{\delta S\over\delta\xi_\mu}=0\,,\qquad
{\delta S\over\delta\xi^5}=0\,.
\label{diffid}
\ee
The first equality reduces to the fact that (\ref{sceq1}) follows from taking the derivative with respect to $r$ of
$e^{3A}$ times equation (\ref{sceq3}).
On the other hand, the second equality in (\ref{diffid}) results in the following non-obvious identity:
\begin{align}
0=&-6e^{3A}\Box(\ref{sceq3})+6(\partial_r-A')\left[e^{3A}(\ref{sceq4})\right]+\nb\\
&-A'\,e^{3A}\left[24(\ref{sceq1})+6\Box(\ref{sceq2})\right]+2\Phi'\,e^{3A}(\ref{sceq5})-\tau'\,(\ref{sceq6})\,,
\label{diffid2}
\end{align}
which can be checked through a lengthy but straightforward computation.

Indeed the diffeomorphism invariance of ${\cal L}_{\rm sg}$ implies that not all the fields in (\ref{fluclag}) are
dynamical. The fluctuations of the metric, dilaton and tachyon making up  ${\cal L}_{\rm sg}$ add up to 17
components, but, as in \cite{glue}, 5 are eliminated by gauge transformations and 5 more due to the non-dynamical
components of the Einstein equations. We are left with 7 propagating degrees of freedom which, as we will see
correspond to a 4D massive spin-2 field and two massive scalars.

We  now focus on the scalars, governed by the
equations (\ref{sceq1} - \ref{sceq6}). We have just seen that Eq. (\ref{sceq4}) follows from Eq. (\ref{sceq1}), and
the identity (\ref{diffid2}) allows us to get rid of eq. (\ref{sceq2}).
Next, we can solve for $\phi$ and $\Box(W-E')$ in terms of $\psi$, $\chi$ and $s$ from equations
(\ref{sceq3}) and (\ref{sceq4}) respectively. Finally, substituting the resulting expressions into the equations
(\ref{sceq5}) and (\ref{sceq6}) we obtain two equations that only depend on the following two linear combinations
of $\psi$, $\chi$ and $s$:
\be
\zeta=\psi-{\Awf'\over\Phi'}\,\chi\, ,\qquad \xi=\psi-{\Awf'\over\tau'}\,s\, ,
\label{xizetdef}
\ee
which are invariant under the gauge transformations (\ref{5ddiff}).
The equations for $\zeta$ and $\xi$ take the form
\begin{align}
\zeta''+k(r)\,\zeta'+p(r)\,\xi'+\Box\zeta+N_1(r) \left(\zeta-\xi\right)&=0\,,\label{zeteq}\\
\xi''+q(r)\,\zeta'+n(r)\,\xi'+t\,\Box\xi+N_2(r) \left(\xi-\zeta\right)&=0\,,\label{xieq}
\end{align}
in terms of the following functions:
\begin{align}
k(r)& =3\Awf'(r)-2{\Awf''(r)\over \Awf'(r)}+2{\Phi''(r)\over\Phi'(r)}\,,\nonumber\\
p(r)& =-\frac12 x\,V_f(\Phi,\t) \Bigg[{3 \h(\Phi,\t)\, \ẗ́'(r)^2\over4\Phi'(r)\, \G(r)}
\frac{\partial_\Phi V_f(\Phi,\t)}{V_f(\Phi,\t)}+\frac{3 \t'(r)^2\,(1+\G(r)^2)}{8\Phi'(r)\, \G(r)^3}
\partial_\Phi\h(\Phi,\t)+\nonumber\\
&\qquad \ +{\h(\Phi,\t)^2\,\tau'(r)^4\over3\Awf'(r)\,e^{2 \Awf(r)}\,\G(r)^3}\Bigg]\,,\label{mpcf}\\
q(r)&=\Phi'(r)\left[{4\Phi'(r)\,\h(\Phi,\t)\,\tau'^2\over9e^{2\Awf(r)}\,\Awf'(r)}+{1+\G(r)^2\over2}\,{\partial_\Phi\h(\Phi,\t)\over\h(\Phi,\t)}+\G(r)^2\,{\partial_\Phi V_f(\Phi,\t)\over V_f(\Phi,\t)}
\right]\,,\nonumber\\
 t(r)& =\G(r)^2\,,\label{qtcf}\\
n(r)& = \left(4\G(r)^2-7\right)\Awf'(r)-2{\Awf''(r)\over \Awf'(r)}+\Phi'(r)\bigg[{\G(r)^2-3\over2}\,{\partial_\Phi\h(\Phi,\t)\over\h(\Phi,\t)}
+\nonumber\\
&\qquad \ -(2-G(r)^2)\,{\partial_\Phi V_f(\Phi,\t)\over V_f(\Phi,\t)}\bigg]+{2e^{2\Awf(r)}\over\h(\Phi,\t)\,\tau'(r)}\,
{\partial_\tau V_f(\Phi,\t)\over V_f(\Phi,\t)}\,.\nb\label{ncf}\\
\end{align}
The expressions for the functions $N_1(r)$ and $N_2(r)$ are very lengthy, we write them down in a somewhat more compact
notation:
\begin{align}
&{N_1(r)\over x}\cdot \mathfrak{D}_1=\rc
&=24\Awf'\,e^{2\Awf}\,\h\,\R\,\tau'\,V_f\bigg\{V_f\,\tau'\left[4x\,e^{4\Awf}\,\R^3\,
V_f+(1+\R^2)\left(e^{4\Awf}\,\R^2\,V_g-2\h\,\tau'^2\,\Phi'^2\right)\right](\pfi \h)+\rc
&+2\h\,\tau'\,\R^2\left[4x\,e^{4\Awf}\,\R\,V_f+e^{4\Awf}\,\R^2\,V_g-2\h\,\tau'^2\,\Phi'^2\right]
(\pfi V_f)+\rc
&+2e^{4\Awf}\,\R^2\,\left[\h\,V_f\,\tau'\,(\pfi V_g)+2\R^2\,\Phi'(\ptt V_f)\right]\bigg\}+
8\h\,\R\,V_f^2\,\Phi'\Big[-8x\,e^{6\Awf}\,\h\,\R^3\,V_f\,\tau'^2+\rc
&-2e^{6\Awf}\,\h\,\tau'^2\,\R^2(1+\R^2)\,V_g+2\h^3\,\tau'^6\,\Phi'^2\Big]+\rc
&+e^{4\Awf}\left(2x\,
e^{2\Awf}\,V_f+e^{2\Awf}\,\R\,V_g-\R\,\Phi'^2\right)\Big[-3(1+\R^2)^2\tau'^2\,V_f^2\Phi'(\pfi \h)^2+\rc
&-12\h\,\R^2(1+\R^2)\,V_f\,\tau'^2\,\Phi'\,(\pfi \h)\,(\pfi V_f)-12\h^2\,\R^4\,\tau'^2\,\Phi'
(\pfi V_f)^2+\rc
&-6\R^2\,V_f^2\,\tau'^3\,(\pfi\h)(\ptt\h)+12e^{2\Awf}\,\R^4\,\tau'\,V_f\,(\pfi\h)(\ptt V_f)+12e^{2\Awf}\,\h\,\tau'\,\R^4(\pfi V_f)
(\ptt V_f)+\rc
&+6\h\,\tau'^3\,\R^2\,V_f^2\,(\partial_{\tau\Phi}^2 \h)+12e^{2\Awf}\,\h\,\tau'\,\R^4\,V_f\,
(\partial_{\tau\Phi}^2 V_f)\Big]\,,
\label{n1}
\end{align}
and
\begin{align}
&N_2(r)\cdot\mathfrak{D}_2=\rc
&=\Awf'\bigg\{-6\Phi'\R\left(2x\,e^{2\Awf}\,V_f+\R\,\Q\right)\bigg[2e^{4\Awf}\R^2\,\h\,
(\partial_\Phi V_f)\,(\partial_\tau V_f)+\rc
&+e^{2\Awf}(1+\R^2)V_f^2\,\Phi'\,\tau'\,\h\,(\pdfi\h)-V_f^2\,e^{2\Awf}\,\tau'^2\,(\pfi\h)(\ptt\h)+V_f^2\,\tau'^2\,e^{2\Awf}\,\h\,(\partial_{\tau\Phi}^2\h)+\rc
&+2e^{2\Awf}\,
V_f\,\Phi'\,\h^2\,\tau'\,\R^2\,(\pdfi V_f)-2e^{4\Awf}V_f\,\h\,\R^2\,(\partial_{\tau\Phi}^2 V_f)+2\,e^{4\Awf}\,V_f(\pfi\h)\,(\ptt V_f)\bigg]+\rc
&-3e^{4\Awf}\h\,\tau'(1+\R^2)\left(2x\,e^{2\Awf}\,\R\,V_f+\R^2\,\Q\right)(\pfi V_g)\,(\pfi\h)+\rc
&-16\h^2\,\tau'\,V_f^2\,\Phi'^2
\Big[x\,e^{2\Awf}\,\R\,V_f\left(4e^{2\Awf}-\h\,\tau'^2(1+\R^2)\right)+2\R^2\left(e^{4\Awf}\,V_g-\h\,\tau'^2\,\Q\right)
\Big]+\rc
&-3V_f^2\,\tau'\Big[2x^2\,e^{4\Awf}\,\h\,\tau'^2\,V_f^2(1+\R^2)-4e^{2\Awf}\,\R^2\,\Phi'^2\,\Q
+x\,\R\,V_f\,e^{2\Awf}\Big(e^{2\Awf}\,\h\,\tau'^2\,V_g\,(1+\R^2)+\rc
&-\Phi'^2\left(8e^{2\Awf}+\h\,\tau'^2(1+\R^2)\right)\Big)\Big]\,(\pfi\h)^2-12\,e^{2\Awf}\,\h\,V_f\,\tau'\,
\Big[2x^2\,e^{8\Awf}\,V_f^2\,\R^4+\rc
&+e^{2\Awf}\,\h\,\tau'^2\,\R^2\,\Phi'^2\,\Q+x\,e^{2\Awf}\,\R\,V_f\left(e^{6\Awf}\,V_g\,\R^4-\left(e^{4\Awf}
+\h^2\,\tau'^4\right)\Phi'^2\right)\Big](\pfi\h)\,(\pfi V_f)+\rc
&-12\h^2\,\tau'\,\R^2\Big[2x^2\,e^{6\Awf}\,V_f^2\,\R^2-e^{2\Awf}\,\R^2\,\Phi'^2\,\Q+\rc
&+x\,e^{4\Awf}\,\R
\,V_f(e^{2\Awf}\,V_g\,\R^2-(2+\R^2)\,\Phi'^2)\Big](\pfi V_f)^2+\rc
&-6 e^{4\Awf}\,\h^2\,\tau'\,\R^2\,(\pfi V_g)\,V_f\,\left(2x\,e^{2\Awf}\,\R\,V_f+\R^2\,\Q\right)
(\pfi V_f)\bigg\}+\rc
&+4e^{4\Awf}\,\R\,\h^2\,\tau'\,V_f^2\,\Phi'\,(\pfi V_g)\left(2x\,e^{2\Awf}\,V_f+
\R\,\Q\right)+\rc
&-2\h\,\tau'\,V_f^2\,\Phi'\,\Big[4x^2\,e^{2\Awf}\,V_f^2\left(2e^{4\Awf}+\h^2\,\tau'^4\right)+
\Q\,V_g\left(2e^{4\Awf}+e^{2\Awf}\,\h\,\tau'^2-\h^2\,\tau'^4\right)+\rc
&+2x\,\R\,V_f\,V_g\,e^{2\Awf}\left(4e^{4\Awf}-e^{2\Awf}\,\h\,\tau'^2+\h^2\,\tau'^4\right)
-2x\,\R\,V_f\left(2e^{4\Awf}+\h^2\,\tau'^4\right)\Phi'^2
\Big](\pfi \h)+\rc
&-4e^{-2\Awf}\,\h^2\,\tau'\,\Phi'\,V_f\,\Big[4x^2\,e^{6\Awf}\,\h\,\tau'^2\,\R^2\,V_f^2
+e^{2\Awf}\,\R^2\,\Q\left(e^{4\Awf}\,\R^2\,V_g-2\,\h\,\tau'^2\Phi'^2\right)+\rc
&+2x\,e^{2\Awf}\,\R\,V_f\Big(e^{6\Awf}\,\R^4\,V_g-\h\,\tau'^2\,\Phi'^2\,e^{2\Awf}(2+\R^2)\Big)
\Big](\pfi V_f)+\rc
&+8e^{2\Awf}\,\h\,V_f\,\Phi'^2\,e^{2\Awf}\left(2x\,e^{2\Awf}\,\R\,V_f+\R^2\,\Q\right)(\ptt V_f)\,,
\label{n2}
\end{align}
with the following definitions:
\begin{align}
\R&=\sqrt{1+e^{-2\Awf}\,\h\,\tau'^2}\, ,\qquad\qquad\quad\;
\Q=e^{2\Awf}\,V_g-\Phi'^2\, ,
\rc
\mathfrak{D}_1&=-144\,\Awf'^2\,e^{4\Awf}R^4\,\h\,V_f\,\Phi'\, ,\qquad
\mathfrak{D}_2=-144\,\Awf'^3\,e^{2\Awf}\,R^2\,\h^2\,\tau'\,V_f^2\,.
\end{align}

\subsubsection{Spin-two fluctuations}
As we have seen in the previous section, the propagating degrees of freedom resulting from the lagrangian
${\cal L}_{\rm sg}$ of eq (\ref{fluclag}) are two scalars (described by Eqs. (\ref{zeteq}) and (\ref{xieq})) and
a massive 4D spin-2 field described by equation (\ref{graveq}). These spin-2 fluctuations correspond to the
$2^{++}$ glueballs, and their equation (\ref{graveq}) is the same as in \cite{glue}:
\be
 \left(h^{\bot\bot}_{\mu\nu}\right)''+3\Awf'\,\left(h^{\bot\bot}_{\mu\nu}\right)'+\Box\,h^{\bot\bot}_{\mu\nu}=0\,.
 \label{tenseom}
\ee

\section{Schr\"odinger form}
\label{app:schro}

In this section, we write the fluctuation equations in the Schr\"odinger
form. A generic quadratic five-dimensional action for a field
$\Psi(x^\mu,r)$ is
\be
S = -\frac12 {\cal K}_\Psi \int d^4 x dr \left(
C_1(r) (\partial_r \Psi)^2 + C_2(r) \eta^{\mu\nu} \partial_\mu \Psi \partial_\nu \Psi +
C_3(r) \Psi \partial_r \Psi + M(r) \Psi^2
\right)\,,
\label{quadracti}
\ee
where we have allowed an arbitrary constant multiplying the action.
We  consider $\Psi = e^{iqx} \psi(r)$ and define as $ m_n^2$ the discrete
set of values of $-q^2$ which satisfy the appropriate normalizability conditions of the
Sturm-Liouville problem.  The 
 solutions satisfy the fluctuation equation
extracted from (\ref{quadracti}):
\be
-\frac{1}{C_2(r)} \partial_r \left( C_1(r) \partial_r \psi_n (r)\right)
+ H(r) \psi_n(r) = m_n^2 \psi_n(r)\, ,
\label{lalala}
\ee
where we have introduced
\be
H(r)\equiv \frac{1}{C_2(r)} \left( M(r) -\frac12 \partial_r C_3(r)\right)\,.
\label{defhofz}
\ee
We can define the orthonormality condition:
\be
 \int dr C_2(r) \psi_n(r) \psi_m (r) = \delta_{mn}\, .
\label{normpsi}
\ee

We now define a new radial variable $u$ (with $u=0$ in the UV), and a rescaled field $\alpha$
in terms of a function $\Xi$ as
\be
du = \sqrt{\frac{C_2(r)}{C_1(r)}} dr\,,\qquad
\alpha = \Xi\, \psi \,,\qquad
\Xi(r) = (C_1(r)C_2(r))^\frac14\,.
\label{defsschrei}
\ee
The Sturm-Liouville problem now takes the Schr\"odinger form:
\be
- \frac{d^2 \alpha_n(u)}{du^2} + V(u) \alpha_n(u) = m_n^2 \alpha_n (u)
\label{schroeqap}
\ee
where the Schr\"odinger potential is
\be
V(u) = \frac{1}{\Xi(u)}\frac{d^2 \Xi(u)}{du^2} + H(u) \, .
\label{schrlike}
\ee
Substituting (\ref{normpsi}) in (\ref{defsschrei}), we find that in the new variables,
the normalization condition is the canonical one:
\be
 \int du \alpha_n(u) \alpha_m(u) = \delta_{mn} \, .
\label{normalpha}
\ee

\section{The UV behavior of the two-point functions} \label{app:2ptf}

\subsection{Vector two-point function}

We calculate the vector two-point function in the bulk for large Euclidean momentum and match to the perturbative QCD result. In this way, we can determine the value  of the function $\gf(\lambda,\tau)$ on the boundary, as it will be explained in the following. The function $\gf(\l,\t)$ multiplies the gauge field strength in Sen's action, Eq.(\ref{Senaction}).

To compute the vector two-point function holographically, we first find the quadratic vector on-shell action and differentiate it with respect to the source of the boundary value of the vector field. As shown in appendix \ref{app:quadfluctdet}, the quadratic action and the equation of motion are
\be
\begin{split}
& S_V  = -  {1 \over 2} M^3 N_c\, {\mathbb Tr} \int d^4x\, dr
V_f(\l,\t) \gf(\l,\t)^2 e^A
\left[ \frac12 G^{-1} V_{\m\n}V^{\m\n} +
G \partial_r V_\m \partial_r V^\m
\right]\,, \\
& \frac{1}{V_f(\l,\t)\, \gf(\l,\t)^2\, e^{\Awf}\,\G}
\partial_r \left( V_f(\l,\t)\, \gf(\l,\t)^2 \,e^{\Awf}\,
\G^{-1}\, \partial_r \psi_V \right)
-q^2\, \psi_V  = 0 \, .
\label{vectoracti2}
\end{split}
\ee
The on-shell action is then found to be an integral on the boundary of space-time:
\be
S_V= {1 \over 2} M^3 N_c\,  {\mathbb Tr}\int_{\partial {\mathcal M}} d^4 x \left. V_f (\l, \t) \gf(\l,\t)^2 e^{\Awf} \G^{-1} V_{\m}(x,r) \partial_r V^{\m}(x,r) \right|_{r=0} \, .
\label{onshellactvec}
\ee

As mentioned above, the holographic two-point function in momentum space is the second derivative of $S_V$ with respect to the boundary value of the vector field:
\be
\Pi^{ab}_{\mu \nu}(q,p)={\delta^2 S_V \over \delta V^{a \, \mu}_{0}
  (q) \delta V^{b\, \nu}_{0} (p)}=-(2 \pi)^4 \delta^4 (p+q) \frac{2\delta^{ab}}{N_f}
\left( q^2 \eta_{\m\n}-{q_{\m} q_{\n}}\right)\Pi_V (q) \, ,
\label{pivadsdef}
\ee
where we have set the fields to be
\be
V^{\m}(x)=\int {d^4 q \over (2\pi)^4} e^{i q x} V^{a\, \m}_{0}(q) t^a \, \psi_V(r)\;,
 \ee
 where $t^a$, $a=1,\ldots, N_f^2-1$ are the $SU(N_f)$, flavor group generators and $V^{a\, \m}_{0}$ is the source of the vector current. The factor of $2/N_f$ is included in order for non-singlet and singlet $\Pi_V$ to have similar normalizations. The bulk fields have been normalized in such a way that the boundary coupling
of the field theory current to the bulk field is
\be
2 {\mathbb Tr} \int d^4x J^{\mu}_V V_{\mu} \, .
\ee

The function $\Pi_V$ is calculated from (\ref{onshellactvec}),
\be
\begin{split}
\Pi_V(q) & =- {1 \over 4}M^3 N_c N_f \left. V_f (\l, \t) \gf(\l,\t)^2 e^{\Awf}
  \G {\psi_V(r) \partial_r \psi_V(r) \over q^2} \right|_{r=\epsilon} \\
  & = -{1 \over 4} M^3
N_c N_f W_0  \gf_0^2 \ell {\partial_r \psi_V(\epsilon) \over q^2 \,\epsilon} \, ,
\end{split}
\label{pivdef}
\ee
where $r=\epsilon$ is the UV cutoff of spacetime, $w_0=w(\l=0)$ and $\psi_V$ is the
IR normalizable solution of the fluctuation equation with UV boundary
condition $\psi_V(0)=1$. We have also used the normalization of the $SU(N_f)$
generators, ${\mathbb Tr} (t^a t^b)= {\delta^{ab} / 2}$. The leading term of the two-point function in the limit of large momentum is determined by the near-boundary solution of the equation of motion. The contribution from the IR is exponentially suppressed in terms of momentum.

Hence, using the UV expansions of the background fields and potentials, given in section \ref{uvstruct} and Appendix \ref{uvasback}, we determine the UV asymptotics of the fluctuation equation (\ref{vectoracti2}) which are reliable for $r \sim 1/q \to 0$, for large $q^2$.
\be
\psi_V^{\prime\prime}- {1 \over r} \left[1+{\mathcal O} \left( {1 \over  \log^2(r \Lambda)}\right)  \right]\psi_V^{\prime}-q^2 \left[1+ {\mathcal O} \left( r^2 \log^2(r \Lambda)\right) \right] \psi_V=0 \, .
\ee
Keeping only the leading terms in the UV expansion of the equation above is enough to determine
the leading large $q^2$ expansion of $\Pi_V$. In this limit, the equation coincides with the corresponding hard-wall AdS/QCD equation.
By normalizing $\psi_V$ such that $\psi_V(0)=1$ we find the solution,
\be
\psi_V(r)=q \, r\, K_{1}\left( q r \right) \, .
\ee

Then, the correlator in the large momentum limit reads
\be
\Pi_V(q) =
- {1\over 8}  M^3 N_c N_f W_0 \gf_0^2 \ell \, \log q^2 +\cdots \, ,
\ee
where the subleading corrections are suppressed by logarithms of $\log q^2$.
By matching to the QCD result, $\Pi_V(q) =-{N_c N_f \over 48 \pi^2} \log q^2$, we find
\be
M^3 N_c N_f W_0 \, \gf_0^2 \ell={N_c N_f \over 6 \pi^2} \, .
\label{matchcon}
\ee


\subsection{Scalar two-point function}

We will also calculate the two-point function of the non-singlet scalar field in order to determine the boundary value of $\h_0$ of $\h(\l,\t)$. The UV expansion of the tachyon, Eq. (\ref{TUVres}), implies that the boundary coupling of the tachyon to the dual field theory operator is proportional to
\be
{\mathbb Tr} \int d^4 x {T(x, \epsilon) \over \ell \, \epsilon} {\bar q(x)} q(x) \, .
\ee
When the tachyon field is expanded as $T=(\tau+s+\mathfrak{s}^a t^a)e^{i\theta+i\,\pi^a t^a}$, the on shell action of the fluctuation is
\be
\begin{split}
S_S&= {1 \over 2} M^3 N_c {\mathbb Tr} \int d^4 x \left. C_1(r)
  \mathfrak{s} \partial_r  \mathfrak{s} \right|_{r=\epsilon} \\
&= {1 \over 2}M^3 N_c {\mathbb Tr} \int d^4 x \left. V_{f}(\l,\t) e^{3 \Awf} \G^{-3} \h(\l,\t) \mathfrak{s} \partial_r  \mathfrak{s} \right|_{r=\epsilon} \, .
\end{split}
\ee
Here, we denoted $\mathfrak{s}=\mathfrak{s}^a t^a$. The scalar-scalar correlator is defined analogously to the vector-vector one in (\ref{pivadsdef})
\be
\Pi^{ab}(q,p) = {\delta^2 S_S \over \delta S^a_{0} (q) \delta S^b_{0} (p)}=(2 \pi)^4 \delta^4 (p+q) \frac{2\delta^{ab}}{N_f} \Pi_S (q) \, ,
\ee
where we have set the fields to be $\mathfrak{s} (x)=\int {d^4 q \over (2\pi)^4} e^{i q x}\, \ell \, S^a_{0}(q) t^a \, \psi_S(r)$, where $t^a$ are the $SU(N_f)$ flavor group generators and the fluctuation $\psi_S$ satisfies the boundary condition $\psi_S (r=\epsilon)=\epsilon$ . The AdS radius appears in the above expression in order for $S^a_{0}(q)$ to have the correct dimension of the source of the boundary operator. 

The scalar fluctuation equation for large Euclidean momentum (i.e., near the boundary) is
\be
 \begin{split}
&  \psi_S^{\prime\prime}-{3 \over r} \left[1+{\mathcal O} \left( {1 \over  \log^2(r \Lambda)}\right)  \right] \psi_S^{\prime}+ {3 \over r^2} \left[ 1- {\mathcal O}\left({1 \over \log (r \Lambda)}\right)  \right]\psi_S \\
&- q^2 \left[1+ {\mathcal O} \left( r^2 \log^2(r \Lambda)\right) \right]  \psi_S=0 \, .
\end{split}
\ee
Solving the above equation we determine the IR normalizable wavefunction
\be
\psi_S(r) = q \, r^2 \, K_{1}\left( q r \right) \, ,
\ee
with the following boundary asymptotics
\be
\psi_S(r)=r+{1\over 4} r^3 q^2 \log (q^2 r^2) +{1 \over 4} \left( (-2
  \rho -1 -\log 4)q^2 \right)r^3  + \cdots \, ,
\ee
where $\rho$ is defined in~\eqref{rhodef}.
The on-shell action is then
\be
S_S=  M^3 N_c W_0 \h_0 \ell^2 {\delta^{a b} \over 4} \int {d^4 q \over (2
  \pi)^4} {\cal S}_{a\,0}(q) {\cal S}_{b\, 0}(-q) \left( {\ell^3 \over \epsilon^2}+
  \ell^3 q^2 \log \epsilon^2+ \ell^3 q^2 \log q^2 +\cdots \right) \, .
\label{sconsh}
\ee
The divergences of the on-shell action are cancelled by adding the
counterterms
\begin{align}
S_{ct1}&=-{M^3 N_c W_0 \h_0 \over 2 \ell} {\mathbb Tr} \int d^{4}x \sqrt{\gamma}  \mathfrak{s}(x,\epsilon)^{2} \, ,
\label{sc2pfct1}\\
S_{ct2}&={M^3 N_c W_0 \h_0 \ell \over 4} {\mathbb Tr} \int d^{4}x \sqrt{\gamma}  \mathfrak{s}(x,\epsilon)\square_{\gamma}  \mathfrak{s}(x,\epsilon) \log\epsilon^{2} \, .
\label{sc2pfct2}
\end{align}
The first counterterm adds a finite part, $-{1 \over 2} \ell^3 q^2 \log q^2$, to the parenthesis in the expression (\ref{sconsh}) for the on-shell action. Hence, the leading order result is
\be
\Pi_S(q)={M^3 N_c N_f W_0 \ell^3 \h_0 \over 8}\, q^2 \log q^2 \, .
\ee
Matching to the field theory result, $\Pi_S(q)={N_c N_f \over 32
\pi^2 }q^2 \log q^2$, we obtain
\be
M^3 N_c N_f W_0 \ell^5 \h_0= {N_c N_f \over 4 \pi^2} \, .
\label{matchscalar}
\ee
Finally, notice that combining (\ref{matchcon}) and (\ref{matchscalar})
one arrives at the expression (\ref{kappawrelation}) which relates $\gf_0$ and $\h_0$.

\section{UV and IR asymptotics of the background} \label{app:bgUVIR}

In this Appendix we calculate the UV and IR asymptotics for rather generic choices of the potentials $V_g$, $V_{f}$, and $\h$ of the action. For reference,  we first repeat here the background equations of motion \cite{jk}:
\begin{align}\nn
6A''+6(A')^2&=-{4\over3}{(\lambda')^2\over\lambda^2}+e^{2A}V_g(\lambda)- x
\,V_f(\lambda, \t)\,e^{2A}\sqrt{1+e^{-2A}\h(\lambda, \t)\ (\t')^2}\,,\\\label{eoms}
12(A')^2&={4\over3}{(\lambda')^2\over\lambda^2}+e^{2A}V_g(\lambda)- x\,
V_f(\lambda, \t)\,{e^{2A}\over\sqrt{1+e^{-2A}\h(\lambda, \t)\ (\t')^2}}\,,\\\nn
\lambda''-{(\lambda')^2\over\lambda}+3A'\,\lambda'&={3\over8}\,\lambda^2
\,e^{2A}\bigg[-{d\,V_g\over d\lambda}+ x\,{\partial V_f\over \partial \lambda}
\,\sqrt{1+e^{-2A}\h\ (\t')^2}\ \\\nn
&\quad + { x \over 2}\ {\partial \h\over \partial \lambda} {e^{-2A}V_f \ (\t')^2\over
\sqrt{1+e^{-2A}\h\ (\t')^2}}\ \bigg]\,,\\\nn
\t''+\ &{e^{-2A}}\left(4 \h \,A'+{\partial \,V_f\over \partial \lambda}\,{\h
\lambda'\over V_f} +{ \lambda'\over 2} {\partial \h \over \partial
\lambda}\right)(\t')^3+\left({1 \over 2\h}{\partial \h \over \partial \t}-{1 \over
V_f}{\partial \,V_f\over \partial \t}\right)(\t')^2+\\
+&\left(3\,A'+{\lambda' \over V_f}{\partial V_f \over \partial \lambda}+{\lambda'
\over \h}{\partial \h \over \partial \lambda}\right)\,\t'-{e^{2A} \over \h \
V_f}\,{\partial V_f \over \partial \t}=0\,. \label{teom}
\end{align}
 We shall use here the Ansatz (see the discussion in Sec.~\ref{sec:constraints})
\be
 V_f(\l,\t) = V_{f0}(\l) e^{-a(\l)\t^2}\,,
\ee
and assume that $\h(\l,\t)$ depends on $\l$ only.

\subsection{UV}
\label{uvasback}

We make here the standard assumption that all potentials are analytic in the UV and are then matched with perturbative QCD, as explained in Sec.~\ref{sec:constraints}.

\subsubsection{Fields $\l $ and $\Awf$}

Setting the tachyon to zero, the equations of motion for $\l$ and $A$ involve the effective potential
\be \label{Videfapp}
 V_{\rm eff}(\l)=V_g(\l)-x V_f(\l,0)=\frac{12}{\ell^2}\left[1 + V_1 \l +V_2
\l^2+\cdots \right] \, .
\ee
Then, the (leading) UV expansions of $\Awf$ and $\l$ can be found by substituting suitable Ans\"atze
in the equations of motion~\eqref{eoms}. The result reads
\begin{align} \label{UVexpsapp}
\Awf(r) \ =& -\log\frac{r}{\ell} + \frac{4}{9 \log(r \Lambda)}  \\
&+ \frac{
  \frac{1}{162} \left[95  - \frac{64 V_2}{V_1^2}\right] +
   \frac{1}{81} \log\left[-\log(r \Lambda)\right] \left[-23 + \frac{64
V_2}{V_1^2}\right]}{
  \log(r \Lambda)^2} +{\cal O}\left(\frac{1}{\log(r\Lambda)^3}\right) \nn \\
  V_1 \l(r)&=-\frac{8}{9 \log(r \Lambda)} + \frac{
   \log\left[-\log(r \Lambda)\right] \left[\frac{46}{81} - \frac{128 V_2}{81
V_1^2}\right]}{\log(r \Lambda)^2}+{\cal
O}\left(\frac{1}{\log(r\Lambda)^3}\right) \, .
\end{align}
Two combinations of the series coefficients of the effective potential appear here. As the potential
is matched with perturbative QCD, they become
\begin{align}
 V_1 &=  \frac{8}{9} b_0  = \frac{88-16x}{216 \pi^2}  \\
 \frac{V_2}{V_1^2} &= \frac{23}{64}+\frac{9 b_1}{16 b_0^2} =\frac{1}{64}
 \left(23+\frac{54 (34-13 x)}{(11-2 x)^2}\right)
\end{align}
where $b_i$ are the coefficients of the perturbative QCD $\beta$-function in the Veneziano limit. Our convention is such that $\beta(\l)\equiv d\l/d\log\mu = -b_0 \l +b_1 \l^2 + b_2\l^3+\cdots$.

\subsubsection{The tachyon}
\label{apptach}

As the tachyon is decoupled near the UV boundary, its UV behavior can be studied by
inserting the expansions calculated above for $\l$ and $\Awf$ into the tachyon EoM~\eqref{teom}.
We also develop the potentials as series in the UV:
\begin{align} \label{Vhexps}
 V_{\rm eff}(\l) &= V_g(\l)-x V_f(\l,0)=\frac{12}{\ell^2}\left[1 + V_1 \l +V_2
\l^2+\cdots \right] \\\nn
 \frac{\h(\l)}{a(\l)} &= \frac{2\ell^2}{3}\left[1 + \h_1 \l +\h_2 \l^2+\cdots
\right] \, .
\end{align}
Here the leading coefficient of $\h/a$
was already fixed in order to have the correct UV mass of the tachyon
\cite{ckp}.
The general solution for $r \to 0$ reads
\begin{align} \label{TUVres}
 \frac{1}{\ell}\tau(r) \ =\ &m_q r
(-\log(r\Lambda))^{-\rho} \left[1+ {\cal
O}\left(\frac{1}{\log(r\Lambda)}\right)\right] \\ \nn
&+\sigma r^3
(-\log(r\Lambda))^{\rho} \left[1+ {\cal
O}\left(\frac{1}{\log(r\Lambda)}\right)\right] \, .
\end{align}
Here matching with the perturbative anomalous dimension of the quark mass in QCD gives
\be \label{rhodef}
 \rho=  -\frac{4}{3}-\frac{4 \h_1}{3
V_1} = \frac{\gamma_0}{b_0} = \frac{9}{22-4 x}
\ee
where $\gamma_0$ is the leading coefficient of the anomalous dimension.

\subsection{IR}
\label{irasback}

We will only repeat here the discussion for the
particular asymptotics of $V_g$ in~\eqref{Vgas} that reproduces well several properties of QCD in the IR, \cite{ihqcd}.
The potential used in this article has this asymptotics.

\subsubsection{$\l$ and $\Awf$}

We  assume that the potential $V_g$ has the asymptotic behavior
\be \label{Vgas}
 V_g(\l) = v_0 \left(\frac{\l}{8\pi^2}\right)^{4/3}\sqrt{\log \frac{\l}{8\pi^2}}\left[1 + {v_1\over\log\left(\frac{\l}{8\pi^2}\right)} +
 {v_2\over \log^2\left(\frac{\l}{8\pi^2}\right)} + \cdots \right] \, .
\ee
Then the asymptotic solution to the equations~\eqref{eoms} reads
\begin{align} \label{IRresA}
 \Awf =& -\frac{r^2}{R^2}+\frac{1}{2}\log\frac{r}{R}-\log R-\frac{1}{2}\log v_0+
 \frac{5}{4}\log 2+\frac{3}{4}\log 3 + \frac{23}{24}+\frac{4 v_1}{3}\nn \\
&+\frac{R^2 \left(-173+512 v_1^2+1024 v_2\right)}{3456 r^2} + \mathcal{O}\left(r^{-4}\right)\\
\log \l &= \frac{3}{2}\frac{ r^2}{R^2}-\frac{23}{16} 
-2 v_1-\frac{R^2
\left(151+512 v_1^2+1024 v_2\right)}{2304 r^2}+ \mathcal{O}\left(r^{-4}\right)\,,
\label{IRresl}
\end{align}
where, for our choice of $V_g$ in~\eqref{potIandIIcommon},
\begin{align}
 v_0 &= (8\pi^2)^2\frac{92 \left(b^{\mathrm{YM}}_0\right)^2-144 b^{\mathrm{YM}}_1}{27\ell(0)^2} = \frac{18476}{243}\\
 v_1 &= \frac{1}{2} \ ; \qquad v_2 = - \frac{1}{8}\,.
\end{align}
Here we set the probe limit AdS radius $\ell(x=0)$ to one.
The IR scale $R=1/\Lambda_\mathrm{IR}$ appears as an integration constant.

\subsubsection{The tachyon}
\label{app:tachyonIR}

The IR expansion of the tachyon depends on the large-$\l$ asymptotics of the potentials
$\h$, $a$, and $V_{f0}$. Power-law asymptotics for the
potentials were analyzed in \cite{jk}.
We consider here a slightly more general case.

We discuss the following asymptotics of the potentials:
 \be \label{potIRas}
 \h(\l) \sim \h_c \l^{-\kappa_p} (\log \l)^{-\kappa_\ell} \ ; \qquad a(\l) \sim a_c \l^{a_p} (\log \l)^{a_\ell} \ ; \qquad
V_{f0}(\l) \sim v_c \l^{v_p} \, ,
\ee
where $\h_c$ and $a_c$ are assumed to be positive.
We also write the asymptotics of $\l$ and $\Awf$ as
\be
 \log \l =  \frac{3}{2}\frac{ r^2}{R^2}+\l_c + \mathcal{O}\left(r^{-2}\right) \ ; \qquad \Awf =  -\frac{ r^2}{R^2} + \frac{1}{2}\log\frac{r}{R}+A_c + \mathcal{O}\left(r^{-2}\right)
\ee
where
\begin{align}
 \l_c &= -\frac{23}{16} -2 v_1 = -\frac{49}{16} \\
 A_c &= -\log R-\frac{1}{2}\log v_0+
 \frac{5}{4}\log 2+\frac{3}{4}\log 3 + \frac{23}{24}+\frac{4 v_1}{3} \nn\\
 &= -\log R+ \frac{13}{8}+\frac{13}{4} \log3 + \frac{1}{4} \log2 - \frac{1}{2} \log4619 \\
 &\equiv -\log R + \bar A_c \, .
\end{align}
The tachyon EoM~\eqref{teom} has the form
\be
 \label{Teomg}
 \tau'' + F_1 \tau' + F_2 \tau'^3 + (1+F_3\tau'^2)(F_4 \tau  + F_5 \tau' \tau^2)  = 0 \, ,
\ee
where
\begin{align}
F_1 &= 3 A' + \l'\frac{d}{d\l}\log(\h(\l)V_{f0}(\l))\,,& F_4 &= \frac{2 a(\l)e^{2A}}{\h(\l)}\,,\\
F_2 &= \h(\l)e^{-2A}\, \left[4 A' + \l'\frac{d}{d\l}\log(\sqrt{\h(\l)}V_{f0}(\l))\right]\,,& F_5 &=-\l'\frac{da(\l)}{d\l} \,,\\
F_3 &=\h(\l)e^{-2A}\, .&
\end{align}
The IR asymptotics of the various factors are
\begin{align} \label{F1as}
 F_1&  \sim -\frac{3  (2+\kappa_p-v_p  ) }{R}\frac{r}{R} \\
 F_2&  \sim  -\frac{2^{\kappa_\ell-2}\h_c  (32-12v_p +6 \kappa_p ) e^{-2 A_c-\kappa_p  \lambda_c}}{3^{\kappa_\ell} R}\left(\frac{r}{R}\right)^{-2 \kappa_\ell } \exp \left[\left(2-\frac{3}{2}\kappa_p
\right)\left(\frac{r}{R}\right)^{2}\right]\\
 F_3 &\sim  \frac{2^{\kappa_\ell}\h_c  e^{-2 A_c-\kappa_p  \lambda_c} }{3^{\kappa_\ell}}\left(\frac{r}{R}\right)^{-1-2 \kappa_\ell }\exp \left[\left(2-\frac{3}{2}\kappa_p
\right)\left(\frac{r}{R}\right)^{2}\right]\\
 F_4 &\sim \frac{3^{a_\ell+\kappa_\ell}  a_c e^{2 A_c+\kappa_p  \lambda _c+a_p \lambda_c}}{2^{\kappa_\ell+a_\ell-1}\h_c}
\left(\frac{r}{R}\right)^{1+2a_\ell+2\kappa_\ell} \exp\left[\left(\frac{3}{2}\kappa_p
+\frac{3}{2} a_p-2\right)\left(\frac{r}{R}\right)^{2} \right]  \\
 F_5 &  \sim  -\frac{3^{a_\ell+1} a_c a_p   e^{a_p \lambda _c} }{2^{a_\ell}R}
\left(\frac{r}{R}\right)^{1+2a_\ell}\exp \left[ \frac{3}{2}a_p
\left(\frac{r}{R}\right)^{2}\right] \,.
\end{align}
For $a_p=0$ the leading behavior of $F_5$ is
\be \label{F5asv2}
  F_5   \sim  -\frac{3^{a_\ell} a_c a_\ell }{2^{a_\ell-1}R}
\left(\frac{r}{R}\right)^{2a_\ell-1} \,.
\ee
If $a(\l)$ is constant, $F_5$ vanishes. All the expressions have subleading corrections suppressed by $1/r^2$.

The most important parameters in the asymptotics of Eq.~\eqref{potIRas} are $a_p$ and $\kappa_p$. If $a_p<0$, the tachyon potential $V_f(\l,\tau)$ does not vanish asymptotically in the IR, which leads to a problem with flavor anomalies \cite{ckp}. Therefore we take $a_p\ge 0$. The special case $a_p=0$ and the region with $a_p>0$ are seen to have qualitatively different asymptotics.
As suggested by the exponential factors in~\eqref{F1as} --~\eqref{F5asv2}, the asymptotics also changes when $\kappa_p$ passes the critical value $4/3$.
Indeed we will have three choices for $\kappa_p$ which lead to different asymptotics: $\kappa_p>4/3$, $\kappa_p=4/3$, and $\kappa_p<4/3$. When either of these parameters takes its critical value, $a_p =0$ or $\kappa_p=4/3$, the logarithmic corrections in~\eqref{potIRas} will also be important. Notice also that when $\h_p=4/3$, the tachyon equation~\eqref{Teomg} can be written asymptotically, up to corrections suppressed by $1/r^2$, as
\be
 \label{Teomgs}
 \tau'' +  (1+F_3\tau'^2)(F_1 \tau' + F_4 \tau  + F_5 \tau' \tau^2)  \simeq 0 \, ,
\ee
which can be used to simplify the analysis.

\begin{figure}
\centerline{
\includegraphics[width=.45\textwidth]{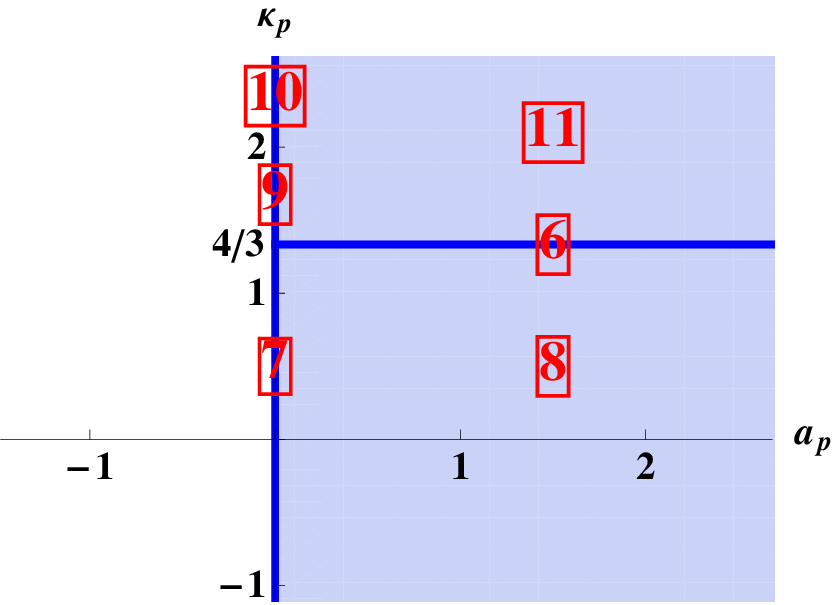}\hspace{0.05\textwidth}%
\includegraphics[width=.45\textwidth]{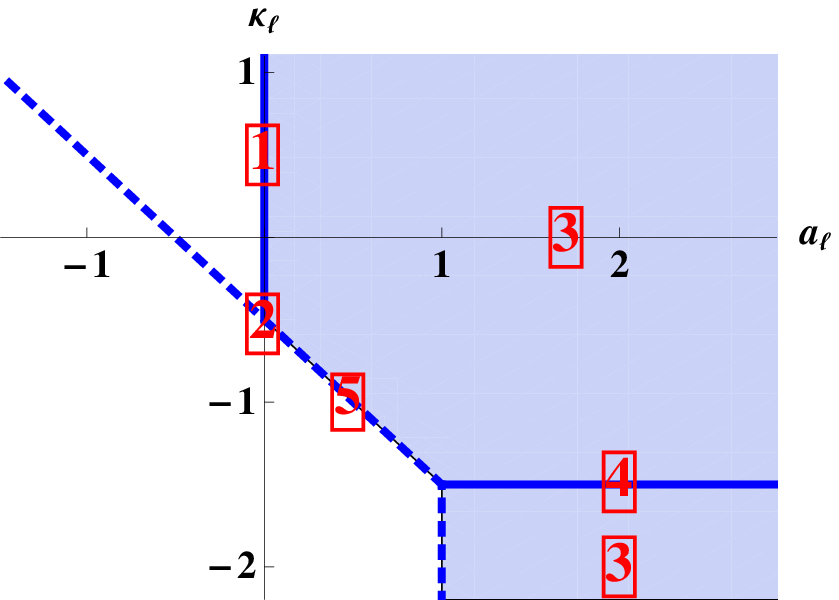}}
\caption{
Map of the acceptable IR asymptotics of the functions $\h(\l)$ and $a(\l)$ (see also Fig.~\protect\ref{fig:IRmaps}). Left: qualitatively different regions of tachyon asymptotics as a function of the parameters $\h_p$ and $a_p$ characterizing the power-law asymptotics of the functions.
Right: regions of tachyon asymptotics at the critical point $\h_p =4/3$, $a_p=0$ as a function of the parameters $\h_\ell$ and $a_\ell$ characterizing the logarithmic corrections to the functions.  The framed numbers refer to the various tachyon asymptotics listed in the text.
}\label{fig:IRmapsApp}
\end{figure}

We list here several potentially acceptable cases, i.e., choices of potentials for which there is a repulsive tachyon solution as $r \to \infty$ such that $V_f(\l,\t)$ goes to zero. We do not discuss the cases that do not satisfy this criterion.

We will require that $V_f(\l,\t)$ vanishes faster than a power of $e^{2A}$ or $1/\l$ in the IR, i.e., that the factor $a(\l)\t^2$ grows faster than $r^2$ in the IR.
This behavior guarantees that the DBI action (as well as its fluctuations) decouple from the glue in the IR. In cases where $V_f(\l,\t)$ vanishes slower than this
(see e.g. item 5 below)\footnote{One can also take $v_p \le 0$. Then  $V_f(\l,\t)$ vanishes in the IR also in many cases which are excluded below, e.g., in the excluded the cases where $0<a_\ell\le 1$.}
the decoupling is more involved. It also turns out that natural asymptotics of the meson
spectra are hard to obtain in such cases.

We will list all results with acceptable tachyon asymptotics.  A map of the asymptotics is shown in Fig.~\ref{fig:IRmapsApp}, where the shaded regions present acceptable solutions. The red framed numbers show where the various asymptotics of the
list lie on the map. The list below contains the results for asymptotics of the tachyon and of the tachyon kinetic term
\be
 K = e^{-2A} \kappa(\l) (\tau')^2\,,
\ee
which will be useful in the next subsection.
 We will also give the equations which the tachyon satisfies in the various cases after the list. In the cases below, the parameters $\kappa_\ell$, $a_\ell$ and $v_p$ can take any real values,
unless stated otherwise.

\bigskip

 \paragraph{1.} $\kappa_p=4/3$, $a(\l)=\mathrm{const.} = a_c$ (so $a_p=0=a_\ell$), $v_p<10/3$ and $\kappa_\ell>-1/2$. This is the case of potentials I,
which have $\kappa_\ell=0$ and $v_p=2$. We find that
 \be
  \tau(r) = \tau_0 \exp\left\{C_\mathrm{I} \left(\frac{r}{R}\right)^{1+2\kappa_\ell}\left[1+\mathcal{O}\left(\frac{1}{r^2}\right)\right]\right\}
 \ee
 where
 \be
  C_\mathrm{I} =  \frac{3^{\kappa_\ell}  a_c e^{2\bar A_c+4 \l_c/3}}{2^{\kappa_\ell-1}(10-3v_p)(1+2\kappa_\ell) \h_c}
 \ee
 and $\tau_0$ is a constant of integration. Here the corrections $\mathcal{O}(1/r^2)$ can be extended to a Taylor series in powers of $1/r^2$, the coefficients of which can be related to the coefficients in the asymptotic expansions of $\l$ and $\Awf$, and which are independent of $\tau_0$. The kinetic term diverges,
\be
 K \sim r^{2\kappa_\ell-1}\tau^2 \sim \exp\left[2 C_\mathrm{I} \left(\frac{r}{R}\right)^{1+2\kappa_\ell}\right]\,,
\ee
as $r \to \infty$.

\bigskip

 \paragraph{2.} $\kappa_p=4/3$, $a(\l)=\mathrm{const.} = a_c$ (so $a_p=0=a_\ell$), $v_p<10/3$ and $\kappa_\ell = -1/2$.
 Notice that this choice is the one singled out  in (\ref{finalchoice}), because
  this is the only set of parameters for which we obtain exactly linear
 trajectories for the meson spectra (as we will show in Appendix \ref{app:Regge}).
 In this case we find a power-law asymptotic for the tachyon,
 \be
  \tau(r) = \tau_0\left(\frac{r}{R}\right)^{C_\mathrm{II}}\left[1+\mathcal{O}\left(\frac{1}{r^2}\right)\right] \, ,
 \ee
where
 \be
  C_\mathrm{II} =  \frac{2\sqrt{2} a_c e^{2\bar A_c+4 \l_c/3}}{\sqrt{3}(10-3v_p) \h_c} \, .
 \ee
Note that requiring $V_f(\l,\t)$ to vanish in the IR gives the additional constraint $C_\mathrm{II}>1$. Taking this into account, the kinetic term diverges again:
\be
 K \sim r^{2\left(C_\mathrm{II}-1\right)}\,,\qquad (r \to \infty)\,.
\ee

\bigskip

 \paragraph{3.} $\kappa_p=4/3$, $a_p=0$, $a_\ell > 0$. In this case, which has the critical choice of $\kappa_p$ and $a_p$, the logarithmic corrections play a major role. The most generic solution of this type is found in two regions on the $(\kappa_\ell,a_\ell)$-plane. The first region is given by the equations $a_\ell>0$, $a_\ell+\kappa_\ell>-1/2$, and $\kappa_\ell>-3/2$, while the second one is limited by $a_\ell>1$ and $\kappa_\ell \le -3/2$.
In these regions the solution may be written as\footnote{We do not write down the order of the dropped subleading terms here, since they are somewhat complicated. There are different terms which are subleading with respect to the leading one by a (possibly noninteger) power of $1/r$. The same applies to the cases 4, 5, and 7 below.}
 \be
 \tau(r) \sim \sqrt{ C_\mathrm{III} \left(\frac{r}{R}\right)^{3+2\kappa_\ell}+\tau_0}\,, \qquad (r \to \infty)\,,
 \ee
 where
 \be
   C_\mathrm{III} = \frac{3^{\kappa_\ell} e^{\bar 2A_c+4 \l_c/3}}{2^{\kappa_\ell-1}a_\ell \h_c (3+2\kappa_\ell)} \, .
 \ee
  For $\kappa_\ell>-3/2$, the tachyon therefore has a power-law divergence. For $\kappa_\ell \le -3/2$, the tachyon tends to a constant. In the latter case, the integration constant $\tau_0$ needs to be positive.

For the kinetic term we find
 \be
  K \sim \mathrm{const}\,,\qquad (r \to \infty)\,,
 \ee
 when $\kappa_\ell>-3/2$. When $\kappa_\ell<-3/2$,  the kinetic term vanishes asymptotically:
 \be
  K \sim r^{3+2\kappa_\ell}\,,\qquad (r \to \infty)\,.
 \ee

On the line with $a_\ell=1$, $\h_\ell<-3/2$, the solution is qualitatively similar and can be written as
\be
\tau(r) \sim \t_0 - \frac{3^{\h_\ell+1}e^{\bar 2A_c+4 \l_c/3 a_c \t_0 }}{2^{\h_\ell}\h_c(-2\h_\ell-3)(10-3v_p+3 a_c \t_0^2)}\left(\frac{r}{R}\right)^{3+2\kappa_\ell} \, .
\ee
The requirement that $V_f(\l,\t)$ vanishes in the IR may be met, but only for large enough values of $\tau_0$, and the convergence to zero is slow (only a power of $1/\l$). Therefore we will exclude this particular case.

\bigskip

\paragraph{4.} $\kappa_p=4/3$, $a_p=0$, $\kappa_\ell =-3/2$. In addition we require that $a_\ell \ge 1$. This is a limiting case of the previous asymptotics as $\kappa_\ell \to -3/2$. Notice however that the region of validity is now extended to $a_\ell = 1$. We find that\footnote{For $a_\ell =1$ there is also a term $\sim \log\log r$ under the square root which is leading with respect to the constant $\tau_0$.}
 \be
 \tau(r) \sim \sqrt{ C_\mathrm{IV} \log\frac{r}{R}+\tau_0}\,,
 \ee
 where
 \be
   C_\mathrm{IV} = \frac{3^{\kappa_\ell}}{2^{\kappa_\ell-1}a_\ell \h_c } e^{\bar 2A_c+4 \l_c/3} \, .
 \ee
The kinetic term behaves as
\be
 K \sim \frac{1}{\log r}\,.
\ee

\bigskip

\paragraph{5.} $\kappa_p=4/3$, $a_p=0$, $a_\ell=-\kappa_\ell -1/2$. In addition we require that $-3/2<\kappa_\ell <-1/2$. Define first
\be
 C_\mathrm{V} = \frac{3^{\kappa_\ell+1/2}(10-3v_p)}{2^{\kappa_\ell+3/2} a_c (-\kappa_\ell-1/2)} \,.
\ee
If $C_\mathrm{V} > \frac{C_\mathrm{IV}}{(3+2\kappa_\ell)}$, we find that
\be
 \tau(r) \sim \tau_0 r^{C_\mathrm{IV}/(2 C_\mathrm{V})} \, .
\ee
Then, however, ${C_\mathrm{IV}/(2 C_\mathrm{V})}< 3/2+\kappa_\ell<1$, and consequently $a(\l)\t^2$ grows in the IR slower than $r^2$, so the tachyon potential does not vanish in the IR.

When  $C_\mathrm{V} < \frac{C_\mathrm{IV}}{(3+2\kappa_\ell)}$ the asymptotics is
\be
 \tau(r) \sim \sqrt{\left(C_\mathrm{III}-C_\mathrm{V}\right) r^{3+2\kappa_\ell}} \, .
\ee
where we used $C_\mathrm{III} =  \frac{C_\mathrm{IV}}{(3+2\kappa_\ell)}$.
In this case we omitted some subleading terms, which involve an integration constant.

Now $a(\l)\t^2 \sim r^2$ in the IR. One can check that $V_f(\l,\t)$ vanishes in the IR if
$C_\mathrm{III}-C_\mathrm{V}$ is large enough.
A further check would be needed to see when this is enough to decouple the tachyon in the IR, as $V_f(\l,\t)$ only vanishes as a power of $1/\l$. Decoupling is, however, guaranteed as $\h_\ell \to -3/2$ since $C_\mathrm{III}$ diverges. In this case the kinetic term $K$ asymptotes to a constant as $r\to\infty$.

\bigskip

 \paragraph{6.} $\kappa_p=4/3$, $a_p>0$.
Potentials II
belong to this class. The tachyon asymptotics is now
 \be \label{tauc6}
  \tau(r)  = \sqrt{ C_\mathrm{VI}\left(\frac{r}{R}\right)^{1+2\kappa_\ell}\left[1+\mathcal{O}\left(\frac{1}{r^2}\right)\right]+\tau_0}
 \ee
 where
 \be
  C_\mathrm{VI} =  \frac{3^{\kappa_\ell-1} e^{2 \bar A_c+4 \l_c/3}}{2^{\kappa_\ell-2}a_p \h_c (1+2\kappa_\ell)}
 \ee
For $\kappa_\ell>-1/2$ the asymptotics is a power law. For $\kappa_\ell<-1/2$, the tachyon tends to a constant value and $\tau_0$ needs to be positive. The case $\kappa_\ell=-1/2$ is obtained by taking the limit $\kappa_\ell \to -1/2$ in~\eqref{tauc6}, so that a logarithm arises under the square root.

We find that the kinetic term vanishes asymptotically:
 \be
  K \sim 1/r^2\,,\qquad (r \to \infty)\,,
 \ee
 when $\kappa_\ell>-1/2$, and
 \be
  K \sim r^{2\kappa_\ell-1}\,,\qquad (r \to \infty)\,.
 \ee
when $\kappa_\ell<-1/2$.

\bigskip

 \paragraph{7.}  $\kappa_p<4/3$, $a_p=0$.
 The tachyon tends to a constant in the IR.
For $a_\ell>1$,
 \be
  \tau(r) \sim \tau_0 - \frac{C_\mathrm{VII}}{\tau_0} \left(\frac{r}{R}\right)^{1+2 \kappa_\ell } \exp\left[\left(\frac{3 \kappa_p }{2} -2\right)\left(\frac{r}{R}\right)^2\right]
 \ee
 where
 \be
   C_\mathrm{VII} = \frac{3^{\kappa_\ell} e^{2\bar A_c + \kappa_p\l_c}}{2^{\kappa_\ell}a_\ell (4-3\kappa_p)\h_c }  \, .
 \ee
The tachyon kinetic term vanishes fast in the IR:
\be
 K \sim \exp\left[-\left(2-\frac{3}{2}\kappa_\ell\right)\left(\frac{r}{R}\right)^2\right]\,.
\ee

For $a_\ell=1$ (assuming $v_p<10/3$)\footnote{It is not enough to require that $v_p<10/3+a_c\tau_0^2$. For $10/3<v_p<10/3+a_c\tau_0^2$ the IR singularity is not repulsive.} the result can be written as
\be
 \tau(r) \sim \tau_0 - \frac{3^{\kappa_\ell+1} a_c e^{2\bar A_c + \kappa_p\l_c} \tau_0}{2^{\kappa_\ell}\h_c (4-3\kappa_p)(10-3v_p+3a_c\tau_0^2) } \left(\frac{r}{R}\right)^{1+2 \kappa_\ell } \exp\left[\left(\frac{3 \kappa_p }{2} -2\right)\left(\frac{r}{R}\right)^2\right] \, .
\ee
Notice, however, that depending on the value of $\tau_0$, the tachyon potential may not vanish in the IR in this case.

\bigskip

 \paragraph{8.}  $\kappa_p<4/3$, $a_p>0$.
 Assuming $v_p<8/3+\kappa_p/2$,
 \be
   \tau(r) = \tau_0 - \frac{C_\mathrm{VIII}}{\tau_0}\left(\frac{r}{R}\right)^{2\kappa_\ell-1}  \exp\left[\left(\frac{3 \kappa_p }{2} -2\right)\left(\frac{r}{R}\right)^2\right]\left[1+\mathcal{O}\left(\frac{1}{r^2}\right)\right]
  \ee
  where
 \be
   C_\mathrm{VIII} = \frac{3^{\kappa_\ell-1} e^{2\bar A_c + \kappa_p\l_c}}{2^{\kappa_\ell-1}  \h_c a_p(4-3\kappa_p)} \, .
 \ee
The kinetic term behaves similarly as in the previous case:
\be
 K \sim \exp\left[-\left(2-\frac{3}{2}\kappa_\ell\right)\left(\frac{r}{R}\right)^2\right]\,.
\ee

\bigskip

 \paragraph{9.}  $\kappa_p>4/3$, $a(\l)=\mathrm{const.}=a_c$ ($a_p=0=a_\ell$).
Then
 \be
  \tau(r) = \tau_0 \exp\left\{ C_\mathrm{IX}\left(\frac{r}{R}\right)^{2\kappa_\ell-1 }  \exp\left[\left(\frac{3 \kappa_p }{2} -2\right)\left(\frac{r}{R}\right)^2\right]\left[1+\mathcal{O}\left(\frac{1}{r^2}\right)\right]\right\}
 \ee
  where
 \be
   C_\mathrm{IX} =  \frac{3^{\kappa_\ell} a_c e^{2\bar A_c+\kappa_p \l_c} }{2^{\kappa_\ell-2} \h_c (16-6v_p+3\kappa_p)(3\kappa_p-4)}\, .
 \ee
The tachyon kinetic term diverges very fast in the IR, essentially as tachyon squared.

\bigskip

 \paragraph{10.}  $\kappa_p>4/3$, $a_p=0$, $a_\ell > 0$.
Then
 \be
  \tau(r) = C_\mathrm{X} \left(\frac{r}{R}\right)^{1/2+\kappa_\ell} \exp\left[\left(\frac{3 \kappa_p }{4} -1\right)\left(\frac{r}{R}\right)^2\right]\left[1+\mathcal{O}\left(\frac{1}{r^2}\right)\right]
 \ee
  where
 \be
   C_\mathrm{X} =   \frac{3^{\kappa_\ell/2}  e^{\bar A_c+\kappa_p \l_c/2} }{2^{\kappa_\ell/2-1/2} \sqrt{\h_c a_\ell(3\kappa_p-4)}}\, ,
 \ee
 and the integration constant appears in exponentially subleading terms which we have not written down. The kinetic term has the asymptotics
\be
 K \sim r^2\,,\qquad(r\to\infty)\,.
\ee

\bigskip

 \paragraph{11.}  $\kappa_p>4/3$, $a_p>0$.
Then
 \be
  \tau(r) = C_\mathrm{XI} \left(\frac{r}{R}\right)^{\kappa_\ell-1/2} \exp\left[\left(\frac{3 \kappa_p }{4} -1\right)\left(\frac{r}{R}\right)^2\right]\left[1+\mathcal{O}\left(\frac{1}{r^2}\right)\right]
 \ee
  where
 \be
   C_\mathrm{XI} =   \frac{3^{\kappa_\ell/2-1/2}  e^{\bar A_c+\kappa_p \l_c/2} }{2^{\kappa_\ell/2-1} \sqrt{\h_c a_p(3\kappa_p-4)}}\, .
 \ee
In this case the kinetic term asymptotes to a constant in the IR.

In summary, the acceptable asymptotics are shown as shaded regions and solid blue lines in Fig.~\ref{fig:IRmapsApp}.

 We briefly comment on the cases which were not covered by the above list. For $a_p<0$, or when $a_p = 0$ and $a_\ell<0$, the tachyon EoM often does not admit a regular, nontrivial solution that extends to $r \to \infty$. In the cases where such solution exists, $V_f$ does not vanish in the IR.
The same applies to those cases with $a_p=0$, $\h_p \le 4/3$, and with $0\le a_\ell \le 1$ which were not covered by the list above. We often required that $v_p$ is below a certain critical value. If this value is exceeded, the tachyon solution is not repulsive at $r \to \infty$. In addition, typically $V_f$ does not vanish in the IR in such cases.

Even though the above list contains many qualitatively different acceptable asymptotics, they can be classified in two classes for which the tachyon satisfies asymptotically two different simple equations. The classes are\footnote{The cases where the potential $V_f(\l,\t)$ does not vanish in the IR, or only vanishes as a power of $1/\l$ (including item 5 and the last expressions of items 3 and 7 in the list), fall outside of this classification.}
\begin{enumerate}
 \item Constant $a(\l)$ (items 1, 2, and 9 in the list). Then the tachyon asymptotics arises from the terms involving the coefficients $F_2$ and $F_4$. Asymptotically the tachyon therefore satisfies
 \be \label{taueqconst}
   e^{-2A}\h(\l) \, \left[4 A' + \l'\frac{d}{d\l}\log(\sqrt{\h(\l)}V_{f0}(\l))\right]\t' + 2 a(\l)\, \t \simeq 0 \, .
 \ee
 \item Non-constant $a(\l)$. Then the tachyon asymptotics arises from the terms involving the coefficients $F_4$ and $F_5$. Asymptotically we find
 \be \label{taueqnonconst}
   \l'\frac{da(\l)}{d\l}\, \t\t' = \frac{2 a(\l)e^{2A}}{\h(\l)}\, .
 \ee
\end{enumerate}

\subsubsection{The tachyon with modified power in the DBI action} \label{app:modDBI}

We have also studied how the tachyon asymptotics change if we allow for different powers from the usual ${1\over 2}$ in the DBI action. More precisely, we replace
\be
 \sqrt{-\det(g_{\mu\nu} + \kappa\, \partial_\mu \tau \partial_\nu \tau)} = \sqrt{-\det g_{\mu\nu}}\sqrt{\det(\delta^\mu_\nu+ \kappa\,  \partial^\mu \tau \partial_\nu \tau)}
\ee
by
\be \label{DBImod}
 \sqrt{-\det g_{\mu\nu}}\,\left[\det(\delta^\mu_\nu+ \kappa\,  \partial^\mu \tau \partial_\nu \tau)\right]^b
\ee
where $b$ is a free parameter. Notice that this replacement maintains the diffeomorphism invariance of the action. As we are interested in the background solutions, we set here the gauge fields to zero and assumed that $T = \tau(r) \mathbb{I}_{N_f}$.

As we require that the tachyon decouples asymptotically in the IR, it is enough to study how the tachyon EoM and asymptotics change when $b \ne 1/2$. The tachyon EoM can be simplified to
\be \label{modteom}
 \begin{split}
 &\left[1+(2b-1)K\right]\tau'' + \frac{a}{b \hat\kappa}\left[1+2(1-b)K-(2b-1)K^2\right]\tau\\
 & +\left[(1+K)\partial_r \log \hat V+(1+bK)\partial_r\log \hat\kappa\right]\tau'- \l'\,\partial_\l a\,(1+K)\tau^2\tau'=0\,,
 \end{split}
\ee
where
\be
 \hat \kappa = e^{-2A} \kappa\,,\qquad \hat V = e^{5A} V_{f0}\,,\qquad K= e^{-2A} \kappa (\tau')^2 = \hat \kappa\,   (\tau')^2\,.
\ee
We immediately notice that cancellations take place in~\eqref{modteom} for $b=1/2$ which suggest that it is a special value.

We now discuss what happens to the tachyon asymptotics which were solved for at $b=1/2$ in the previous subsection. Depending on the behavior of the factor $K$, the various solutions can be divided into three classes:
\begin{itemize}
 \item[a)] Solutions where $K$ vanishes asymptotically in the IR. These include the cases 3 (when $\kappa_\ell<-3/2$), 4, 6, 7, and 8 in Sec.~\ref{app:tachyonIR}. As $K$ vanishes, the modification in~\eqref{DBImod} is unimportant in the IR, and the solutions of  Sec.~\ref{app:tachyonIR} remain valid for $b \ne 1/2$ (the dependence on $b$ needs to be included in some coefficients).
 \item[b)] Solutions where $K$ asymptotes to a constant in the IR. These include the cases 3 (when $\kappa_\ell>-3/2$), 5, and 11 in Sec.~\ref{app:tachyonIR}.
These solutions are unchanged at qualitative level for $b\ne 1/2$, at least when $b$ is not too large.\footnote{For large values of $b$ it may not be possible to find such solution to~\eqref{modteom} that the various coefficient factors asymptote to constants having the same signs as for the $b=1/2$ solutions. In such cases the tachyon asymptotic must change qualitatively.} We have verified this numerically in several cases.
 \item[c)] Solutions where $K$ diverges asymptotically in the IR. These include the cases 1, 2, 9, and 10 in Sec.~\ref{app:tachyonIR}. For large $K$ and $b\ne 1/2$, factors which vanish at $b=1/2$ may dominate the EoM in~\eqref{modteom}. Therefore drastic changes are possible when $b$ is moved away from its critical value $1/2$.
\end{itemize}

Interestingly, the class c) includes potentials~I as well as the tachyon asymptotics which was singled out in Sec.~\ref{sec:constraints} as the only one having exactly linear meson trajectories (cases 1 and 2 above). Recall that these choices are also motivated by the relation to string theory, as pointed out in Sec.~\ref{sec:string}. We have checked what happens for $b\ne 1/2$ for these potentials.

For $b<1/2$ we could not find any regular solutions, numerically or analytically. When $b>1/2$, there is a regular solution for which the tachyon diverges as
\be \label{spectasympt}
 \tau(r) \sim r^{\kappa_\ell+3/2}\,.
\ee
(Notice that $\kappa_\ell\ge -1/2$ for these potentials.) Therefore,  the asymptotics indeed change qualitatively whenever $b$ deviates from its critical value. For $\kappa_\ell=-1/2$ (case 2) the tachyon divergence is only linear which is not enough to guarantee the decoupling of the tachyon as discussed in Sec.~\ref{app:tachyonIR}, so this case is ruled out.

Interestingly, the solution~\eqref{spectasympt} is smoothly connected to the case 3 of Sec.~\ref{app:tachyonIR}, where similar power-law asymptotics were found for $a_\ell>0$ (and $b=1/2$). It is easy to see that the analysis of asymptotic meson trajectories for the case 3 from Appendix~\ref{app:Regge} also applies to the present case. Therefore,  the trajectories are not linear, but
\be
 m_n^2 \sim n^{4\frac{\kappa_\ell+1}{2\kappa_\ell+3}}\,.
\ee

In conclusion, we did not find any extra solutions, which would satisfy all constraints from QCD, by generalizing to $b \ne 1/2$. To the contrary, the single choice of potentials which has linear trajectories, only works for $b=1/2$.

\section{Asymptotic spectrum and radial trajectories} \label{app:Regge}

\subsection{Flavor non-singlet trajectories}

We now calculate how meson masses scale at large excitation number $n$. The scaling of the decay constants is discussed in Appendix~\ref{app:asdecconn}
We start from the flavor non-singlet sector where the fluctuation equations can be transformed into the Schr\"odinger form.
The IR asymptotics of Schr\"odinger potentials are computed for all excitations in the different cases of tachyon asymptotics, which are given in section~\ref{app:tachyonIR}.

For ease of reference, we repeat here the results for the Schr\"odinger potential from Appendix~\ref{app:quadfluctdet}. The potential is defined by
\be \label{VSchRegge}
 V(u) = \frac{1}{\Xi(u)}\frac{d^2\Xi}{du^2} + H(u)\,.
\ee
Here the functions $\Xi$ and $H$ are for the vector, axial vector, pseudoscalar, and scalar sectors
\begin{align} \label{XisHs}
\Xi_V&=V_f(\l,\t)^{1/ 2} \gf(\l,\t) e^{A / 2}\,, &H_V&=0\, ,\nn \\
\Xi_A&=\Xi_V \,, &H_{A}&= 4 e^{2 A} \t^2 {\kappa(\l,\t) \over
  \gf(\l,\t)^2}\, ,\nn\\
\Xi_{P}&=V_f(\l,\t)^{-1/2} \t^{-1} e^{-3 A / 2}
\kappa(\l,\t)^{-1/2} \,,  &H_{P}&=H_A\, ,\nn\\
 \Xi_{S}&=V_f(\l,\t)^{1\over 2} e^{3 A / 2} G^{-1}\kappa(\l,\t)^{1/2} \,, & H_{S}&=-2 e^{2 A}
{a(\l) \over \kappa(\l,\t)}\, ,
\end{align}
respectively, and the coordinate $u$ is defined by
\be
 \frac{du}{dr} = G(r) = \sqrt{1 + e^{-2A(r)}\h(\l,\t)(\t')^2 }\,.
\ee

The WKB approximation is then used
to find meson masses in terms of $n$.
The Schr\"odinger potential have the UV asymptotics
\be
V(u) \sim {v_\mathrm{UV} \over u^2} \, ,\qquad  u \simeq r\,,\qquad (u \to 0)\,.
\label{uvexppot}
\ee
where $v_\mathrm{UV}$ equals $15/4$ for the pseudoscalars (assuming zero quark mass) and $3/4$ for the other towers.

As we shall see shortly, the UV asymptotics of the potential is not relevant in the calculation.
All tachyon asymptotics which we shall consider, are covered by two qualitatively different
cases for the IR asymptotics of the potential:
\begin{align}
\hspace{0.1\linewidth} &\hspace{0.15\linewidth} & \hspace{0.05\linewidth} &\hspace{0.15\linewidth} &\hspace{0.05\linewidth} & \hspace{0.5\linewidth} \nonumber \\[-\baselineskip]
&\textbf{I}& V(u) &\sim v_{s}^2 u^{p} \log^{q} u\,,& (u &\to \infty)\,,
\label{VdefI} \\
&\textbf{II}& V(u) &\sim v_s^2 e^{2 p u^2}\,,& (u &\to \infty)\,, \label{VdefII}
\end{align}
where $v_s$, $p$ and $q$ can be determined in terms of the potentials of the action separately for each case.

For large excitation number the mass eigenvalues satisfy
\be
{d (m_n^2) \over dn} \sim 2 \pi \left(\int_{u_1}^{u_2} {du \over \sqrt{m_n^2-V(u)}} \right)^{-1}\,,\qquad (n \to\infty)\,,
\label{wkbf}
\ee
where $u_1$ and $u_2$ are turning points of the potential.
The contribution to the integral from the regime close to the UV turning point $u_1$ is $\propto 1/m_n$ and will be subleading at large $n$. Therefore the exact value of $u_1$ is unimportant.

For large $m_n$, and for the two different cases for the potential the IR turning point $u_2$ satisfies
\begin{align}
\hspace{0.1\linewidth} &\hspace{0.05\linewidth} & \hspace{0.05\linewidth} &\hspace{0.8\linewidth}\nonumber \\[-\baselineskip]
&\textbf{I}& &u_2^p \log^q u_2 \,\simeq \left({m_n\over v_s}\right)^{2} \Rightarrow u_2 \simeq \left({m_n\over v_s}\right)^{2 \over p} \left({2 \over p} \log{m_n \over v_s} \right)^{-{q \over p}}\,,
\\
&\textbf{II}& &u_2\simeq\sqrt{{1\over p} \log \left({m_n \over v_s}\right)}\,.
\end{align}
In each case, the integral in Eq.(\ref{wkbf}) is dominated by its IR limit as $m_n \rightarrow
\infty$. Therefore,
\begin{align}
\hspace{0.1\linewidth} &\hspace{0.05\linewidth} & \hspace{0.25\linewidth} &\hspace{0.6\linewidth}\nonumber \\[-\baselineskip]
&\textbf{I}&
\int^{u_2}\!\! {du \over m_n \sqrt{1-{v_s^2 \over m_n^2} u^p \log^{q} u}}&={u_2
  \over m_n} \int^1 {dx \over \sqrt {1-\left(1+{\log x \over \log u_2}
    \right)^q x^p}} \nn\\
&&&\simeq {m_n^{{2\over p}-1}\over v_s^{2/p}}  \left({2 \over p} \log{m_n \over v_s} \right)^{-{q \over p}} \int^1 \!\!{dx \over \sqrt {1-x^p}}\,,
\label{wkbint1} \\
&\textbf{II}&
\int^{u_2}\!\! {du \over m_n \sqrt{1-{v_s^2 \over m_n^2} e^{2 p u^2}}}&={u_2
  \over m_n} \int^1 \!\!{dx \over \sqrt {1-{v_s^2 \over m_n^2} e^{\log(m_n^2/v_s^2) x^2}}}\nn\\
&&&\simeq {1\over m_n}\sqrt{{1\over p} \log \left({m_n \over v_s}\right)} \int^1 \!\! dx\,.
\label{wkbint2}
\end{align}
Consequently, Eq.~(\ref{wkbf}) becomes,
\begin{align}
\hspace{0.1\linewidth} &\hspace{0.1\linewidth} & \hspace{0.1\linewidth}&\hspace{0.05\linewidth} &\hspace{0.05\linewidth}&\hspace{0.65\linewidth}\nonumber \\[-\baselineskip]
&\textbf{I}& {d (m_n^2) \over dn}&\sim m_n^{1- {2 \over
    p}} \left({2 \over p} \log {m_n \over v_s} \right)^{{q \over p}}\,,& (n &\to \infty)\,,
\\
&\textbf{II}& {d (m_n^2) \over dn}&\sim {m_n \over \sqrt{\log\left({m_n \over v_s}\right)}}\,,&  (n &\to \infty)\,.
\end{align}
Finally, we find the mass in terms of $n$ as
\begin{align}
\hspace{0.1\linewidth} &\hspace{0.1\linewidth} & \hspace{0.1\linewidth}&\hspace{0.05\linewidth} &\hspace{0.05\linewidth}&\hspace{0.65\linewidth}\nonumber \\[-\baselineskip]
&\textbf{I}& m_n^2 &\sim v_s^\frac{4}{2+p} n^{2p \over 2+p} \left( \log n \right)^{{2 q \over 2
      +p}} \,,& (n &\to \infty)\,,
\label{masI} \\
&\textbf{II}& m_n^2 &\sim {n^2 \over \log n}\,,& (n &\to \infty)\,. \label{masII}
\end{align}
Hence, the spectrum follows linear trajectories only for $p=2$ and $q=0$ in the first case. Notice that the proportionality constant in~\eqref{masI} depends on $v_s$, as we have stressed by writing down the dependence explicitly, while the leading term in~\eqref{masII} is independent of $v_s$.\footnote{The leading result in~\eqref{masII} also remains unchanged if the include power-like corrections in~\eqref{VdefII}.}

We now proceed to the analysis of the different potentials which were
presented in the Appendix (\ref{apptach}). In each case, we need to analyze
the IR asymptotics of the two terms
 of the Schr\"odinger potential in~\eqref{VSchRegge}.
The first term, $\Xi''(u)/\Xi(u)$, will usually have the same IR asymptotics for the
 different excitation towers (but will be dependent on the tachyon asymptotics).

 As pointed out above, the Schr\"odinger coordinate $u$ will also be tower independent. For the second term in~\eqref{VSchRegge}, we have two nontrivial functions, given as $H_A$ and $H_S$ in~\eqref{XisHs}, and we need to check separately if they contribute to the potential.
As we shall see, $H_S$ is always suppressed in the IR, but $H_A$ may contribute depending on our choice of $w(\l)$. Usually $H_A$ contributes, if $w(\l)$ vanishes faster in the IR than $\kappa(\l)$.

We parametrize the infrared asymptotics of the potential functions as above:
 \begin{align} \label{potiras}
 \h(\l) &\sim \h_c \l^{-\kappa_p} (\log \l)^{-\kappa_\ell} \, , & a(\l) &\sim a_c \l^{a_p} (\log \l)^{a_\ell} \, ; \nonumber\\
V_{f0}(\l) &\sim v_c \l^{v_p}  \, , &\gf(\l) &\sim \l^{-\nw} (\log \lambda)^{-w_\ell} \, .
\end{align}

We  then discuss the behavior in the various cases of tachyon asymptotics. As above, the numbers in the list are mapped to values of the coefficients of~\eqref{potiras} as shown in Fig.~\ref{fig:IRmapsApp}. We shall fix the units such that the IR scale $R=1$ for simplicity.

\bigskip
\paragraph{1.}
$\kappa_p=4/3$, $a(\l)=\mathrm{const.} = a_c$ (so
$a_p=0=a_\ell$), $v_p<10/3$ and $\kappa_\ell>-1/2$.
The asymptotics of the tachyon is in this case
\be
  \tau(r) \sim \tau_0 \exp\left[C_\mathrm{I} r^{1+2\kappa_\ell}\right] \ ; \qquad (r \to \infty)\,.
 \ee

In this case the Schr\"odinger coordinate behaves in the IR as
\be
u\sim r^{-\kappa_\ell-{1\over 2}}\, \tau \sim  e^{C_\mathrm{I} r^{1+2\kappa_\ell} } \, .
\ee
The various terms of the Schr\"odinger potential behave as
\begin{align}
{1\over \Xi} \label{Xic1}
{d^2 \Xi \over du^2} &= v_s^2 u^2 \log^2 u \left[1+ \morder{\frac{1}{\log u}}\right]\,,\\
\label{Hc1}
H_A &= c_1 u^2 (\log u)^{{2+4 w_\ell \over 1+ 2 \kappa_\ell}} \left[ 1+ {\mathcal O}\left({1 
\over \log u}\right)\right]\, \nn\\
 &\quad\times \exp\left\{(3 \nw -4)\left(
    {\log u \over C_\mathrm{I}} \right)^{2 \over 1+2 \kappa_\ell} \left[  1+ {\mathcal O} \left({1\over \log u} \right)  \right]\right\}\,,\\ \nonumber
H_S &\sim \log u\,,
\end{align}
as $u \to\infty$. Here $c_1$ and $v_s$ can be calculated in terms of the potentials, and $v_s$ is the same for all excitation towers.

For $w_p \le {4\over 3}$, $H_A(u)$ in~\eqref{Hc1} is suppressed in the IR with respect to $\Xi''(u)/\Xi(u)$ in~\eqref{Xic1}. Comparing~\eqref{Xic1} to~\eqref{VdefI}, the asymptotics of the masses in~\eqref{masI} becomes
\be \label{case1res}
m_n^2\sim n  \log n\,.
\ee
Because $v_s$ in~\eqref{Xic1} was the same for all excitation towers, also the slopes of the various towers, i.e., the proportionality constants in~\eqref{case1res}, are the same. Therefore,  the trajectories are almost, but not exactly, linear.
For $w_p={4 \over 3}$, and if  $w_\ell<\kappa_\ell$, the extra term $H_A(u)$ is again suppressed with respect to $\Xi''(u)/\Xi(u)$ ,~\eqref{case1res} holds, and the slopes are the same. In other cases the slopes, or even the asymptotic behavior of $m_n$, are different among the various towers due to extra contribution from $H_A$.

\bigskip


\paragraph{2.} $\kappa_p=4/3$, $a(\l)=\mathrm{const.} = a_c$ (so
   $a_p=0=a_\ell$), $v_p<11/4$ and $\kappa_\ell =-
   1/2$. The asymptotics of the tachyon is in this case
 \be
  \tau(r) \sim \tau_0r^{C_\mathrm{II}}\,.
 \ee
For $C_\mathrm{II}>1$ we find that
\be
u \sim c_1 \t \sim c_1\,\t_0\,r^{C_\mathrm{II}} \,,\qquad{1\over \Xi}{d^2 \Xi \over du^2} = v_s^2 u^2\left[1+ \mathcal{O} \left( u^{ -\epsilon}\right) \right]
\ee
as $u \to\infty$, where $v_s$ is the same for all towers, $c_1$ is a coefficient which is determined in terms of the parameters of the potentials, and $\epsilon = \min(2/C_\mathrm{II},2-2/C_\mathrm{II})$. We also find that
\begin{align}
H_A &= c_2\exp\left\{ (3 \nw-4) \left(\frac{u}{c \t_0}\right)^{2 \over C_\mathrm{II}} \left[1+ \mathcal{O} \left( u^{ -\epsilon}\right) \right]\right\} u^{2 +{2 \over C_\mathrm{II}}+4 {w_\ell \over C_\mathrm{II}}} \left[1+ \mathcal{O} \left( u^{ -\epsilon}\right) \right] \,,\nn\\
H_S &\sim \textrm{const.}
\end{align}
For $w_p < 4/3$, $H_A$ is suppressed and the spectrum is linear:
\be
m_n^2 \sim n
\ee
for large $n$, and all towers have the same slopes. In the critical case $w_p=4/3$ we find the same result, if also $w_\ell<-1/2 (=\kappa_\ell)$. If ($w_p=4/3$ and) $w_\ell=-1/2$ the slopes of the axial vectors and pseudoscalars are larger than to those of the vectors and scalars due to the contribution from $H_A$. If $w_\ell>-1/2$ the trajectories of axials and pseudoscalars are no longer linear.

When $C_\mathrm{II}\le 1$, the tachyon potential does not vanish in the IR, so these potentials are not acceptable. If we anyhow repeat the above calculation, we find quite similar results as above. In particular,
\begin{align}
 {1\over \Xi}{d^2 \Xi \over du^2} = v_s^2 u^2\left[1+ \mathcal{O} \left( u^{ -\epsilon}\right) \right]
\end{align}
but now the coefficient $v_s$ may vary between the different towers. Therefore,  the spin-one and spin-zero trajectories have, in general, different slopes (if $H_A$ is asymptotically suppressed). However, if $w_p$ has the critical value $4/3$, the slopes turn out to be the same.


\bigskip

\paragraph{3.} $\kappa_p=4/3$, $a_p=0$, and with $a_\ell$ and $\kappa_\ell$ constrained as depicted in Fig.~\ref{fig:IRmapsApp} (right). The
  tachyon has the asymptotics
 \be
 \tau(r) \sim \sqrt{ C_\mathrm{III} r^{3+2\kappa_\ell}+\tau_0}\,.
 \ee

This case of tachyon asymptotics appears in two distinct regions.
In the first region ($a_\ell+\kappa_\ell>-1/2$ and $\kappa_\ell>-3/2$), the coordinate $u\sim r$, and the terms in the Schr\"odinger potential read
\begin{align}
 {1\over \Xi} {d^2 \Xi \over du^2}  &= v_s^2 u^{4 a_\ell +4\kappa_\ell +4}\left[1+ \mathcal{O} \left( u^{-\epsilon}\right) \right]\,,\nn\\
H_A &= c_1\exp\left\{c_2 (3 \nw -4)u^2\left[1+ \mathcal{O} \left( u^{-\epsilon}\right) \right]\right\} u^{4 +4 w_\ell}\left[1+ \mathcal{O} \left( u^{-\epsilon}\right) \right] \,,\nn\\
H_S &\sim u^{2 a_\ell +2
  \kappa_\ell+1}\,,
\end{align}
where the coefficients $c_i$ and $\epsilon>0$ are calculable in terms of the potentials of the action.
Hence, when $H_A$ is suppressed in large $n$ limit, the masses asymptote as
\be \label{masc1}
m_n^2 \sim n^{4 {a_\ell +\kappa_\ell+1 \over 2 a_\ell + 2 \kappa_\ell +3}}\,.
\ee
Therefore, to have linear spectrum we should have $a_\ell +\kappa_\ell
=-{1\over 2}$, but then the tachyon asymptotic changes (see the case 5 below). With the constraints given above, we actually find that the power in~\eqref{masc1} satisfies
\be
 1<4 {a_\ell +\kappa_\ell+1 \over 2 a_\ell + 2 \kappa_\ell +3}<2\,.
\ee

In the second region ($a_\ell > 1$ and $\kappa_\ell \le - 3/2$), again $u\sim r$, and the terms in the potential become
\begin{align}
  {1\over \Xi} {d^2 \Xi \over du^2}  &= v_s^2 u^{4 a_\ell-2}\left[1+ \mathcal{O} \left( u^{2 \kappa_\ell+3}\right) \right]\,,\nn\\
 H_A &= \exp\left\{c (3 w_p -4)u^2\left[1+ \mathcal{O} \left( u^{2 \kappa_\ell+3}\right) \right]\right\} u^{1-2 \kappa_\ell+4 w_\ell}\left[1+ \mathcal{O} \left( u^{2 \kappa_\ell+3}\right) \right] \,,\nn\\
 H_S &\sim u^{2 a_\ell+2\kappa_\ell+1}\,.
\end{align}
When $H_A$ is suppressed, the trajectories are given by
\be
 m_n^2 \sim n^\frac{2 a_\ell-1}{a_\ell}\,.
\ee
Linear trajectories would therefore require $a_\ell=1$, but in this case the tachyon potential $V_f(\l,\t)$ might not vanish in the IR, so that the potential would not be acceptable as explained in Appendix~\ref{app:tachyonIR}.

\bigskip


 \paragraph{4.} $\kappa_p=4/3$, $\kappa_\ell=3/2$, $a_p=0$,
   $a_\ell \ge 1$.
The asymptotics of the tachyon is in this case
\be
 \tau(r) \sim \sqrt{ C_\mathrm{IV} \log r+\tau_0}
\ee
The Schr\"odinger coordinate behaves as $u \simeq r$ and the potential terms asymptote to
\begin{align}
 {1\over \Xi}{d^2 \Xi \over du^2}  &= v_s^2 u^{4 a_\ell-2} \log^2u \nn\\
  H_A &= c_1\exp\left\{(3 w_p -4)u^2\left[1+ \mathcal{O} \left({1 \over \log u}\right) \right]\right\} u^{4+4 w_\ell}\log u \left[1+ \mathcal{O} \left({1 \over \log u}\right) \right] \,,\nn\\
 H_S &\sim u^{2 a_\ell-2}
\end{align}
When $H_A$ is suppressed, the masses behave as
\be
m_n^2 \sim n^{2 a_\ell -1 \over a_\ell} \log^{1 \over a_\ell} n\,.
\ee
The spectrum is therefore linear with logarithmic corrections at the endpoint $a_\ell=1$.

\bigskip

 \paragraph{5.}  $\kappa_p=4/3$, $a_p=0$,
   $a_\ell=-\kappa_\ell-1/2$, $-1/2>\kappa_\ell >-3/2$.
In this case we find two subregions with slightly different tachyon asymptotics.

First, if $C_\mathrm{V}>{C_\mathrm{IV} \over 3+2\kappa_\ell}$,
the tachyon asymptotics reads
\be
 \tau(r) \sim \tau_0 r^{C_\mathrm{IV}/(2 C_\mathrm{V})} \, .
\ee
The coordinate $u\simeq r$, and the terms in the potential behave as
\begin{align}
 {1\over \Xi}{d^2 \Xi \over du^2} &= v_s^2 u^{2}\left[1+ \mathcal{O} \left( u^{-\epsilon}\right) \right]\,,\nn\\
H_A &= c_1\exp\left\{ (3 \nw -4)u^2\left[1+ \mathcal{O} \left( u^{-\epsilon}\right) \right]\right\} u^{C_\mathrm{IV}/C_\mathrm{V}-2
  \kappa_\ell+4 w_\ell+1}\left[1+ \mathcal{O} \left( u^{-\epsilon}\right) \right]\,,\nn\\
   H_S &\sim u^{C_\mathrm{IV}/C_\mathrm{V}+2 \kappa_\ell-3}\,.
\end{align}
The coefficient $v_s$ is the same for spin-zero and spin-one excitations only if $w_p=4/3$. Assuming that $H_A$ is suppressed asymptotically in the IR, the large $n$ trajectories are linear:
\be
m_n^2 \sim n\,,
\ee
and the slopes are the same for the various towers if in addition $w_p=4/3$. However, as pointed out in Appendix~\ref{app:tachyonIR}, the tachyon potential does not vanish in the IR and therefore this solution is not acceptable.

Second, for $C_\mathrm{V}<{C_\mathrm{IV} \over 3-2\kappa_\ell}$  we have
\be
 \tau(r) \sim \sqrt{\left(\frac{C_\mathrm{VI}}{(3+2\kappa_\ell)}-C_\mathrm{V}\right) r^{3-2\kappa_\ell}} \, .
\ee
The coordinate $u\sim r$, and the terms in the potential behave as
\begin{align}
 {1\over \Xi}{d^2 \Xi \over du^2} &= v_s^2 u^{2} \left[1+ \mathcal{O} \left( u^{-\epsilon}\right) \right]\nn\\
H_A &= c_1\,\exp\left\{c_2 (3 \nw -4)u^2\left[1+ \mathcal{O} \left( u^{-\epsilon}\right) \right]\right\} u^{4+ 4 w_\ell}\left[1+ \mathcal{O} \left( u^{-\epsilon}\right) \right]\,,\nn\\
H_S &\sim \textrm{const.}
\end{align}
where again $c_i$ and $\epsilon$ can be calculated. When $H_A$ is subleading in the IR
the spectrum is linear:
\be
m_n^2 \sim n\,.
\ee
All excitation towers have the same slopes if $w_p=4/3$. As discussed in Appendix~\ref{app:tachyonIR}, the solution is not acceptable in general, but for $\kappa_\ell$ sufficiently close to $-3/2$ the tachyon potential will vanish in the IR. Therefore, if in addition the tachyon decouples from the other fields asymptotically in the IR as we have assumed for the background solutions, this asymptotics may be acceptable.


\bigskip

 \paragraph{6.}  $\kappa_p=4/3$, $a_p>0$.
The asymptotic of the tachyon is in this case
\be
  \tau(r)  \sim \sqrt{ C_\mathrm{VI}r^{1+2\kappa_\ell}+\tau_0}
 \ee
In this case $u\simeq r$ and the various terms asymptote as
\begin{align}
  {1\over \Xi}{d^2 \Xi \over du^2} &\sim  e^{3 a_p u^{2}\left[1+ \mathcal{O} \left( u^{-\epsilon}\right) \right]} \nn\\
H_A &\sim e^{ (3 w_p -4)u^2\left[1+ \mathcal{O} \left( u^{-\epsilon}\right) \right]}
\,,\qquad
H_S\sim e^{{3 \over 2} a_p u^{2}}\,,
\end{align}
where $\epsilon=2$ for $\kappa_\ell>-{1 \over 2}$ and $\epsilon=1-2 \kappa_\ell$ for $\kappa_\ell < -{1\over 2}$. The asymptotic spectrum in this case is not linear, the masses behave as
\be
m_n^2\sim {n^2 \over \log n}
\ee
for large $n$.

\bigskip

 \paragraph{7.}  $\kappa_p<4/3$, $a_p=0$, $a_\ell>1$.
The asymptotics of the tachyon is in this case
\be \label{tasc7}
  \tau(r) \sim \tau_0 - \frac{C_\mathrm{VII}}{\tau_0} r^{1+2 \kappa_\ell } \exp\left[\left(\frac{3 \kappa_p }{2} -2\right)r^2\right]\,.
 \ee
Asymptotically in the IR $u \simeq r$, and the Schr\"odinger functions are found to be
\begin{align}
  {1\over \Xi}{d^2 \Xi \over du^2} & = u^{4 a_\ell-2}\left[1+ \mathcal{O} \left( e^{\left({3 \over 2} \kappa_p-2 \right)u^2}\right) \right]\,,\nn\\
H_A &=c_1 e^{ (3 w_p -{3\over 2}\kappa_p-2)u^2 \left[1+ \mathcal{O} \left( e^{\left({3 \over 2} \kappa_p-2 \right)u^2}\right) \right]} u^{3+2 \kappa_\ell + 4 w_\ell}\left[1+ \mathcal{O} \left( e^{\left({3 \over 2} \kappa_p-2 \right)u^2}\right) \right]\,, \nonumber\\
H_S&\sim u^{2 a_\ell-1}\,.
\end{align}
When $H_A$ is asymptotically suppressed, we find that
\be
m_n^2 \sim n^\frac{2a_\ell-1}{a_\ell}\,.
\ee
Linear trajectories would require again that $a_\ell=1$, but this case is not acceptable as explained in Appendix~\ref{app:tachyonIR}.


\bigskip

\paragraph{8.}  $\kappa_p<4/3$, $a_p>0$. In this case we have
\be
   \tau(r) \sim \tau_0 - \frac{C_\mathrm{VIII}}{\tau_0}r^{-1+2\kappa_\ell}  \exp\left[\left(\frac{3 \kappa_p }{2} -2\right)r^2\right]\,.
  \ee
The coordinate behaves as $u \simeq r$, and the terms of the Schr\"odinger potential are
\begin{align}
 {1\over \Xi}{d^2 \Xi \over du^2} &\sim e^{3 a_p u^2\left[1+ \mathcal{O} \left( e^{\left({3 \over 2} \kappa_p -2 \right)u^2}\right) \right]}\,,\nn\\
H_A &\sim e^{(3 \nw -{3\over 2}\kappa_p-2)u^2\left[1+ \mathcal{O} \left( e^{\left({3 \over 2} \kappa_p -2 \right)u^2}\right) \right]}
\,, \nonumber\\
H_S&\sim e^{({3\over 2} a_p +{3\over 2}\kappa_p-2)u^2}\,.
\end{align}
The asymptotic spectrum is again not linear:
\be
m_n^2 \sim {n^2 \over \log n}\,.
\ee


\bigskip

\paragraph{9.}  $\kappa_p>4/3$, $a_p=\hat a_\ell=0$.
We now have
\be
  \tau(r) \sim \tau_0 \exp\left\{ C_\mathrm{IX}r^{2\kappa_\ell-1 }  \exp\left[\left(\frac{3 \kappa_p }{2} -2\right)r^2\right]\right\}\,.
 \ee
The coordinate $u$ behaves as
\be
u \sim r^{-\kappa_\ell-1/2}  e^{
\left(1-\frac{3}{4}\kappa_p\right)r^2}
\tau \, .
\ee
Consequently, the terms in the Schr\"odinger potential become
\begin{align}
{1\over \Xi}{d^2 \Xi \over du^2} &= v_s^2 u^2 (\log u)^2 (\log\log u)^2\left[1+ \mathcal{O} \left(\frac{1}{\log\log u}\right) \right]\nn\\
H_A &=c_1 u^2 (\log u)^{2(3 w_p-4) \over 3 \kappa_p
  -4} (\log\log u)^{2 w_\ell+1-(2\kappa_\ell-1)\frac{3 w_p-4}{3 \kappa_p
  -4}}\left[1+ \mathcal{O} \left(\frac{1}{\log\log u}\right) \right] \,,\nn\\
 H_S &\sim \log u
\end{align}
In this case the trajectories are not exactly linear due to the logarithmic
corrections. When $H_A$ is asymptotically subleading in the IR, the mass asymptotics are
\be
m_n^2 \sim n\, \log n\, \log \log n\,,
\ee
and the slopes of different towers are the same.


\bigskip

\paragraph{10.}  $\kappa_p>4/3$, $a_p=0$, $a_\ell>0$. Now the asymptotics of the tachyon is
\be
  \tau(r) \sim C_\mathrm{X} r^{\kappa_\ell +1/2} \exp\left[\left(\frac{3 \kappa_p }{4} -1\right)r^2\right]
 \ee
 In this case $u \sim r^2$, and the Schr\"odinger functions are
\begin{align}
 {1\over \Xi}{d^2 \Xi \over du^2} &\sim e^{c_1(3 \kappa_p-4) u\left[1+{\mathcal O} \left({1\over \sqrt{u}} \right) \right]} \nn\\
H_A &\sim
e^{c_1(3 w_p -4) u \left[1+{\mathcal O} \left({1\over \sqrt{u}} \right) \right]} \,,\qquad H_S \sim e^{c_1\left({3 \over 4}
  \kappa_p-1\right) u}\,.
\end{align}
The spectrum is not linear, the masses scale as
\be
m_n^2 \sim {n^2 \over \log^2 n}\,.
\ee

\bigskip

\paragraph{11.}  $\kappa_p>4/3$, $a_p>0$. In the last case, the tachyon behaves as
\be
  \tau(r) \sim C_\mathrm{XI} r^{\kappa_\ell -1/2} \exp\left[\left(\frac{3 \kappa_p }{4} -1\right)r^2\right]\,.
 \ee
We now find that $u\sim r$ in the IR, and the Schr\"odinger functions read
\begin{align}
 {1\over \Xi}{d^2 \Xi \over du^2} &\sim u^{4 a_\ell +4 \kappa_\ell}e^{(3 \kappa_p +3
  a_p-4) u^2\left[1+{\mathcal O} \left({1\over u^2} \right) \right]} \nn\\
H_A &\sim
e^{(3 \nw -4) u^2\left[1+{\mathcal O} \left({1\over u^2} \right) \right]} \,,\qquad H_S \sim e^{\left({3 \over
      2}\kappa_p+{3 \over 2}a_p-2\right) u^2}\,.
\end{align}
In the limit of large $n$, the masses behave as
\be
m_n^2 \sim {n^2 \over \log n}\,.
\ee
The slopes are the same for $w_p<\kappa_p+a_p$.

\subsection{Flavor singlet trajectories}

Analyzing the asymptotic behavior of flavor singlet scalars and pseudoscalars is more involved than in the flavor non-singlet sector. This is the case because gluonic and mesonic degrees of freedom are coupled, and the resulting system of equations cannot be written in Schr\"odinger form in general.
We shall now show that for high excitations the glueballs and mesons are, however, decoupled. Then mesons have tower by tower the same trajectories as in the flavor non-singlet case, whereas all glueballs have linear trajectories with common slope.

\subsubsection{Axial vectors}

The flavor singlet axial vectors contain no glueballs, but there is an extra term
\be \label{tildeHdef}
 \widetilde H_A = \frac{4 x\,e^{2\Awf}\, Z(\l)\, V_a(\l,\t)^2 }{G\, \gf(\l,\t)^2\, V_f(\l,\t) }
\ee
in the Schr\"odinger potential with respect to the flavor non-singlet axials, which arises from the CP-odd action $S_a$. We expect that both $V_a$ and $V_f$ are exponentially suppressed in the IR, $\log V_a(\l,\t)\sim -a(\l)\t^2 \sim \log V_f(\l,\t)$.
Assuming that the function $a(\l)$ is the same in both potentials, $\widetilde H_a$ is suppressed as well since it is proportional to $V_a(\l,\t)^2/V_f(\l,\t)$. The exponential factor is enough to make sure that  $\widetilde H_a$ is a subleading contribution asymptotically in the IR for all acceptable potentials. Therefore, the asymptotics of the spectrum are identical to the flavor non-singlet case.

\subsubsection{Pseudoscalars}

The flavor singlet pseudoscalar fluctuations satisfy the coupled pair of equations~\eqref{cpsys1} and~\eqref{cpsys2}. They do no admit such simple Schr\"odinger form as the equations of the flavor non-singlet sector. The results for the asymptotic spectra are, however, expected to arise from the IR behavior of the various coefficients of the equations also in this case.

We  consider the second equation
\be \label{PSQeq}
\partial_r\left[e^{3\Awf}\,Z\,\left(
4x\,e^{2\Awf}\,{V_a\,V_f\,\h\,\t^2\over N_a+N_b}\,P'+x\,{V_a'\,N_a\over N_a+N_b}\,P
+{N_a\over N_a+N_b}Q'\right)\right]+m^2\,e^{3\Awf}\,Z\,Q=0\,,
\ee
where $N_a$ and $N_b$ were defined in~\eqref{NaNbdef}.
In the UV, the coefficients behave qualitatively in the same way as in the flavor non-singlet sector, and therefore the UV region again does not contribute to the asymptotic trajectories.\footnote{The fields $P$ and $Q$ have the standard power-like normalizable and non-normalizable solutions in the UV, which signals the fact that the UV structure cannot give rise to any modes.} In the intermediate region, with $r=\mathcal{O}(1)$, all coefficients have regular behavior.
Therefore, as $m^2 \to \infty$, the only way to balance with the mass term in~\eqref{PSQeq} is that the derivatives of the modes grow large, $d/dr \sim m$, and the dominant terms are the mass term and the terms with second order derivatives. This leads to rapidly oscillating solutions where the number of nodes is linked to the number of states as in the Schr\"odinger picture. With growing $m^2$, this approximation holds until larger and larger values of $r$.
For all acceptable potentials the coefficient of the mass term turns out to have a rather mild dependence on $r$, such
that for high enough $m^2$ the counting of the number of nodes (and therefore the number of states) is dominated by the region of large $r$ where IR asymptotics works.

In conclusion, in order to extract the trajectories, it is safe to use the IR expansions of the background in~\eqref{PSQeq}. Doing so, we notice that the coefficients of $P$ and $P'$ in the square brackets are exponentially suppressed in the same way as the extra mass term~\eqref{tildeHdef} of the flavor singlet axials. Therefore $P$ decouples from~\eqref{PSQeq} at high $m^2$, and the spectra for $Q$ is determined by the equation
\be \label{PSQeq2}
\frac{1}{e^{3\Awf}\,Z}\partial_r\left[e^{3\Awf}\,Z\,Q'\right]+m^2\,Q=0\,,
\ee
where we used the fact that $N_b$ is exponentially suppressed with respect to $N_a$ in the IR. This matches precisely the fluctuation equation for pseudoscalar glueballs in IHQCD \cite{ihqcd}, which is known to produce linear trajectories.

The remaining task is to solve the trajectories for $P$. As $Q$ was solved from an independent equation we can safely set it to zero\footnote{A more precise argument is as follows. When solving for $P$ in the IR from~\eqref{cpsys1}, we should also include the solution for $Q$ which is sourced by $P$ in~\eqref{PSQeq}.
But as the couplings of $P$ are exponentially suppressed in~\eqref{PSQeq}, so is the sourced solution for $Q$. Therefore it can be neglected in~\eqref{cpsys1}.} in~\eqref{cpsys1}. Further dropping terms which are exponentially suppressed in the IR, the equation becomes
\be
\partial_r\left[\frac{V_f\, G^{-1}\,e^{3\Awf}\,\h\,\t^2}{m^2-H_A}P'\right]+V_f\,e^{3\Awf}\,G\,\h\,\t^2\, P = 0 \, ,
\ee
where
\be
 H_A= \frac{4e^{2\Awf}\,\h\,\t^2}{w^2}\, .
\ee
Defining
\be
 \hat P =\frac{V_f\, G^{-1}\,e^{3\Awf}\,\h\,\t^2}{m^2-H_A}P'\, ,
\ee
the equation can be rewritten as
\be
 V_f\, G^{-1}\,e^{3\Awf}\,\h\,\t^2\ \partial_r\left[\frac{1}{V_f\,e^{3\Awf}\,G\,\h\,\t^2} \hat P'\right] +m^2\, \hat P-H_A\,\hat P = 0\,.
\ee
This matches exactly the fluctuation equation of the flavor non-singlet pseudoscalars~\eqref{PSflucts}.
Therefore,  the pseudoscalar meson trajectories are similar in the flavor singlet and non-singlet sectors.

\subsubsection{Scalars}

The asymptotics of the flavor singlet scalar modes can be solved similarly as for the pseudo-scalars. Now the coupled fluctuation
equations are~\eqref{zeteq} and~\eqref{xieq}. It is not difficult to check that the functions $p(r)$ and $N_1(r)$
in~\eqref{zeteq} are exponentially suppressed in the IR. Therefore the relevant equation for $\zeta$ is
\be
 \zeta'' + \left[3 \Awf' - 2 \frac{\Awf''}{\Awf'} + 2 \frac{\l''}{\l'}-2\frac{\l'}{\l}\right]\zeta' + m^2 \zeta = 0 \, .
\ee
Here the coefficient of $\zeta'$ can be written as
\be
 \frac 12 k = \partial_r\left[\frac{3}{2}\Awf+ \log \left(\frac{\l'}{3\l\Awf'}\right)^2\right]
\ee
and therefore the equation matches the scalar glueball fluctuation equation of IHQCD~\cite{ihqcd}. The scalar glueball trajectory is linear with the same slope as in the pseudoscalar case.

Setting $\zeta$ to zero in~\eqref{xieq}, and assuming\footnote{This is plausible because the $N_2$ term does not involve factors of $m$ or derivatives, which become large as $m$ grows. We have also verified the validity of this assumption explicitly for the potentials~I and~II which were used in the numerical analysis.} that the last term involving $N_2$ can be neglected in the limit of large $m$, we find that
\be
 \xi'' + n\, \xi' +m^2\,t\,\xi = 0\, ,
\ee
where
\begin{align}\label{ncf2}
 t &= G^2 = 1+e^{-2A}\,\h(\l,\t)\, (\t')^2 \, ,\nonumber \\
 n &= \left(4\G^2-7\right)\Awf'-2{\Awf''\over \Awf'}+\l'\bigg[{\G^2-3\over2}\,{\partial_\l\h(\l,\t)\over\h(\l,\t)}
+\nonumber\\
&\quad -(2-G^2)\,{\partial_\l V_f(\l,\t)\over V_f(\l,\t)}\bigg]+{2e^{2\Awf}\over\h(\l,\t)\,\tau'}\,{\partial_\tau V_f(\l,\t)\over V_f(\l,\t)}\,.
\end{align}

Apparently the connection to the flavor non-singlet equation is not as simple as for the pseudoscalars. The flavor non-singlet scalar fluctuations satisfy~\eqref{flscalflucts}.  We notice that the form of $t(r)$ is such that the Schr\"odinger coordinate is the same as in flavor non-singlet case.
In the analysis of the flavor non-singlet trajectories, the asymptotically leading contribution to the Schr\"odinger potential arose from the exponential factor $\exp[-a(\l)\t^2/2]$ in $\Xi$ for all acceptable potentials. In order to prove that this factor is the same in the flavor singlet case, it is enough to show that
\be
 n(r) \simeq - \partial_r\left[a(\l)\,\t(r)^2\right]\,,\qquad (r\to \infty)\, .
\ee
We check this separately for the two classes of potentials having either constant or nonconstant $a(\l)$.

For all acceptable potentials with constant $a(\l)$, from the results of Appendix~\ref{app:tachyonIR} it follows that
\be
 G^2 \sim K = e^{-2A} \h(\l) (\t')^2
\ee
and this factor diverges (at least as a power of $r$) as $r \to\infty$. The terms $\propto G^2$ in~\eqref{ncf2} can be written as
\begin{align}
 &\quad\ 4\G^2\Awf'+\l'\bigg[\frac{\G^2}{2}\,{\partial_\l\h(\l,\t)\over\h(\l,\t)}+G^2\,{\partial_\l V_{f0}(\l)\over V_{f0}(\l)}\bigg] \nonumber\\
&\simeq  e^{-2A} \h(\l) (\t')^2\, \left[4 A' + \l'\frac{d}{d\l}\log(\sqrt{\h(\l)}V_{f0}(\l))\right] \simeq -2 a(\l)\, \t\t'
\end{align}
where we used~\eqref{taueqconst} at the last step. Notice that the last term in~\eqref{ncf2} can be written as
\be
 {2e^{2\Awf}\over\h(\l,\t)\,\tau'}\,{\partial_\tau V_f(\l,\t)\over V_f(\l,\t)} \simeq -\frac{4 a(\l)\,\t\t'}{G^2}
\ee
so it is suppressed. The other additional terms in~\eqref{ncf2} are also suppressed, by $1/G^2$ at least. Therefore, as $a(\l)$ is constant,
\be
  n(r) \simeq - \partial_r\left[a(\l)\,\t(r)^2\right]\,,\qquad (r\to\infty)\, ,
\ee
with corrections suppressed at least by powers of $r$.

We  then consider the case where $a(\l)$ depends on $\l$. The logarithmic derivative $\partial_\l V_f(\l,\t)/V_f(\l,\t)$ then involves the term $-da(\l)/d\l \t^2$. For the potentials to be acceptable, we required that $a(\l)\t^2$ diverges faster than $r^2$ in order to make $V_f(\l,\t)$ vanish in the IR. Therefore
\be
 \l'(r)\, \frac{da(\l)}{d\l}\, \t(r)^2 = \frac{d a(\l(r))}{dr}\, \t(r)^2
\ee
diverges faster than linearly in $r$. The factors $\l'\, \partial_\l \log \h$,  $\l'\, \partial_\l \log  V_{f0}$, and $A'$ only grow linearly in $r$. Dropping subleading terms we therefore find\footnote{In a special case, $a_\ell=1$ in item 4 of the list of asymptotics of Appendix~\ref{app:tachyonIR}, the subleading terms are only suppressed by $1/\log r$. This is still enough for the trajectory to be fully determined by~\eqref{nproofnc}.}
\be \label{nproofnc}
 n \simeq (2-G^2)\,\l'\, \frac{da(\l)}{d\l}\, \t^2 - {4e^{2\Awf}\,a(\l)\,\t\over\h(\l,\t)\,\tau'} \simeq -\partial_r\left[a(\l)\,\t^2\right]\,,\qquad (r\to \infty)\, ,
\ee
where the last step follows by using~\eqref{taueqnonconst}.

We conclude that the masses of the flavor singlet scalar mesons have the same asymptotics as the flavor non-singlet scalars (as well as all other meson towers) for all acceptable potentials.

\section{Decay constants} \label{app:fn}
In this Appendix, it is shown how a two-point function of a generic operator is expressed as an infinite sum over the normalizable fluctuations of the dual bulk field. The decay constants of the states, created by this operator, can then be extracted from this decomposition. We follow the analysis of \cite{ikp} and we use  the action of a massless bulk vector field in a general gravitational background. This is the simplest case which captures the most important aspects of this analysis.

The decay constants are defined as the residues of the two point functions at $q^2=-M_n^2$, where $M_n^2$ is the mass of the $n\text{th}$ excited state. Therefore, $\Pi_V$, defined in \eqref{pivdef}, may be rewritten as
\be
\Pi_V=\sum{ F_n^2 \over q^2 +M_n^2 -i\epsilon} \, ,
\label{sumrule}
\ee
see \eqref{2}.
To determine the decay constants we start from a general action for the vector fluctuations
\be
S_V = -  {x M^3 N_c^2 \delta^{ab} \over 2} \int d^4x\, dr
\left[{1\over 2}C_2(r) V_{\m\n\,\,a} V^{\m\n}_b + C_1(r) \partial_r
  V_{\m \,\, a} \partial_r V^{\m}_b\right]\, .
\label{vecacgen}
\ee
Changing variables to Schr\"odinger form (see Appendix \ref{app:schro}), the action can be  written as
\be
S_V= -  {x M^3 N_c^2  \over 2}\int du d^4 x \left[ \Xi(u)^2\partial_u
  V_{\m}^a \partial_u V^{\m}_a+ {1\over 2}\Xi(u)^2 V_{\m\n\,\,a} V^{\m\n}_b \right] \, ,
\label{actksi}
\ee
where $u$ and $\Xi(u)$ are defined in Eq. (\ref{defsschrei}).

The fluctuation equation becomes
\be
\partial_u(\Xi(u)^2\partial_u \psi_V)-q^2 \Xi(u)^2\psi_V=0 \, ,
\label{eomu}
\ee
 We also define $\phi=\Xi \psi_V$. Then, Eq. (\ref{eomu}) becomes
\be
\phi''-\left[{\Xi''\over \Xi}+q^2\right]\phi=0 \, .
\label{phieq}
\ee
The boundary condition, $\psi_V(\epsilon)=1$, implies that
\be
\phi(\e)=\Xi(\e) \, .
\label{bcphi}
\ee
The equation for the spectrum is
\be
\phi_n''-\left[{ \Xi'' \over \Xi}-m_n^2\right]\phi_n=0 \, ,
\label{phieqm}
\ee
where $\phi_n$ satisfy
\be
\int_{\e}^{\infty} du~\phi^*_n(u)\phi_m(u)=\delta_{n,m} \,,\qquad \sum_{n}\phi^*_n(u)\phi_n(u')=\delta(u-u') \, .
\label{norm}
\ee
The propagator of Eq. (\ref{phieq}) reads
\be
G(u,u')=-\sum_n{\phi^*_n(u)\phi_n(u')\over q^2+m_n^2}\sp G(u',u)=G^*(u,u')
\label{prop}
\ee
and satisfies
\be
\left[\pa_u^2-\left({ \Xi'' \over \Xi}+q^2\right)\right]G(u,u')=\delta(u-u') \, .
\label{eqprop}
\ee

We define a new function $g$ by
\be
\phi(u)=\Xi(u)+g(u) \, .
\label{gdef}
\ee
This is useful because the leading non-normalizable terms of $\phi$ are included in $\Xi$ on the left hand side, so that $g(u)$ is normalizable in the UV.  
Substituting the definition in~\eqref{phieq}, we find
\be
\left[\partial_u^2-\left({\Xi'' \over \Xi}+q^2\right)\right]g=q^2 \Xi \, .
\label{geq}\ee
In this case, the solution in terms of the propagator is
\be
g(u)=\int_{\e}^{\infty} G(u,u') q^2 \Xi(u') du'\, .
\label{qsol}
\ee
Consequently, $\psi_V$ reads
\be
\psi_V(u)=1+{g(u) \over \Xi(u)}=1+\int_{\e}^{\infty} \frac{G(u,u')}{\Xi(u)} q^2 \Xi(u') du' \, .
\label{solforphi}
\ee

Eq.(\ref{pivdef}), for the two point function, reads
\be
\Pi_V=-x M^3 N_c^2 \Xi(u)^2 \left.  {\psi_{V}^* \partial_u \psi_V \over
q^2} \right|_{u=\epsilon} \, ,
\label{pivq}
\ee
where the wave function obeys $\psi_V(u=\epsilon)=1$. Inserting Eq. \eqref{solforphi} in the above definition we find
\be
\Pi_V=-{x M^3 N_c^2 \over q^2} \Xi(\epsilon)^2 \int_{\e}^{\infty}
\left[\partial_u\frac{G(u,u')}{\Xi(u)} \right]_{u=\e} q^2 \Xi(u')du' \, .
\label{pivG}
\ee
From Eq. (\ref{prop}) we have
\be
\pa_u{G(u,u')\over \Xi(u)}=-\sum_n{\partial_u \psi^*_n(u)\phi_n(u')\over q^2+m_n^2} \, ,
\label{Gder}
\ee
where $\psi_n = {\phi_n \over \Xi}$. Therefore, Eq. \eqref{pivG} becomes
\be
\begin{split}
\Pi_V&=x M^3 N_c^2 \Xi(\epsilon)^2 \sum_n {\partial_u \psi_n^*(\epsilon)
  \over q^2 +m_n^2} \int_{\e}^{\infty} \phi_n(u')\Xi(u')du' \\
&=x M^3 N_c^2 \Xi(\epsilon)^2 \sum_n {1\over m_n^2} {\partial_u \psi_n^*(\epsilon)
  \over q^2 +m_n^2} \int_{\e}^{\infty} m_n^2\psi_n(u')\Xi(u')^2du' \, .
\label{pivG2}
\end{split}
\ee
Using the fluctuation equation of $\psi_n$ to replace the integrand, we conclude that
\be
\Pi_V=x M^3 N_c^2 \Xi(\epsilon)^4 \sum_n {1\over m_n^2} {|\partial_u
  \psi_n(\epsilon)|^2 \over q^2 +m_n^2} \,.
\label{pivfin}
\ee
Hence, by comparing to (\ref{sumrule}), we find the decay constants:\footnote{The decay constants of the axial vectors are given by analogous
formula, where $\psi_n$ will be the normalizable solutions of Eq. \eqref{axvectoreom}. The formula for $f_{\pi}$ is different from the above since it is proportional to the coefficient of the mass term, $H_A$, \eqref{XiHA}, see \cite{ikp}. }
\be
\left(F_n^{(V)}\right)^2=x M^3 N_c^2 {\Xi(\epsilon)^4 |\partial_u  \psi_n|^2 \over m_n^2}\, .
\ee

Going back to the $A$ coordinate, and using the UV expansion of the normalizable wave-functions
\be
\psi_n (r) =c_2^{(n)} r^2 +\cdots= c_2^{(n)} \ell^2 e^{-2 A}+\cdots
\ee
 we obtain
\be
\left(F_n^{(V)}\right)^2= 4 x M^3 N_c^2 C_1(A)^2 \left. \left( {d r \over
  dA}(A)\right)^2 \right|_{A=\infty} {\left(c_2^{(n)}\right)^2 \over m_n^2} \, .
  \label{fneq}
\ee
Using also the matching condition (\ref{matchcon}), the final result may be written as
\be
\left(F_n^{(V)}\right)^2= {N_c N_f \over 3 \pi^2 W_0 \ell} C_1(A)^2 \left. \left({d r \over
  dA}(A)\right)^2 \right|_{A=\infty} {\left(c_2^{(n)}\right)^2 \over m_n^2} \, .
\ee

\subsection{Asymptotic decay constants for large excitation number} \label{app:asdecconn}

 Similarly to appendix \ref{app:Regge}, we now compute how the vector and axial vector decay constants depend on the excitation number, $n$, for large $n$. In the WKB approximation (for $n \to \infty$), the solution of \eqref{schroeqap} reads
 \be
\a (u)= {c_0 \over  \left( m_n^2-V(u) \right)^{1 \over 4}} \cos\left(\int^u \sqrt{m_n^2-V(u)} \, du +\theta \right) \, ,
\label{wf1}
\ee
where $V(u)$ is the Schr\"odinger potential of the vector or axial vector fluctuation. The constant $c_0$ is determined by the normalization of $\a$, see \eqref{normalpha}. We find that
\be
c_0 = \sqrt{2} \left(\int_{u_1}^{u_2} {du \over \sqrt{m_n^2-V(u)}} \right)^{-{1 \over 2}} \, ,
\ee
where $u_1$ and $u_2$ are the turning points of the potential.
For the two different cases of IR potentials which are shown in \eqref{VdefI} and \eqref{VdefII}, the above integral is given by Eqs. \eqref{wkbint1}, and \eqref{wkbint2}. The result for $c_0$ is
\begin{align}
\hspace{0.05\linewidth} &\hspace{0.05\linewidth} & \hspace{0.1\linewidth}&\hspace{0.05\linewidth} &\hspace{0.05\linewidth}&\hspace{0.7\linewidth}\nonumber \\[-\baselineskip]
&\textbf{I}& c_0 & \sim v_s^{1 \over p} m_n^{{1 \over 2}- {1 \over
    p}} \left({2 \over p} \log {m_n \over v_s} \right)^{q \over 2 p} \sim v_s^{2 \over 2+p} n^{{1 \over 2}-{2 \over p+2}} \left( \log \, n \right)^{q \over p+2} \,,& (n &\to \infty)\,
\\
&\textbf{II}& c_0 &\sim { \sqrt{m_n} \over \left( \log {m_n \over v_s} \right)^{1 \over 4}} \sim \sqrt{ n \over \log \, n} \,,&  (n &\to \infty)\,.
\end{align}

To compute the decay constants we need the expansion of the normalizable wavefunction in the UV, see \eqref{fneq}.
Keeping only the leading UV asymptotics of the potential, \eqref{uvexppot}, the UV normalizable solution of \eqref{fneq} is
\be
\tilde \a (u) = c_1 \sqrt{u} \,  J_{1} (m_n u)\, .
\label{uvsola}
\ee
In the limit of large $n$, this solution agrees with that of~\eqref{wf1} in the regime $\Lambda_\mathrm{UV}^{-1} \ll u \ll u_2$,  where both approximations are accurate.
Therefore, the wave-function \eqref{wf1} (setting $V(u)=0$) must match the IR expansion of $\tilde \a$. Indeed we find that
\be
\begin{split}
& \a (u)\big|_{V(u)=0}={c_0 \over \sqrt{m_n}} \cos \left( m_n \, u + \theta \right) \, , \\
& \tilde \a (u) \simeq - {c_1 \over \sqrt{m_n}} \sqrt{2 \over \pi} \cos \left( m_n \, u + {\pi \over 4}\right) \, , \qquad (u\to\infty)\,,
\end{split}
\label{iratil}
\ee
from which we obtain
\be
c_1 =-\sqrt{\pi \over 2} c_0 \, , \qquad \theta = {\pi \over 4} \,.
\ee
The UV expansion of the wavefunction $\psi_{UV}(r) = { \tilde \alpha(r) \over \Xi(r)} $, \eqref{defsschrei}, is
\be
\psi_{UV} (r) \sim c_0 m_n \,  r^2 + \cdots \, .
\ee
Therefore, the coefficient of the normalizable solution is $c_2^{(n)} \sim c_0 \, m_n$. Finally, \eqref{fneq} for the decay constant yields
\begin{align}
\hspace{0.05\linewidth} &\hspace{0.1\linewidth} & \hspace{0.1\linewidth}&\hspace{0.05\linewidth} &\hspace{0.05\linewidth}&\hspace{0.65\linewidth}\nonumber \\[-\baselineskip]
&\textbf{I}& F_n & \sim v_s^{2 \over 2+p} n^{{1 \over 2}-{2 \over p+2}} \left( \log \, n \right)^{q \over p+2} \,,& (n &\to \infty)\,
\\
&\textbf{II}& F_n & \sim \sqrt{ n \over \log \, n} \,,&  (n &\to \infty)\,.
\end{align}

Notice that in the case I, the $v_s$ dependence of the proportionality constants, as well as the leading logarithmic corrections, cancel in the ratio $F_n/m_n$, see~\eqref{masI}. This ensures asymptotic cancellation between the contributions from the vector and axial states, and consequently convergence of the series, in the spectral decomposition of the S-parameter in~\eqref{Sdef}. When meson trajectories are linear ($p=2$ and $q=0$ in the case I), the decay constants asymptote to a constant,
\be
F_n  \sim v_s^{1/2} \,,\qquad (n \to \infty)\,.
\ee

\section{Asymptotics of the meson wave functions} \label{app:wfasympt}
In this appendix we will compute the IR and UV asymptotic form of the wavefunctions for the singlet and non-singlet mesons. As we have seen above, the knowledge of the asymptotic behavior of the wavefunctions is essential to compute the mass spectra of mesons.

\subsection{Non-singlet flavor fluctuations}

 We begin with the analysis of the non-singlet fluctuations (presented in appendix \ref{app:SUNfluct}). These include the vector and axial vector modes plus the pseudoscalar and scalar fluctuations. We will study them first in the UV ($r\to0$), and then in the IR ($r\to \infty$) region.

\subsubsection{UV}
In order to solve the fluctuation equations in the $r\to0$ limit we need to read the UV expansions of the potentials
and background fields from sections \ref{uvstruct} and \ref{uvasback}.
We collect them here written in a form that will be useful for the asymptotic analysis of the equations:
\begin{align}
&A(r)\sim-\log(r/\ell)+{4\over 9\log(\Lambda\, r)}\,,\qquad
\log\l(r) = \Phi(r) \sim \log\left({-8\over9V_1\,\log(\Lambda\,r)}\right)\,,\label{uvback1}\\
\nonumber \\
&\tau(r)\sim \ell \left(m_q\,r\,(-\log(\Lambda\,r))^{-\rho}+\sigma\,r^3(-\log(\Lambda\,r))^{\rho}\right)\,,
\\
\nonumber \\
&V_{f0}(\l) \sim W_0\left(1+W_1\,\l\right)\,,\qquad
V_g(\l)-xV_{f0}(\l)\sim \frac{12}{\ell^2}\left(1+V_1\,\l\right)\,, \qquad  \kappa(\l)\sim \kappa_0  \,,\\
\nonumber \\
&w(\l) \sim w_0 \left(1+w_1 \lambda \right)\,,\qquad \frac{\kappa(\l)}{a(\l)} \sim \frac{2\ell^2}{3}\left(1+\kappa_1\l\right)\,, \qquad a(\l) \sim a_0 \left( 1+a_1 \l \right)
\end{align}
where
\be
\rho= -{4\over3}\left(1+{\kappa_1\over V_1}\right)\,.
\label{uvback4}
\ee

We first consider the vector and axial vector modes described by Eq. (\ref{vecaxeommain}).
Near the boundary ($r\rightarrow 0$) the equation of motion for the vector fluctuations reads:
\be
\psi_V^{\prime\prime}- {1 \over r} \left[1+ {4 (V_1-4w_1-2 W_1) \over 9 V_1 \log^2(r \Lambda)} +{\mathcal O} \left( {1 \over  \log^3(r \Lambda)}\right)  \right]\psi_V^{\prime}+m^2 \left[1+ {\mathcal O} \left( r^2 \log^2(r \Lambda)\right) \right] \psi_V=0.
\label{nsvfUV}
\ee
The solution of the above equation close to the boundary is given by
 \be \label{wfVUV}
 \psi_V(r)=c_1 \left( 1 -{1 \over 2} m^2 r^2 \log(r \Lambda)+\cdots \right)+ c_2 r^2 \left(1-{4 (V_1-4w_1-2 W_1) \over 9 V_1 \log (r \Lambda)} +\cdots \right)
 \ee
The equation for the axial vectors involves an additional mass term with respect to the one for the vectors. Near
the boundary ($r\to 0$) it takes the form:
 \be
 \begin{split} \label{wfAUV}
&\psi_A^{\prime\prime}- {1 \over r} \left[1+ {4 (V_1-4w_1-2 W_1) \over 9 V_1 \log^2(r \Lambda)} +{\mathcal O} \left( {1 \over  \log^3(r \Lambda)}\right) \right]]
\psi_A^{\prime} \\
&-4 {\kappa_0 \over w_0^2}m_q^2 \ell^4 \log^{-2 \rho}(r \Lambda) \left[1+{\mathcal O}
\left({1\over \log (r \Lambda)} \right) \right] \psi_A+ m^2 \left[1+ {\mathcal O} \left( r^2 \log^2(r \Lambda)\right) \right] \psi_A=0\,.
\end{split}
\ee
 The UV asymptotic solution is then
  \be
  \begin{split}
 \psi_A(r)&=c_1 \left( 1 -{9 \over 10} m^2 r^2 \log(r \Lambda)+ 2 {\kappa_0 m_q^2 \ell^4 \over w_0^2 \left(1+2 \gamma \right)} r^2 \log^{2 \rho+1} (r \Lambda) +\cdots \right) \\
 & + c_2 r^2 \left(1-{4 (V_1-4w_1-2 W_1) \over 9 V_1 \log (r \Lambda)} +\cdots \right)
\end{split}
 \ee

As for the non-singlet pseudoscalar fluctuations, their equation of motion is Eq. (\ref{PSflucts}).
When solving it asymptotically in the UV we have to distinguish two cases: zero, and nonzero quark mass $m_q$.
First, when the quark mass is nonzero ($ m_q \not= 0$) the fluctuation equation takes the asymptotic form:
 \be
 \begin{split}
& \hat \psi_P^{\prime\prime}+ {1 \over r} \left[1+{2 \rho \over \log(r \Lambda)}+{\mathcal O}
\left( {1 \over  \log^2(r \Lambda)}\right)  \right]\hat \psi_P^{\prime}\\
&-4 {\kappa_0 \over w_0^2}\,m_q^2\, \ell^4
\log^{-2 \rho}(r \Lambda) \left[1+{\mathcal O}\left({1\over \log (r \Lambda)} \right) \right] \hat \psi_P + m^2 \left[1+ {\mathcal O} \left( r^2 \log^2(r \Lambda)\right) \right] \hat \psi_P=0\,,
\end{split}
\ee
and therefore the wavefunction as $r \rightarrow 0$ is
\be
 \begin{split}
\hat \psi_P\,=\,&c_1 \left(1- {m^2 \over 4} r^2 + {\kappa_0\, m_q^2\, \ell^4 \over w_0^2}\, r^2
\log^{-2 \rho} (r \Lambda) +\cdots \right)\\
+&\, c_2\, \log^{-2 \rho +1} (r \Lambda) \left( 1+
{\kappa_0\, m_q^2\, \ell^4 \over w_0^2}\, r^2 \log^{-2 \rho} (r \Lambda) +\cdots \right) \,.
\end{split}
\ee
When $m_q =0$, the asymptotic equation reads:
 \be
 \begin{split}
& \hat \psi_P^{\prime\prime}-{3 \over r} \left[1-{{4 \over 9} - {2 \over 3} \rho \over  \log(r \Lambda)}+{\mathcal O} \left( {1 \over  \log^2(r \Lambda)}\right)  \right]\hat \psi_P^{\prime}+{\mathcal O}\left(r^2 \right)\hat \psi_P \\
&+ m^2 \left[1+ {\mathcal O} \left( r^2 \log^2(r \Lambda)\right) \right] \hat \psi_P=0\,,
\end{split}
\ee
and then the asymptotic wavefunction for the pseudoscalars when $m_q=0$ is given by
\be
\hat \psi_P=c_1 \left(1+ {m^2 \over 4} r^2 +\cdots \right) + c_2\left( r^4 \log^{2 \rho } (r \Lambda) +\cdots\right)
\ee

Finally, the equation of motion for the non-singlet scalar fluctuations is Eq. (\ref{flscalflucts}), which in the $r\to0$
limit becomes:
\be
 \begin{split}
&  \psi_S^{\prime\prime}-{3 \over r} \left[1+{\mathcal O} \left( {1 \over  \log^2(r \Lambda)}\right)  \right] \psi_S^{\prime}+ {3 \over r^2} \left[ 1- {p_s \over \log (r \Lambda)}  \right]\psi_S \\
&+ m^2 \left[1+ {\mathcal O} \left( r^2 \log^2(r \Lambda)\right) \right]  \psi_S=0\,,
\end{split}
\ee
where $p_s={8 \over 9} \left( {a_1 \over V_1} -{\kappa_1 \over V_1} -1 \right)$.
The UV asymptotic wavefunction is then
\be
\psi_S= c_1\left( r \log^{-{ 3 \over 2} p_s} (r \Lambda ) +\cdots\right)
+c_2\left( r^{3}\log^{{3 \over 2} p_s} (r \Lambda ) +\cdots\right)\,.
\ee

\subsubsection{IR}
We shall now determine the asymptotic form of the wavefunction for the non-singlet modes in the IR region
($r\to\infty$). We will consider two different scenarios
corresponding to backgrounds generated by type I and type II potentials respectively
(see section \ref{sec:potentials}).

\paragraph{IR -- Potentials I}
The asymptotics of the background fields are presented in appendix \ref{irasback}.
For the type I potentials the IR ($r\to \infty$) asymptotic form of the background fields is given by
 \begin{align} \label{potirasIwav}
&V_g(\l)\sim \tilde v_0\,\l^{4/3}\sqrt{\log \l}\left(1 + {v_1\over\log\l} \right)\,,  \quad
V_{f0}(\l) \sim v_c \l^{2}  \,, \quad
a(\l) \sim a_c  \,, \nn \\
& \h(\l) \sim \h_c\, \l^{-{4 \over 3}} \,, \quad
\gf(\l) \sim \l^{-\nw} (\log \lambda)^{-w_\ell}\,;\qquad (\l\to\infty)\,,
\end{align}
and
\be\label{potirasItau}
  \tau(r) \sim \tau_0 \exp \left[ C_\mathrm{I}\frac{r}{R}\right] \, ; \qquad (r\to\infty)\,.
 \ee
We will set here $R=1$ for notational simplicity so that dimensionful quantities are given in units of $1/R$. As before let us start with the vector fluctuations, whose equation of motion takes the following form in the IR:
\be
\psi_V^{\prime\prime}-p_1 e^{2 C_\mathrm{I}r\left[1+{\mathcal O}
\left( {1 \over r^2} \right) \right]}  \psi_V^{\prime} +
m^2 p_2 { e^{2 C_\mathrm{I}r} \over r}\left[1
+{\mathcal O} \left( {1 \over r^2} \right) \right] \psi_V=0\,,
\ee
where $p_1={2 a_c \tau_0^2 C_\mathrm{I}}$,
$p_2={2 \h_c e^{-2 A_c-{4 \over 3}\lambda_c} C_\mathrm{I}^2 \tau_0^2}$.
The solution in the $r \rightarrow \infty$ limit is
\be
\psi_V = c_1  r^{-m^2 {p_2 \over p_1}} e^{{p_1 \over 2 C_{\mathrm{I}}}
e^{2 C_{\mathrm{I}} r}-2 C_{\mathrm{I}} r }\left[ 1+{\mathcal O}
\left( e^{-2 C_\mathrm{I}r}\right) \right] +c_2 r ^{m^2 {p_2 \over p_1}}
\left[ 1+{\mathcal O}\left( {e^{-2 C_\mathrm{I}r} \over r^2}\right) \right]\,.
\label{nsvfIR}
\ee
The extra mass term present in the equation of motion for the axial vector fluctuations (see Eq.
(\ref{vecaxeommain})) is
subleading  in the IR region, and therefore the asymptotic ($r\to\infty$) solution for the axial vector wavefunction
is also Eq. (\ref{nsvfIR}).

In the IR limit, the pseudoscalar equation (\ref{PSflucts}) becomes (the term proportional to the quark mass is
subleading in the IR):
\be
\hat \psi_P^{\prime\prime}+p_1 e^{2 C_\mathrm{I}r\left[1
+{\mathcal O} \left( {1 \over r^2} \right) \right]} \hat \psi_P^{\prime}
+ m^2 p_2 { e^{2 C_\mathrm{I}r} \over r}\left[1
+{\mathcal O} \left( {1 \over r^2} \right) \right] \hat \psi_P=0\,,
\ee
with solution
\be
\hat \psi_P = c_1 r^{m^2 {p_2 \over p_1}}
e^{-{p_1  \over 2 C_{\mathrm{I}}} e^{2 C_{\mathrm{I}} r}-2 C_{\mathrm{I}} r }
\left[ 1+{\mathcal O}\left( e^{-2 C_\mathrm{I}r}\right) \right] +c_2 r^{-m^2 {p_2 \over p_1}}
\left[ 1+{\mathcal O}\left( {e^{-2 C_\mathrm{I}r} \over r^2}\right) \right]\,.
\ee
For the scalar fluctuations, in the IR Eq. (\ref{flscalflucts}) reduces to
\be
\psi_S^{\prime \prime}-p_1 e^{2 C_{\mathrm{I}} r\left[1
+{\mathcal O} \left( {1 \over r^2} \right) \right]} \psi_S^{\prime}
+  \frac{p_2 m^2}{r} e^{2 C_{\mathrm{I}} r\left[1
+{\mathcal O} \left( {1 \over r^2} \right) \right]}  \psi_S + p_3 e^{2 C_{\mathrm{I}} r
\left[1+{\mathcal O} \left( {1 \over r^2} \right) \right]} \psi_S = 0\,,
\ee
where $p_3= {4 a_c e^{{4 \over 3} \lambda_c}C_{\mathrm{I}} \tau_0^2 }$. The asymptotic solution is
\be
\psi_S=c_1 r ^{-m^2 {p_2 \over p_1}}e^{{p_1 \over 2 C_\mathrm{I}} e^{2 C_\mathrm{I}r}
-\left(2 C_\mathrm{I} +{p_3 \over p1} \right) r } \left[ 1+{\mathcal O}\left( e^{-2 C_\mathrm{I}r}\right) \right]
+ c_2 r ^{m^2 {p_2 \over p_1}}e^{{p_3 \over p_1} r }\left[ 1
+{\mathcal O}\left( e^{-2 C_\mathrm{I}r}\right) \right]\,.
\ee

\paragraph{IR -- Potentials II}
We now study the asymptotics for backgrounds corresponding to type II potentials. The IR asymptotics of the
background fields are:
\begin{align} \label{potirasIIwav}
&V_g(\l)\sim \tilde v_0\,\l^{4/3}\sqrt{\log \l}\left(1 + {v_1\over\log\l} \right)\,,  \quad
V_{f0}(\l) \sim v_c\, \l^{2}  \,, \quad
a(\l) \sim a_c\, \lambda^{2/3}  \,, \nn \\
& \h(\l) \sim \h_c\, \l^{-{4 \over 3}} \,, \quad
\gf(\l) \sim \l^{-\nw} (\log \lambda)^{-w_\ell}\,;\qquad (\l\to\infty)\,,
\end{align}
and
\be\label{potirasIItau}
  \tau(r) \sim \sqrt{C_\mathrm{VI}\frac{r}{R} + \tau_0} \,;\qquad (r\to\infty)\,.
 \ee
 Again we set $R=1$, and start with the vectors whose equation of motion becomes:
\be
\psi_V^{\prime\prime}-p_1 {r^2}  e^{{r^2}}\left[1+{\mathcal O} \left( {1 \over r^2} \right) \right]  \psi_V^{\prime} + m^2 \left[1+{\mathcal O} \left( {1 \over r^2} \right) \right] \psi_V = 0 \, ,
\ee
where $p_1=2 a_c e^{2 \lambda_c \over 3} C_{\mathrm VI} $.
The solution in the IR reads:
\be
\psi_V= c_1\, e^{- e^{-{r^2 }} {m^2 \over 2 p_1 r^3}
\left[ 1+{\mathcal O}\left( {1 \over r^2}\right) \right]} +c_2\, e^{ {p_1 \over 2} r e^{r^2}
\left[ 1+ { \mathcal O}\left( {1 \over r^2}\right) \right]} \, ,
\ee
which is also the IR asymptotic wavefunction of the axial vector fluctuations (as before the additional mass
term present in the equation of the axial vector modes is subleading in the IR).

The pseudoscalar equation in the IR region $r\to\infty$ reduces to
\be
\hat \psi_P^{\prime\prime}+p_1 {r^2 } e^{{r^2}}\left[1
+{\mathcal O} \left( {1 \over r^2} \right) \right]  \hat \psi_P^{\prime}
+ m^2 \left[1+{\mathcal O} \left( {1 \over r^2} \right) \right] \hat \psi_P = 0 \, ,
\ee
with solution
\be
\hat \psi_P= c_1\, e^{e^{-{r^2} } {m^2 \over 2 p_1 r^3}  \left[ 1
+{\mathcal O}\left( {1 \over r^2}\right) \right]} +c_2\,e^{ -{p_1 \over 2} r e^{r^2} \left[ 1
+ { \mathcal O}\left( {1 \over r^2}\right) \right]} \, .
\ee
Finally, the equation of motion for the scalar modes takes the following IR asymptotic form:
\be
\psi_S^{\prime\prime}-p_1 {r^2} e^{r^2}\left[1+{\mathcal O} \left( {1 \over r^2} \right) \right]  \psi_S^{\prime} + m^2 \left[1+{\mathcal O} \left( {1 \over r^2} \right) \right] \psi_S + p_2 r e^{r^2 } \psi_S = 0 \, ,
\ee
where $p_2 = {2 e^{2 \lambda_c} a_c \over \kappa_c}$. Its solution is
\be
\psi_S= c_1\,r^{p_2 \over p_1} \left[ 1+{\mathcal O}\left( {e^{-r^2} \over r^3 }\right) \right]
+c_2\, e^{ {p_1 \over 2} r e^{r^2} \left[ 1+ { \mathcal O}\left( {1 \over r^2}\right) \right]} \, .
\ee

\subsection{Singlet flavor fluctuations}
We will now analyze the asymptotic behavior of the fluctuations in the singlet flavor sector.
We will only consider the case of the singlet scalar modes, since it is for these that we computed numerically the
mass spectrum in section~\ref{sec:results}.

\subsubsection{Scalar fluctuations}
The scalar singlet fluctuations are described by the system of coupled equations ({\ref{zeteq}-\ref{xieq}}). We
will solve this system first in the UV ($r \to0$) and then in the IR ($r\to\infty$) limit.

\paragraph{UV}
The UV asymptotic forms of the background fields can be read from sections \ref{uvstruct} and \ref{uvasback},
and is summarized in Eqs. (\ref{uvback1} - \ref{uvback4}).
Then, the functions (\ref{mpcf} - \ref{ncf}) take the following UV form:
\begin{align}
&k\sim-{3\over r}-{2\over r\,\log(\Lambda\,r)}\,,\qquad
q\sim{8K\over 3W_0\,\ell^2}\,{1\over r\,(\log(\Lambda\,r))^2}\, ,\\ \nonumber\\
&n(m_q\neq0)\sim-{1\over r}-{2\rho\over r\,\log(\Lambda\,r)}\,,\qquad
n(m_q=0)\sim{3\over r}+{2\rho\over r\,\log(\Lambda\,r)}\,,\nonumber\\ \\
&p\sim -x\,m_q^2\,r\,(-\log(\Lambda\,r))^{-2\rho}\,K-
x\,\sigma^2\,r^5\,(-\log(\Lambda\,r))^{2\rho}\,9K+\cdots\,,\label{puv}
\qquad\\ \nonumber\\
&t\sim1+\mathcal{O}(r^2)\, ;\qquad{\rm with}\quad K= {W_0 \ell^2(\kappa_1+W_1)\over 3 V_1}\, ,
\end{align}
where in (\ref{puv}) we have omitted a term $\sim m_q\,\sigma\,r^3$.
The functions $N_1$ and $N_2$ take the following form in the UV:
\begin{align}
&N_1(m_q\neq0)\sim n_1\left(- \log(\Lambda\,r)\right)^{-2\rho}\, ,\qquad
N_1(m_q=0)\sim \tilde n_1\,r^4\left(- \log(\Lambda\,r)\right)^{2\rho}\,,\\
&n_1=x\,m^2\,W_0\,\ell^2\left(1-\frac{2W_1}{3V_1}+\frac{\kappa_1}{3V_1}\right)\,,\\
&\tilde n_1=3x\,\sigma^2\,W_0\,\ell^2\left(1+\frac{2W_1}{V_1}+\frac{3\kappa_1}{V_1}\right)\,,\\
&N_2\sim {n_2\over r^2  (\log(\Lambda\,r))^2}\, ,\qquad
n_2= {8\over3}\left(1+\frac{\kappa_1}{W_1}\right)\,.
\end{align}
Therefore, in the UV the system (\ref{zeteq} - \ref{xieq}) reduces to
\begin{align}
&\zeta''+{1\over r}\left(-3-{2\over \log(\Lambda\,r)}\right)\zeta'+m^2\,\zeta=x\,\ell^2\left[m_q^2\,K\,r\,(-\log(\Lambda\,r))^{-2\rho}\right]\xi'+\nonumber \\
&+\left[n_1\,(-\log(\Lambda\,r))^{-2\rho}\right]\,(\xi-\z)\;;\qquad (m_q\neq0)\,,\label{zetaequv1}\nonumber \\ \\
&\zeta''+{1\over r}\left(-3-{2\over \log(\Lambda\,r)}\right)\zeta'+m^2\,\zeta=x\,\ell^2\left[\sigma^2\,K\,r^5\,
(-\log(\Lambda\,r))^{2\rho}\right]\xi'+\nonumber \\
&+\left[\tilde n_1\,r^4\,(-\log(\Lambda\,r))^{2\rho}\right]\,(\xi-\z)\;;\qquad (m_q=0)\,,\label{zetaequv2}\nonumber \\ \\
&\xi''+{1\over r}\left(-1-{2\rho\over \log(\Lambda\,r)}\right)\xi'+\left({n_2\over r^2  (\log(\Lambda\,r))^2}
+m^2\right)\xi=\nonumber \\
&\qquad=-{8K\over27W_0}\,{1\over r (\log(\Lambda\,r))^2}\,\zeta'+{n_2\over r^2  (\log(\Lambda\,r))^2}\,\zeta
\;;\qquad (m_q\neq0)\label{xiequv1}
\,,\nonumber \\ \\
&\xi''+{1\over r}\left(3+{2\rho\over \log(\Lambda\,r)}\right)\xi'+\left({n_2\over r^2  (\log(\Lambda\,r))^2}+m^2\right)\xi=
\nonumber \\
&\qquad=-{8K\over27W_0}\,{1\over r (\log(\Lambda\,r))^2}\,\zeta'+{n_2\over r^2  (\log(\Lambda\,r))^2}\,\zeta
\;;\qquad (m_q=0)\label{xiequv2}\,.
\end{align}
Neglecting the mixing terms in (\ref{zetaequv1} - \ref{xiequv2}) we obtain the following solutions as $r \to 0$:
\begin{align}
&\zeta\sim B_1(1+{\mathcal O}(r))+B_2\, r^4(\log(\Lambda\,r))^2\,(1+{\mathcal O}((\log(\Lambda\,r))^{-1}))\,,\label{zetauv}\\
&\xi(m_q\neq0)\sim E_1(1+{\mathcal O}(r))+E_2\, r^2\,(-\log(\Lambda\,r))^{2\rho}\,(1+{\mathcal O}((\log(\Lambda\,r))^{-1})\,,\label{xiuv1}\\
&\xi(m_q=0)\sim F_1\, r^{-2}\,(-\log(\Lambda\,r))^{-2\rho}\,(1+{\mathcal O}((\log(\Lambda\,r))^{-1})+F_2(1+{\mathcal O}(r))\,.\label{xiuv2}
\end{align}
For the massless quark case it is consistent to take the subdominant normalizable solutions
with  $B_1=E_1=F_1=0$, since the mixing terms in (\ref{zetaequv1} - \ref{xiequv2}) are clearly
subleading.
However, when $m_q\neq0$, even when taking the subdominant solution for $\xi$ ($E_1=0$), the
mixing terms in (\ref{zetaequv1}) are not subleading. A consistent solution for $\z$ is then:
\be
\z\sim B_2\, r^4\,(\log(\Lambda\,r))^2\,(1+{\mathcal O}((\log(\Lambda\,r))^{-1}))-
{E_2\over4}\left(2x\,\ell^2\,m_q^2\,K+n_1\right)r^4 \log(\Lambda\,r)\,,
\label{xiuvmq}
\ee
as $r \to 0$, where the second term
is the one sourced by $\xi$.

According to these UV asymptotics the operators dual to $\zeta$ and $\xi$ have dimension $\Delta=4$ and $\Delta=3$
respectively.


\paragraph{IR -- Potentials I}
We now move on to the analysis of the IR behavior of the flavor singlet scalar modes. We will first consider the
scenario given by the type I potentials. The asymptotic form of the background fields is presented in appendix
\ref{irasback} (see also Eqs. (\ref{potirasIwav}-\ref{potirasItau}) above).

In the IR limit the functions determining the system (\ref{zeteq} - \ref{xieq}) are given by
\begin{align}
&k\sim-6{r}+{3\over 2r}+{\mathcal O}(r^{-3})\, ,\qquad p\sim e^{-a_c\,\t^2}\, ,\qquad
q\sim{C_\mathrm{I}^2\,\kappa_c\,\tilde v_0\,\tau_0^2\over 6\sqrt{6} }\,e^{2C_\mathrm{I}r}\,,
\\ \rc
&n\sim-{C_\mathrm{I}^2\,\kappa_c\,\tilde v_0\,\tau_0^2\over 3\sqrt{6} }\,e^{2C_\mathrm{I}r}\, ,\qquad\qquad
t\sim{C_\mathrm{I}^2\,\kappa_c\,\tilde v_0\,\tau_0^2\over 12\sqrt{6}}\,{1\over r}\,e^{2C_\mathrm{I}r}\,, \\ \rc
&N_1\sim e^{-a_c\,\t^2}\, ,\qquad\qquad
N_2\sim{C_\mathrm{I}^2\,\kappa_c\,\tilde v_0\,\tau_0^2(\lambda_c+2v_1+6)\over 12\sqrt{6}}\,{e^{2C_\mathrm{I}r}\over r}
\;; \;\qquad (r\to\infty)\,,
\label{cfirtype1}
\end{align}
where, as for the non-singlet case, we are setting $R=1$.
Assuming that the exponential suppression of $p(r)$ and $N_1(r)$ in the IR kills the mixing in (\ref{zeteq})
we can solve for $\zeta$ in the IR:
\be
\zeta\sim C_1\,r^{m^2\over6}(1+ {\mathcal O}(r^{-2}))+C_2\,e^{3r^2}\,
r^{-{5\over2}-{m^2\over6}}(1+{\mathcal O}(r^{-2}))\, ,\qquad (r\to\infty)\,.
\label{zetairsol}
\ee
We  now write the IR form of Eq. (\ref{xieq}):
\begin{align}
\xi''+{\kappa_c\,C_\mathrm{I}^2\,\t_0^2\,\tilde v_0\over\sqrt{6}}\,e^{2C_\mathrm{I}r}
\bigg[&-{\xi'\over3}+{1\over12 r}\left(\lambda_c+2v_1+6+m^2\right)\xi+\rc
&+{1\over6}\,\zeta'-{\lambda_c+2v_1+6\over 12}\,{1\over r}\,\zeta\bigg]=0\,.
\label{xiireq}
\end{align}
When looking for a regular solution we shall take $\zeta_\mathrm{IR}$ with $C_2=0$ and neglect the second
derivative term. We then obtain:
\be \label{xiirnormI}
\xi \sim D_1\, r^{{1\over4}(m^2+6+2v_1+\l_c)}-C_1{m^2-3(6+2v_1+\l_c)\over m^2
+3(6+2v_1+\l_c)}\,r^{{m^2\over6}}\, ,\qquad (r\to\infty)\,.
\ee
On the other hand, the leading (divergent) solution for $\xi$ in the IR is given by:
\be
\xi\sim D_2\,\exp\left({C_\mathrm{I}\,\t_0^2\,\kappa_c\,\tilde v_0\over 6\sqrt{6}}\,e^{2C_\mathrm{I}r}\right)
\left(e^{-2C_\mathrm{I}r}+{\mathcal O}(e^{-4C_\mathrm{I}r})\right)\, ,\qquad (r\to\infty)\,.
\ee

\paragraph{IR -- Potentials II}
One can read the corresponding asymptotic form of the background fields in appendix
\ref{irasback} (see also Eqs. (\ref{potirasIIwav} - \ref{potirasIItau})).
For the type II potentials the functions appearing in the system (\ref{zeteq} - \ref{xieq}) take the following form in
the IR:
\begin{align}
&k\sim-6{r}+{3\over 2r}+{\mathcal O}(r^{-3})\,\qquad
p\sim \exp\left(-a_c\,\l^{2/3}\,\t^2\right)\, ,\qquad q\sim-q_0\,{r^2}\,e^{r^2}\, , \\ \rc
&n\sim -q_0\,{r^2}\,e^{r^2}\, ,\qquad\quad
t\sim 1+\mathcal{O}(r^{-2})\, ,\qquad\quad N_1\sim \exp\left(-a_c\,\l^{2/3}\,\t^2\right)\, , \\ \rc
& N_2\sim q_0\,{r}\,e^{r^2}\, ,\qquad{\rm with}\quad
q_0={48\sqrt{6}\,a_c\over \kappa_c\, \tilde v_0}\,
e^{{2\over3}\l_c}\;; \;\qquad (r\to\infty)\,,
\label{cfirtype2}
\end{align}
where again we have set $R=1$.
Assuming that the exponential suppression of $p(r)$ and $N_1(r)$ in the IR kills the mixing in (\ref{zeteq})
the solution for $\zeta$ in the IR is again (\ref{zetairsol}). On the other hand, equation (\ref{xieq}) now reads:
\be
\xi''+m^2\,\xi-q_0\,{r^2}\,e^{r^2}\left( \xi'+\zeta'+{1\over r}(\zeta-\xi)\right)\,.
\label{xiireqt2}
\ee
Proceeding as before we find a dominant solution diverging exponentially and a subdominant one which
results from solving (\ref{xiireqt2}) after discarding the first two terms. The result is:
\begin{align}
\xi\sim \tilde D_2\exp\left({24\sqrt{6}a_c\over\kappa_c\,\tilde v_0}\,e^{{2\over3}\l_c}\,e^{r^2}\,r
\right) +\tilde D_1\,r-C_1\,{6+m^2\over m^2-6}\,r^{m^2\over6}\,,\qquad (r \to \infty)\,,
\label{xiirt2}
\end{align}
where $C_1$ is the constant appearing in (\ref{zetairsol}).

\section{Numerical methods} \label{app:nummet}

We shall discuss here the most involved steps of the numerical computations, which were done to derive the results of sections~\ref{sec:results} and~\ref{sec:Sparam}.

\subsection{Computing the flavor singlet scalar spectrum}

The starting point of the numerical analysis for the flavor singlet scalars is the coupled pair of equations~\eqref{zeteq} and~\eqref{xieq}. Following~\cite{jk}, we first make the coordinate transformation from $r$ to $A$, because this helps to resolve the UV behavior of the various functions. For $x^\mu$ dependence we insert the usual plane wave Ansatz such that $-\Box$ is replaced by the square of the four momentum $q^2$.
We first solve the background as explained in~\cite{jk} and evaluate the coefficients of the fluctuation equation on the background. The $A$ dependence of the fluctuation wave functions is then obtained by solving the resulting coupled ordinary differential equations by shooting from the IR. We require normalizability in the IR, and choose the starting values at an IR cutoff according to the asymptotic expansions of Appendix~\ref{app:wfasympt}.

The scalar meson (and glueball) masses are found as follows. We choose a basis for the IR normalizable solutions by setting $D_1=\pm C_1$ or $\tilde D_1 =\pm C_1$ in~\eqref{zetairsol},~\eqref{xiirnormI}, and~\eqref{xiirt2} (and by setting $C_2=D_2=\tilde D_2=0$). We denote the solutions obtained by using the boundary conditions with plus [minus] signs by $(\zeta_1,\xi_1)$ [$(\zeta_2,\xi_2)$], respectively.
Notice that these solutions will not be UV normalizable, in general. We then study the determinant
\be
 \left.\left(\begin{array}{cc}
  \zeta_1 & \xi_1 \\
  \zeta_2 & \xi_2
 \end{array}\right)\right|_{A=A_\mathrm{UV}}
\ee
evaluated at an UV cutoff $A_\mathrm{UV}$ as $q^2$ is varied. The value of the determinant is dominated by the non-normalizable term of the solutions in the UV, and approximately at its nodes the two solutions become linearly dependent asymptotically in the UV. One can form a linear combination where the non-normalizable terms cancel, which is Therefore normalizable both in the IR and in the UV. Therefore, the nodes of the determinant mark the values of the meson masses, $q^2=-m_n^2$.

Notice that the dependence of the coefficients of the fluctuation equations~\eqref{mpcf}-\eqref{n2} on the background is complicated. Consequently, numerical evaluation of the coefficients is rather time consuming, which slows down the analysis of the fluctuation equations. Therefore we have tabulated the values of the coefficients for each background, and used interpolation from the tabulated values when solving the fluctuation equations.
Since the coefficients are independent of $q^2$, the tabulation procedure needs to be done only once for each background. This method speeds up the scans over $q^2$ considerably.

It is important to check that the UV and IR cutoffs are far enough so that the values of the masses are not sensitive to small changes in their values. This can be particularly tricky for small $x_c-x$, as the system develops separate UV and IR scales, which both need to be included in the range of solutions.
One also needs to check that the  grid used when tabulating the coefficients of the fluctuation equations is dense enough, and that the numerical precision of the background solutions is sufficient, in order to obtain reliable solutions to the fluctuation equations.

\subsection{Computing the S-parameter}

The S-parameter can in principle computed in a straightforward manner by applying the definition of Sec.~\ref{sec:Sparam}:
\be \label{Sdefapp}
S=4 \pi {d \over dq^2}\left[q^2 (\Pi_V - \Pi_A)\right]_{q=0} =-{N_c N_f \over 3 \pi} {d \over dq^2}\left. \left( {\partial_r \psi^V (r) \over
    r}-{\partial_r \psi^A (r) \over r} \right) \right|_{r=\epsilon,\, q=0}\,,
\ee
with the numerically evaluated vector and axial fluctuation wave functions $\psi_V$ and $\psi_A$. However, this often produces imprecise or unreliable results due to issues with numerical precision, in particular when $x_c-x$ is small: On one hand, the S-parameter arises from the normalizable parts of the wave functions which are highly suppressed in the UV.
On the other hand, the formula~\eqref{Sdefapp} has corrections which are only suppressed by logarithms of $r$, and the cutoff $\epsilon$ needs to be exponentially close to the boundary in order to suppress these. We present here one possible method\footnote{We present here a solution which is rather straightforward, but technically involved.
Another possibility could be to use the formalism of Appendix~\ref{app:Pidiff}.} to overcome such issues and to obtain $S$ at high precision. Analogous methods can be used to compute $f_\pi$ or $S'$ in~\eqref{Spdef} reliably.

We first introduce some notation. We will take the quark mass to be zero. Let $\psi_V^{(\mathrm{IR})}(A,q^2)$ and $\psi_A^{(\mathrm{IR})}(A,q^2)$ be the ``standard'' numerical solutions, which are obtained by matching with the normalizable asymptotics in the IR, shooting from the IR towards the UV, and normalized by  $\psi_V^{(\mathrm{IR})}(A_\mathrm{UV},q^2) = 1 = \psi_A^{(\mathrm{IR})}(A_\mathrm{UV},q^2)$ at the UV cutoff.
Notice that $\psi_V^{(\mathrm{IR})}(A,0) \equiv 1$ for all $A$. We also denote by $\psi_V^{(\mathrm{UV})}(A,q^2)$ and $\psi_A^{(\mathrm{UV})}(A,q^2)$ the analogous solutions which are normalizable in the UV instead, matched with the UV expansions of the normalizable modes in~\eqref{wfVUV},~\eqref{wfAUV}, and obtained by shooting from the UV towards the IR. We do not need to normalize them at this point.

We will need one more set of solutions: the ones corresponding to the subleading terms in the UV-non-normalizable solutions for the wave functions. Therefore we write $\psi_{V,A} = 1 + \delta\psi_{V,A}$. Inserting this in the fluctuation equations~\eqref{vectoreom} and~\eqref{axvectoreom} we find that
\begin{align} \label{dpsiVA}
 &\delta\psi_{V}''(r) + \partial_r \log C_1(r)\ \delta\psi_{V}'(r) - q^2 G(r)^2 \simeq 0 \nn\\
 &\delta\psi_{A}''(r) + \partial_r \log C_1(r)\ \delta\psi_{A}'(r) + G^2(r) H_A(r) - q^2 G(r)^2 \simeq 0 \,,
\end{align}
where we approximated that $\delta\psi_{V,A}$ are small. The functions appearing in the coefficients are defined in~\eqref{Gdef},~\eqref{ABdefs}, and~\eqref{XiHA}. We denote  by $\delta \psi_{V,A}(A,q^2)$ the solutions of~\eqref{dpsiVA}, matched to the subleading terms of the non-normalizable modes in~\eqref{wfVUV},~\eqref{wfAUV}, and shot from the UV towards the IR.
Notice that these solutions necessarily also contain terms proportional to the normalizable modes in~\eqref{wfVUV} and~\eqref{wfAUV}, because the normalizable modes dominate when we shoot towards the IR.

 We now discuss the actual calculation. First,  we choose a (small) value $q^2$, and construct the vector wave function  $\psi_V^{(\mathrm{IR})}(A,q^2)$,
for all coordinate values between the cutoffs. The high precision UV solution is then given by the combination
\be \label{psiVUV}
 1 + \delta \psi_V(A,q^2) + C_V \psi_V^{(\mathrm{UV})}(A,q^2)\,,
\ee
where $C_V$ needs to be determined by matching to the ``standard'' solution $\psi_V^{(\mathrm{IR})}(A,q^2)$ in the region where the two last terms of~\eqref{psiVUV} are much smaller than one, but not too small, so that they can be extracted from  $\psi_V^{(\mathrm{IR})}(A,q^2)$ as well. In practice we study the ratio
\be \label{Vmatch}
\frac{\psi_V^{(\mathrm{IR})}(A,q^2) -1 - \delta \psi_V(A,q^2)}{\psi_V^{(\mathrm{UV})}(A,q^2)}
\ee
as a function of $A$. It takes a very precisely constant value in the UV region (up to a value of $A$ where numerical noise sets in), and this value is identified as $C_V$.

We repeat a similar procedure for the axial wave function. There is, however, complication as the axial sector couples to the Goldstone mode, which induces extra contributions at $q=0$.
Therefore, it is convenient to subtract the $q=0$ solutions and match\footnote{Notice that analogously subtracting $q=0$ contributions in the vector sector would give exactly the matching discussed after~\eqref{psiVUV}, because due to $\psi_V^{(\mathrm{IR})}(A,0)=1$ and $\delta \psi_V(A,0)=0$. In the matching of the axial wave functions, we could also approximate $\psi_A^{(\mathrm{IR})}(A,0) \simeq 1$, but using the exact solution leads to slightly better results.}
\be \label{psiAUV}
  \delta \psi_A(A,q^2)-\delta \psi_A(A,0) + C_A \psi_A^{(\mathrm{UV})}(A,q^2)\,,
\ee
to $\psi_A^{(\mathrm{IR})}(A,q^2)-\psi_A^{(\mathrm{IR})}(A,0)$.

The subleading terms of $\psi_V-\psi_A$ in the UV, which will determine the S-parameter trough~\eqref{Sdefapp}, are now included in
\be \label{SUVsubl}
 \Psi(A,q^2)\equiv \delta \psi_V(A,q^2) + C_V \psi_V^{(\mathrm{UV})}(A,q^2)-\left[\delta \psi_A(A,q^2)-\delta \psi_A(A,0) + C_A \psi_A^{(\mathrm{UV})}(A,q^2) \right]\,.
\ee
This is a big improvement with respect to the solutions $\psi_{V,A}^{(\mathrm{IR})}$, because all terms here are obtained by shooting from the UV and can be evaluated essentially arbitrary close to the boundary (or arbitrary high values of $A$) with good precision.\footnote{In the end the fact that $\psi_{V,A}^{(\mathrm{UV})}$ is suppressed with respect to $\delta \psi_{V,A}$ by $-\log r \sim A$ sets a limit on how close to the boundary we can get. We have been using the cutoff value $A=300$.}
The subleading terms of the non-normalizable modes in~\eqref{wfVUV},~\eqref{wfAUV} cancel up to highly suppressed contributions due to the difference between the vector and axial wave functions in the definition of $\Psi(A,q^2)$.\footnote{Notice, however, that we cannot leave out the functions $\delta \psi_{V,A}$ in the definition, because
 they also contain nontrivial contributions to the normalizable modes.} Therefore, it has the UV asymptotics of the normalizable modes:
\be
 \Psi(A,q^2) = c_2(q^2)\, \ell^2 e^{-2 A} \left[1 + \frac{C_n}{A} + \morder{\frac{1}{A^2}}\right]\,,
\ee
where
\be
 C_n=-\frac{4(V_1+4w_1+2W_1)}{9V_1}
\ee
with the coefficients defined in~\eqref{uvback1}.
Since $\Psi(A,q^2)$ can be evaluated very close to the boundary, it is easy to find $c_2(q^2)$ by matching with the UV expansion.

Finally, we can compute the S-parameter from the formula~\eqref{Sdefapp}, which depends on the wave functions through the combination
\be
 \lim_{r \to 0}\left[ {\partial_r \psi^V (r) \over  r}-{\partial_r \psi^A (r) \over r} \right] = -\ell^{-2} \lim_{A\to\infty} e^{2A} \partial_A \Psi(A,q^2)+\frac{12\pi^2f_\pi^2}{N_cN_f} = 2 c_2(q^2) +\frac{12\pi^2f_\pi^2}{N_cN_f} \,.
\ee
By construction, $\Psi(A,0)=0=c_2(0)$. Therefore we find that
\be \label{Sfinal}
 S = -{2 N_c N_f \over 3 \pi} \lim_{q\to 0} \frac{c_2(q^2)}{q^2}\,.
\ee
In practice, we cannot evaluate $c_2(q^2)$ at arbitrarily small $q^2$ because some of the functions, e.g., in the ratio~\eqref{Vmatch} vanish at small $q^2$ and we will face issues with numerical precision. Therefore, we pick a small value of $q^2$ where matching still works, evaluate $c_2(q^2)$ also at the negative value $-q^2$, and approximate the limit in~\eqref{Sfinal} by the ``averaged'' ratio $(c_2(q^2)-c_2(-q^2))/(2q^2)$.
The precision of the interpolation to $q=0$ could be improved by computing $c_2$ at even larger set of values of $q^2$.

\section{Spectrum in the limit $x \to x_c$} \label{app:xtoxc}

In this Appendix we discuss in more detail how the spectrum behaves as $x \to x_c$ from below.
We start by an argument~\cite{alho} which explains why  all mass ratios tend to constants in this limit. The argument is sketched without proofs, and the details may be checked numerically.

First recall from Sec.~\ref{sec:bg} that as $x \to x_c$, the walking behavior of the background is linked to the IR fixed point at $\l=\l_*$, which is present when $x_c \le x < 11/2$. As $x \to x_c$, the coupling remains approximately constant with $\l \simeq \l_*$ for a large range of $r$,  but in the end,  the IR fixed point is screened by the tachyon solution, which is nonzero but exponentially suppressed in the UV.
The idea is that the background can be divided into two regions, the UV one having $\l<\l_*$, and the IR one having $\l>\l_*$, which will become distinct in the limit $x \to x_c$.

Also recall from Sec.~\ref{sec:bg} that we can define the characteristic UV and IR scales $\Lambda_\mathrm{UV}$ and $\Lambda_\mathrm{IR}$ by using the UV and IR expansions of the background solutions, respectively.
If we take $x \to x_c$ keeping $r \Lambda_\mathrm{UV}$ fixed, we expect that the background approaches a nontrivial limit, which has $\l<\l_*$. The limiting functions are identified as the UV piece of the background.
Similarly, if we take $x \to x_c$ keeping $r \Lambda_\mathrm{IR}$ fixed instead, the background approaches a nontrivial limit with $\l>\l_*$, which is the IR piece. We can define explicitly, for example, the limiting functions of the coupling:
\begin{align} \label{lambdalims}
 \l(r) &\to \l_\mathrm{UV}(r)\,,& x &\to \xcfb\quad\textrm{with}\quad r\Lambda_\mathrm{UV}\quad\textrm{fixed}\,;\nonumber\\
 \l(r) &\to \l_\mathrm{IR}(r)\,,& x &\to \xcfb\quad\textrm{with}\quad r\Lambda_\mathrm{IR}\quad\,\textrm{fixed}\,.
\end{align}
The convergence of the former and latter limits is pointwise in $r\Lambda_\mathrm{UV}$ and in $r\Lambda_\mathrm{IR}$, respectively. The UV solution is the standard solution at $x=x_c$ with an IR fixed point: $\l_\mathrm{UV} \to \l_*$ as $r \to \infty$. The IR solution diverges in the IR and satisfies $\l_\mathrm{IR} \to \l_*$ as $r \to 0$. The UV and IR solutions have essentially one characteristic scale each, given by $\Lambda_\mathrm{UV}$ and $\Lambda_\mathrm{IR}$, respectively.

We can now understand why the ratios of any two masses and decay constants tend to fixed values in the limit $x \to x_c$: these quantities are essentially functions of the IR piece of the solution only, and therefore take fixed values in units of $\Lambda_\mathrm{IR}$ in this limit.
It is indeed easy to check that masses in the non-singlet flavor towers (except for the scalar case which will be discussed below) depend on the IR dynamics only,
because the fluctuation equations can be cast into the Schr\"odinger from, see Appendix~\ref{app:schro}. The bottom of the  potential is located around $r \sim \Lambda_\mathrm{IR}^{-1}$, and for $r\ll \Lambda_\mathrm{IR}^{-1}$ the potential diverges as $V(r) \sim r^{-2}$.
Therefore the effect of any structure of the potential at $r \sim \Lambda_\mathrm{UV}^{-1}$ on the masses is highly suppressed as $x \to x_c$, and the limiting values of the masses can be calculated solely in terms of the IR piece of the background at $x=x_c$.

For the decay constants the situation is slightly more involved, as their values are
given by the UV asymptotics of the wave functions of the fluctuation modes (see Appendix~\ref{app:fn}). The wave functions take their largest values in the region $r \sim \Lambda_\mathrm{IR}^{-1}$, and their shape in this region determines their normalization up to highly suppressed contributions. Therefore the scale of the coefficients of the UV asymptotics, and consequently that of the decay constants, is fixed to $\Lambda_\mathrm{IR}$.
There is, however, some structure at $r \sim \Lambda_\mathrm{UV}^{-1}$ in the Schr\"odinger potential, which may modify the values of these coefficients even as $x \to x_c$. The resulting $\morder{1}$ correction term may be calculated explicitly for the vector and axial decay constants as we now demonstrate. In the walking regime and in the UV, only the first terms in the fluctuation equations~\eqref{vectoreom} and~\eqref{axvectoreom} are relevant. Therefore,  we find that
\be \label{fluctUV}
 C_1(r) \partial_r \psi_{V/A}(r) = \mathrm{const.}
\ee
where $C_1$ is defined in~\eqref{ABdefs}. In the UV we have $G(r)\simeq 1$ and $V_f(\l,\t) \simeq V_{f0}(\l)$. At UV/IR fixed points the metric reads
\be
 e^A \simeq \frac{\ell}{r} = \frac{\sqrt{12}}{\sqrt{V_\mathrm{eff}(\l)}\, r}
\ee
so that equating the values of the constant in~\eqref{fluctUV} at the fixed points leads to
\be \label{UVmod}
 \left.\frac{\partial_r \psi_{V/A}(r)}{r}\right|_{r=0} = \frac{V_{f0}(\l_*)}{V_{f0}(0)}\left(\frac{w(\l_*)}{w(0)}\right)^2\sqrt{\frac{V_\mathrm{eff}(0)}{V_\mathrm{eff}(\l_*)}}\left.\frac{\partial_r \psi_{V/A}(r)}{r}\right|_{\l=\l_*}\,.
\ee
The decay constants depend on the fluctuation wave function through the factor on the left hand side, whereas its value at the IR fixed point, the last term on the right hand side, only depends on the IR solutions in the limit $x \to x_c$. The correction from the UV dynamics may therefore be found by inserting~\eqref{UVmod} in the definition of the decay constants.

Notice that the above arguments do not exclude the possibility of a light dilaton: there could in principle be a scalar state the mass of which behaves as $\Lambda_\mathrm{IR}$ in general, but becomes exactly zero as $x \to x_c$. And indeed there is something interesting happening in the scalar sector. It is not difficult to check that the IR piece of the Schr\"odinger potential for the non-singlet scalars
at $x=x_c$
has the critical behavior~\cite{son}
\be
 V_S(r) \sim -\frac{1}{4r^2}\,,\qquad (r \to 0)\,.
\ee
Therefore, the system is close to becoming unstable: if the coefficient was even slightly less than $-1/4$, the scalar sector would have an infinite tower of tachyons as $x \to x_c$.

We may elaborate on this by considering the region with $x<x_c$ and $x_c-x\ll 1$. We find that (see~\eqref{VSUVFP} and~\eqref{VSIRFP} in Appendix~\ref{app:SUNfluct})
\begin{align} \label{Vschsc}
 V_S(r) &\sim \frac{3}{4r^2} & (0<\ &r \ll \Lambda_\mathrm{UV}^{-1}) \nonumber\\
 V_S(r) &\sim \left[\frac{15}{4}-\Delta(4-\Delta)\right]\frac{1}{r^2}& (\Lambda_\mathrm{UV}^{-1} \ll\ &r \ll \Lambda_\mathrm{IR}^{-1})\,,
\end{align}
where $\Delta$ is the dimension of the chiral condensate at the fixed point $\l=\l_*$. When $x<x_c$ the BF bound is violated at the fixed point, so that $\Delta(4-\Delta)>4$ and the coefficient of $1/r^2$ in~\eqref{Vschsc} is smaller than $-1/4$. This means that there is an instability, if the range of the approximation in~\eqref{Vschsc} (with walking) is long enough.

In order to check this explicitly, we can analyze the wave functions for the potential~\eqref{Vschsc} a bit further. In order to see if there is an instability in the non-singlet scalar sector, it is enough to check if the fluctuation wave function at zero mass has nodes. By recalling that in terms of the mass of the tachyon and the IR AdS radius we have $\Delta(4-\Delta)=-m_\mathrm{IR}^2\ell_\mathrm{IR}^2$, and using the relation~\eqref{hatKres}, we find that
\be
 \frac{15}{4}-\Delta(4-\Delta) \simeq - \frac{1}{4} -\frac{\pi^2}{\hat K^2}(x_c-x)\,.
\ee
Consequently, in the walking region, the Schr\"odinger wave function $\alpha(r)$ behaves as
\be \label{scalflu}
 \alpha(r) \propto \sqrt{r} \sin\left(\frac{\pi}{\hat K}\sqrt{x_c-x}\log r + \phi\right)\,,
\ee
where $\phi$ is a constant. Recalling the ratio $\Lambda_\mathrm{UV}/\Lambda_\mathrm{IR}$ from~\eqref{scalescal}, we observe that the length of the walking region is half of the period of the sine function in~\eqref{scalflu}, i.e., the region has exactly the largest possible length which is still consistent with $\alpha(r)$ having no nodes.
However, since the length is exactly critical, unstable modes or a light dilaton cannot be excluded by this argument, and numerics is needed to check the stability in the end.

One way to understand the above results is to notice that when $r \ll \Lambda_\mathrm{IR}^{-1}$, the background tachyon is much smaller than one, and therefore the tachyon and its zero-mass fluctuations (singlet and non-singlet) solve the same equations.\footnote{This is nontrivial for the singlet scalars due to mixing of the glue and flavor fluctuation modes, but may be checked by a straightforward calculation.}
The ``critical'' scalar fluctuations therefore just reflects similar behavior of the tachyon background, which is linked to the BKT transition and Miransky scaling. Since the standard background tachyon does not have nodes, the fluctuations should not have those either for $r \ll \Lambda_\mathrm{IR}^{-1}$.

In summary, the negative dip of~\eqref{Vschsc} in the Schr\"odinger potentials of the scalars has therefore just the critical width, so that is not clear without doing the numerics if there are unstable states or possibly a light dilaton state. While we did the above analysis mostly for flavor non-singlet scalars, for which the Schr\"odinger picture is well-defined, it is possible to check that the behavior of the wave function is similar to~\eqref{scalflu} for the flavor singlet scalars.
In the end, numerical analysis  reveals that the dilaton (as well as tachyonic states) are absent from the spectrum. In fact, as seen from Fig.~\eqref{fratios}, the lightest flavor singlet meson is heavier than its non-singlet counterpart, and it is also heavier than the spin-one states for potentials~II.

\section{$\Pi_V-\Pi_A$ at large $q^2$} \label{app:Pidiff}

Here we shall go through the discussion of Sec.~\ref{sec:Sparamxxc} in more detail. To start with, the fluctuation equations for vectors and axial vectors are
\begin{align}
  &\frac{1}{C_2(r)}\partial_r\left[C_1(r)\psi_V'(r)\right]- q^2 \psi_V(r) = 0\nn\\
  &\frac{1}{C_2(r)}\partial_r\left[C_1(r)\psi_A'(r)\right]- H(r)\psi_A(r) - q^2 \psi_A(r) = 0\,,
\end{align}
where the coefficients can be found in Appendix~\ref{app:SUNfluct} (we denote again $H=H_A$). Defining $\phi = (\psi_A-\psi_V)/\psi_A$, the exact counterpart of~\eqref{feq} reads
\be \label{phieqapp}
 \frac{1}{C_2(r)}\partial_r\left[C_1(r)\phi'(r)\right] + 2\partial_r \log \psi_A(r)\, \phi'(r) = (1-\phi(r))H(r)\,.
\ee
Assuming that $\phi(r)$ is small (which can be checked from the result), we drop the term proportional to $\phi(r)$ on the right hand side. Defining
\be \label{phi2g}
g(r) = C_1(r) \phi'(r)\,,
\ee
 we therefore find
\be
 g'(r) + 2 G(r)^2\,\partial_r \log \psi_A(r)\,g(r) \simeq M(r)\,,
\ee
where $G$ and $M$ are given in Eqs.~\eqref{Gdef} and~\eqref{MdefA}, respectively. This equation is solved by
\be\begin{split}\label{gsol}
 g(r) \simeq&\ \exp\left[-2\int_0^r dr'\,G^2(r')\, \partial_r \log \psi_A(r') \right] \\
&\times \left\{-\int_r^\infty dr'\,M(r')\,\exp\left[2\int_0^{r'} dr''\,G^2(r'')\, \partial_r \log \psi_A(r'') \right] +\mathrm{const.}\right\}\,.
\end{split}\ee

So far we did not use the fact that $q^2$ is large. This is necessary to rule out the constant piece in the wavy brackets of~\eqref{gsol}. When $q\gg \Lambda_\mathrm{IR}$, the IR-normalizable solution behaves as $\psi_A(r) \sim e^{-qr}$ for $1/q \ll r \ll 1/\Lambda_\mathrm{IR}$, so that $ \partial_r \log \psi_A(r) \simeq -q$.
We see that the constant piece corresponds to the IR-non-normalizable solution for $g(r)$, and must be exponentially suppressed as $q \to \infty$. Setting the constant to zero in~\eqref{gsol} and by using the relation~\eqref{phi2g}, we can verify that the qualitative properties of the solution match with~\eqref{phiappsol}: $g(r)$ goes to a constant as $r \to 0$, and behaves as
\be \label{gappsol}
 g(r) \sim -e^{2 qr} \int_r^\infty M(r') e^{-2 q r'} \simeq -\frac{1}{2 q} M(r) \sim \frac{H(r)}{q r}
\ee
for $1/q \ll r \ll 1/\Lambda_\mathrm{IR}$. We also see that indeed $\phi(r)\ll 1$ for $r \ll 1/\Lambda_\mathrm{IR}$ even in the presence of walking dynamics, so the approximation done above is valid.

Finally, we write down the result for $\Pi_V-\Pi_A$. By using the definition in~\eqref{Sdef},
\begin{align} \label{Pidiffexact}
 q^2\left[\Pi_A(q^2)-\Pi_V(q^2)\right] &= \frac{N_cN_f}{12 \pi^2}\,\lim_{r\to 0}\frac{\partial_r\left(\psi_V(r)-\psi_A(r)\right)}{r} \nn\\
 &= -\frac{N_cN_f}{12 \pi^2}\,\lim_{r\to 0} \frac{g(r)}{r C_1(r)} \nn\\
 &\simeq \frac{N_cN_f}{12\pi^2 W_0w_0^2\ell}\int_0^\infty dr\,M(r)\,\exp\left[2\int_0^{r} dr'\,G^2(r')\, \partial_r \log \psi_A(r') \right]\,,
\end{align}
where the coefficients are as defined in Sec.~\ref{sec:constraints}. By its derivation, this result holds for all $q\gg \Lambda_\mathrm{IR}$ also for small $x_c-x$, i.e., when the coupling constant walks and $\Lambda_\mathrm{UV}\gg \Lambda_\mathrm{IR}$.
Corrections to this formula can be computed iteratively by taking into account the extra term which we dropped in~\eqref{phieqapp}. By using, e.g.,~\eqref{gappsol}, we see that the corrections are suppressed by $1/q^6$ for $q\gg \Lambda_\mathrm{UV}$, and by $1/q^4$ in the regime governed by walking dynamics, $\Lambda_\mathrm{UV} \gg q \gg \Lambda_\mathrm{IR}$.

We finish the discussion by simplifying~\eqref{Pidiffexact} for $q \gg \Lambda_\mathrm{UV}$. In this regime, up to logarithmically suppressed corrections, $\psi_A(r) = qr K_1(qr)$, where $K$ is the modified Bessel function of the second kind, and we may also set $G(r)=1$. Therefore,
\be\begin{split}
 q^2\left[\Pi_A(q^2)-\Pi_V(q^2)\right] &\simeq \frac{N_cN_f}{12\pi^2 } \int_0^\infty dr\,\frac{(qr)^2 \left(K_1(qr)\right)^2}{r}\,H(r) \\
&= \frac{N_cN_f}{12\pi^2 } \int_0^\infty dr\,r\, \left(K_1(r)\right)^2\,H\left(\frac{r}{q}\right)\,.
\end{split}\ee
Therefore,  at leading order the result is completely explicit and matches with the standard AdS one (see, e.g.,~\cite{Son:2010vc}), but the corrections are only logarithmically suppressed, while~\eqref{Pidiffexact} has power suppressed corrections.


\addcontentsline{toc}{section}{References}

\begin{thebibliography}{99}


\bibitem{veneziano}
  G.~Veneziano,
  {\em ``Some Aspects of a Unified Approach to Gauge, Dual and Gribov Theories,''}
  \href{http://www.sciencedirect.com/science/article/pii/0550321376904120}{Nucl.\ Phys.\ B {\bf 117} (1976) 519}.





\bibitem{vu1}
  G.~Veneziano,
  {\em ``U(1) Without Instantons,''}
  \href{http://dx.doi.org/10.1016/0550-3213(79)90332-8}{Nucl.\ Phys.\ B\ {\bf
159} (1979) 213}.



  \bibitem{bankszaks}
  T.~Banks, A.~Zaks,
  {\em ``On the Phase Structure of Vector-Like Gauge Theories with Massless
Fermions,''}
  \href{http://dx.doi.org/10.1016/0550-3213(82)90035-9}{Nucl.\ Phys.\  {\bf B196
} (1982)  189}.

\bibitem{jk}
  M.~J\"arvinen and E.~Kiritsis,
  {\em ``Holographic Models for QCD in the Veneziano Limit,''}
  JHEP {\bf 1203}, 002 (2012)
  \hri{1112.1261}{[hep-ph]}.


\bibitem{bkt}
  J.~M.~Kosterlitz and D.~J.~Thouless,
  {\em ``Ordering, metastability and phase transitions in two-dimensional systems,''}
  J.\ Phys.\ C {\bf 6} (1973) 1181.

\bibitem{miransky}
  V.~A.~Miransky and K.~Yamawaki,
  {\em ``Conformal phase transition in gauge theories,''}
  Phys.\ Rev.\ D {\bf 55} (1997) 5051
   [Erratum-ibid.\ D {\bf 56} (1997) 3768]
  \hre{hep-th}{9611142}.

\bibitem{son}
  D.~B.~Kaplan, J.~-W.~Lee, D.~T.~Son, M.~A.~Stephanov,
  {\em ``Conformality Lost,''}
  Phys.\ Rev.\  {\bf D80}, 125005 (2009)
  \hri{0905.4752}{[hep-th]}.





\bibitem{ds}
  T.~Appelquist, K.~D.~Lane and U.~Mahanta,
  {\em ``On The Ladder Approximation For Spontaneous Chiral Symmetry
Breaking,''}
  \href{http://dx.doi.org/10.1103/PhysRevLett.61.1553}{Phys.\ Rev.\ Lett.\ \
{\bf 61}, 1553  (1988)};\\
  A.~G.~Cohen and H.~Georgi,
  {\em ``Walking Beyond The Rainbow,''}
  \href{http://dx.doi.org/10.1016/0550-3213(89)90109-0}{Nucl.\ Phys.\ B\ {\bf
314}, 7  (1989)}.

\bibitem{cw}
  E.~Gardi and G.~Grunberg,
  {\em ``The Conformal window in QCD and supersymmetric QCD,''}
  JHEP\ {\bf 9903}, 024  (1999)
  \hre{hep-th}{9810192};\\
  T.~Appelquist, A.~G.~Cohen and M.~Schmaltz,
  {\em ``A new constraint on strongly coupled field theories,'' }
  Phys.\ Rev.\  D {\bf 60}, 045003 (1999)
  \hre{hep-th}{9901109};\\
  F.~Sannino and K.~Tuominen,
  {\em ``Orientifold theory dynamics and symmetry breaking,'' }
  Phys.\ Rev.\  D {\bf 71}, 051901 (2005)
  \hre{hep-ph}{0405209};\\
  H.~Gies and J.~Jaeckel,
  {\em ``Chiral phase structure of QCD with many flavors,''}
  Eur.\ Phys.\ J.\ C\ {\bf 46}, 433  (2006)
  \hre{hep-ph}{0507171};\\
  T.~A.~Ryttov and F.~Sannino,
  {\em ``Supersymmetry Inspired QCD Beta Function,'' }
  Phys.\ Rev.\  D {\bf 78}, 065001 (2008)
  \hri{0711.3745}{[hep-th]};\\
  E.~Poppitz and M.~Unsal,
  {\em ``Conformality or confinement: (IR)relevance of topological
excitations,'' }
  JHEP {\bf 0909}, 050 (2009)
  \hri{0906.5156}{[hep-th]};\\
  A.~Armoni,
  {\em ``The Conformal Window from the Worldline Formalism,'' }
  Nucl.\ Phys.\  B {\bf 826}, 328 (2010)
  \hri{0907.4091}{[hep-ph]};\\
  M.~T.~Frandsen, T.~Pickup, M.~Teper,
  {\em ``Delineating the conformal window,'' }
  Phys.\ Lett.\  {\bf B695}, 231-237 (2011)
  \hri{1007.1614}{[hep-ph]}.

\bibitem{antipin}
  O.~Antipin and K.~Tuominen,
  {\em ``Constraints on Conformal Windows from Holographic Duals,''}
  Mod.\ Phys.\ Lett.\ A\ {\bf 26}, 2227  (2011)
  \hri{0912.0674}{[hep-ph]}.


\bibitem{lattice}
  T.~Appelquist, G.~T.~Fleming and E.~T.~Neil,
  {\em ``Lattice Study of Conformal Behavior in SU(3) Yang-Mills Theories,''}
  Phys.\ Rev.\ D\ {\bf 79}, 076010  (2009)
  \hri{0901.3766}{[hep-ph]};\\
  M.~Hayakawa, K.~-I.~Ishikawa, Y.~Osaki, S.~Takeda, S.~Uno and N.~Yamada,
  {\em ``Running coupling constant of ten-flavor QCD with the Schr\'odinger
functional method,''}
  Phys.\ Rev.\ D\ {\bf 83}, 074509  (2011)
  \hri{1011.2577}{[hep-lat]};\\
  Z.~Fodor, K.~Holland, J.~Kuti, D.~Nogradi, C.~Schroeder, K.~Holland, J.~Kuti
and D.~Nogradi {\it et al.},
  {\em ``Twelve massless flavors and three colors below the conformal window,''}
  Phys.\ Lett.\ B\ {\bf 703}, 348  (2011)
  \hri{1104.3124}{[hep-lat]}; \\
  E.~T.~Neil,
  {\em ``Exploring Models for New Physics on the Lattice,''}
  PoS LATTICE {\bf 2011} (2011) 009
  \hri{1205.4706}{[hep-lat]}.



 \bibitem{holdom}
  B.~Holdom,
  {\em ``Techniodor,''}
  \href{http://dx.doi.org/10.1016/0370-2693(85)91015-9}{Phys.\ Lett.\ B\ {\bf
150} (1985) 301};\\
 T.~W.~Appelquist, D.~Karabali and L.~C.~R.~Wijewardhana,
  {\em ``Chiral Hierarchies and the Flavor Changing Neutral Current Problem in
Technicolor,''}
  \href{http://dx.doi.org/10.1103/PhysRevLett.57.957}{Phys.\ Rev.\ Lett.\ \ {\bf
57} (1986) 957}.



\bibitem{walk2}
K.~Yamawaki, M.~Bando and K.~-i.~Matumoto,
  {\em ``Scale Invariant Technicolor Model and a Technidilaton,''}
  \href{http://dx.doi.org/10.1103/PhysRevLett.56.1335}{Phys.\ Rev.\ Lett.\ \
{\bf 56} (1986) 1335}.


\bibitem{tech}
  S.~Dimopoulos and L.~Susskind,
  {\em ``Mass Without Scalars,''}
  Nucl.\ Phys.\ B {\bf 155} (1979) 237.


\bibitem{review}
  K.~D.~Lane,
  {\em ``An Introduction to technicolor,''}
  In *Boulder 1993, Proceedings, The building blocks of creation* 381-408, and
Boston U. - BU-HEP-94-02 (94,rec.Jan.) 32 p
  \hre{hep-ph}{9401324};\\
 C.~T.~Hill and E.~H.~Simmons,
  {\em ``Strong dynamics and electroweak symmetry breaking,''}
  Phys.\ Rept.\ \ {\bf 381} (2003) 235
   [Erratum-ibid.\ \ {\bf 390} (2004) 553]
  \hre{hep-ph}{0203079};\\
  F.~Sannino,
  {\em ``Conformal Dynamics for TeV Physics and Cosmology,''}
  Acta Phys.\ Polon.\ B\ {\bf 40}, 3533  (2009)
  \hri{0911.0931}{[hep-ph]}.


\bibitem{Holdom2}
  B.~Holdom,
  {\em ``Raising Condensates Beyond The Ladder,''}
  \href{http://dx.doi.org/10.1016/0370-2693(88)91776-5}{Phys.\ Lett.\  {\bf
B213}, 365 (1988)}.






\bibitem{fukano}
  H.~S.~Fukano and F.~Sannino,
  {\em ``Conformal Window of Gauge Theories with Four-Fermion Interactions and
Ideal Walking,''}
  Phys.\ Rev.\ D\ {\bf 82}, 035021  (2010)
  \hri{1005.3340}{[hep-ph]}.


  \bibitem{peskin}
  M.~E.~Peskin and T.~Takeuchi,
  {\em ``Estimation of oblique electroweak corrections,''}
  \href{http://dx.doi.org/10.1103/PhysRevD.46.381}{Phys.\ Rev.\ D\ {\bf 46}
(1992) 381}.



  \bibitem{as}
  T.~Appelquist and F.~Sannino,
  {\em ``The Physical spectrum of conformal SU(N) gauge theories,''}
  Phys.\ Rev.\ D {\bf 59} (1999) 067702
  \hre{hep-ph}{9806409}.



 \bibitem{topdown}
   J.~Babington, J.~Erdmenger, N.~J.~Evans, Z.~Guralnik and I.~Kirsch,
  {\em``Chiral symmetry breaking and pions in non-supersymmetric gauge /  gravity
  duals,''}
  Phys.\ Rev.\  D {\bf 69}, 066007 (2004).
  \hre{hep-th}{0306018}; \\
N.~J.~Evans and J.~P.~Shock,
  {\em``Chiral dynamics from AdS space,''}
  Phys.\ Rev.\  D {\bf 70}, 046002 (2004)
  \hre{hep-th}{0403279}; \\
  M.~Kruczenski, D.~Mateos, R.~C.~Myers and D.~J.~Winters,
  {\em ``Towards a holographic dual of large-N(c) QCD,''}
  JHEP {\bf 0405}, 041 (2004)
  \hre{hep-th}{0311270}; \\
  T.~Sakai and S.~Sugimoto,
  {\em ``Low energy hadron physics in holographic QCD,''}
  Prog.\ Theor.\ Phys.\  {\bf 113}, 843 (2005)
  \hre{hep-th}{0412141}.


\bibitem{Bigazzi:2005md}
  F.~Bigazzi, R.~Casero, A.~L.~Cotrone, E.~Kiritsis and A.~Paredes,
  {\em ``Non-critical holography and four-dimensional CFT's with fundamentals,''}
  JHEP {\bf 0510} (2005) 012
  \hre{hep-th}{0505140}.

\bibitem{Casero:2006pt}
  R.~Casero, C.~Nunez and A.~Paredes,
  {\em ``Towards the string dual of N=1 SQCD-like theories,''}
  Phys.\ Rev.\ D {\bf 73} (2006) 086005
  \hre{hep-th}{0602027};\\
  C.~Nunez, A.~Paredes and A.~V.~Ramallo,
  {\em ``Unquenched Flavor in the Gauge/Gravity Correspondence,''}
  Adv.\ High Energy Phys.\  {\bf 2010} (2010) 196714
  \hri{1002.1088}{[hep-th]}.



\bibitem{Kehagias:1998gn}
  A.~Kehagias,
  {\em ``New type IIB vacua and their F-theory interpretation,''}
  Phys.\ Lett.\  B {\bf 435}, 337 (1998)
  \hre{hep-th}{9805131};\\
  O.~Aharony, A.~Fayyazuddin and J.~M.~Maldacena,
  {\em ``The large N limit of N = 2,1 field theories from three-branes in
  F-theory,''}
  JHEP {\bf 9807}, 013 (1998)
  \hre{hep-th}{9806159}.




\bibitem{Itzhaki:1998uz}
  N.~Itzhaki, A.~A.~Tseytlin and S.~Yankielowicz,
  {\em ``Supergravity solutions for branes localized within branes,''}
  Phys.\ Lett.\  B {\bf 432}, 298 (1998)
  \hre{hep-th}{9803103};\\
O.~Pelc and R.~Siebelink,
  {\em ``The D2-D6 system and a fibered AdS geometry,''}
  Nucl.\ Phys.\  B {\bf 558}, 127 (1999)
  \hre{hep-th}{9902045}.



\bibitem{Nastase:2003dd}
  H.~Nastase,
  {\em ``On Dp-Dp+4 systems, QCD dual and phenomenology,''}
  \hre{hep-th}{0305069}.


\bibitem{Burrington:2007qd}
  B.~A.~Burrington, V.~S.~Kaplunovsky and J.~Sonnenschein,
  {\em ``Localized Backreacted Flavor Branes in Holographic QCD,''}
  JHEP {\bf 0802}, 001 (2008)
  \hri{0708.1234}{[hep-th]}.


  \bibitem{Erlich:2005qh}
  J.~Erlich, E.~Katz, D.~T.~Son and M.~A.~Stephanov,
  {\em ``QCD and a holographic model of hadrons,''}
  Phys.\ Rev.\ Lett.\  {\bf 95} (2005) 261602
  \hre{hep-ph}{0501128};\\
 L.~Da Rold and A.~Pomarol,
  {\em ``Chiral symmetry breaking from five dimensional spaces,''}
  Nucl.\ Phys.\  B {\bf 721} (2005) 79
  \hre{hep-ph}{0501218}.






 \bibitem{butc}
  J.~Hirn and V.~Sanz,
  {\em ``A Negative S parameter from holographic technicolor,''}
  Phys.\ Rev.\ Lett.\ \ {\bf 97} (2006) 121803
  \hre{hep-ph}{0606086};
  {\em ``The Fifth dimension as an analogue computer for strong interactions at
the LHC,''}
  JHEP\ {\bf 0703} (2007) 100
  \hre{hep-ph}{0612239};\\
   M.~Piai,
  {\em ``Precision electro-weak parameters from AdS(5), localized kinetic terms
and anomalous dimensions,''}
  \hre{hep-ph}{0608241};\\
  C.~D.~Carone, J.~Erlich and J.~A.~Tan,
  {\em ``Holographic Bosonic Technicolor,''}
  Phys.\ Rev.\ D\ {\bf 75} (2007) 075005
  \hre{hep-ph}{0612242};\\
  M.~Fabbrichesi, M.~Piai and L.~Vecchi,
  {\em ``Dynamical electro-weak symmetry breaking from deformed AdS: Vector
mesons and effective couplings,''}
  Phys.\ Rev.\ D\ {\bf 78} (2008) 045009
  \hri{0804.0124}{[hep-ph]};\\
  J.~Hirn, A.~Martin and V.~Sanz,
  {\em ``Describing viable technicolor scenarios,''}
  Phys.\ Rev.\ D\ {\bf 78} (2008) 075026
 \hri{0807.2465}{[hep-ph]};\\
  D.~D.~Dietrich and C.~Kouvaris,
  {\em ``Constraining vectors and axial-vectors in walking technicolour by a
holographic principle,''}
  Phys.\ Rev.\ D\ {\bf 78}, 055005  (2008)
  \hri{0805.1503}{[hep-ph]}; 
  {\em ``Generalised bottom-up holography and walking technicolour,''}
  Phys.\ Rev.\ D\ {\bf 79} (2009) 075004
  \hri{0809.1324}{[hep-ph]};\\
   D.~D.~Dietrich, M.~J\"arvinen and C.~Kouvaris,
  {\em ``Mixing in the axial sector in bottom-up holography for walking
technicolour,''}
  JHEP\ {\bf 1007}, 023  (2010)
  \hri{0908.4357}{[hep-ph]}.



\bibitem{hong}
  D.~K.~Hong and H.~-U.~Yee,
  {\em ``Holographic estimate of oblique corrections for technicolor,''}
  Phys.\ Rev.\ D {\bf 74}, 015011 (2006)
  \hre{hep-ph}{0602177}; \\
  K.~Agashe, C.~Csaki, C.~Grojean and M.~Reece,
    {\em ``The S-parameter in holographic technicolor models,''}
  JHEP {\bf 0712}, 003 (2007)
  \hri{0704.1821}{[hep-ph]}; \\
  K.~Haba, S.~Matsuzaki and K.~Yamawaki,
 {\em ``S Parameter in the Holographic Walking/Conformal Technicolor,''}
  Prog.\ Theor.\ Phys.\  {\bf 120}, 691 (2008)
  \hri{0804.3668}{[hep-ph]}.



 \bibitem{nunez}
  C.~Nunez, I.~Papadimitriou and M.~Piai,
  {\em ``Walking Dynamics from String Duals,''}
  Int.\ J.\ Mod.\ Phys.\ A\ {\bf 25} (2010) 2837
  \hri{0812.3655}{[hep-th]} ; \\
  S.~Prem Kumar, D.~Mateos, A.~Paredes and M.~Piai,
  {\em ``Towards holographic walking from N=4 super Yang-Mills,''}
  JHEP\ {\bf 1105} (2011) 008
  \hri{1012.4678}{[hep-th]}; \\
  L.~Anguelova, P.~Suranyi and L.~C.~R.~Wijewardhana,
  {\em ``Holographic Walking Technicolor from D-branes,''}
  Nucl.\ Phys.\ B\ {\bf 852} (2011) 39
  \hri{1105.4185}{[hep-th]}.


  \bibitem{kutasov}
  D.~Kutasov, J.~Lin and A.~Parnachev,
  {\em ``Conformal Phase Transitions at Weak and Strong Coupling,''}
  Nucl.\ Phys.\ B {\bf 858}, 155 (2012)
  \hri{1107.2324}{[hep-th]};
  {\em ``Holographic Walking from Tachyon DBI,''}
  Nucl.\ Phys.\ B {\bf 863} (2012) 361
  \hri{1201.4123} {[hep-th]}.







   \bibitem{ihqcd}
  U.~Gursoy and E.~Kiritsis,
  {\em ``Exploring improved holographic theories for QCD: Part I,''}
  JHEP {\bf 0802} (2008) 032
  \hri{0707.1324}{[hep-th]}; \\
  U.~Gursoy, E.~Kiritsis, F.~Nitti,
  {\em ``Exploring improved holographic theories for QCD: Part II,''}
  JHEP {\bf 0802}, 019 (2008)
  \hri{0707.1349}{[hep-th]}; \\
  E.~Kiritsis,
  {\em ``Dissecting the string theory dual of QCD,''}
  Fortsch.\ Phys.\  {\bf 57} (2009) 396
  \hri{0901.1772}{[hep-th]};\\
  U.~Gursoy, E.~Kiritsis, L.~Mazzanti, G.~Michalogiorgakis and F.~Nitti,
  {\em ``Improved Holographic QCD,''}
  Lect.\ Notes Phys.\ \ {\bf 828} (2011) 79
  \hri{1006.5461}{[hep-th]}.


   \bibitem{ihqcd2}
  U.~Gursoy, E.~Kiritsis, L.~Mazzanti, F.~Nitti,
  {\em ``Deconfinement and Gluon Plasma Dynamics in Improved Holographic QCD,''}
  Phys.\ Rev.\ Lett.\  {\bf 101 } (2008)  181601
  \hri{0804.0899}{[hep-th]}; \\
  {\em ``Holography and Thermodynamics of 5D Dilaton-gravity,''}
  JHEP {\bf 0905}, 033 (2009)
  \hri{0812.0792}{[hep-th]}.

  \bibitem{data}
    U.~Gursoy, E.~Kiritsis, L.~Mazzanti, F.~Nitti,
  {\em ``Improved Holographic Yang-Mills at Finite Temperature: Comparison with Data,''}
  Nucl.\ Phys.\  {\bf B820 } (2009)  148-177
  \hri{0903.2859}{[hep-th]}.

\bibitem{Lucini:2012gg}
  B.~Lucini and M.~Panero,
  {\em ``SU(N) gauge theories at large N,''}
  Phys.\ Rept.\  {\bf 526} (2013) 93
  \hri{arXiv:1210.4997}{[hep-th]}.


\bibitem{Jarvinen:2009fe}
  M.~J\"arvinen, F.~Sannino,
  {\em ``Holographic Conformal Window - A Bottom Up Approach,''}
  JHEP {\bf 1005}, 041 (2010)
  \hri{0911.2462}{[hep-ph]};\\
  J.~Alanen and K.~Kajantie,
  {\em ``Thermodynamics of a field theory with infrared fixed point from
gauge/gravity duality,''}
  Phys.\ Rev.\ D\ {\bf 81}, 046003  (2010)
  \hri{0912.4128}{[hep-ph]};\\
  J.~Alanen, K.~Kajantie and K.~Tuominen,
  {\em ``Thermodynamics of Quasi Conformal Theories From Gauge/Gravity
Duality,''}
  Phys.\ Rev.\ D\ {\bf 82}, 055024  (2010)
  \hri{1003.5499}{[hep-ph]};\\
  J.~Alanen, T.~Alho, K.~Kajantie and K.~Tuominen,
  {\em ``Mass spectrum and thermodynamics of quasi-conformal gauge theories from
gauge/gravity duality,''}
  Phys.\ Rev.\ D\ {\bf 84}, 086007  (2011)
  \hri{1107.3362}{[hep-th]}.



  \bibitem{ckp}
  R.~Casero, E.~Kiritsis and A.~Paredes,
  {\em ``Chiral symmetry breaking as open string tachyon condensation,''}
 Nucl.\ Phys.\ B {\bf 787} (2007) 98
  \hri{hep-th/0702155}{[hep-th]}.




  \bibitem{sen}
  A.~Sen,
  {\em ``Tachyon dynamics in open string theory,''}
  Int.\ J.\ Mod.\ Phys.\ A\ {\bf 20} (2005) 5513
  \hre{hep-th}{0410103}.




  \bibitem{ikp}
   I.~Iatrakis, E.~Kiritsis and A.~Paredes,
  {\em ``An AdS/QCD model from Sen's tachyon action,''}
  Phys.\ Rev.\ D {\bf 81} (2010) 115004
  \hri{1003.2377}{[hep-ph]}; 
  {\em ``An AdS/QCD model from tachyon condensation: II,''}
  JHEP {\bf 1011} (2010) 123
\hri{1010.1364}{[hep-ph]}.



\bibitem{sstachyon}
  O.~Bergman, S.~Seki and J.~Sonnenschein,
  {\em ``Quark mass and condensate in HQCD,''}
  JHEP {\bf 0712}, 037 (2007)
  \hri{0708.2839}{[hep-th]};\\
  A.~Dhar and P.~Nag,
  {\em ``Sakai-Sugimoto model, Tachyon Condensation and Chiral symmetry Breaking,''}
  JHEP {\bf 0801}, 055 (2008)
  \hri{0708.3233}{[hep-th]};
  {\em ``Tachyon condensation and quark mass in modified Sakai-Sugimoto model,''}
  Phys.\ Rev.\ D {\bf 78}, 066021 (2008)
  \hri{0804.4807}{[hep-th]};\\
  N.~Jokela, M.~J\"arvinen and S.~Nowling,
  {\em ``Winding effects on brane/anti-brane pairs,''}
  JHEP {\bf 0907}, 085 (2009)
  \hri{0901.0281}{[hep-th]}.

\bibitem{Bali:2013kia}
  G.~S.~Bali, F.~Bursa, L.~Castagnini, S.~Collins, L.~Del Debbio, B.~Lucini and M.~Panero,
  {\em ``Mesons in large-N QCD,''}
  JHEP {\bf 1306} (2013) 071
  \hri{arXiv:1304.4437}{[hep-lat]}.



\bibitem{Alvares:2012kr}
  R.~Alvares, N.~Evans and K.~-Y.~Kim,
  {\em ``Holography of the Conformal Window,''}
  Phys.\ Rev.\ D {\bf 86}, 026008 (2012)
  \hri{1204.2474}{[hep-ph]};\\
  T.~Alho, N.~Evans and K.~Tuominen,
  {\em ``Dynamic AdS/QCD and the Spectrum of Walking Gauge Theories,''}
  \hri{1307.4896}{[hep-ph]}.

  \bibitem{alho}
  T.~Alho, M.~J\"arvinen, K.~Kajantie, E.~Kiritsis and K.~Tuominen,
  {\em ``On finite-temperature holographic QCD in the Veneziano limit,''}
  \hri{1210.4516}{[hep-ph]}.

\bibitem{letter}
  D.~Arean, I.~Iatrakis, M.~J\"arvinen and E.~Kiritsis,
  {\em ``V-QCD: Spectra, the dilaton and the S-parameter,''}
  Phys.\ Lett.\ B {\bf 720}, 219 (2013)
  \hri{1211.6125}{[hep-ph]}.


\bibitem{dilaton}
   D.~Elander, C.~Nunez and M.~Piai,
  {\em ``A Light scalar from walking solutions in gauge-string duality,''}
  Phys.\ Lett.\ B {\bf 686}, 64 (2010)
  \hri{0908.2808}{[hep-th]}; \\
 K.~Haba, S.~Matsuzaki and K.~Yamawaki,
  {\em ``Holographic Techni-dilaton,''}
  Phys.\ Rev.\ D {\bf 82}, 055007 (2010)
  \hri{1006.2526}{[hep-ph]}; \\
  D.~Elander and M.~Piai,
  {\em ``Light scalars from a compact fifth dimension,''}
  JHEP {\bf 1101}, 026 (2011)
  \hri{1010.1964}{[hep-th]}; \\
  L.~Anguelova, P.~Suranyi and L.~C.~R.~Wijewardhana,
  {\em ``Scalar Mesons in Holographic Walking Technicolor,''}
  Nucl.\ Phys.\ B {\bf 862}, 671 (2012)
  \hri{1203.1968}{[hep-th]};
  {\em ``Glueball Spectrum in a Gauge Theory with Two Dynamical Scales,''}
  JHEP {\bf 1305}, 003 (2013)
  \hri{1212.1176}{[hep-th]};\\
  D.~Elander and M.~Piai,
  {\em ``On the glueball spectrum of walking backgrounds from wrapped-D5 gravity duals,''}
  Nucl.\ Phys.\ B {\bf 871}, 164 (2013)
  \hri{1212.2600}{[hep-th]};\\
  N.~Evans and K.~Tuominen,
  {\em ``Holographic Modelling of a Light Techni-Dilaton,''}
  \hri{1302.4553}{[hep-ph]};\\
  B.~Bellazzini, C.~Csaki, J.~Hubisz, J.~Serra and J.~Terning,
  {\em ``A Naturally Light Dilaton and a Small Cosmological Constant,''}
  \hri{1305.3919}{[hep-th]}.




\bibitem{cobi}
  O.~Mintakevich and J.~Sonnenschein,
  {\em ``Holographic technicolor models and their S-parameter,''}
  JHEP {\bf 0907} (2009) 032
  \hri{0905.3284}{[hep-th]}.


\bibitem{rubakov}
  D.~G.~Levkov, V.~A.~Rubakov, S.~V.~Troitsky and Y.~A.~Zenkevich,
  {\em ``Constraining Holographic Technicolor,''}
  Phys.\ Lett.\ B {\bf 716} (2012) 350
  \hri{1201.6368}{[hep-ph]}.


\bibitem{sv}
  M.~Shifman and A.~Vainshtein,
  {\em ``Highly Excited Mesons, Linear Regge Trajectories and the Pattern of the Chiral Symmetry Realization,''}
  Phys.\ Rev.\ D {\bf 77} (2008) 034002
  \hri{0710.0863}{[hep-ph]}.


  \bibitem{glo}
  L.~Y.~.Glozman,
  {\em ``QCD symmetries in excited hadrons,''}
  eConf C {\bf 070910} (2007) 140
  \hri{0710.0978}{[hep-ph]}.


  \bibitem{espriu}
  A.~A.~Andrianov and D.~Espriu,
  {\em ``Parity doubling from Weinberg sum rules,''}
  Phys.\ Lett.\ B {\bf 671} (2009) 275
\hri{0803.4104}{[hep-ph]}.



 \bibitem{sannino}
    F.~Sannino,
  {\em ``Mass Deformed Exact S-parameter in Conformal Theories,''}
  Phys.\ Rev.\ D {\bf 82}, 081701 (2010)
  \hri{1006.0207}{[hep-lat]};
  {\em ``Magnetic S-parameter,''}
  Phys.\ Rev.\ Lett.\  {\bf 105}, 232002 (2010)
  \hri{1007.0254}{[hep-ph]}; \\
  S.~Di Chiara, C.~Pica and F.~Sannino,
  {\em ``Flavor Dependence of the S-parameter,''}
  Phys.\ Lett.\ B {\bf 700}, 229 (2011)
  \hri{1008.1267}{[hep-ph]}.



\bibitem{Barbieri:2004qk}
  R.~Barbieri, A.~Pomarol, R.~Rattazzi and A.~Strumia,
  {\em ``Electroweak symmetry breaking after LEP-1 and LEP-2,''}
  Nucl.\ Phys.\ B {\bf 703}, 127 (2004)
  \hre{hep-ph}{0405040}.


\bibitem{blaise}
C.~Charmousis, B.~Gouteraux, B.~S.~Kim, E.~Kiritsis and R.~Meyer,
  {\em ``Effective Holographic Theories for low-temperature condensed matter systems,''}
  JHEP {\bf 1011} (2010) 151
  \hri{1005.4690}{[hep-th]};\\
  B.~Gouteraux and E.~Kiritsis,
  {\em ``Generalized Holographic Quantum Criticality at Finite Density,''}
  JHEP {\bf 1112} (2011) 036
\hri{1107.2116}{[hep-th]}.





\bibitem{Gursoy}
  M.~Caselle, L.~Castagnini, A.~Feo, F.~Gliozzi, U.~Gursoy, M.~Panero and A.~Schafer,
  {\em ``Thermodynamics of SU(N) Yang-Mills theories in 2+1 dimensions II. The Deconfined phase,''}
  JHEP {\bf 1205} (2012) 135
  \hri{1111.0580}{[hep-th]};\\
  U.~Gursoy,
  {\em ``Continuous Hawking-Page transitions in Einstein-scalar gravity,''}
  JHEP {\bf 1101} (2011) 086
  \hri{1007.0500}{[hep-th]}.

\bibitem{2}
  M.~Panero,
  {\em ``Thermodynamics of the QCD plasma and the large-N limit,''}
  Phys.\ Rev.\ Lett.\  {\bf 103} (2009) 232001
  \hri{0907.3719}{[hep-lat]};\\
  G.~Boyd, J.~Engels, F.~Karsch, E.~Laermann, C.~Legeland, M.~Lutgemeier and B.~Petersson,
  {\em ``Thermodynamics of SU(3) lattice gauge theory,''}
  Nucl.\ Phys.\ B {\bf 469} (1996) 419
  \hre{hep-lat}{9602007};\\
  P.~N.~Meisinger, T.~R.~Miller and M.~C.~Ogilvie,
  {\em ``Phenomenological equations of state for the quark gluon plasma,''}
  Phys.\ Rev.\ D {\bf 65} (2002) 034009
  \hre{hep-ph}{0108009};\\
  R.~D.~Pisarski,
  {\em ``Fuzzy Bags and Wilson Lines,''}
  Prog.\ Theor.\ Phys.\ Suppl.\  {\bf 168} (2007) 276
  \hre{hep-ph}{0612191}.


\bibitem{parnachev}
  M.~Goykhman and A.~Parnachev,
  {\em ``S-parameter, Technimesons, and Phase Transitions in Holographic Tachyon DBI Models,''}
  Phys.\ Rev.\ D {\bf 87} (2013) 026007
  \hri{arXiv:1211.0482}{hep-th]}.


\bibitem{cs}
  U.~G\"ursoy, I.~Iatrakis, E.~Kiritsis, F.~Nitti and A.~Obannon,
  {\em ``The Chern-Simons Diffusion Rate in Improved Holographic QCD,''}
  JHEP {\bf 1302} (2013) 119
  \hri{1212.3894}{[hep-th]}.





\bibitem{gubser}
  S.~S.~Gubser,
  {\em``Curvature singularities: The Good, the bad, and the naked,''}
  Adv.\ Theor.\ Math.\ Phys.\  {\bf 4}, 679 (2000)
  \hre{hep-th}{0002160}.

\bibitem{Harada:2003dc}
  M.~Harada, M.~Kurachi and K.~Yamawaki,
  {\em ``Meson masses in large N(f) QCD from Bethe-Salpeter equation,''}
  Phys.\ Rev.\ D {\bf 68}, 076001 (2003)
  \hre{hep-ph}{0305018};\\
  M.~Kurachi and R.~Shrock,
  {\em ``Study of the Change from Walking to Non-Walking Behavior in a Vectorial Gauge Theory as a Function of N(f),''}
  JHEP {\bf 0612}, 034 (2006)
  \hre{hep-ph}{0605290}.

\bibitem{dilaton2}
  W.~A.~Bardeen, C.~N.~Leung and S.~T.~Love,
  {\em ``The Dilaton and Chiral Symmetry Breaking,''}
 \href{http://dx.doi.org/10.1103/PhysRevLett.56.1230}{Phys.\ Rev.\ Lett.\  {\bf 56}, 1230 (1986)};\\
  B.~Holdom and J.~Terning,
  {\em ``A Light Dilaton in Gauge Theories?,''}
  \href{http://dx.doi.org/10.1016/0370-2693(87)91109-9}{Phys.\ Lett.\ B {\bf 187}, 357 (1987)};
  {\em ``No Light Dilaton In Gauge Theories,''}
  \href{http://dx.doi.org/10.1016/0370-2693(88)90783-6}{Phys.\ Lett.\ B {\bf 200}, 338 (1988)};\\
  D.~D.~Dietrich, F.~Sannino and K.~Tuominen,
  {\em ``Light composite Higgs from higher representations versus electroweak precision measurements: Predictions for CERN LHC,''}
  Phys.\ Rev.\ D {\bf 72}, 055001 (2005)
  \hre{hep-ph}{0505059};\\
  T.~Appelquist and Y.~Bai,
  {\em ``A Light Dilaton in Walking Gauge Theories,''}
  Phys.\ Rev.\ D {\bf 82}, 071701 (2010)
  \hri{1006.4375}{[hep-ph]};\\
  M.~Hashimoto and K.~Yamawaki,
  {\em ``Techni-dilaton at Conformal Edge,''}
  Phys.\ Rev.\ D {\bf 83}, 015008 (2011)
  \hri{1009.5482}{[hep-ph]};\\
  B.~Grinstein and P.~Uttayarat,
  {\em ``A Very Light Dilaton,''}
  JHEP {\bf 1107}, 038 (2011)
  \hri{1105.2370}{[hep-ph]};\\
  O.~Antipin, M.~Mojaza and F.~Sannino,
  {\em ``Light Dilaton at Fixed Points and Ultra Light Scale Super Yang Mills,''}
  Phys.\ Lett.\ B {\bf 712}, 119 (2012)
  \hri{1107.2932}{[hep-ph]}.


\bibitem{SinDS}
  M.~Harada, M.~Kurachi and K.~Yamawaki,
  {\em ``The pi+ - pi0 mass difference and the S parameter in large N(f) QCD,''}
  Prog.\ Theor.\ Phys.\  {\bf 115}, 765 (2006)
  \hre{hep-ph}{0509193};\\
  M.~Kurachi and R.~Shrock,
  {\em ``Behavior of the S Parameter in the Crossover Region Between Walking and QCD-Like Regimes of an SU(N) Gauge Theory,''}
  Phys.\ Rev.\ D {\bf 74}, 056003 (2006)
  \hre{hep-ph}{0607231}.



\bibitem{walkingS}
  T.~Appelquist and G.~Triantaphyllou,
  {\em ``Precision tests of technicolor,''}
  \href{http://dx.doi.org/10.1016/0370-2693(92)90204-H}{Phys.\ Lett.\ B {\bf 278}, 345 (1992)};\\
  R.~Sundrum and S.~D.~H.~Hsu,
  {\em ``Walking technicolor and electroweak radiative corrections,''}
  Nucl.\ Phys.\ B {\bf 391}, 127 (1993)
  \hre{hep-ph/9206225};\\
  M.~Harada and Y.~Yoshida,
  {\em ``QCD S parameter from inhomogeneous Bethe-Salpeter equation,''}
  Phys.\ Rev.\ D {\bf 50}, 6902 (1994)
  \hre{hep-ph/9406402};\\
  S.~R.~Ignjatovic, L.~C.~R.~Wijewardhana and T.~Takeuchi,
  {\em``ACD estimation of the S parameter,''}
  \href{http://dx.doi.org/10.1103/PhysRevD.61.056006}{Phys.\ Rev.\ D {\bf 61}, 056006 (2000)}.




\bibitem{Son:2010vc}
  D.~T.~Son and N.~Yamamoto,
  {\em``Holography and Anomaly Matching for Resonances,''}
  \hri{1010.0718}{[hep-ph]};\\
  I.~Iatrakis and E.~Kiritsis,
  {\em ``Vector-axial vector correlators in weak electric field and the holographic dynamics of the chiral condensate,''}
  JHEP {\bf 1202} (2012) 064
  \hri{1109.1282}{[hep-ph]}.

\bibitem{glue}
  E.~Kiritsis and F.~Nitti,
  {\em ``On massless 4D gravitons from asymptotically AdS(5) space-times,''}
  Nucl.\ Phys.\ B {\bf 772} (2007) 67
  \hre{hep-th}{0611344}.







\end{thebibliography}
\end{document}